%% file: paper.tex
\documentclass[10pt]{iopart}
\usepackage{cite}
\usepackage{epsfig}
\usepackage{index} 

\begin{document}
\include{command}
\eqnobysec

\title[Quark-model study of few-baryon systems]
{Quark-model study of few-baryon systems}

\author{A Valcarce\dag, H Garcilazo\dag\footnote[3]{Permanent address:
Escuela Superior de F\' \i sica y Matem\'aticas,
Instituto Polit\'ecnico Nacional, Edificio 9,
07738 M\'exico D.F., Mexico}, F Fern\'andez\dag and P. Gonz\'alez\ddag}

\address{\dag\ Grupo de F\' \i sica Nuclear,
Universidad de Salamanca, E-37008 Salamanca, Spain}

\address{\ddag\ Dpto. de F\' \i sica Te\'orica - IFIC,
Universidad de Valencia - CSIC, E-46100 Burjassot, Valencia, Spain}

\eads{\mailto{valcarce@usal.es}, \mailto{humberto@esfm.ipn.mx},
\mailto{fdz@usal.es}, \mailto{pedro.gonzalez@uv.es}}

\begin{abstract}
We review the application of non-relativistic
constituent quark models to
study one, two and three non-strange baryon systems. 
We present results for the baryon spectra, potentials and observables of the
NN, N$\Delta$, $\Delta\Delta$ and NN$^*(1440)$ systems, and also for the 
binding energies of three non-strange baryon systems. We make emphasis
on observable effects related to quark antisymmetry and its interplay
with quark dynamics.
\end{abstract}

\submitto{\RPP}

\maketitle

\addcontentsline{toc}{part}{}

\include{ch0rev}
\include{ch1rev}
\include{ch2rev}
\include{ch3rev}

\include{ch4rev}

\include{ch5rev}
\include{ch6rev}
\include{ch7rev}

\include{ref}
\end{document}

%% file: command.tex
\newcommand{\vecprod}[2]{\vec{#1} \cdot \vec{#2}}    
\newcommand{\bra}[1]{\left \langle {#1} \right|}  
\newcommand{\ket}[1]{\left | {#1} \right \rangle}  
\newcommand{\ares}[3]{Y_{#1#2}(\hat{#3})}
\newcommand{\expect}[1]{\langle {#1} \rangle}
\def \PRP{{\it Phys. Rep.} }
\def \ANP{{\it Adv. Nucl. Phys.} }
\def \PPNP{{\it Prog. Part. Nucl. Phys.} }
\def \PTP{{\it Prog. Theor. Phys.} }
\def \PTPS{{\it Prog. Theor. Phys. Supp.} }
\def \FB{{\it Few-Body Syst.} }
\def \FBS{{\it Few-Body Syst. Supp.} }
\def \PH{{\it Physica} }
\def \PP{{\it Preprint} }
\def \YF{{\it Yad. Fiz.} }
\def \PAN{{\it Phys. At. Nucl.} }
\def \JMP{{\it J. Math. Phys.} }
\def \LNC{{\it Lett. Nuov. Cim.} }
\def \IJMP{{\it Int. J. Mod. Phys.} }
\def \EPJ{{\it Eur. Phys. J.} }
\def \IRNP{{\it Int. Rev. Nucl. Phys.} }
\def \SPJ{{\it Sov. Phys.-JETP} }

%% file: ch0rev.tex
\section{Introduction}

Hadron physics, ranging from particle to nuclear physics, aims to a precise
and consistent description of hadronic structure and interactions. This is a
formidable task that can only be ideally accomplished within the framework
of the Standard Model of the strong and electroweak interactions. The
strong interaction part, Quantum Chromodynamics (QCD), should give account
of the main bulk of hadronic data but in its current state of development
this objective seems far from being attainable. This has motivated the
formulation of alternative descriptions, based on QCD and/or phenomenology,
which restrict their study to a particular set of hadronic systems and are
able to reproduce the data and to make useful testable predictions. In this
sense non-relativistic constituent quark models incorporating 
{\it chiral} potentials provide undoubtedly the most
successful, consistent and universal microscopic description of the baryon
spectrum and the baryon-baryon interaction altogether. Therefore they are
ideal frameworks to analyze few-baryon systems. 
The purpose of this review article is to report the progress
made in the last years in this analysis.
But first to put these models in perspective we shall start by recalling
the main aspects of the different approaches in hadron physics. Then we
shall present the historical development of chiral constituent quark models
emphasizing their connection to phenomenology and their plausible relation
to the basic theory.

QCD \cite{FRI73} is nowadays accepted as the basic theory to describe the
hadrons and their strong interactions. QCD is a renormalizable quantum field
theory of quarks and gluons based on a local gauge principle on the group 
$SU(3)_{Colour}$. Apart from this local symmetry the QCD Lagrangian has also,
for massless quarks, a global $SU(n)\times SU(n)$ ($n$: number of quark
flavours) chiral symmetry that appears to be spontaneously broken in nature.
Other relevant properties of QCD are related to the running of the
quark-gluon coupling constant: for high momenta (i.e., high momentum
transfers as compared to the QCD scale, $\Lambda _{\rm QCD}$) the coupling tends
to vanish and the quarks are essentially free, a property named asymptotic
freedom; for low momenta the coupling becomes strong, a property known as
infrared slavery. As a consequence of this behaviour a
solution of the theory is only attainable, perturbatively, at high momenta.
In the low-momentum regime one has to resort to non-perturbative calculation
methods (sum rules, instanton based calculations,...) of limited
applicability or to reformulations of the theory from which to approach the
exact solution. Thus lattice QCD has been developed in the hope of getting
from it the exact solution by a limiting procedure 
from the discrete lattice space to the
continuum. However progress in making precise, detailed
predictions of the physical states of the theory is slow (due in part to the
enormous computer capacity required) \cite{MON94}.

Alternatively effective field theory approaches have been proposed: from the
action and measure of QCD one can integrate out the irrelevant degrees of
freedom in the momentum region under consideration to obtain a more
tractable field theory with the same $S$ matrix. This procedure, used with
notable precision at high momenta where the smallness of the coupling
constant makes feasible to calculate the effective Lagrangian,
cannot be pursued with the same precision at low momenta. Such a difficulty
can be overcome by the application of a celebrated, though unproven, theorem
by Weinberg \cite{WEI79} which states that one should write the most general
Lagrangian, constructed from the accessible degrees of freedom in the
momentum region under consideration, which satisfies the relevant symmetries
of the theory. This is the scheme used in Chiral Perturbation Theory
\cite{GAS84} which has been successful in the study of low momentum hadron (in
particular meson) physics.

Other approaches take limits of QCD and generate from them systematic
expansions to get corrections. Among them we shall briefly comment on
the Heavy Quark Effective Theory
(HQET) \cite{ISG89}, the Non-Relativistic QCD (NRQCD) \cite{THA91} and the
$1/N_c$ approach \cite{THO74,WIT79}.

For physical systems with a heavy quark, interacting with light quarks and
gluons carrying a momentum of order $\Lambda _{\rm QCD}$, an appropriate limit
of QCD involves taking the heavy quark mass going to infinity, 
$m_{Q}\rightarrow \infty$ (HQET).
In this limit the strong interactions of the
heavy quark become independent of its mass and spin. These heavy flavour and
spin symmetries, not present in QCD, lead to model independent
relations that allow for instance the description of exclusive decays in
terms of a few parameters. Corrections are obtained from an expansion in
terms of $1/m_{Q}$, the inverse of the heavy quark mass.

When more than one heavy quark is present in the system (for example in
quarkonia) the heavy quark kinetic energy, treated as a small correction in
HQET, cannot be treated as a perturbation. For such systems the appropriate
limit of QCD to examine is the $c\rightarrow \infty $ limit 
(or $v/c \rightarrow 0$ in quarkonia, being 
$v$ the relative heavy quark velocity and
$c$ the speed of light). The resulting field theory is called
Non-Relativistic QCD. An expansion in terms of $1/c$ is done. NRQCD has
improved the understanding of the decays and production of quarkonia.

The $1/N_{c}$ approach is the limit of QCD when the number of
colours $N_{c}$ tends to infinity. An expansion in terms of $1/N_{c}$,
the inverse of the number of colours, is performed. Remarkably, large 
$N_{c}$ QCD reduces smoothly to an effective field theory of non-interacting
mesons to leading order. Baryons emerge as solitons in this weakly coupled
phase of mesons. This is the base of topological models where baryons are
constructed from non-linear meson fields in chiral Lagrangians 
(Skyrme model \cite{ZAH86}, chiral soliton model \cite{DIA97},...)

At a much more phenomenological level model building has been extremely
useful to provide a classification of hadronic data and to make predictions.
Though the connection with QCD is not clearly established and there is not a
sound systematics to obtain corrections, models provide simple physical
pictures which connect the phenomenological regularities observed in the
hadron data with the underlying structure.

Historically the first model of hadron structure, a non-relativistic quark
model, appeared in the nineteen sixties right after Gell-Mann and Zweig
\cite{GEL64,ZWE64} introduced the quarks as the members of the triplet
representation of $SU(3)_{Flavour}$ from which the hadron representations
could be constructed. The lack of observation of free quarks in nature
seemed to indicate their strong binding inside hadrons. Dalitz \cite{DAL65}
performed a quark model analysis of hadronic data based on flavour. This
analysis seemed to point out some possible inconsistencies with the
Spin-Statistic theorem. The introduction of colour \cite{GRE64,HAN65}, a new
degree of freedom for the quarks, saved the consistency and led to the
formulation of QCD. The other way around QCD provided from its infrared
slavery property some justification for coloured quark confinement inside
colour singlet hadrons. Moreover an interquark colour potential was
derived by de R\'{u}jula, Georgi and Glashow \cite{RUJ75} from the one-gluon
exchange (OGE) diagram in QCD, and applied in an effective manner to reproduce
energy splittings in hadron spectroscopy.

In the 70's the potentialities of quark model calculations in hadronic
physics were established. The non-relativistic quark model was
formulated under the assumption that the hadrons are colour singlet non
relativistic bound states of constituent quarks with phenomenological
effective masses and interactions.

Non-relativistic quark models incorporating effective OGE and confinement
potentials provided in the heavy meson sector a very good description of the
charmonium and bottomonium spectra \cite{APP75,EIC75}. The reasonable results
obtained when applied to the light baryon sector \cite{ISG78} were somewhat
surprising since according to the size of the baryons ($\sim$ 0.8 fm) and the
mass of the light constituent quarks employed ($m_u=m_d\simeq$ 300 MeV)
the quark velocities were close to $c$ and the non-relativistic treatment
could hardly be justified. In the spirit of quark potential model
calculations this pointed out that the effectiveness of the parameters of
the potential could be taking, at least partially into account in an
implicit form, some relativistic corrections. 
Years later the 
low-lying states of the meson \cite{BHA81} and baryon spectra \cite{SIL85}
were reasonably reproduced with a unique set of
parameter values.

On the same line relativistic quark models started to be developed. 
The bag model \cite{CHO74} considered asymptotically free quarks
confined into a space-time tube through field boundary conditions. Chiral
symmetry was incorporated through current conservation at the boundary
associated to the presence of a meson cloud \cite{CHO75}.
Hadron spectra and properties were calculated. However, despite the very
intuitive image the bag model provided of a nucleon as a quark core
surrounded by a meson cloud, calculations had to face the technical
difficulty of separating the center of mass motion, an endemic problem in
all relativistic quark models.

Mid-way between non-relativistic and relativistic treatments
semirelativistic quark models, where the relativistic expression for the
kinetic energy was used in the Schr\"{o}dinger equation, were explored
\cite{GOD85}.

At the same time there were attempts to study the hadron-hadron interaction
within the non-relativistic quark model framework \cite{LIB77}. It was soon
realized that the colour structure of the OGE interquark potential could
provide an explanation for the nucleon-nucleon (NN) short-range repulsion
\cite{NEU77}. This encouraged several groups to undertake the ambitious project
of describing the NN interaction from the quark-quark (confinement+OGE)
potential \cite{TOK80,RIB80,OKA80a,OKA80b,FAE82,CVE83,HARV81} by using 
the resonating group method or
Born-Oppenheimer techniques. According to the expectations a quantitative
explanation for the short-range NN repulsion was found. 
However, for the medium and
long-range parts of the NN interaction, though $(q\overline{q})$ and 
$(q \overline{q})^{2}$ excitations generated by off-shell terms of the
Fermi-Breit piece of the OGE were considered so that the NN potential
became attractive in the $0.8-1.5$ fm range, the attraction was too weak
to bind the deuteron or to fit the extreme low-energy $S$-wave scattering
\cite{FUJ86}.

To remedy this {\it hybrid quark models} \cite{OKA83,FUJ86b}
containing both, interquark (confinement+OGE) and interbaryon long-range 
one-pion and medium range one-sigma, or two-pion exchange potentials, were
introduced. Although these models were efficient to reproduce scattering and
bound state data it was compelling, for the sake of consistency,
to find a justification for these
interbaryon potentials at the quark level. By considering that they could be
due to (chiral) meson clouds surrounding the nucleons quark core, and
pursuing the philosophy initiated in \cite{RUJ75} of trying to incorporate into
the quark potential model the dynamics and symmetries of QCD, an implementation of
chiral symmetry at the quark potential level was needed.

To accomplish this task the progress in the understanding of the connection
between quark potential models and the basic theory was of great help.
Manohar and Georgi \cite{MAN84} argued that the scale associated to confinement
in a hadron ($\Lambda _{\rm CON}\sim$ 100$-$300 MeV) being smaller than the one
associated to chiral symmetry breaking $(\Lambda \sim$ 1 GeV) would drive to
a picture where quarks, gluons and pions coexisted in a region of momentum.
Using an effective field theory approach they obtained a model where
constituent quarks (with a mass generated through chiral symmetry breaking)
and gluons interacted via conventional colour couplings while quarks and
pions did via a non-linear sigma model. By the same time Diakonov, Petrov
and Yu got similar results concerning chiral symmetry breaking from a
picture of the QCD vacuum as a dilute gas of instantons \cite{DIA84,DIA96}.

The effect of incorporating to the simple (confinement+OGE) model a one-pion
exchange (OPE) potential at the quark level started to be analyzed. It was
realized that the OPE potential gave rise to a NN short-range repulsion
\cite{SHI84} to be added to the OGE one. As a consequence the quark-gluon
coupling constant needed to fit the data got nicely reduced from its former
effective value ($>$1) to a value ($<$1) much more
according to the perturbative derivation of the OGE from QCD. The same
conclusion came out from the OPE contribution to the $\Delta -$N mass
difference \cite{BRA85}. On the other hand the OPE potential generated, at large
internucleonic distances, a NN conventional pion exchange interaction as
needed. Nonetheless at the medium range it did not provide enough 
NN attraction to fit the data. 

The introduction of sigma as well as pion exchanges between
the quarks, in the form dictated by a Nambu-Goldstone realization of chiral
symmetry within the linear sigma model, allowed to overcome 
this problem. Thus, the first non-relativistic constituent quark model
of the NN interaction incorporating a chiral $SU(2)\times SU(2)$
potential came out \cite{OBU90,FER93a}.
NN phase shifts (with the exception of $P$-waves) and deuteron
properties were satisfactorily described. Furthermore, reasonable, though
not precise, hadron spectra were predicted with the very same model (i.e.
with the same set of parameters fitted from the NN interaction)
\cite{OBU90,VAL96a}. 
This model has been usually referred in the literature as chiral
constituent quark model (CCQM). We shall maintain this logo hereforth.

Later on a $SU(3)\times SU(3)$ chiral quark model where constituent quarks
interact only through pseudoscalar Goldstone bosons (GBE) was developed to describe very
successfully the baryon spectra \cite{GLO96}. However its application to the NN
interaction revealed a too strong tensor force, generated from 
the $\pi$-exchange, and the absence of the necessary medium-range 
attraction \cite{NAK98}. 
When properly implemented with the one-gluon exchange and 
scalar $SU(3)\times SU(3)$ Goldstone boson
interactions, these models were successfully applied to the NN and
nucleon-hyperon interactions \cite{FUJ96,FUJ04}.

The significant role played by the semirelativistic kinematics in the 
GBE model to fit the spectroscopy encouraged the implementation of a
relativistic treatment in the CCQM. When done
\cite{GAR03b} a pretty nice description 
of the non-strange baryon spectra was obtained as well.

The success in the description of the non-strange baryon spectroscopy and the NN
interaction in a consistent manner makes the CCQM
to be a powerful tool to treat, in a parameter-free way and on the same
footing, other baryon-baryon interactions. This consistent treatment is
mandatory in the study of few-baryon systems due to the intertwined role of
nucleons and resonances in them. 
In this article we review the application of non-relativistic
constituent quark models to obtain the energy spectra of hadrons
and the baryon-baryon effective potentials (we will not consider
baryon form-factors or amplitudes for weak, electromagnetic or strong 
decays since they are very sensitive to relativity \cite{DES04}).
Though most of the results will refer to the CCQM, results from other
models will be included for completeness. The order of the presentation is
the following. In \sref{ch2} the derivation of the 
{\it chiral} part of the quark-quark potential from an
effective Lagrangian, incorporating spontaneous chiral symmetry breaking, is
detailed. A one-gluon exchange potential, giving account of the residual
(perturbative) colour interaction, completes the quark-quark potential whose
parameters are fitted from the NN interaction and the non-strange baryon
spectrum. In order to be able to generate a baryon-baryon potential from the
quark-quark one and to deepen the understanding of the role played by
quark antisymmetry, a variational two-baryon wave function is introduced in
\sref{ch3}. A detailed explanation of the NN short-range repulsion based on
the interplay between quark antisymmetry and dynamics is presented. The
baryon-baryon potential for NN, N$\Delta$, $\Delta\Delta$ and NN$^*$(1440)
channels are constructed in \sref{ch4} and \sref{ch5}. 
Comparison to experimental
data, when available, and to other model results, are shown. A coupled channel
calculation for the NN system above the pion threshold is reported. In
\sref{ch6} a thorough study of the non-strange baryon spectrum is carried
out. Some related comments on the first experimental candidate for an exotic baryon,
the $\Theta ^{+}$ resonance, are also included. Consistency of the baryon
spectrum wave functions with the variational baryon-baryon wave functions is
shown. Section \ref{ch7} is devoted 
to the search for unstable two- and three-baryon
resonances. A more complete calculation for the triton binding
energy is presented. Finally in \sref{ch9} we summarize the main
results and conclusions.

%% file: ch1rev.tex
\section{The chiral constituent quark model.}
\label{ch2}
\subsection{The quark-quark potential}
\label{ch2.1}

In non-relativistic quark models quark colour confinement inside
colour singlet hadrons is taken for granted. Though confinement has
not been rigorously derived from QCD, lattice calculations show
in the so-called quenched approximation (only valence quarks) an 
interquark potential 
linearly increasing with the interquark distance \cite{BAL01}.
This potential can be physically interpreted in 
a picture in which the quarks are linked
with a one-dimensional colour flux tube or string 
and hence the potential is proportional to the 
distance between the quarks,
\begin{equation}
V_{\rm{CON}} ({\vec r}_{ij}) =
-a_c \, {\vec \lambda}_i \cdot {\vec
\lambda}_j \, r_{ij}  ,
\end{equation}
where $a_c$ is the confinement strength, the ${\vec \lambda}$'s are 
the $SU(3)$ colour matrices, and the colour structure
prevents from having confining interaction 
between colour singlets \cite{SHI89}. 
Hadron sizes correspond to a scale of confinement $\Lambda_{\rm CON}
\sim$ 100$-$300 MeV. On the other hand low-lying hadron masses
are much bigger than light quark (up and down) current masses in 
QCD. So when dealing with hadrons one can reasonably assume as a 
good approximation the light quarks to be massless.

In the limit of zero light-quark masses the QCD Lagrangian is invariant under the
chiral transformation $SU(2)_{L}\otimes SU(2)_{R}$.
This symmetry would imply 
the existence of chiral partners, that is, for each low-lying 
hadron there would exist another one with equal mass and opposite parity 
\footnote{There are two ways in which a symmetry of a Lagrangian 
manifests itself in nature \cite{YND99}. The first 
one is the standard Wigner-Weyl realization 
when the generators of the symmetry group annihilate the vacuum. In this case 
nature exhibits the symmetry in the form of degenerate multiplets.
The second one is the Goldstone realization and it 
corresponds to the case of a vacuum that is not annihilated by
all the generators of the group.
The symmetry of the Lagrangian is not evident in nature, one says it is hidden,
and this is referred to as the spontaneous breaking of
the symmetry. It is essentially the case for QCD \cite{YND99,PIC95}.}.
This is not observed in nature 
\footnote{The splitting between the vector $\rho$ and the axial
$a_1$ mesons is about 500
MeV (2/3 of the $\rho$ mass) and the splitting between the
nucleon and its chiral partner is even larger ($940-1535$) MeV.}
what points out to a {\it spontaneous chiral symmetry breaking} in QCD.
As a consequence, the current quarks get dressed becoming constituent quarks
and Goldstone bosons are generated \cite{GOD61}. 
Would the whole process be exact, one would end up with 
massless Goldstone bosons exchanged between the constituent 
quarks. In the real world chiral symmetry is only an approximate 
symmetry so one ends up with low-mass bosons (with masses
related to the quark masses) exchanged between the constituents.

The picture of the QCD vacuum as a dilute medium of
instantons \cite{DIA84,DIA96} explains nicely the 
spontaneous chiral symmetry breaking, which is
the most important non-perturbative phenomenon for hadron structure at low
momenta. Quarks interact with fermionic zero modes of the individual
instantons in the medium and therefore the propagator of a light quark gets
modified and quarks acquire a momentum dependent mass which drops to zero
for momenta higher than the inverse of the average instanton size 
$\overline{\rho }$. The momentum dependent quark mass acts as a natural 
cutoff of the theory. In the domain of momenta $q<1/\overline{\rho }$, 
a simple chiral invariant Lagrangian can be derived as \cite{DIA84}
\begin{equation}
L=\overline{\psi}(\rmi \gamma^\mu \partial_\mu -MU^{\gamma _{5}})\psi
\label{eq1}
\end{equation}
where $U^{\gamma _{5}}=\exp (i\vec{\pi} \cdot \vec{\tau} \gamma
_{5}/f_{\pi })$. $\vec{\pi}$ denotes a Goldstone pseudoscalar field,
$f_\pi$ is the pion decay constant,
and $M$ is the constituent quark mass. The momentum dependence of 
the constituent quark mass can be obtained from the theory.
It has been effectively parametrized as 
$M(q^{2})=m_{q}F(q^{2})$ \cite{BRA85,FER93a} with
\begin{equation}
F(q^{2})=\left[ \frac{\Lambda^{2}}{\Lambda^{2}+q^{2}} \right] ^{\frac{1}{2}}
\label{fofa}
\end{equation}
where $\Lambda$ is the chiral symmetry breaking scale, $\Lambda 
\sim 1/\overline{\rho}$ and $m_q=M(q^2=0)\approx$ 350 MeV.
We shall call hereforth $m_q$ the constituent quark mass.
$U^{\gamma _{5}}$ can be expanded in terms of boson fields as,
\begin{equation}
U^{\gamma _{5}}=1+\frac{\rmi}{f_{\pi }}\gamma ^{5}\vec{\tau}\cdot \vec{\pi} -\frac{1}{%
2f_{\pi }^{2}}\vec{\pi} \cdot \vec{\pi} +...
\end{equation}
The first term generates the constituent quark mass and the second gives
rise to a pseudoscalar (PS) one-boson exchange interaction between quarks. The main
contribution of the third term comes from the two-pion exchange which is 
usually simulated by means of a scalar (S) exchange. From the
non-relativistic approximation of the Lagrangian
one can generate in the static approximation the quark-meson
exchange potentials:
\begin{eqnarray}
\fl V_{\rm{PS}} ({\vec r}_{ij}) = {1 \over 3}
\, \alpha_{\rm{ch}} {\Lambda^2  \over \Lambda^2 -
m_{\rm{PS}}^2} \, m_{\rm{PS}} \, \Biggr\{ \left[ \,
Y (m_{\rm{PS}} \, r_{ij}) - { \Lambda^3
\over m_{\rm{PS}}^3} \, Y (\Lambda \,
r_{ij}) \right] {\vec \sigma}_i \cdot
{\vec \sigma}_j + \nonumber \\
 \left[ H( m_{\rm{PS}} \, r_{ij}) - {
\Lambda^3 \over m_{\rm{PS}}^3} \, H( \Lambda \,
r_{ij}) \right] S_{ij} \Biggr\} \,
{\vec \tau}_i \cdot {\vec \tau}_j \, ,
\label{PS}
\label{OPE}
\end{eqnarray}
\begin{equation}
\fl V_{\rm{S}} ({\vec r}_{ij}) = - \alpha_{\rm{ch}} \,
{4 \, m_q^2 \over m_{\rm{PS}}^2}
{\Lambda^2 \over \Lambda^2 - m_{\rm{S}}^2}
\, m_{\rm{S}} \, \left[
Y (m_{\rm{S}} \, r_{ij})-
{\Lambda \over {m_{\rm{S}}}} \,
Y (\Lambda \, r_{ij}) \right] \, ,
\label{OSE}
\end{equation}
where the $i$ and $j$ indices are associated with $i$ and $j$
quarks respectively, ${\vec r}_{ij}$ stands for the interquark
distance, $\alpha_{\rm{ch}}$ is the chiral coupling constant,
the ${\vec \sigma}$'s (${\vec \tau}$'s)
are the spin (isospin) quark Pauli matrices. 
$m_{\rm PS}$ and $m_{\rm S}$ are the masses of the
pseudoscalar and scalar Goldstone bosons, respectively.
$S_{ij} \, = \, 3 \, ({\vec \sigma}_i \cdot
{\hat r}_{ij}) ({\vec \sigma}_j \cdot  {\hat r}_{ij})
\, - \, {\vec \sigma}_i \cdot {\vec \sigma}_j$
is the quark tensor operator
and $Y(x)$ and $H(x)$ are the standard Yukawa functions
defined by $Y(x)=\rme^{-x}/x$ and $H(x)=(1+3/x+3/x^2)Y(x)$.
 
For $q^2 \gg \Lambda^2$ one expects QCD perturbative effects 
playing a role. They mimic the gluon fluctuations 
around the instanton vacuum and are taken into account
through the OGE potential \cite{RUJ75}.
From the non-relativistic reduction of the one-gluon-exchange
diagram in QCD for point-like
quarks one gets 

\begin{equation}
\fl V_{\rm{OGE}} ({\vec r}_{ij}) =
{1 \over 4} \, \alpha_s \, {\vec
\lambda}_i \cdot {\vec \lambda}_j
\Biggl \lbrace {1 \over r_{ij}} -
{1 \over {4 \, m^2_q}} \, \biggl [ 1 + {2 \over 3}
{\vec \sigma}_i \cdot {\vec
\sigma}_j \biggr ] \,\,
{{\rme^{-r_{ij}/r_0}} \over
{r_0^2 \,\,r_{ij}}}
- {3 \over {4 m^2_q \, r^{3}_{ij}}}
\, S_{ij} \Biggr \rbrace \, ,
\label{reg}
\end{equation}
where $\alpha_s$ is an effective strong coupling constant.
Let us realize that the contact term involving a
Dirac $\delta(\vec{r})$ that comes out in the deduction of the potential has been
regularized in the form
\begin{equation}
\delta(\vec{r} \,) \, \rightarrow \,
{1 \over {4 \pi r_0^2}} \,\,
{{\rme^{-r/r_0}} \over r} \, ,
\label{regu}
\end{equation}
where $r_0$ is a regularization parameter giving rise 
to the second term of \eref{reg}. This avoids to get
an unbound baryon spectrum when solving the Schr\"odinger equation
\cite{BHA80}. 

Thus the quark-quark interaction has the form,
\begin{equation}
V_{qq}(\vec r_{ij})= V_{\rm{CON}} (\vec r_{ij}) +
V_{\rm{OGE}} (\vec r_{ij}) + V_{\rm{PS}} (\vec r_{ij}) +
V_{\rm{S}} (\vec r_{ij}) \, .
\label{inte}
\end{equation}
Such a model has an immediate physical interpretation.
In the intermediate region,
between the scale at which the chiral flavour symmetry is
spontaneously broken, $\Lambda \sim$ 0.8 GeV, and
the confinement scale, $\Lambda_{\rm CON} \sim$ 0.2 GeV, QCD is 
formulated in terms of an effective theory of constituent
chiral quarks interacting through the
Goldstone modes associated with the spontaneous breaking of chiral 
symmetry. For $q^2 \gg \Lambda^2$ gluon exchange is also
important.

It is worthwhile to note that vector
meson-exchange potentials ($\rho$, $\omega$) are not considered. 
The problem of unifying
the quark exchange and meson exchange in the nuclear force
has been a matter of discussion \cite{YAZ90}. It has been
shown that the pseudoscalar ($\pi$) and
scalar ($\sigma$) meson-exchange terms can be simply added to the
quark-exchange terms without risk of double counting. However,
the vector-meson exchanges ($\rho$, $\omega$), which play an important role
in meson-exchange models at the baryon level need some care.
In baryonic one-boson exchange models, the $\omega$-meson
provides the short-range repulsion of the NN interaction.
In chiral constituent quark models this task is taken 
over by the antisymmetrization effects on the pseudoscalar exchange 
combined with OGE. Besides, the $\rho$-meson
reduces the strength of the tensor pseudoscalar
interaction, the same effect that is obtained from the quark-exchange terms
of the pseudoscalar potential
as has been checked in charge-exchange reactions \cite{FER94}.

\subsection{Model parameters}
\label{ch2.2}

In the spirit of quark model calculations the parameters
in the potentials have an effective character.
A rough estimate of the values of the parameters can be
made based on general arguments.
It is well established that the NN interaction at
long-range is governed by the one-pion exchange.
Therefore, to reproduce accurately this
piece of the NN interaction, one is forced to identify the
mass of the pseudoscalar field with the physical pion mass. The
mass of the scalar field, the one-sigma exchange (OSE),
is obtained by the PCAC relation \cite{SCA93}
\begin{eqnarray}
m_{\rm{PS}}^2 &=& m_\pi^2 \\ \nonumber
m_{\rm{S}}^2 &=& m^2_{\rm{PS}}+4m^2_q \, .
\end{eqnarray}
As the pseudoscalar field is identified with the
pion, the $\alpha_{\rm ch}$ coupling constant should
reproduce the long-range OPE interaction. If two nucleons are
separated enough, the central part of $V_{\rm PS}$, 
the pseudoscalar interaction between quarks,
generates an interaction between nucleons
of the form,
\begin{equation}
\fl  V_{\rm{PS}} (R)=\frac{1}{3}\,\alpha_{\rm{ch}}\, G \,\,
  {\Lambda^2 \over {\Lambda^2 - m_\pi^2}} 
  \frac{\rme^{-m_\pi R}}{R}
  \left (\frac{5}{3}\right)^2
   (\vec{\sigma}_N\cdot\vec{\sigma}_N)
   (\vec{\tau}_N\cdot\vec{\tau}_N)
\end{equation}
where $R$ is the interbaryon distance and $G$
depends on the nucleon wave function. Comparing with the
standard one-pion-exchange internucleon potential,
\begin{equation}
\fl  V_{\rm{OPE}} (R)=\frac{1}{3}\,\frac{f_{\pi \rm{NN}}^2}{4\pi}\,
  {\Lambda^2 \over {\Lambda^2 - m_\pi^2}} 
  \frac{\rme^{-m_\pi R}}{R}
   (\vec{\sigma}_N\cdot\vec{\sigma}_N)
   (\vec{\tau}_N\cdot\vec{\tau}_N)
\end{equation}
where the same form factor has been used at the quark
and baryon levels, and
using a harmonic oscillator wave function for the nucleon
in terms of quarks (see next section), 
one finally obtains \cite{LIU93},
\begin{eqnarray}
  \alpha_{\rm{ch}}=
  \left (\frac{3}{5}\right)^2
  \frac{f_{\pi \rm{NN}}^2}{4\pi}\,\,
  \rme^{-\frac{b^2 m_\pi^2}{2}}\,=\,
\left( 3 \over 5 \right)^2
{ g_{\pi \rm{NN}}^2 \over {4 \pi}} { m_{\pi}^2
\over {4 m_N^2}} \,
  \rme^{-\frac{b^2 m_\pi^2}{2}} \, .
\end{eqnarray}
This gives the chiral coupling constant $\alpha_{\rm ch}$
in terms of the $\pi$NN
coupling constant, taken to be
$f_{\pi \rm{NN}}^2/4\pi=0.078$ \cite{ERI92}, and $b$.
Usual values in the literature for $b$ 
range between 0.4 and 0.6 fm. The most stringent determination,
$b=$ 0.518 fm, was done
in reference \cite{VAL94a} by means of a simultaneous study of $S$-wave
NN phase shifts and deuteron properties (see \sref{ch5}). 
As it will be discussed, this value turns out
to be consistent with the solution of the baryon spectra as a
three-body problem in terms of the interaction \eref{inte} \cite{VAL96a}.
The value of $\alpha_{\rm ch}$ is then $\alpha_{\rm ch}=$0.0269.

The tensor force is mainly due to the pion interaction. Therefore,
the value of $\Lambda$ can be fixed
examining a process dominated by the tensor term. Such a reaction
could be the $pp \to n \Delta^{++}$ because, at high
momenta, more than $90\,\% $ of the interaction corresponds to the
tensor part. The calculation of reference \cite{FER94}
suggests for $\Lambda$ a value close to 4.2 fm $^{-1}$.

The value of $\alpha_s$ is estimated from
the $\Delta-$N mass difference.
In the chiral constituent quark model there are
contributions not only from spin-spin term of the OGE but also
from the pseudoscalar interaction,
the latter contributing approximately half of the
total mass difference. The rest is attributed to the
OGE, and the value of $\alpha_s$ is adjusted
to reproduce the experimental
$\Delta-$N mass difference. This gives
$\alpha_s \sim 0.4-0.5$ for $r_0=0.25$ fm,
a standard value of $r_0$ within the stability
region for the $\Delta-$N mass difference 
(see \sref{sec624}).

The constituent quark mass $m_q$ is an effective parameter 
whose value is conventionally chosen in the range 
of 300 MeV, using in general smaller
values for the study of the NN interaction and bigger ones for
one-baryon properties. From the proton and
neutron magnetic moments in the impulse approximation 
one gets $m_q=350$ MeV. From NN scattering 
the quark mass value should be close to one third
of the nucleon mass $m_q \sim 1/3 m_N \sim 313$ MeV. 
In this way a correct relation 
between the momentum and kinetic energy 
is guaranteed \cite{TOK80}.
\begin{table}[b]
\caption{\label{table1}Quark model parameters.}
\begin{tabular}{@{}lllllllll}
\br
$m_q$ & $b$ & $\alpha_s$ & $a_c$ & $\alpha_{\rm{ch}}$ & $r_0$ 
& $m_{\rm{S}}$  & $m_{\rm{PS}}$ & $\Lambda$  \\
(MeV) & (fm) & & (MeV fm$^{-1}$) & & (fm) & (fm$^{-1}$) & (fm$^{-1}$) &
(fm$^{-1}$)  \\ 
\mr
313 & 0.518 & 0.485 & 67.0 & 0.0269  & 0.25  & 3.42 & 0.7& 4.2 \\
\br
\end{tabular}
\end{table}

Finally, regarding $a_c$ we should first notice that the
contribution of confinement to the force between two baryons
is very small (zero for quadratic confining, see \sref{ch3.2}).
Hence its value is only constrained by the requirement of
having a confining ($a_c > 0$) and not a deconfining ($a_c < 0$)
interaction. This can be guaranteed through the nucleon mass
stability condition
$\left. \partial M_N(b)/\partial b \right|_{b=0.518} =0$. Concerning
its specific value we can resort to the baryon spectrum which is 
strongly dependent on it (see \sref{ch6}). We quote in
table \ref{table1} the standard value derived from the baryon 
spectrum analysis. 
Some caution is necessary when comparing the strength of the
confining potential to other values given in
the literature. First, one should be aware of the specific form used
for the confining interaction, if the colour Gell-Mann matrices are
used or not (a factor $-8/3$ is in the way). Second, when scalar potentials
between quarks are used, as it is the case of the chiral constituent quark
model, smaller values of $a_c$ than in pure OGE models are obtained.

The standard values of the parameters used in the chiral constituent 
quark model is resumed in table \ref{table1}. 

%% file: ch2rev.tex
\section{The non-strange two-baryon system}
\label{ch3}
\subsection{The two-baryon wave function: quark Pauli blocking}
\label{ch3.1}

The calculation of the interaction between two baryons 
requires the knowledge of the two-baryon wave function and therefore 
that of a single baryon in terms of quarks. 
Single baryon wave functions 
have been calculated using different methods available in the literature 
to solve the three-body problem (see \sref{ch6}).
Although the resulting wave functions have an involved 
structure, it has been shown that for the baryon-baryon 
interaction they can be very well approximated by 
harmonic oscillator eigenfunctions \cite{VAL96a}.
For the non-strange baryons we are going to consider: N, $\Delta$ and N$^*$(1440), 
the total wave function of a single baryon can be explicitly written
as the product of three wave function components: spatial, 
spin-isospin, and colour space respectively,
\footnote{This is not, for example, the case for 
the N$^*$(1535) where the total spin of the particle is 
the result of coupling the intrinsic spin 
and the relative orbital angular momenta of the quarks.} 
\begin{equation}
| \Phi_B \rangle = \phi(\vec{r}_1,\vec{r}_2,\vec{r}_3;\vec{R}) 
\otimes \chi_{ST} \otimes \xi_{C} \, ,
\end{equation}
where $\vec{r}_i$ is the position of quark $i$ and 
$\vec{R}$ denotes the center of mass coordinate of the baryon.
The symmetrization postulate requires the wave function to be
antisymmetric. Explicitly,
\begin{equation}
\fl \Phi_{\rm N}({\vec r}_1,{\vec r}_2,{\vec r}_3;\vec{R}) = 
\prod_{n=1}^{3} \left({\frac{1 
}{\pi b^{2}}} \right)^{3/4} \rme^{-{\frac{ 
( \vec{r}_{n}- \vec{R})^2 }{2 b^{2} }}
} \otimes [3]_{S=1/2,T=1/2} \otimes [1^{3}]_{C} \, ,
\label{nwf}
\end{equation}
\begin{equation}
\fl \Phi_{\Delta}({\vec r}_1,{\vec r}_2,{\vec r}_3;\vec{R}) = 
\prod_{n=1}^{3} \left({\frac{1 
}{\pi b^{2}}} \right)^{3/4} \rme^{-{\frac{ 
( \vec{r}_{n}- \vec{R})^2 }{2 b^{2} }}
} \otimes [3]_{S=3/2,T=3/2} \otimes [1^{3}]_{C} \, ,
\label{dwf}
\end{equation}
\begin{equation}
\fl \Phi_{{\rm N}^*(1440)}({\vec r}_1,{\vec r}_2,{\vec r}_3;\vec{R}) = 
\left ( \sqrt{2\over3} \phi_1- \sqrt{1\over3}\phi_2 \right )
\otimes[3]_{S=1/2,T=1/2}\otimes[1^3]_{C}  \, ,
\label{ropw}
\end{equation}
where $[3]_{ST}$ stands for a completely symmetric $SU(4)$ spin-isospin
wave function and $[1^{3}]_C$ for a completely antisymmetric
$SU(3)$ colour wave function \cite{STA96}, and
\begin{equation}
\fl \phi_1({\vec r}_1,{\vec r}_2,{\vec r}_3;\vec{R}) =  {\sqrt{2} \over 3}
\left( {\frac{1 }{\pi b^{2} }} \right)^{9/4}
\sum_{k=1}^3 \left[ {\frac{3 }{2}} - {\frac{ ( \vec{r}_k
- \vec{R})^2 }{b^{2} }}\right] \prod_{i=1}^3 
\rme^{-{\frac{ ( \vec{r}_{i}- \vec{R})^2 }{2b^{2} }} } ,
\label{fi1}
\end{equation}
\begin{equation}
\fl \phi_2({\vec r}_1,{\vec r}_2,{\vec r}_3;\vec{R}) =  -{2 \over 3}
\left( 1\over \pi^{9\over4} b^{13\over2}  \right)  
\sum_{j<k=1}^3 (\vec{r}_j-\vec{R})\cdot (\vec{r}_k-\vec{R})
 \prod_{i=1}^3 \rme^{-{\frac{ ( \vec{r}_{i}- \vec{R})^2 }{2
b^{2} }} } .
\label{fi2}
\end{equation}

Once single baryon wave functions have been constructed, one
can proceed to study the two-baryon wave functions.
Again the symmetrization postulate forces 
the wave function of a system
of $n$ identical quarks to be totally antisymmetric
under the exchange of any two of them. 
As single baryon wave functions are already antisymmetric,
there is an important simplification to construct
two-baryon wave functions: one needs the antisymmetrizer for a 
system of six identical particles but already clustered in two 
antisymmetric groups. The antisymmetrizer can then be
written as \cite{HOL84}
\begin{equation}
{\cal A} = {1\over N}\left (1- \sum_{i=1}^{3} \sum_{j=4}^{6} 
P_{ij}\right)(1-{\cal P}) \, ,
\label{anti}
\end{equation}
where $P_{ij}$ is the operator that exchanges particles 
$i$ and $j$, ${\cal P}$ exchanges the two clusters, and 
$N$ is a normalization constant. $P_{ij}$ can be explicitly written 
as the product of permutation operators in colour ($C$), 
spin-isospin ($ST$) and spatial ($O$) spaces,
\begin{equation}
P_{ij}= P^{C}_{ij} \; P^{ST}_{ij} \; P^{O}_{ij} \, .
\end{equation}
Taking into account that any two-baryon state can be decomposed
in a symmetric plus an antisymmetric part under the exchange of the baryon
quantum numbers, one can write for 
a definite symmetry (specified by $f$ even or odd) and projecting
onto a partial wave to make clear the effect of the exchange 
operator \cite{SUZ84b,VAL95b}:
\begin{eqnarray}
\fl \Psi_{B_1 B_2}^{LST}({\vec R}) & = & {\frac{{\cal A} }{\sqrt{1 + \delta_{B_1
B_2}}}} \sqrt{\frac{1 }{2}} \Biggr\{ \left[ \Phi_{B_1} \left( 123;{-{\frac{{\vec R} 
}{2}}} \right) \Phi_{B_2} \left( 456; {\frac{{\vec R} }{2}} \right) \right]_{LST} 
\nonumber \\
& + & (-1)^{f} \,  \left[ \Phi_{B_2} \left( 123;{-{\frac{{\vec R} }{2}}}
\right) \Phi_{B_1} \left( 456; {\frac{{\vec R} }{2}} \right) \right]_{LST} \Biggr \} 
\, ,  \label{Gor}
\end{eqnarray}
where $S$, $T$ and $L$ correspond to the total spin, 
isospin and orbital angular momentum
of the two-baryon system. ${\cal A}$ is the six-quark 
antisymmetrizer described above. The action of ${\cal P}$, appearing in 
the antisymmetrizer, on a state with definite 
quantum numbers $LST$ is given by,
\begin{eqnarray}
\fl {\cal P} \;
         \left[ \Phi_{B_1} \left( 123;{-{\frac{{\vec R}}{2}}} \right) 
                \Phi_{B_2} \left( 456;  {\frac{{\vec R} }{2}} \right) 
 +    (-1)^{f} \, 
                \Phi_{B_2} \left( 123;{-{\frac{{\vec R} }{2}}}\right) 
                \Phi_{B_1} \left( 456;  {\frac{{\vec R} }{2}} \right) 
\right]_{LST} \nonumber  \\
\fl  =\left[\Phi_{B_1} \left( 456;{-{\frac{{\vec R}}{2}}} \right) 
          \Phi_{B_2} \left( 123;  {\frac{{\vec R} }{2}} \right) 
 +    (-1)^{f} \, 
          \Phi_{B_2} \left( 456;{-{\frac{{\vec R} }{2}}}\right) 
          \Phi_{B_1} \left( 123;  {\frac{{\vec R} }{2}} \right) 
\right]_{LST} \nonumber  \\
\lo =(-)^{L+S_1+S_2+T_1+T_2-S-T+f}
         \left[ \Phi_{B_1} \left( 123;{-{\frac{{\vec R}}{2}}} \right) 
                \Phi_{B_2} \left( 456;  {\frac{{\vec R} }{2}} \right) \right.\nonumber \\
\lo  +\left. (-1)^{f} \, 
                \Phi_{B_2} \left( 123;{-{\frac{{\vec R} }{2}}}\right) 
                \Phi_{B_1} \left( 456;  {\frac{{\vec R} }{2}} \right) 
\right]_{LST} \, . 
\end{eqnarray}
Then, due to the $(1- {\cal P})$ factor in equation \eref{anti},
the wave function vanishes unless:
\begin{equation}
L+S_1+S_2-S+T_1+T_2-T+f = {\rm odd} \, .
\label{LST}
\end{equation}
For non-identical baryons this relation indicates the symmetry $f$ 
associated to a given set of values $LST$. The non-possible symmetries 
correspond to forbidden states. For identical baryons, $B_1=B_2$, 
such as nucleons
(note that $f$ has to be even in order to have a non-vanishing wave function),
one recovers the well-known selection rule $L+S+T = {\rm odd}$.

Certainly, the effect of quark substructure goes beyond the
$(1-{\cal P})$ factor appearing in the antisymmetrizer and
it also appears through the quark permutation 
operator $P_{ij}$, whose effect can be analyzed in part in
a simple way through the 
norm of the two-baryon system. This is 
a measure of the overlapping between the 
two baryons and it shows out the 
consequences of the Pauli principle. 
The norm of a two-baryon system $B_1B_2$ is defined as,
\begin{equation}
{\cal N}^{LSTf}_{B_1B_2}(R)=\left \langle 
\Psi_{B_1 B_2}^{L\,
S\, T} ({\vec{R}}) \mid 
\Psi_{B_1 B_2}^{L\, S\, T} 
({\vec{R}}) \right \rangle\,.
\label{norm}
\end{equation}
Making use of the wave function (\ref{Gor}) one obtains: 
\begin{equation}
{\cal N}^{LSTf}_{B_1B_2}(R)= {\cal N}_L^{\rm di}(R) -
C(S,T,f;B_1B_2)\; {\cal N}^{\rm ex}_L(R)\, ,
\label{gnorm}
\end{equation}
where ${\cal N}_L^{\rm di}(R)$ and ${\cal N}^{\rm ex}_L(R)$ 
refer to the direct and exchange 
kernels, respectively. The direct kernel corresponds to 
the identity operator appearing in the antisymmetrizer,
while the exchange kernel arises from the $P_{ij}$ operator.
$C(S,T,f;B_1B_2)$ is a spin-isospin coefficient defined 
as follows, 
\begin{eqnarray}
\fl C(S,T,f;B_1B_2) = {3\over 1+\delta_{B_1B_2}}\; 
\left[ \expect{B_1(123)B_2(456)|P_{ij}^{ST}|B_1(123)B_2(456)}_{ST} 
\right.\nonumber \\
\lo +\left. (-1)^f \expect{B_1(123)B_2(456)|P_{ij}^{ST}
|B_2(123)B_1(456)}_{ST} \right] \,.
\label{cst}
\end{eqnarray}
These spin-isospin coefficients, summarized in table \ref{cstf},
determine the degree of
Pauli attraction or repulsion as we will see later on.
Let us note that for the NN$^*$(1440) system, being the 
N$^*$(1440) spin-isospin wave function completely symmetric, 
the spin-isospin coefficients
are the same as for the NN case.
The explicit expressions of the direct and 
exchange kernels depend on the baryons considered. 
For the NN, N$\Delta$ and $\Delta\Delta$ cases
(the NN$^*$(1440) system is much more involved \cite{JUL01} 
and will be discussed below) the norm kernels are given by,
\begin{equation}
\eqalign{ {\cal N}_L^{\rm di}(R)&= 4 \pi \,
\exp{\left({-{3\over4}{R^2\over b^2}} \right)} \; 
\imath_L\left({3\over4}{R^2\over b^2}\right ) \\
{\cal N}_L^{\rm ex}(R)&= 4 \pi \, 
\exp{\left({-{3\over4}{R^2\over b^2}} \right)} \; 
\imath_L\left({R^2\over4b^2}\right )} 
\end{equation}
where $\imath_L$ are the modified spherical Bessel functions.
\begin{table}
\caption{\label{cstf}Spin, isospin coefficients.
``+'' (``$-$'') refers to $f$ even (odd).}
\begin{indented}
\item[]\begin{tabular}{@{}llll}
\br
$(S,T,f)$     &  $C(S,T,f;{\rm NN})$ & $C(S,T,f;{\rm N}\Delta)$ & 
$C(S,T,f;\Delta\Delta)$ \\
\mr
(0,0,+)               &  $7/9$   &          &  ${1/3}$  \\
(0,1,+),(1,0,+)       &  $-1/27$ &          &  ${1/9}$  \\
(0,2,+),(2,0,+)       &          &          &  $-{1/3}$ \\
(0,3,+),(3,0,+)       &          &          &  ${-1}$   \\
(1,1,+)               &  $31/81$ &${31/27}$ &  ${1/27}$ \\
(1,1,$-$)             &          & ${1}$    &           \\
(1,2,+),(2,1,+)       &          & ${1/9}$  &  $-1/9$   \\ 
(1,2,$-$),(2,1,$-$)   &          & ${-1/3}$ &           \\ 
(1,3,+),(3,1,+)       &          &          &  $-{1/3}$ \\
(2,2,+)               &          & ${7/3}$  &  ${1/3}$ \\
(2,2,$-$)             &          &  ${1}$   &          \\
(2,3,+),(3,2,+)       &          &          &   ${1}$  \\
(3,3,+)               &          &          &   ${3}$  \\ 
\br
\end{tabular}
\end{indented}
\end{table}

To examine the physical content of $\cal{N}$, it is convenient
to take the limit 
of the distance between the baryons approaching zero, $R\to0$, 
\begin{eqnarray}
\fl {\cal N}^{LSTf}_{B_1B_2}(R) \rightarrow 
4\pi \left( 1 - {3\over4}{R^2\over b^2}\right){1\over 1\cdot3\cdots(2L+1)} 
\left({R^2\over 4b^2}\right)^L 
\times \Biggr \{ [3^L-C(S,T,f;B_1B_2)] \nonumber \\
\lo +{1\over 2(2L+3)} \left({R^2\over 4b^2}\right)^2 
\times [3^{L+2}-C(S,T,f;B_1B_2)]+\cdots \Biggr \} \, .
\end{eqnarray}
Of significant interest are those cases where 
\begin{equation}
3^L=C(S,T,f;B_1B_2) \,,
\label{csteq}
\end{equation}
because it implies that the overlapping of 
the two-baryon wave function behaves as $R^{2L+4}$ instead 
of the centrifugal barrier behaviour $R^{2L}$, 
indicating that quark Pauli blocking occurs, the available
spin-isospin-colour degrees of freedom saturate and then some quarks are
Pauli expelled to higher orbits.
This suppression in the overlapping
of the two-clusters, that may be a source of short-range
repulsion, is not present at baryonic level
in those cases where $B_1 \ne B_2$,
because of the distinguishability of baryons.
Looking at table \ref{cstf} we differentiate the following
cases:
\begin{itemize}
\item In the NN system there are not Pauli blocked channels.

\item In the N$\Delta$ system some partial waves 
present quark Pauli blocking, those corresponding 
to $(S,T)=(1,1)$ and $(2,2)$, with orbital angular 
momentum $L=0$ ($f=$ odd). Pauli blocking will translate 
into a strong short-range repulsion that can be 
checked experimentally looking at $\pi d$ elastic scattering \cite{FERR89}.
We will return to this point in \sref{ch4.2.2}.

\item In the $\Delta\Delta$ system, the spin-isospin 
coefficients fulfill equation \eref{csteq} for the 
cases $(S,T)=(2,3)$ and $(3,2)$ both with orbital 
angular momentum $L=0$ ($f=$ even). It is also important to mention the 
existence of quark Pauli blocking for a channel 
with $L\neq 0$ which is a characteristic feature of the 
$\Delta\Delta$ interaction, this corresponds to 
$(S,T)=(3,3)$ with orbital angular momentum $L=1$ ($f=$ even).
In a group theory language,
\begin{equation}
\eqalign{\Psi_{\Delta \Delta} [^5S_2 (T=3)]  =  \,
& [ 2^3 ]_{C} \, [ 42 ]_{O} \, [51]_{ST} \, = \Psi_{\Delta \Delta}
[^7S_3 (T=2)]  \\
\Psi_{\Delta \Delta} [^7P_{2,3,4} (T=3)] & = \,
[ 2^3 ]_{C} [ 33 ]_{O} [6]_{ST}}
\end{equation}
showing the absence of the lower energy
spatial states ($[6]_O$ for $(S,T)=(2,3)$
and $[51]_O$ for $(S,T)=(3,3)$) forbidden by the symmetrization postulate.
\end{itemize}
\begin{figure}[t]
\vspace*{-1.8cm}
\centerline{\psfig{figure=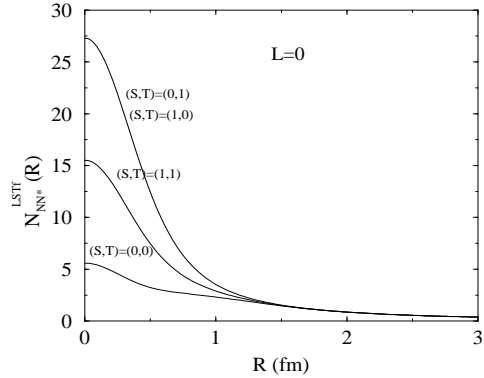,height=5.1in,width=3.8in}}
\vspace*{-5.9cm}
\caption{NN$^*$(1440) overlapping for $L=0$ partial waves.}
\label{f1c2}
\end{figure}
 
As previously said, for the NN$^*$(1440) system the calculation
is much more involved. For the most interesting cases, those where there
is no centrifugal barrier, and in the limit
$R\rightarrow 0$ one obtains \cite{JUL01}:
\begin{equation}
\fl {\cal N}_{\rm{NN}^{*}(1440)}^{L=0,STf}(R) \rightarrow 4 \pi \left \{ 1 -{1\over3}
\left[ 5+
2 (-)^{f} \right] C(S,T,f;\rm{NN}) \right \} + {\cal O}(R^4) \, .
\end{equation}
Quark Pauli blocked channels would correspond to
$f$=odd and $C(S,T,f;\rm{NN})$=1, or $f$=even 
and $C(S,T,f;\rm{NN})$=3/7. From the values given
in table \ref{cstf} it is clear that there are no Pauli blocked
channels. However, looking at figure \ref{f1c2} 
one can see how in those $S$-wave channels forbidden in the NN case, 
$(S,T)=(0,0),(1,1)$, the overlapping gets suppressed as compared
to the allowed channels $(S,T)=(1,0),(0,1)$. 
This is a remnant of the near to identity
similarity of N and N$^*$(1440), and it will 
have an important influence when deriving the NN$^*$(1440) interaction.
 
\subsection{NN short-range repulsion}
\label{ch3.2}

As said above, there is no quark Pauli blocking in the NN system.
Therefore the short-range repulsion experimentally observed requires a
different microscopic explanation. 
More than twenty years ago a NN short-range repulsion was derived from
an interplay between quark antisymmetry and OGE quark 
dynamics \cite{NEU77,OKA80b,TOK80,FAE82,CVE83}.
The origin of the NN short-range repulsion was understood
in a simple and intuitive manner in terms of the energy degeneracy 
induced by the color magnetic hamiltonian between the different
spatial symmetries in the two-baryon wave function \cite{MYH86}.
Later on, the need to incorporate chiral potentials (OPE and OSE)
in the description of the NN interaction forced a revision of the
role played by the OGE dynamics and brought forth a new understanding
of the short-range repulsion at the microscopic level.

To settle this process
let us first revise the OGE based explanation
to establish the notation and physical arguments.
In a group theory language, a completely antisymmetric
six-quark state asymptotically
describing two free nucleons in relative $S$-wave is given
[for spin-isospin $(S,T)=(0,1)$ or $(1,0)$] by \cite{HARV81},
\begin{equation}
\Psi_{6q} (R \to \infty) = \sqrt {1 \over 9} \, \Psi_{\{ 6 \}} +
\sqrt {8 \over 9} \, \Psi_{ \{ 42 \} } \, ,
\label{320}
\end{equation}
\noindent
where
\begin{eqnarray}
\Psi_{\{ 6 \}} \, & \equiv & \, [ 2^3 ]_{C} 
\,\, [ 6 ]_{O} \,\, [ 33 ]_{ST}
\, , \nonumber \\
\Psi_{\{ 42 \}} \, & \equiv & \, {1 \over \sqrt{2}} 
\,\, [ 2^3 ]_{C} \,\,
[ 42 ]_{O} \, \left( [ 33 ]_{ST} \, - 
\, [ 51 ]_{ST} \right) \, ,
\label{l1}
\end{eqnarray}
\noindent
the subindex $C$, $O$ or $ST$ 
indicating the colour, spatial or spin-isospin representation respectively. 
This six-quark wave function expressed in terms of the baryon-baryon basis
would contain two-nucleon, two-delta and two-baryon coloured-octet
states. Concerning the spatial part, 
$L=0$ may be obtained not only from six quarks 
in $0s$ states ($[6]_{O}$),
but also from four quarks in $0s$ states plus two quarks in
excited $p$ waves ($[42]_O$). When the distance 
goes to zero the $\{ 42 \} $ configuration
represents an excited state. In fact, if we assume all quarks in a
harmonic oscillator potential of angular frequency $\omega$, their
wave functions will be given by
\begin{eqnarray}
\eta_{0s} ({\vec r}) \, & = & \, \left(
{1 \over {\pi b^2}} \right)^{3/4}
\,\, \rme^{-r^{\, 2}/2 b^2} \, , \nonumber \\
\eta_{0p} ({\vec r} \, ) \, & = & \, {1 \over \sqrt{3}} \,
\sum_{m=-1}^{1}
\, \left( {8 \over {3 \sqrt{\pi}}}
\right)^{1/2} {1 \over {b^{5/2}}} \, r \, \rme^{-r^2/2 b^2}
\, Y_{1m} ({\hat r}) \, ,
\end{eqnarray}
where we have assumed equal probability for all the third angular
momentum components of the $\ell = 1$ quark orbital
excited state $\eta_{0p} ({\vec r} \, )$.
The excitation energy is then given by,
\begin{equation}
\Delta_{\rm ho} \, \equiv \,
\left[ E_{\{ 42 \} } - E_{ \{ 6 \} } \right]_{\rm ho}
\, =  \,
2 \, \hbar \, \omega \, \equiv  \, {{2 \, \hbar^2} \over 
{3 \, b^2 \, m_q}} \, ,
\label{f2}
\end{equation}
the $[6]_O$ being the lowest in energy. This situation
may be changed due to a particular dynamics. Let us for example analyze
the case of the interaction via the OGE of reference \cite{OKA80a} ,
\begin{equation}
V_{\rm OGE} ({\vec r}_{ij}) =
{1 \over 4} \, \alpha_s \, {\vec
\lambda}_i \cdot {\vec \lambda}_j
\Biggl \lbrace {1 \over r_{ij}} -
{{2 \, \pi} \over {3 \, m^2_q}} \,
{\vec \sigma}_i \cdot {\vec
\sigma}_j \, \delta({\vec r}_{ij})
\Biggr \rbrace \, ,
\label{OGEOYA}
\end{equation}
where the values of the parameters are taken to reproduce the 
$\Delta-$N mass difference: $\alpha_s = 1.39$,
$m_q = 300 \,\,{\rm MeV}$ and $b=0.6 \,\,{\rm fm}$. 
One can estimate the contribution of this interaction to the energy of the
$\Psi_{\{ 42 \}}$ or $\Psi_{\{ 6 \}}$ configurations. First of all,
taken into account that
\begin{equation}
\fl \sum_{i < j }
{\vec \lambda}_i \cdot {\vec \lambda}_j \, = \,
{1 \over 2} \left[ \, \left(
\sum_{i} \vec \lambda_i \right)^2 \, - \,
\sum_{i} \vec \lambda_i^2 \, \right] \, = \, 
{1 \over 2} \left(
\sum_{i} \vec \lambda_i \right)^2 \, - \,
{8 \over 3} \, n \, ,
\end{equation}
\noindent
where $n$ is the number of particles, it is clear that quite
approximately ($1/r_{ij}$ has been substituted by an average
value) the Coulomb-like term of $V_{\rm OGE}$
does not contribute to the $\{42\}-\{6\}$ splitting since 
$\left( \sum_{i} \vec \lambda_i \right)^2$ vanishes for a colour
singlet. Hence, only the colour magnetic part (the term depending
on the $\delta$ function) has to be evaluated. 
Following reference \cite{MYH86}, 
the matrix element for the $\{ 6 \}$ configuration [for $(S,T)=(1,0)$] is,
\begin{equation}
\fl < \Psi_{ \{ 6 \} } \mid - \, \sum_{i < j} c_{ij} \,
({\vec \lambda}_i \cdot {\vec \lambda}_j)
({\vec \sigma}_i \cdot {\vec \sigma}_j) 
\mid \Psi_{ \{ 6 \} } >  = 
{{8 c_1} \over 3} = \, {D \over 6} \,
= \, 48.83 \,\, {\rm MeV} \, ,
\label{a1}
\end{equation}
\noindent
where the radial function $c_{ij}$ is easily identified from 
(\ref{OGEOYA}) and $c_1$ is calculated trough
\begin{equation}
\fl c_1 \equiv c_{ij}^{0s,0s} =
< \eta_{0s} ({\vec r}_i) \eta_{0s} ({\vec r}_j)
\mid \, c_{ij} \mid 
\eta_{0s} ({\vec r}_i) \eta_{0s} ({\vec r}_j) > = 
{\alpha_s \over {12 \sqrt{2 \pi} b^3 m_q^2}} =
{D \over 16} ,
\label{e2}
\end{equation}
$D$ standing for the $\Delta-$N mass difference
($D = 293 \,\, {\rm MeV}$).
Similarly, for the $\{ 42 \}$ configuration
\begin{equation}
\fl < \, \Psi_{ \{ 42 \} } \, \mid \, - \, \sum_{i < j} \, c_{ij} \,
({\vec \lambda}_i \cdot {\vec \lambda}_j)
({\vec \sigma}_i \cdot {\vec \sigma}_j) \,
\mid \, \Psi_{ \{ 42 \} } \, > \, = \, -{{55 c_2} \over 3} \,
= \, - 235.63 \,\, {\rm MeV} \, ,
\end{equation}
where $c_2$ is the average two-quark interaction strength for the
$[42]_O$ state,
\begin{equation}
c_2 = {1 \over 15} \left[ 6 \, c_{ij}^{0s,0s} 
+ 8 \, c_{ij}^{0s,0p} + c_{ij}^{0p,0p} \right]
= {379 \over 540} \, c_1 \, ,
\label{a2}
\end{equation}
6, 8 and 1 denoting the number of the corresponding
pairs. Finally one obtains $\Delta_{\rm OGE} \equiv
\left[ E_{\{ 42 \} } - E_{ \{ 6 \} } \right]_{\rm OGE}
\, = \, - 284.46 \,\, {\rm MeV}$.
The different character of the colour magnetic Hamiltonian for
both configurations makes this difference to compensate the
harmonic oscillator energy 
difference, $\Delta_{\rm ho} = 240.36 \,\,{\rm MeV}$, the
$[42]_{O}$ spatial symmetry becoming the lowest in energy
and almost degenerate with the $[6]_O$. 
If the two spatial symmetry states are energy degenerate
then for two free
nucleons ($R \to \infty$ in Eq. \eref{320}) the $[42]_O$ dominates.
To make the physics clear let us for a moment argue as if we had 
only the spatial $[42]_{O}$ symmetry (our results
are not modified if the $[6]_{O}$ symmetry
is included). At zero distance between the two nucleons
the spatial $[42]_{O}$ symmetry implies a $2 \hbar \omega$
excitation in an oscillator basis. If the two nucleons are moved apart,
in each nucleon the three quarks are in a spatially symmetric state
and thus the configuration corresponds to a $0 \hbar \omega$ excitation.
The $2 \hbar \omega$ energy has therefore to be in the relative motion.
This means that for an $S$-wave the relative motion must have a node
and thus cannot be described by a $0s$ but instead by a $1s$ relative wave function.
In the asymptotic part of the wave function such a zero implies 
a phase shift as the one given by a hard-core potential \cite{SAIT68}.

Similar results are obtained with other OGE potentials. 
For example, using the parameters of reference \cite{FAE82} one obtains 
$- 291.08 \,\,{\rm MeV}$ for the energy difference between the
$\{ 42 \}$ and $\{ 6 \}$ configurations, while the harmonic oscillator
energy gap is in this case $296.68 \,\,{\rm MeV}$.
Once again the colour magnetic interaction produces a
mixing of the symmetry states and a short-range repulsion.

This explanation had to be revised in
chiral constituent quark models where, in addition to the OGE,
there are pseudoscalar and scalar Goldstone-boson exchanges between quarks.
As a consequence the value of $\alpha_s$, which drives
the OGE energy gap between the $\{ 42 \}$ and $\{ 6 \}$ 
configurations, is significantly reduced (due to the
pseudoscalar contribution to the $\Delta-$N mass difference)
and correspondingly an 
energy degeneracy from the OGE is not attained.

As a matter of fact if we recalculate the contribution
of the OGE with the parameters of table \ref{table1}, we obtain
\begin{eqnarray}
\fl < \, \Psi_{ \{ 6 \} } \, \mid \, - \, \sum_{i < j} \, c_{ij} \,
({\vec \lambda}_i \cdot {\vec \lambda}_j)
({\vec \sigma}_i \cdot {\vec \sigma}_j) \,
\mid \, \Psi_{ \{ 6 \} } \, > \,  =  \, 24.21 \,\, {\rm MeV} \, ,
\nonumber \\
\fl < \, \Psi_{ \{ 42 \} } \, \mid \, - \, \sum_{i < j} \, c_{ij} \,
({\vec \lambda}_i \cdot {\vec \lambda}_j)
({\vec \sigma}_i \cdot {\vec \sigma}_j) \,
\mid \, \Psi_{ \{ 42 \} } \, > \,  =  \, -116.78 \,\, {\rm MeV} \, .
\label{c1}
\end{eqnarray}
Again, the colour magnetic interaction 
reduces the energy difference between the
$\{ 42 \}$ and $\{ 6 \}$ configurations 
$\Delta_{{\rm OGE}} \, = \, - 140.99 \,\, {\rm MeV}$,
but the reduction of the energy gap is much smaller
than the harmonic oscillator energy difference
$\Delta_{\rm ho} \, = \, 309.10 \,\, {\rm MeV}$ (note that the
value of $b$ in the CCQM is different than in pure OGE models).
To go further let us analyze the OPE and OSE contributions.

For the OPE potential \eref{OPE},
the calculation of the corresponding spin-isospin matrix elements for a $SU(4)$
irreducible representation is done in references \cite{VAL97,HET69}.
The contribution of the OPE to the $\{ 6 \}$ configuration is,
\begin{equation}
\fl < \Psi_{ \{ 6 \} } \mid \sum_{i < j} d_{ij}
({\vec \sigma}_i \cdot {\vec \sigma}_j)
({\vec \tau}_i \cdot {\vec \tau}_j)
\mid \Psi_{ \{ 6 \} } > = 11 \, d_1 =
- {22 \over 3}
{\alpha_{\rm ch} \over {\sqrt{2 \pi} b^3 m_\pi^2}} \, ,
\end{equation}
where $d_1 = d_{ij}^{0s,0s}$ and 
the radial function $d_{ij}$ can be identified from \eref{OPE}.
For the $\{ 42 \}$ component we have,
\begin{eqnarray}
\fl < \, [ 42]_{O} \,\, [ 33 ]_{ST} 
\, \mid \, \sum_{i < j} \, d_{ij} \,
({\vec \sigma}_i \cdot {\vec \sigma}_j)
({\vec \tau}_i \cdot {\vec \tau}_j) \,
\mid \, [ 42 ]_{O} \,\, [ 33 ]_{ST} \, > 
\,  =  \, 11 \,\,  d  \, ,
\nonumber \\
\fl < \, [ 42 ]_{O} \,\, [ 51 ]_{ST} \, \mid 
\, \sum_{i < j} \, d_{ij} \,
({\vec \sigma}_i \cdot {\vec \sigma}_j)
({\vec \tau}_i \cdot {\vec \tau}_j) \,
\mid \, [ 42 ]_{O} \,\, [ 51 ]_{ST} \, > 
\,  =  \, 35 \,\,  d  \, ,
\end{eqnarray}
where 
\begin{equation}
d = {1 \over 15} \left[ 6 \, d_{ij}^{0s,0s}
+ 8 \, d_{ij}^{0s,0p}
+ d_{ij}^{0p,0p} \right] \, . 
\label{b1}
\end{equation}
Then, from equation (\ref{l1})
\begin{equation}
\fl < \, \Psi_{\{ 42 \} } \, \mid \, \sum_{i < j} \, d_{ij} \,
({\vec \sigma}_i \cdot {\vec \sigma}_j)
({\vec \tau}_i \cdot {\vec \tau}_j) \,
\mid \, \Psi_{\{ 42 \} } \, > \, = \, 23 \,\,  d \, = 
\, -335.34 \,\, {\rm MeV}.
\end{equation}
It is clear from the above expressions that differently than in the
OGE case, the OPE contribution has the same sign for both configurations.
Regarding the energy gap between the 
spatial symmetries, one gets $\Delta_{\rm OPE} \equiv
\left[ E_{ \{ 42 \} } - E_{ \{ 6 \} }
\right]_{\rm OPE} \, = - 68.62 \,\, {\rm MeV}$.
For the scalar potential \eref{OSE} we get 
$\Delta_{\rm OSE} \equiv \left[ E_{ \{ 42 \} } - E_{ \{ 6 \} }
\right]_{\rm OSE} \, = 133.11 \,\, {\rm MeV}$,
its effect being just the opposite to the OGE one.

Therefore, in chiral constituent quark models three effects 
conspire against 
the energy degeneracy of the spatial symmetry states.
First, the small value
of $\alpha_s$ which lowers the contribution of the OGE.
Second, the partial cancellation between the OPE contributions to 
the $\{ 42 \}$ and $\{ 6 \}$ configurations,
and third the cancelling effect of the OPE+OSE potential
with respect to the OGE. Putting
all together one obtains 
$\Delta_{\rm OGE} + \Delta_{\rm OPE} + \Delta_{\rm OSE} \, 
= - 76.56 \,\, {\rm MeV}$
which, differently than in the OGE models of references \cite{OKA80a} and 
\cite{FAE82}, is much smaller than the harmonic oscillator energy 
difference, $\Delta_{\rm ho} = 309.10 \,\, {\rm MeV}$.

Although the perturbative separate one-channel calculation
carried out should be only considered as a valuable hint (see
for instance Ref. \cite{CVE83} for configuration mixing effects)
the results obtained make plausible to conclude that
in chiral constituent quark models
there is not enough energy degeneracy
to account for the NN
short-range strong repulsion as a node produced by the
$[42]_{O}$ spatial symmetry. 

Certainly this result depends on the value of the regularization 
cut-off mass $\Lambda$, that as has been explained 
controls the pion/gluon rate. 
The dependence on the cut-off mass $\Lambda$ of the energy gap
$\left[ E_{ \{ 42 \} } - E_{ \{ 6 \} } \right]$ generated by the OPE and
OSE is presented in table \ref{ta3}. One finds a small dependence
of the results on small variations of $\Lambda$ around reasonable values.
It is worth to point out that the strong correlation among
all the parameters does not allow for independent variations
of them \cite{VAL95c}. The 
strong coupling constant $\alpha_s$ has to be recalculated
for each value of $\Lambda$ to reproduce the 
$\Delta-$N mass difference. The new contribution of the OGE 
is also given in table \ref{ta3}. 

\begin{table}[b]
\caption{\label{gap2} Variation of the energy difference in MeV between the
$\{ 42 \}$ and $\{ 6 \}$ configurations. The harmonic
oscillator energy gap is $\Delta_{\rm ho} = 309.10 \,\, {\rm MeV}$.}
\label{ta3}
\begin{indented}
\item[]\begin{tabular}{@{}llllc}
\br
$\Lambda$ (fm$^{-1}$)  &
$\Delta_{\rm OSE}$ &
$\Delta_{\rm OPE}$ &
$\Delta_{\rm OGE}$ &
$\Delta_{\rm OGE}+\Delta_{\rm OPE}+\Delta_{\rm OSE}$  \\
\mr
 3.7  &  123.72  & $-$64.21 &  $-$152.40 & $-$92.89  \\
 4.2  &  133.11  & $-$68.62 &  $-$140.99 & $-$76.50  \\
 4.7  &  140.98  & $-$71.98 &  $-$131.13 & $-$62.13  \\
 5.2  &  147.49  & $-$74.77 &  $-$122.39 & $-$49.67  \\
 5.7  &  152.98  & $-$77.07 &  $-$115.98 & $-$40.07  \\
 6.2  &  157.65  & $-$79.07 &  $-$110.15 & $-$31.57  \\
\br
\end{tabular}
\end{indented}
\end{table}

Since explicit NN calculations have shown that chiral constituent 
quark models have enough short-range repulsion to reproduce 
the experimental data \cite{FER93a,BRA85,YUZ95} the question that
immediately arises is where does the short-range repulsion come
from?. To look for the origin of
this repulsive character of the interaction one should go beyond 
the energy difference and calculate the
specific contribution of the interaction for each symmetry.
In order to see the repulsive or attractive character of each 
term of the potential in both spatial configurations one has to 
subtract twice (one for each nucleon) the corresponding 
nucleon self-energy, given by
\begin{equation}
\eqalign{
{\cal E}_{\rm OGE} \, & =  \, 
\left\langle N | V_{\rm OGE} | N \right\rangle
= \, -8 c_1 \, = \, 
-72.63 \,\, {\rm MeV}\, ,  \\
{\cal E}_{\rm OPE} \, & =  \, 
\left\langle N | V_{\rm OPE} | N \right\rangle
= \, 15 d_1 \, = \,
-311.40 \,\, {\rm MeV} \, , \\ 
{\cal E}_{\rm OSE} \, & =  \,
\left\langle N | V_{\rm OSE} | N \right\rangle
= \, 3 e_1 \, = \,
-66.90 \,\, {\rm MeV} \, ,} 
\end{equation}
where $e_1=e_{ij}^{0s,0s}$, and $e_{ij}$ is easily identified
from Eq. (\ref{OSE}). One then obtains
for the $\{ 6 \}$ configuration,
\begin{equation}
\fl {\rm E}_{ \{ 6 \} }^{{\rm OGE} + {\rm OPE} + {\rm OSE}} \, = \,
( 169.47  \, + \, 486.95 \, - \, 200.70 ) \,\, {\rm MeV} \, = \, 
455.72 \,\, {\rm MeV} \, , 
\label{e1}
\end{equation}
and for the $\{ 42 \}$,
\begin{equation}
\fl {\rm E}_{ \{ 42 \} }^{{\rm OGE} + {\rm OPE} + {\rm OSE}} \, = \,
( 28.48  \, + \, 418.33 \, - \, 67.65 ) \,\, {\rm MeV} \, = \, 
379.16 \,\, {\rm MeV} \, . 
\end{equation}
\begin{figure}[t]
\vspace*{-0.3cm}
\centerline{\psfig{figure=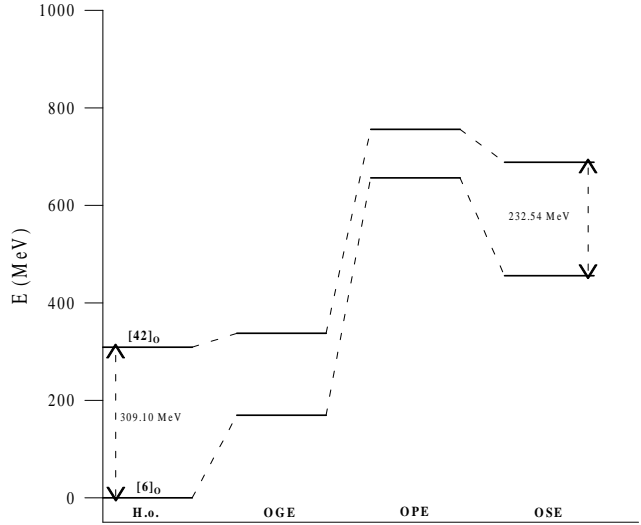,height=3.2in,width=4.8in}}
\vspace*{-0.6cm}
\caption{Contribution of the different pieces of the interaction to the
energies of the $[42]_{O}$ and $[6]_{O}$ spatial symmetries. Energies are 
given at R=0.}
\label{figm32}
\end{figure}
The chiral potential produces strong repulsion in both cases 
due mainly to the OPE.
In figure \ref{figm32} it can be seen the effect of the
different terms of the interaction.
The OGE, repulsive in both configurations,
reduces the energy difference whereas the 
OSE, attractive in both configurations, increases it.
The OPE produces a strong repulsion in both symmetries. 
The net effect is an energy difference of about the same value obtained
in the harmonic oscillator but with an additional repulsion in both
symmetries originated mainly by quark antisymmetry on the OPE.
Therefore, in this type of models the NN $S$-wave hard-core like behaviour
should be mainly
attributed to the strong repulsion in the 
$[6]_{O}$ spatial configuration
(more precisely there is a compromise between repulsion and
kinetic energy difference  in both configurations giving
rise to some configuration mixing).

These simple images of the origin of the NN short-range repulsion
are confirmed by means of explicit RGM or Born-Oppenheimer calculations
of the NN relative motion based on OGE \cite{NEU77,OKA80b,TOK80,FAE82,CVE83}
or OGE plus Goldstone boson exchanges \cite{OBU90,FER93a,YUZ95,FUJ96}. 
Such calculations have been done considering explicitly the NN and $\Delta \Delta$
components (explicit hidden color-hidden color states were 
considered in Refs. \cite{FAE82,CVE83}) and the mixing induced
between them by the different interactions.

%% file: ch3rev.tex
\section{The baryon-baryon potential}
\label{ch4}

It has become clear in the last years the major role played by baryonic
resonances, in particular the low-lying nucleonic resonances $\Delta $(1232)
and N$^*$(1440), in many electromagnetic and strong
reactions that take place in nucleons and nuclei. This justifies the current
experimental effort to study nucleon 
resonances in several facilities: TJNAF with a
specific experimental program of electroexcitation of resonances, WASA in
Uppsala to study ${\rm NN}\rightarrow {\rm NN}\pi \pi $ reactions, etc.
In this context the knowledge of the interaction involving resonances
derived in a consistent way is of great relevance.
The interaction between the nucleon and a resonance has been usually
written as a straightforward extension of some pieces of the NN potential
modifying the coupling constants extracted from their
decay widths. Though this procedure can be appropriate for the very
long-range part of the interaction, it is under suspicion at least for the
short-range part for which the detailed structure of baryons may
determine to some extent the form of the interaction. 
It seems therefore convenient to proceed to a derivation, besides the NN
potential which will serve to fix the quark potential parameters,
of the $\rm{NN}\rightarrow \rm{NR}$ (R : resonance),
$\rm{NR}\rightarrow \rm{NR}$, and $\rm{RR}\rightarrow \rm{RR}$ interactions 
based on the more elementary quark-quark interaction. 
The main comparative advantage of the quark treatment comes out from the fact
that as all the basic interactions are at the quark level, the parameters
of each vertex (coupling constants, cut-off masses,...) are independent 
of the baryon to which the quarks belong, 
what makes its generalization to any other non-strange
baryonic system straightforward. The other way around, the comparison of its
predictions to the experimental data available serves
as a stringent test of the quark potential model.

The derivation of the dynamics of a two-baryon 
system from the dynamics of its constituents is a 
tough six-body problem whose solution cannot be
exactly obtained even for the non-relativistic case. 
This forces the use of approximate calculation methods. 
In the literature 
two methods have been mainly used to get baryonic interactions
from the dynamics of the constituents: the resonating group method (RGM)
and the Born-Oppenheimer (BO) approximation. 
We resume their most relevant aspects. 

\subsection{Calculation methods}
\label{ch4.1}
\subsubsection{Resonating group method potential}
\label{ch4.1.1}

The RGM \cite{WHE37}, widely used in 
nuclear physics to study the nucleus-nucleus 
interaction, can be straightforwardly 
applied to study the baryon-baryon interaction
in the quark model. It allows, once the Hilbert space for 
the six-body problem has been fixed, to treat the 
inter-cluster dynamics in an exact way.

The formulation of the RGM for a system of 
two baryons, $B_1$ and $B_2$, starts from the wave function
of the six-quark system expressed in terms of the Jacobi coordinates
of the baryons. Then,
the spatial part factorizes in a product of two 
three-quark cluster wave functions and the relative motion wave function
of the two clusters so that, 
\begin{equation}
\fl \Psi_{B_1 B_2}=
{\cal A}\; [\varphi(\vec{P})\; \phi_{B_1}(\vec{p}_{\xi_{B_1}}) \; 
\phi_{B_2}(\vec{p}_{\xi_{B_2}}) \;
\chi_{B_1 B_2}^{ST} \xi_{C} [2^3] \,] ,
\end{equation}
where ${\cal A}$ is the six-quark antisymmetrizer, $\varphi(\vec{P})$ is 
the relative motion wave function of the two clusters, 
$\phi_{B_i}(\vec{p}_{\xi_{B_i}})$ is the internal momentum wave function 
of baryon $B_i$, and $\xi_{B_i}$ are the Jacobi 
coordinates of the baryon $B_i$. 
$\chi_{B_1 B_2}^{ST}$ denotes the spin-isospin 
wave function of the two-baryon system coupled to 
spin $S$ and isospin $T$, and, finally, $\xi_{C} [2^3]$ 
is the product of two colour singlets. 

The dynamics of the system is governed by the Schr\"odinger equation:
\begin{equation}
({\cal H} -E_T)| \Psi>=0 \ \ \ \Rightarrow \ \ \ <\delta \Psi | ({\cal H} -E_T)
| \Psi>=0, 
\label{variations}
\end{equation}
where 
\begin{equation}
{\cal H} = \sum_{i=1}^6 {\vec{p_i}^2 \over 2 m_q} +\sum_{i<j=1}^6 V_{qq}(\vec r_{ij})
-T_{\rm CM} \, ,
\end{equation}
$T_{\rm CM}$ being the center of mass kinetic 
energy, $V_{qq}$ the quark-quark interaction,
$\vec{p_i}$ the trimomentum of quark $i$,
and $m_q$ the constituent quark mass. 
In equation \eref{variations} the variations are performed 
on the unknown relative motion wave function $\varphi(\vec{P})$.
Assuming harmonic oscillator wave functions for $\phi_{B_i}$
equation (\ref{variations}), 
after the integration of the internal  
degrees of freedom of both clusters,
can be written in the following way \cite{TAN77},
\begin{equation}
\left ( {\vec{P}^2 \over 2\mu} -E \right ) \varphi(\vec{P}) +
\int \left( V_D (\vec{P}, \vec{P}_i) +
W_{\rm EX} (\vec{P}, \vec{P}_i) \right)
\varphi(\vec{P}) d\vec{P}_i =0  \, ,
\label{ALS}
\end{equation}
where $E=E_T-E_{B_1}-E_{B_2}=E_T-E_{in}$, $E_{in}$ being the internal
energy of the two-body system and $V_D (\vec{P}, \vec{P}_i)$ 
and $W_{\rm EX} (\vec{P}, \vec{P}_i)$ are the direct and exchange RGM 
kernels, respectively. $V_D$ contains the effect of the interaction
between baryonic clusters whereas $W_{\rm EX}$ gives account 
of the quark exchanges between clusters coming from the identity
of quarks. They are evaluated in detail in reference \cite{KAM78}. 

Note that if we do not mind how 
$V_D$ and $W_{\rm EX}$ 
were derived microscopically, equation (\ref{ALS}) 
can be regarded as a general single channel 
equation of motion including an energy-dependent non-local 
potential given by the sum of $V_D$ and $W_{\rm EX}$.
This is the RGM baryon-baryon potential. 

\subsubsection{Born-Oppenheimer potential}
\label{ch4.1.2}

The BO method, also known as adiabatic approximation,
has been frequently employed for the study
of the nuclear force from the microscopic degrees of freedom \cite{LIB77,TAR78}.
It is based on the assumption that quarks move inside the clusters
much faster than the clusters themselves. Then one can
integrate out the 
fast degrees of freedom assuming a fixed position for 
the center of each cluster obtaining in this way a 
local potential depending on the distance 
between the centers of mass of the clusters. 
The potential is defined in the following way \cite{OKA84},
\begin{equation}
V_{B_1 B_2 (L \, S \, T) \rightarrow B_3 B_4 
(L^{\prime}\, S^{\prime}\, T)} (R)
= {\cal V}_{L \,S \, T}^{L^{\prime}\, S^{\prime}\, T} 
(R) \, - \, {\cal V}_{L \,S \,
T}^{L^{\prime}\, S^{\prime}\, T} (\infty) \, ,  
\label{PotBO}
\end{equation}
where
\begin{equation}
\fl {\cal V}_{L \, S \, T}^{L^{\prime}\, S^{\prime}\, T} (R) \, 
= \,\frac{{\left \langle \Psi_{B_1 B_2}^{L^{\prime}\, S^{\prime}\, T} 
(\vec{R}) \mid
\sum_{i<j=1}^{6} V_{qq}({\vec{r}_{ij}})  \mid 
\Psi_{B_3 B_4}^{L \, S \, T} (\vec{R}) \right \rangle }}
{\sqrt{\left \langle \Psi_{B_1 B_2}^{L^{\prime}\,S^{\prime}\, T} (\vec{R}) \mid 
\Psi_{B_1 B_2}^{L^{\prime}\, S^{\prime}\, T}(\vec{R}) 
\right \rangle} 
\sqrt{\left \langle \Psi_{B_3 B_4}^{L \, S \, T} (\vec{R}) \mid 
\Psi_{B_3 B_4}^{L \, S \, T} (\vec{R}) 
\right \rangle}} \, ,
\label{BODEF}
\end{equation}
with $\Psi_{B_i B_j}^{L \, S \, T} (\vec{R})$ given by equation \eref{Gor}.
The subtraction of ${\cal V}_{L \,S \,
T}^{L^{\prime}\, S^{\prime}\, T} (\infty)$ assures that no internal
cluster energies enter in the baryon-baryon interacting potential.

\subsection{Results}
\label{ch4.2}

Both methods permit to evaluate the influence of the Pauli 
principle at the quark level on properties of the baryon-baryon interaction. 
The main conceptual difference between the resulting interactions
is that the BO potential is local while the RGM one is non-local,
\begin{equation}
\eqalign{
V^{\rm RGM}_{B_1B_2\to B_3B_4} & \equiv V^{\rm RGM}_{B_1B_2\to B_3B_4}(R,R')\nonumber\\
V^{\rm BO}_{B_1B_2\to B_3B_4} & \equiv V^{\rm BO}_{B_1B_2\to B_3B_4}(R) \, .}
\end{equation}
This means that the $T$ matrix calculated solving 
a Lippmann-Schwinger equation has a different off-shell behaviour
\footnote{The on-shell behaviour is very similar, 
in fact one can almost achieve on-shell equivalence by fine 
tuning the quark model parameters \protect\cite{JUL02}.}
and thus will give different results,
the larger the difference the more the particles of the system
under consideration explore the off-shell region. 

\begin{figure}[t]
\begin{center}
\mbox{\epsfxsize=120mm\epsfysize=70mm\epsffile{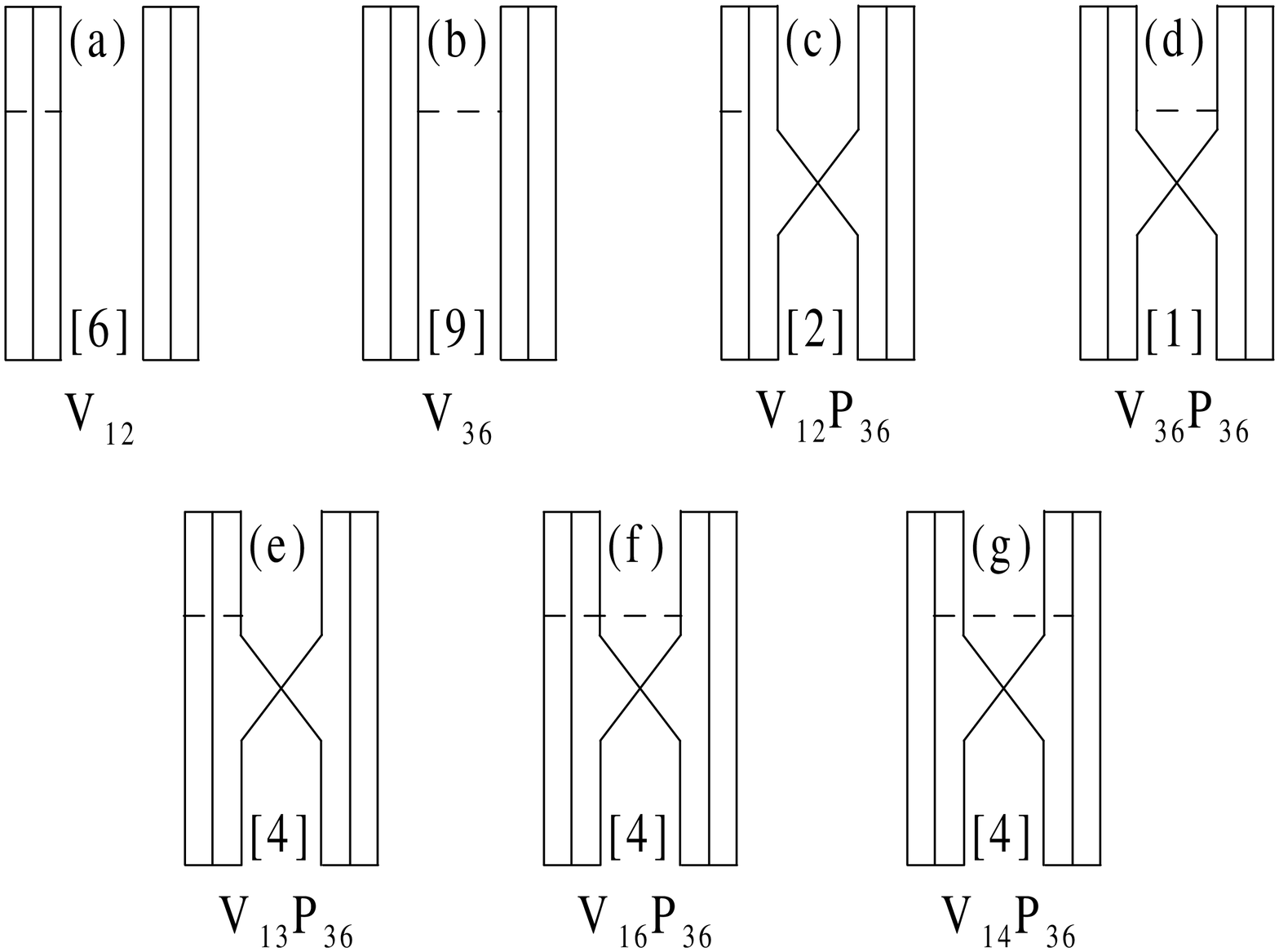}}
\end{center}
\vspace*{-0.5cm}
\caption{Different diagrams contributing to the NN interaction.} 
\label{fig0}
\end{figure}

In both cases the calculation of a baryonic potential from the quark
dynamics involves, due to the antisymmetry operator, the calculation
of many different diagrams that are depicted in figure \ref{fig0}
for the case of the NN interaction.
The direct terms are represented by diagrams (a) and (b), whereas
quark-exchange terms correspond to diagrams (c)$-$(g).
The number in square brackets corresponds to the number of equivalent
diagrams that can be constructed
(it counts the number of quark pairs which are equivalent to the
pair $ij$ singled out in the figure).
Diagram (a) cancels almost exactly with the self-energy
term ${\cal V}_{L \, S \, T}^{L' \, S' \, T}(\infty)$,
it gives a small contribution at short-range.
Diagram (b) generates the
asymptotic behaviour of the interaction. The relevance of
the quark-exchange terms, diagrams (c)$-$(g), 
depends on the overlap of the baryon wave functions.
They are responsible for the short-range structure of the 
quark-model-based potential and they vanish when the two baryons do not overlap.
Next we present results for the NN, N$\Delta$, $\Delta \Delta$, and
NN$^*$(1440) systems.

\subsubsection{The NN interaction}
\label{ch4.2.1}
\begin{figure}[t]
\mbox{\psfig{figure=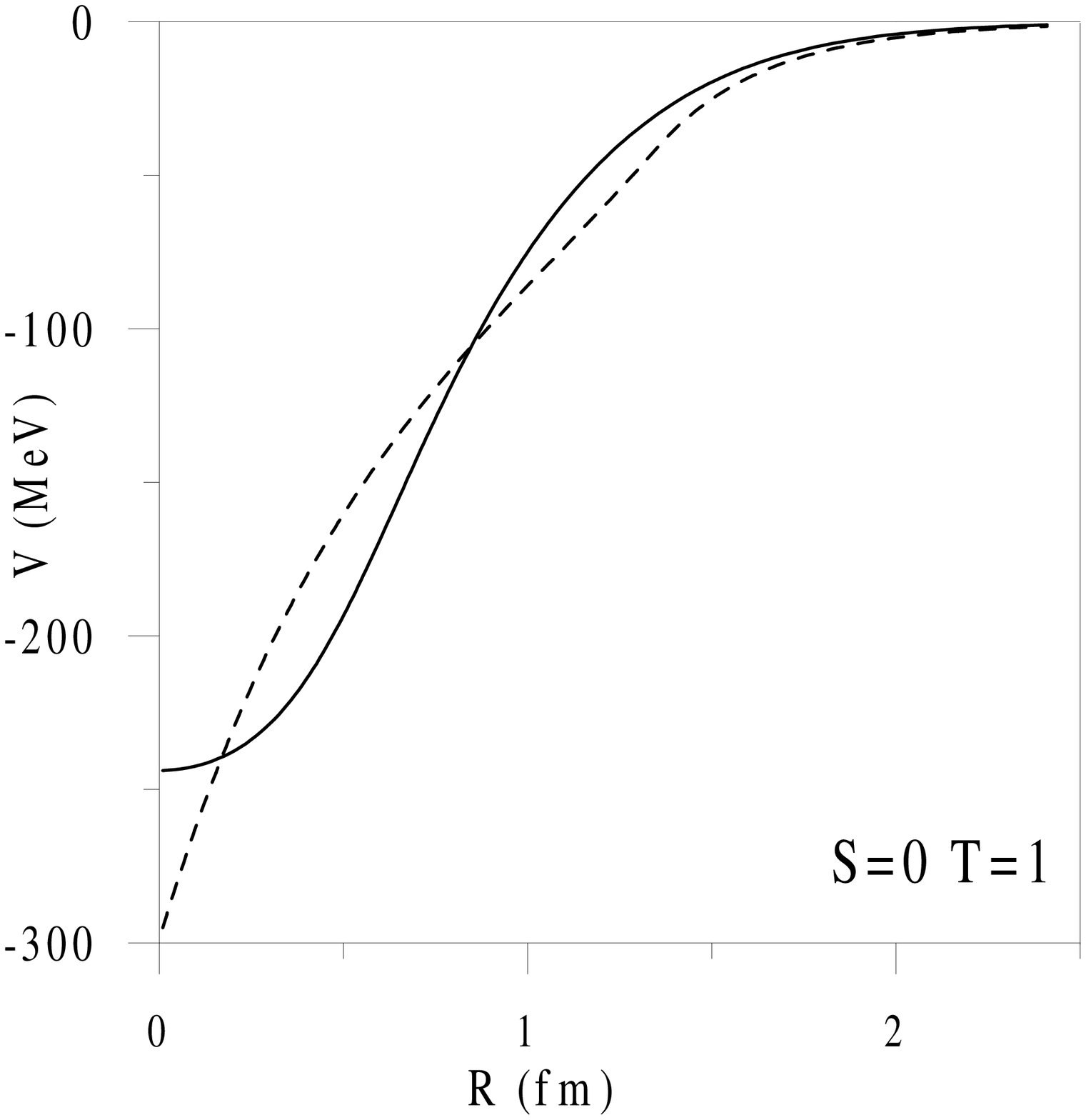,height=2.8in,width=2.2in}}
\hspace*{0.6cm}
\mbox{\psfig{figure=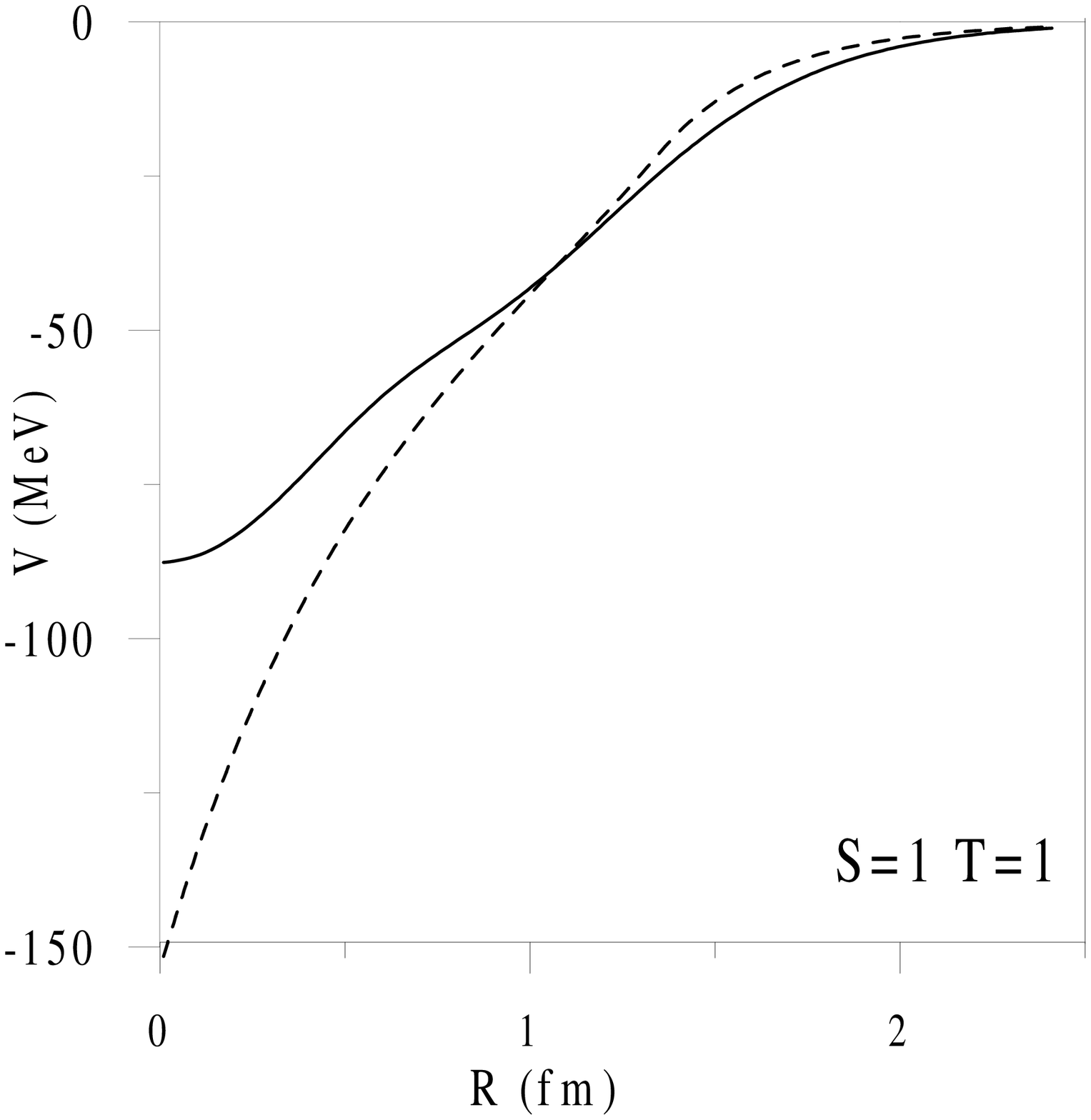,height=2.8in,width=2.2in}}
\vspace*{-2.8cm}
\caption{NN OSE potential for two different spin-isospin
channels. The dashed line represents the parametrization 
used in reference \cite{BRA85}.
The solid line represents the chiral 
constituent quark model result \cite{FER93a}.}
\label{fig1}
\end{figure}

Given the huge amount of experimental data available on the 
NN system, a detailed calculation of the scattering phase shifts 
and bound state properties will be
presented in \sref{ch5}. We use this section to 
discuss qualitative important aspects of the NN interaction 
that are naturally described in chiral constituent quark models.
Let us first mention that the identity of quarks gives rise to the
well-known selection rule L+S+T=odd (see \sref{ch3.1})

By construction the chiral quark pion potential reproduces the
NN long-range interaction (\sref{ch2.2}).
It has been already discussed in \sref{ch3.2}
how the quark substructure of the nucleon allows
to understand the short-range behaviour of the $S$-wave NN interaction.
Concerning the medium-range attraction,
figure \ref{fig1} shows the scalar potential obtained 
in the chiral constituent quark model \cite{FER93a}
as compared to a parametrization at baryonic level used in 
reference \cite{BRA85} to fit the NN experimental data. 
While at baryonic level different coupling constants are used:
$g^2_{\sigma \rm{NN}}/4\pi=$ 3.7 for $(S,T)=(0,1)$ and
$g^2_{\sigma \rm{NN}}/4\pi=$ 1.9 for $(S,T)=(1,1)$,
the quark model result is obtained in both cases from the
same chiral quark coupling constant.
Moreover, the deuteron binding energy
is also reproduced \cite{VAL94a}, while 
the scalar coupling constant used at baryonic level 
in this case is once more a different one,
$g^2_{\sigma \rm{NN}}/4\pi=$ 2.55 \cite{BUC89}.

\begin{figure}[t]
\mbox{\psfig{figure=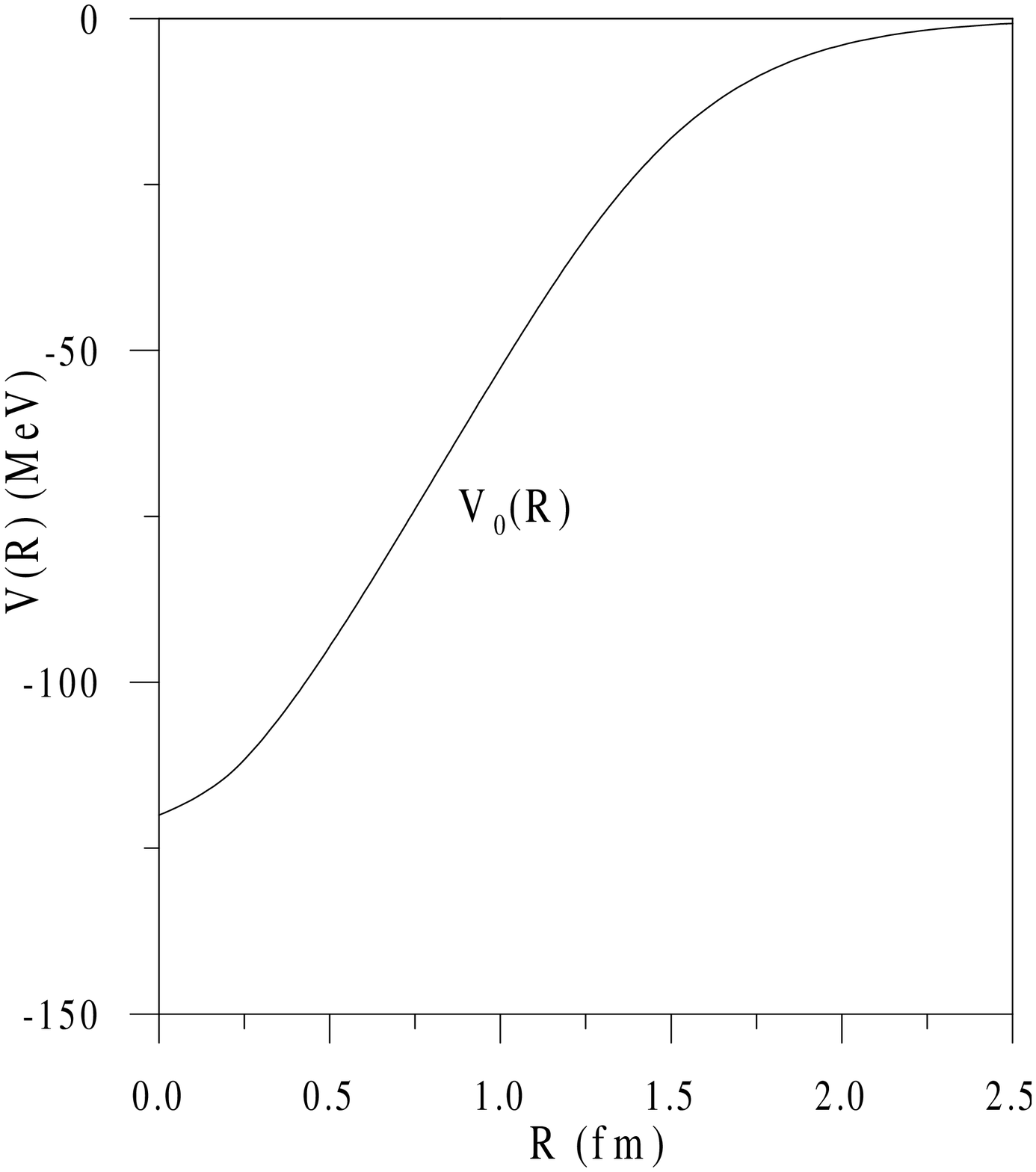,height=2.6in,width=2.2in}}
\hspace*{0.6cm}
\mbox{\psfig{figure=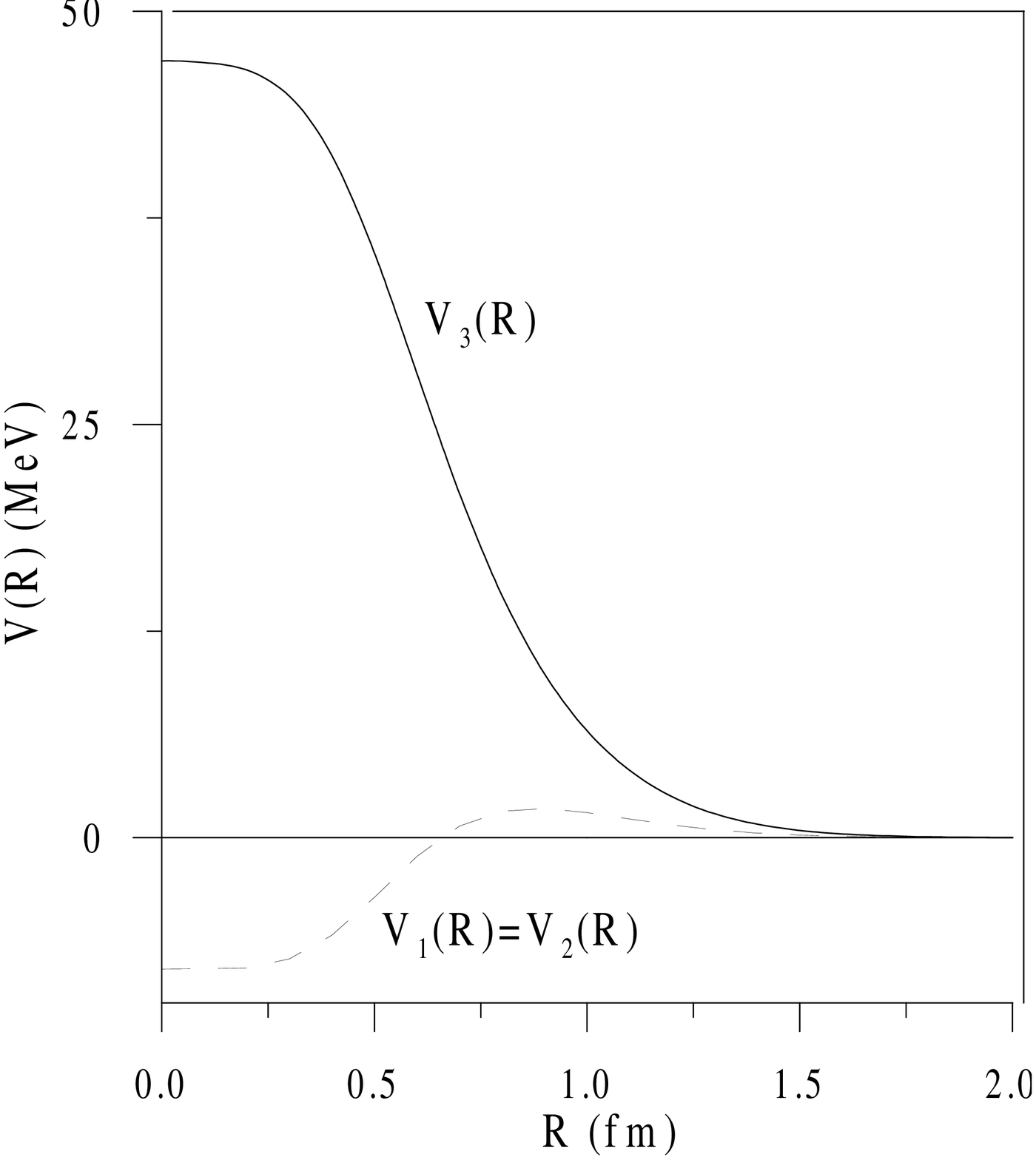,height=2.6in,width=2.2in}}
\vspace*{-1.2cm}
\caption{Spin-isospin independent part ($V_0$) and spin-isospin
dependent parts ($V_1$, $V_2$, and $V_3$) of the OSE potential.}
\label{fig1b}
\end{figure}

Another important feature of chiral constituent quark models is that
although asymptotically the spin-isospin structure of the different
terms of the baryon-baryon potential is the same as the corresponding
terms of the quark-quark one, at short distances
quark exchange generates a rather involved spin-isospin structure. This is
due to the antisymmetrization operator, that 
can be factorized as ${\cal A} \equiv (1- 9 P_{36})(1-{\cal P})$. 
When the quark spin-isospin
operators are transformed algebraically to find the corresponding
baryonic operators different structures are generated \cite{HOL84}. 
Taking into account that
$P^{ST}_{36}={1 \over 4} (1 + \vec{\sigma}_3 \cdot \vec{\sigma}_6) 
(1 + \vec{\tau}_3 \cdot \vec{\tau}_6)$,  
the effective NN interaction can be decomposed as follows, 
\begin{equation}
\fl V_{\rm NN}^{(S,T)} = V_0(R) + 
V_1(R)\, {\vec \sigma}_{B_1} \cdot {\vec \sigma}_{B_2} +
V_2(R)\, {\vec \tau}_{B_1} \cdot {\vec \tau}_{B_2} +
V_3(R)\, {\vec \sigma}_{B_1} \cdot {\vec \sigma}_{B_2} \,
{\vec \tau}_{B_1} \cdot {\vec \tau}_{B_2} \, ,
\label{ttgg}
\end{equation}
where $V_i(R)$ are functions of the interbaryon distance $R$, and
$\vec{\sigma}_{B_i}$ ($\vec{\tau}_{B_i}$)
are the spin (isospin) baryonic operators. 
The $V_i(R)$ functions can be obtained from the
calculated potential for the different $(S,T)$ channels,
\begin{equation}
\fl \eqalign{
V_0(R) &= {1\over 16} \left [V_{\rm NN}^{(0,0)}(R)+3V_{\rm NN}^{(0,1)}(R)
+3V_{\rm NN}^{(1,0)}(R)+9V_{\rm NN}^{(1,1)}(R) \right ] \nonumber \\
V_1(R) &= {1\over 16} \left [-V_{\rm NN}^{(0,0)}(R)+V_{\rm NN}^{(0,1)}(R)
-3V_{\rm NN}^{(1,0)}(R)+3V_{\rm NN}^{(1,1)}(R) \right ] \nonumber \\
V_2(R) &= {1\over 16} \left [-V_{\rm NN}^{(0,0)}(R)-3V_{\rm NN}^{(0,1)}(R)
+V_{\rm NN}^{(1,0)}(R)+3V_{\rm NN}^{(1,1)}(R) \right ] \nonumber \\
V_3(R) &= {1\over 16} \left [V_{\rm NN}^{(0,0)}(R)-V_{\rm NN}^{(0,1)}(R)
-V_{\rm NN}^{(1,0)}(R)+V_{\rm NN}^{(1,1)}(R) \right ]} \, .
\end{equation}
The decomposition \eref{ttgg} comes out from any spin-isospin structure 
of the quark-quark potential. In figure \ref{fig1b} the spin-isospin
independent, $V_0$, and the spin-isospin dependent,
$V_1$, $V_2$ and $V_3$, terms
generated by the OSE potential are shown. The main component of the interaction
is, as expected, scalar and attractive, 
however a small spin-isospin dependence appears.
This dependence is a completely new feature with 
respect to the usual scalar exchange
at baryon level and may play a significant role in the understanding
of different reactions such as the $p(\alpha,\alpha')$
or $p(d,d')$ \cite{JUL03}.

The same decomposition can be applied to the OPE. In this case, the 
relative strength of the spin-isospin
dependent and the tensor terms at the baryonic level
is largely reduced with respect to the 
quark level case, equation \eref{PS}. 
This allows for a simultaneous explanation of the $p(p,\Delta^{++})n$ and
$p(n,p)n$ reactions \cite{FER94} what is difficult to get when using 
baryonic meson-exchange potentials \cite{JAI93}.

\subsubsection{The N$\Delta$ interaction}
\label{ch4.2.2}

The inclusion of N$\Delta$ intermediate states
has been considered as a possible improvement
of NN interaction models at intermediate energies
for a long time \cite{SAU86,LOM82,PEN92}. 
The N$\Delta$ interaction has been
usually described by means of baryonic meson-exchange models \cite{PEN90}
or parametrized by phenomenological potentials \cite{ALE90,FERR89,TAKA88}
with coupling constants and cut-off masses not well determined due to
the lack of sufficient experimental
information about the N$\Delta$ system.
\begin{figure}[t]
\vspace*{-0.2cm}
\mbox{\psfig{figure=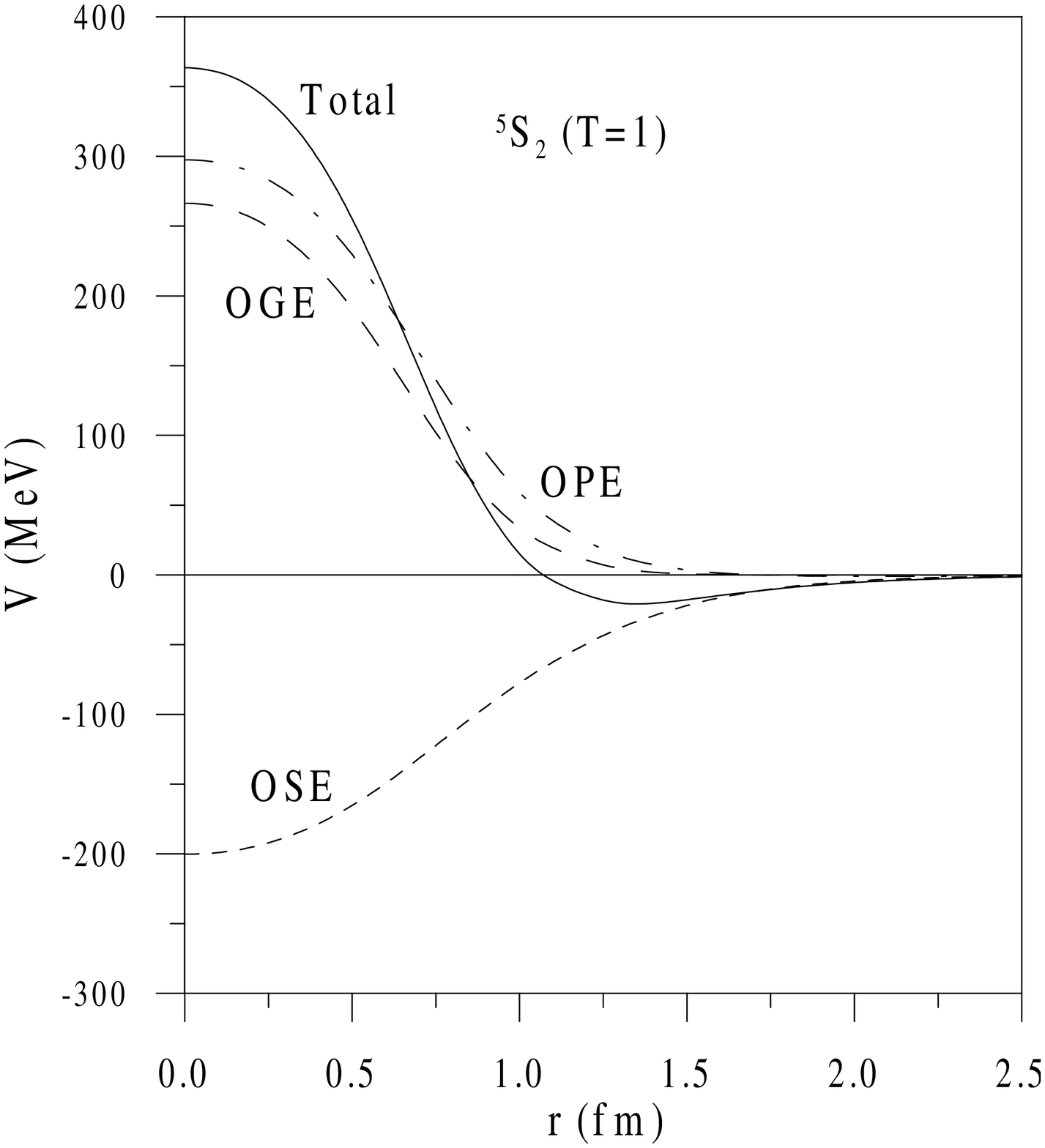,height=2.8in,width=2.4in}}
\hspace*{0.2cm}
\mbox{\psfig{figure=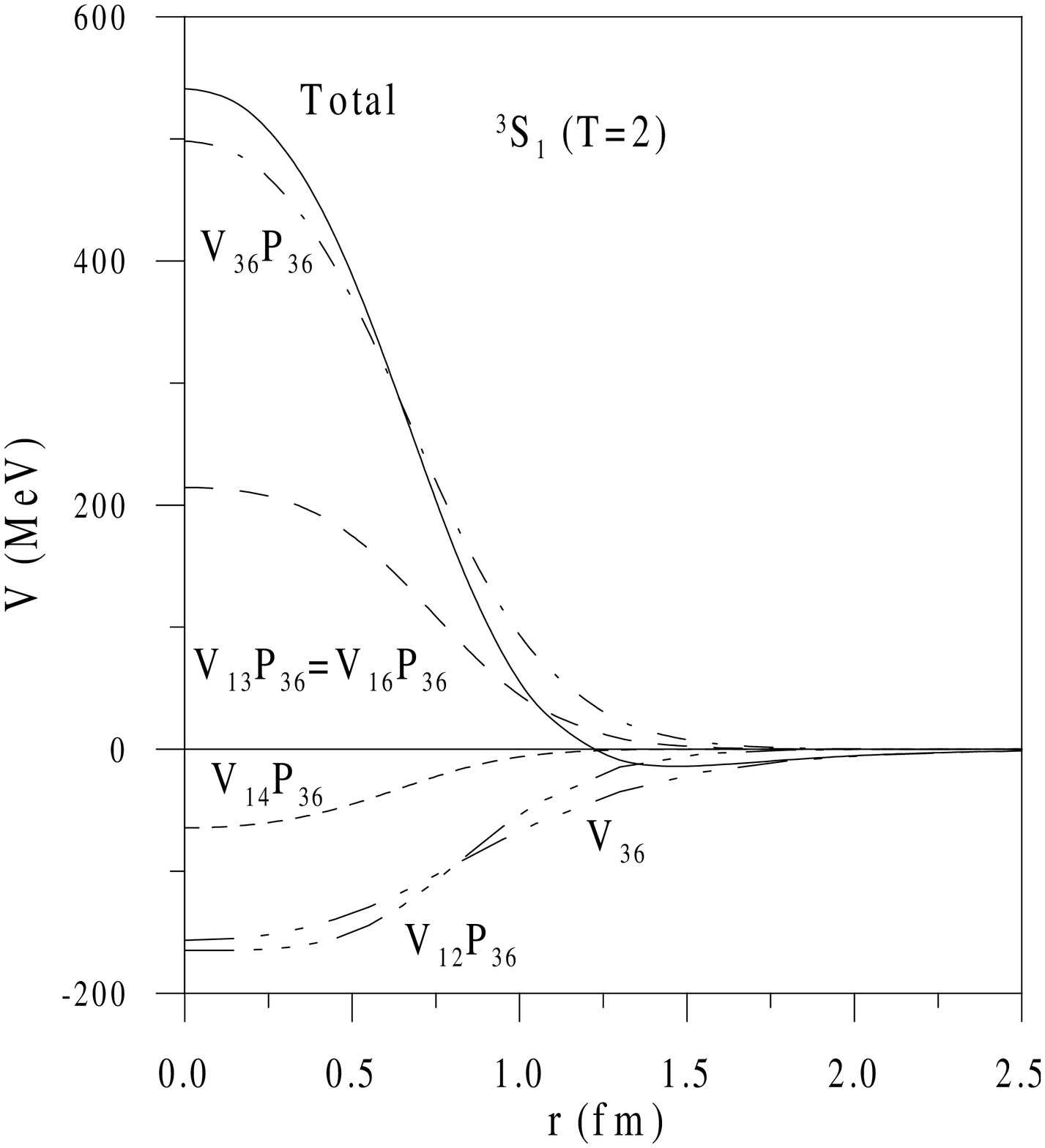,height=2.8in,width=2.4in}}
\vspace*{-1.6cm}
\caption{$^5S_2(T=1)$ and $^3S_1(T=2)$ N$\Delta$ potentials.}
\label{fig2}
\end{figure}
\begin{figure}[b]
\vspace*{-0.2cm}
\mbox{\psfig{figure=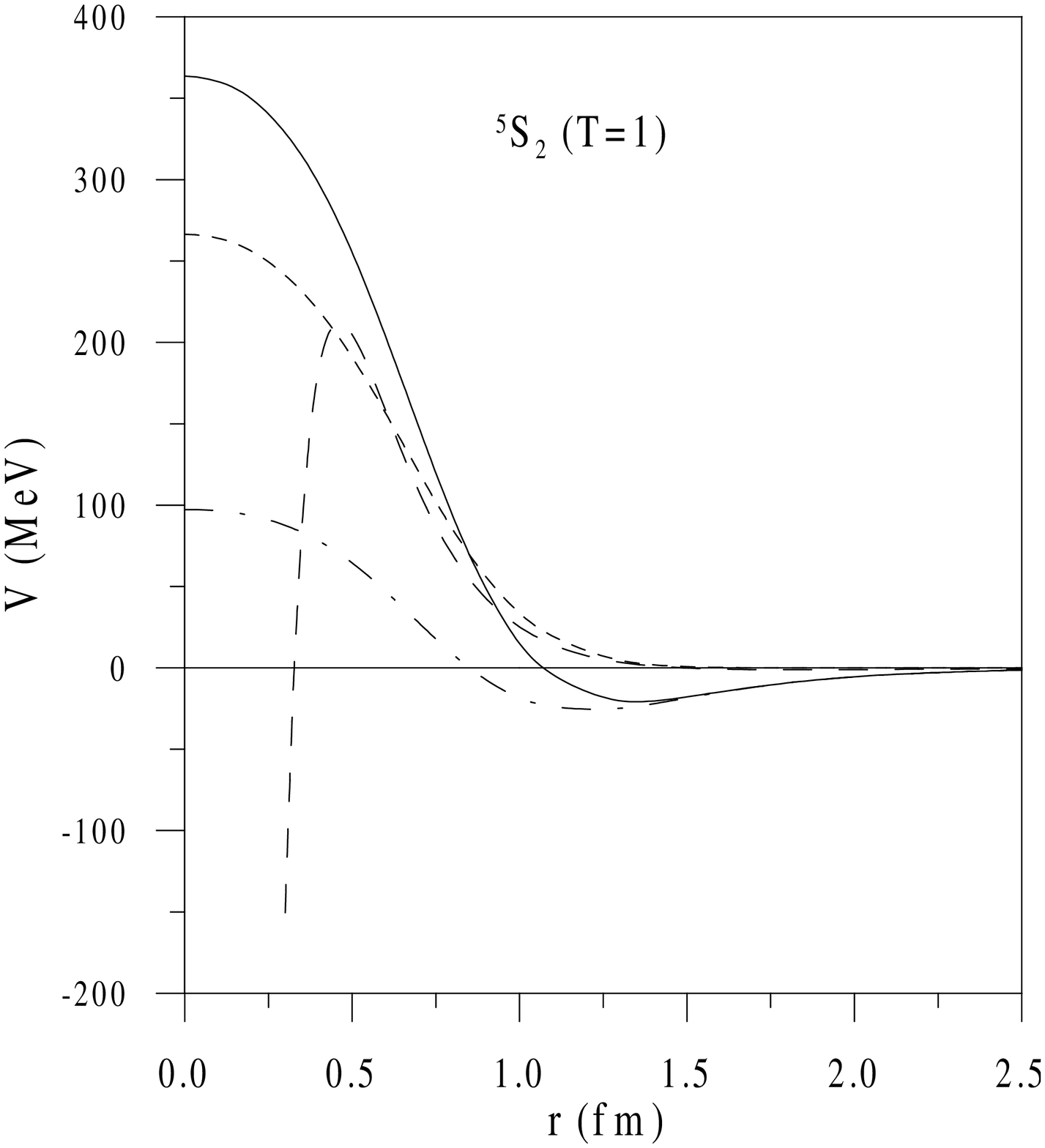,height=2.8in,width=2.4in}}
\hspace*{0.2cm}
\mbox{\psfig{figure=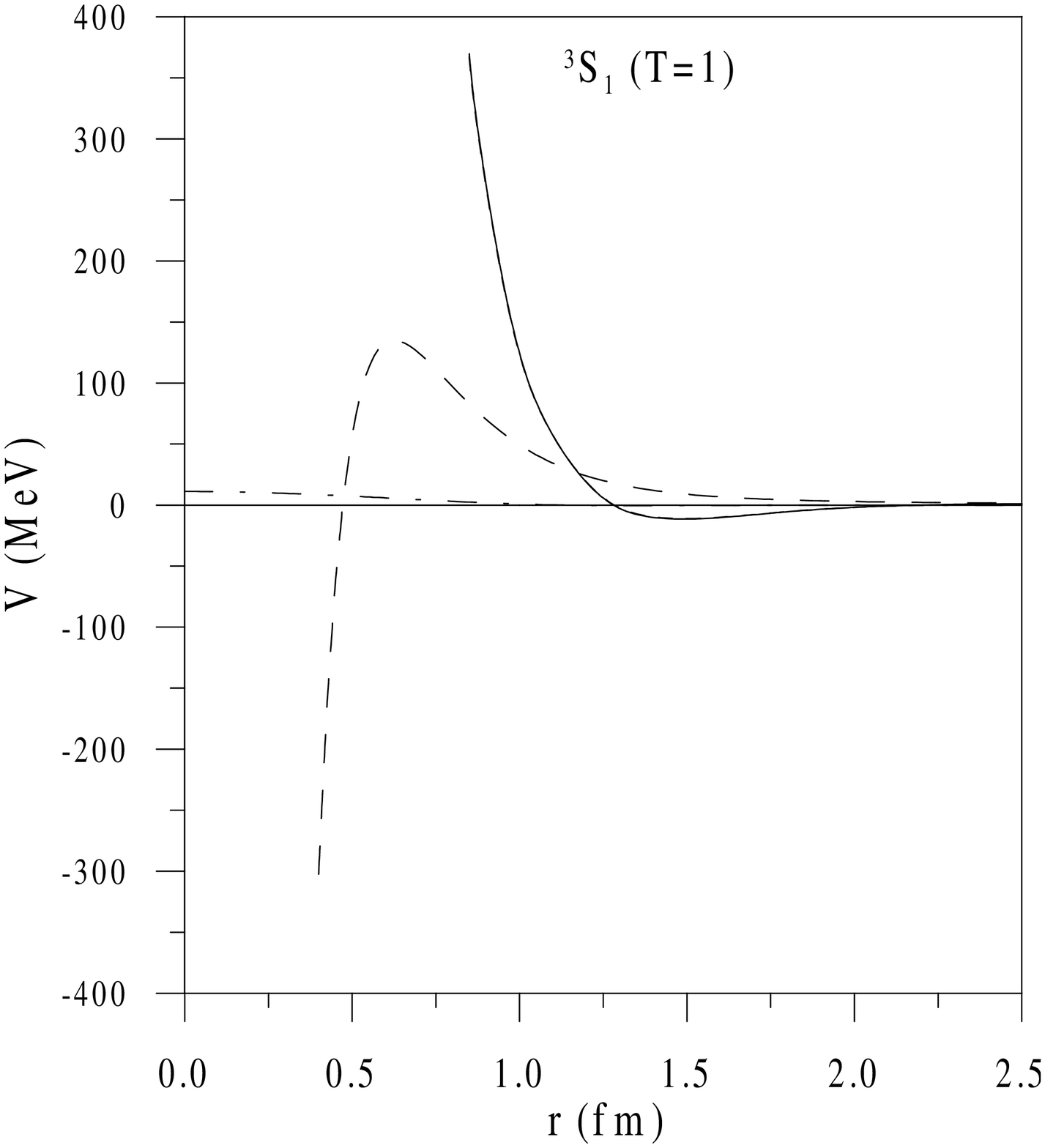,height=2.8in,width=2.4in}}
\vspace*{-1.6cm}
\caption{$^5S_2(T=1)$ and $^3S_1(T=1)$ N$\Delta$ potentials. The solid line
represents the quark-model based result, the dashed-dotted line
the contribution of the OGE and the dotted line that of 
the OPE+OSE (indistinguishable of the total potential
for the $^3S_1(T=1)$ partial wave). The dashed line 
corresponds to the baryonic meson-exchange potential
of reference \protect\cite{PEN90}.}
\label{fig3}
\end{figure}

Since the $\Delta$ is not a stable particle but rather
a $\pi$N resonance, one must establish what it is understood by
the N$\Delta$ interaction. The $\Delta$ is considered as an elementary particle,
being the coupling to the $\pi$N continuum the responsible
for its width \cite{GAR90b}. The vicinity of a nucleon modifies
the properties of the $\Delta$, because they can exchange particles 
between each other, eventually a virtual boson.
This exchange, which looks like the interaction between two
stable particles is what is considered as the N$\Delta$ interaction.
The modification by the coupling to the continuum ($\Delta$ width)
should be implemented when treating any particular problem.

At the quark level, the N$\Delta$ interaction 
has been derived in the chiral constituent
quark model \cite{VAL95b} and used to study the NN system
above the pion threshold \cite{VAL94b}.
Figure \ref{fig2} shows the N$\Delta$ potential calculated 
for two partial waves of different isospin, $^5S_2 \, (T=1)$ and
$^3S_1 \, (T=2)$. In one case the contribution
of the different terms of the potential has been separated. 
As in the other case this separation is qualitatively similar,
the contribution of the 
different diagrams of figure \ref{fig0} is presented. The effect
of quark antisymmetrization can be extracted by comparing
the total potential with the term $V_{36}$ [diagram (b) in 
figure \ref{fig0}],
the only significant one that does not include quark exchanges.
All the exchange diagrams do not appreciably contribute beyond
1.5 fm, where the overlap of the nucleon and $\Delta$ wave functions is negligible. 
Above this distance the interaction is driven by the $V_{36}$ term and it 
equals the total interaction. In general, we see how the
$V_{12} \, P_{36}$ term [diagram (c)] generates additional attraction
and it is the $V_{36} \, P_{36}$ term [diagram (d)] the main responsible
for the short-range repulsion. The behaviour of the other diagrams
depends on the partial wave considered.
\begin{figure}[t]
\vspace*{-0.1cm}
\mbox{\psfig{figure=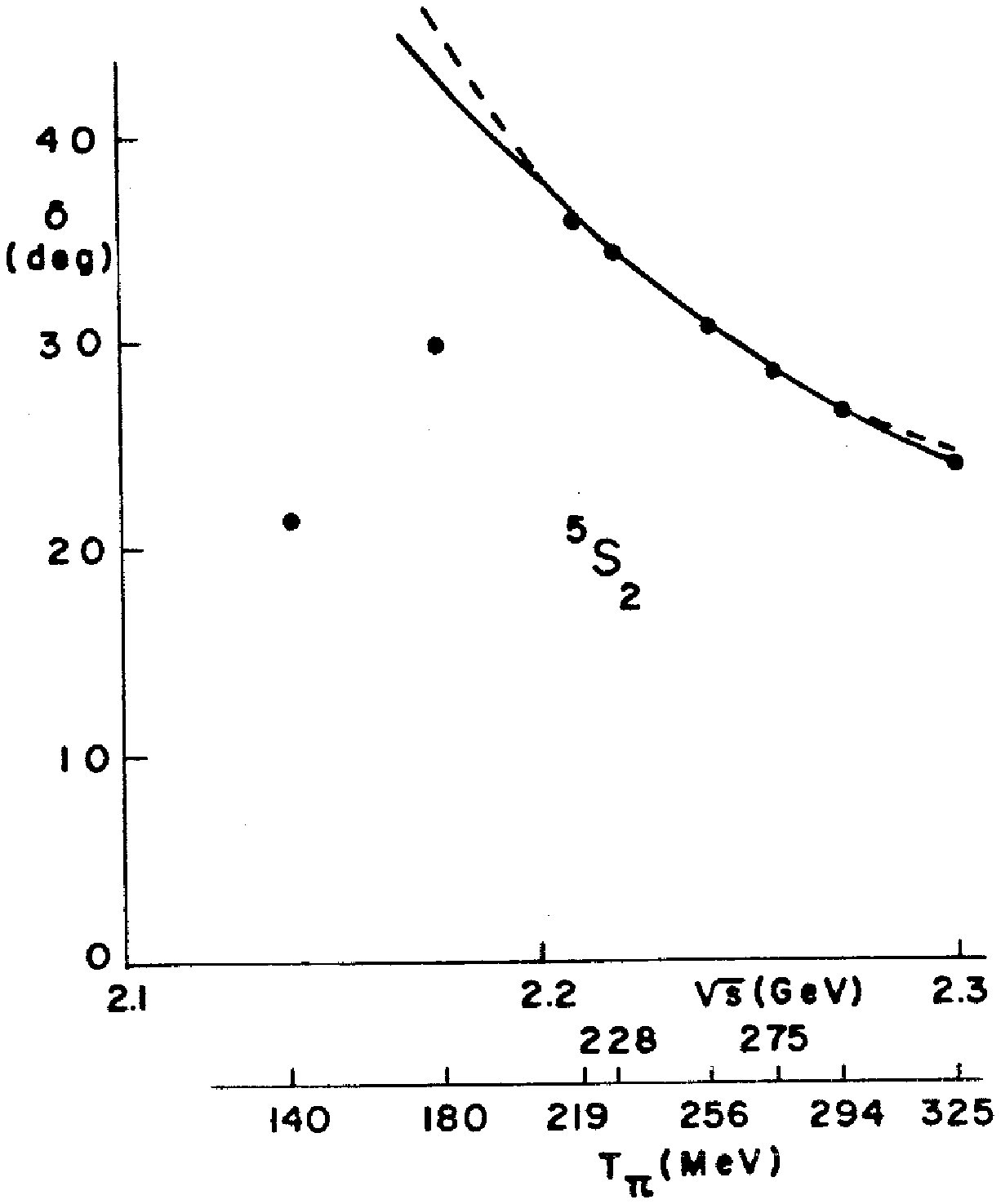,height=2.5in,width=2.in}}
\hspace*{0.1cm}
\mbox{\psfig{figure=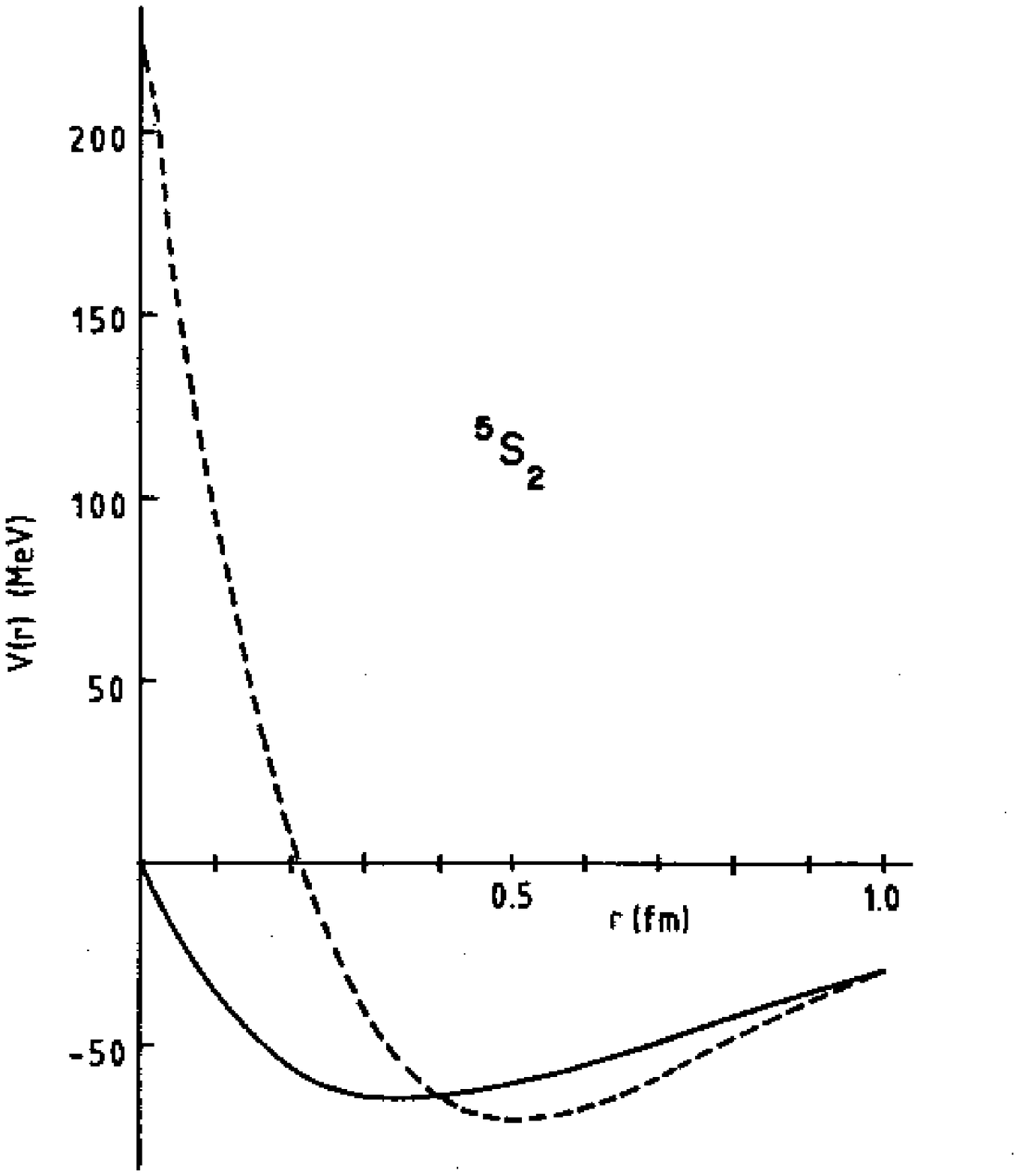,height=2.5in,width=2.in}}
\mbox{\psfig{figure=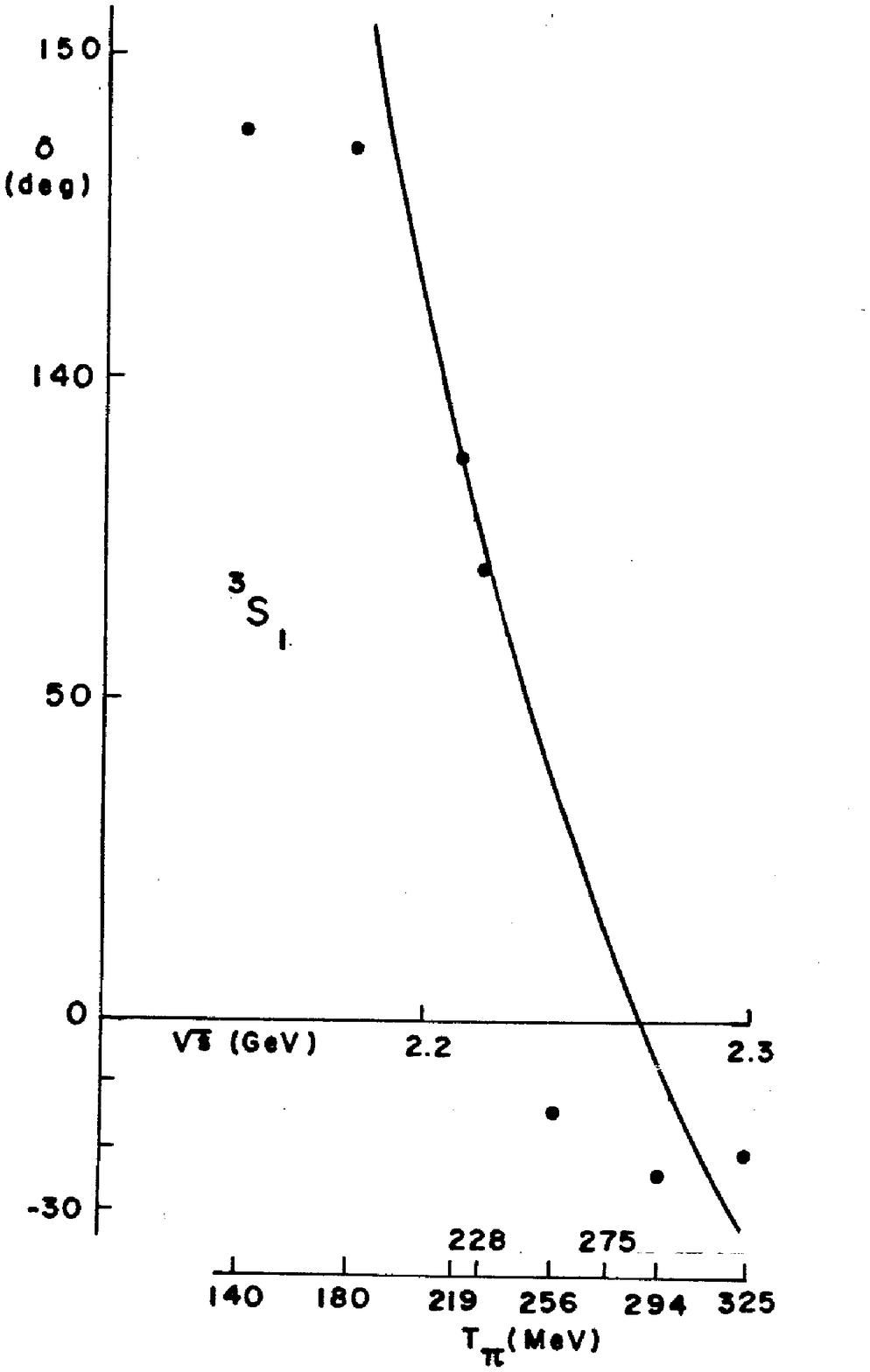,height=2.5in,width=2.in}}
\hspace*{1.5cm}
\mbox{\psfig{figure=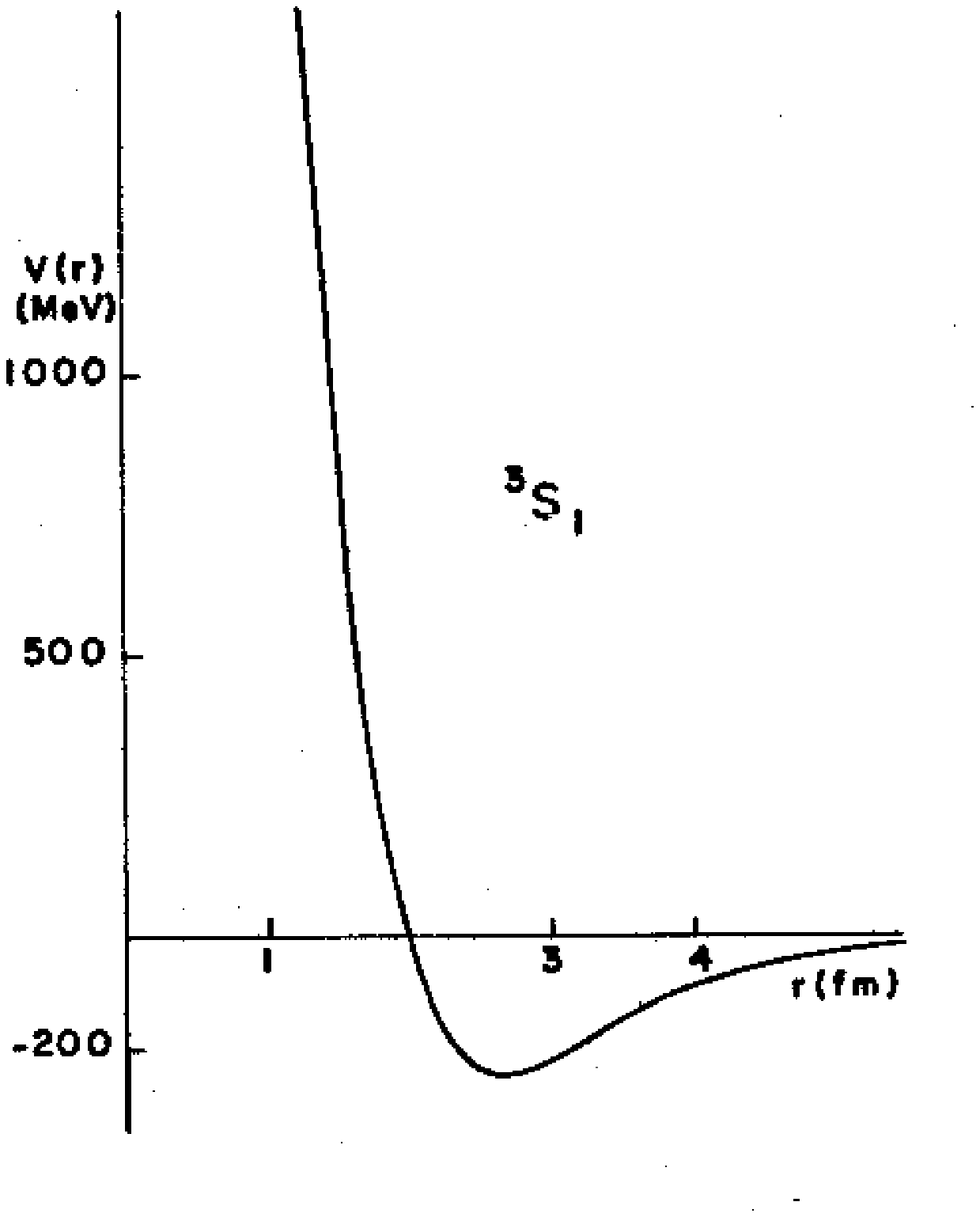,height=2.5in,width=2.in}}
\vspace*{-0.3cm}
\caption{$^3S_1(T=1)$ and $^5S_2(T=1)$ N$\Delta$ phase shifts
and separable potentials of reference \protect\cite{FERR89}.}
\label{fig2c}
\end{figure}
\begin{figure}[t]
\vspace*{-0.2cm}
\mbox{\psfig{figure=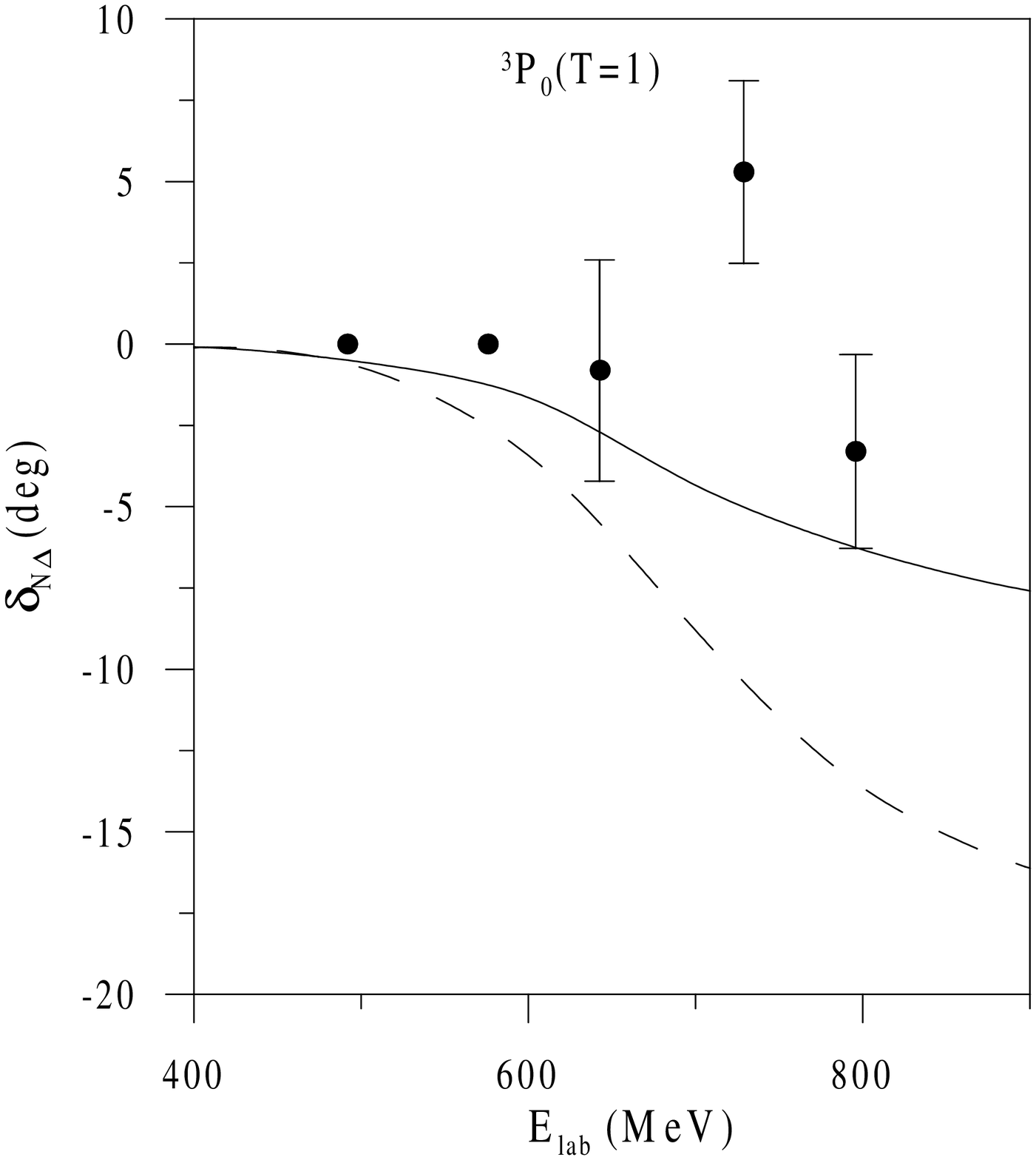,height=3.15in,width=2.50in}}
\hspace*{0.1cm}
\mbox{\psfig{figure=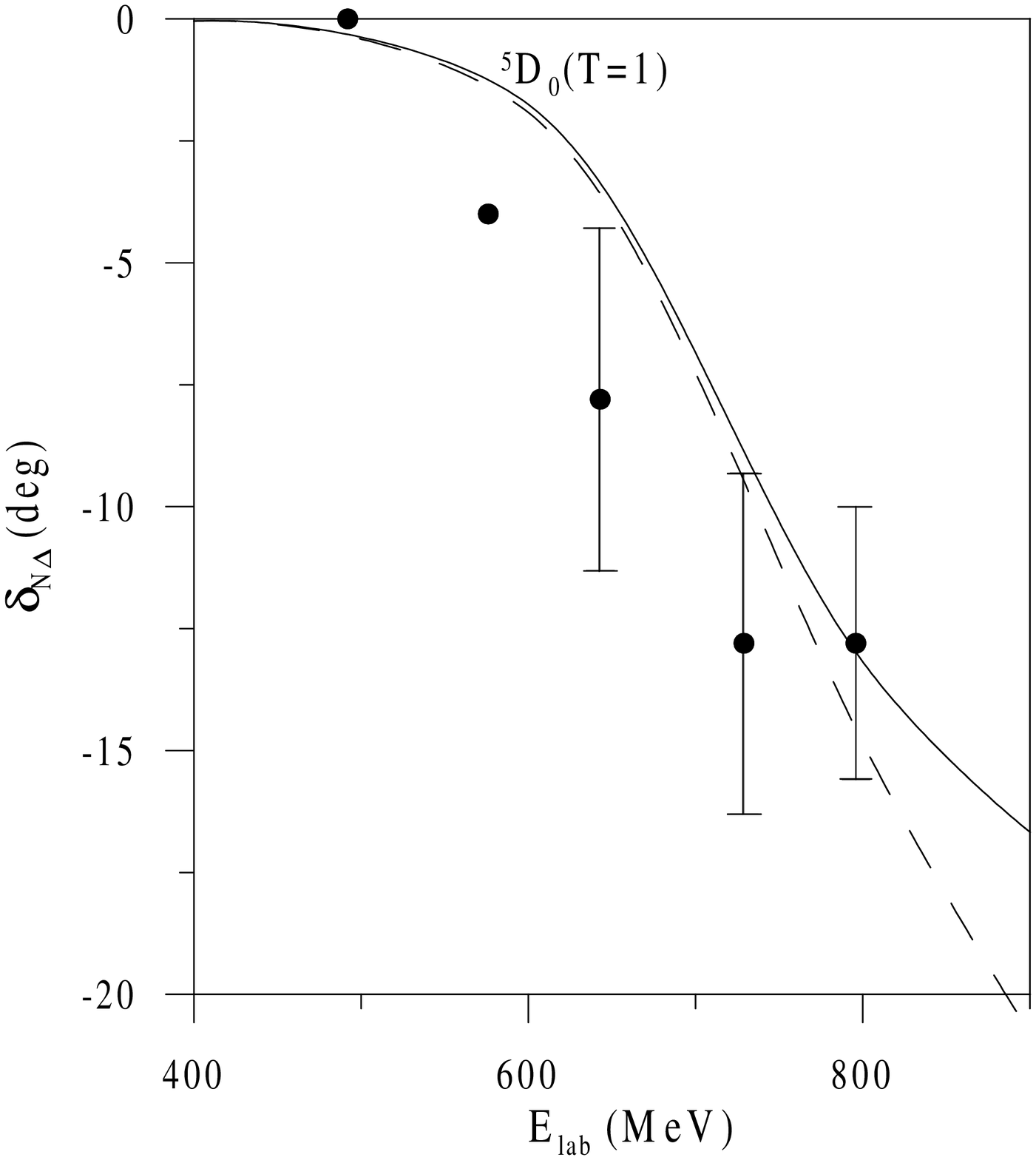,height=3.15in,width=2.45in}}
\vspace*{-2.2cm}
\caption{$^3P_0(T=1)$ and $^5D_0(T=1)$ N$\Delta$ phase shifts. The solid 
line represents the quark model results and
the dashed line those of the baryonic 
meson-exchange model of reference \cite{PEN90}.
Experimental data are from reference \cite{SHY89}.}
\label{fig4}
\end{figure}

In \sref{ch3.1} the presence of
N$\Delta$ quark Pauli blocked channels, specifically the
$^3S_1(T=1)$ and $^5S_2(T=2)$ was discussed. In figure \ref{fig3} we can see 
how quark Pauli blocking, existing in the $^3S_1(T=1)$ partial wave,
translates into a strong short-range repulsion, while this behaviour
is not observed for non-Pauli blocked partial waves as it is the case for 
the $^5S_2(T=1)$ channel. The strong repulsion is due to the 
combined effect of the fast decrease of the 
norm of the six-quark wave function when $R \to 0$ with 
the presence of non-vanishing direct terms in the potential. 
Actually, the repulsion manifests itself through the
direct contribution of the OPE and OSE potentials.
Let us note that if there were no direct contributions, i.e.,
only quark-exchange diagrams contributed
then the resulting behaviour would be quite different since the
quark-exchange terms go to zero with the same power
of $R$ as the norm. 

For the sake of comparison we have also plotted
in figure \ref{fig3} the result for the baryonic meson-exchange model 
of reference \cite{PEN90}: the strong short-range repulsion does not appear.
The only possibility to simulate it would be the use
of large cut-off masses, but this would produce
instabilities in the short-range part of the interaction.
Moreover, the baryonic model gives the same
structure in both partial waves,
the only difference between them being 
a spin-isospin factor. 

In figure \ref{fig2c}, a separable N$\Delta$ 
potential reproducing the phase shifts (dots in figure \ref{fig2c}) 
obtained from the
experimental $\pi d$ elastic cross section is shown \cite{FERR89}.
One can infer a hard-core in the $^3S_1(T=1)$ partial wave
from the rapidly varying phase shifts changing sign for 
a pion energy of 219 MeV and a smooth behaviour
for the $^5S_2(T=1)$ partial wave.
This behaviour is quite similar to the one predicted by the
chiral constituent quark model (see figure \ref{fig3}).
When not considering the quark substructure
a baryon-baryon hard-core has to be introduced by hand to reproduce the
experimental data \cite{ALE90,TAKA88}. 

Nevertheless, potentials are not observable quantities and one should use them 
to calculate phase shifts to be compared to data,
whose extraction is still a matter of controversy. There are a 
few N$\Delta$ channels which are susceptible of being parametrized 
in terms of phase shifts. In figure \ref{fig4} 
we plot the quark-model result for
two uncoupled isospin one N$\Delta$ partial waves, 
$^3P_0$ and $^5D_0$. Although the error bars are still big,
quark-model results agree reasonably well with
the experimental data, whereas the baryonic meson-exchange
predictions, due to its unsmooth character at short
distances \cite{VAL94b}, are far from the data in the $^3P_0$ case.

\subsubsection{The $\Delta\Delta$ interaction}
\label{ch4.2.3}
\begin{figure}[t]
\vspace*{-0.2cm}
\mbox{\psfig{figure=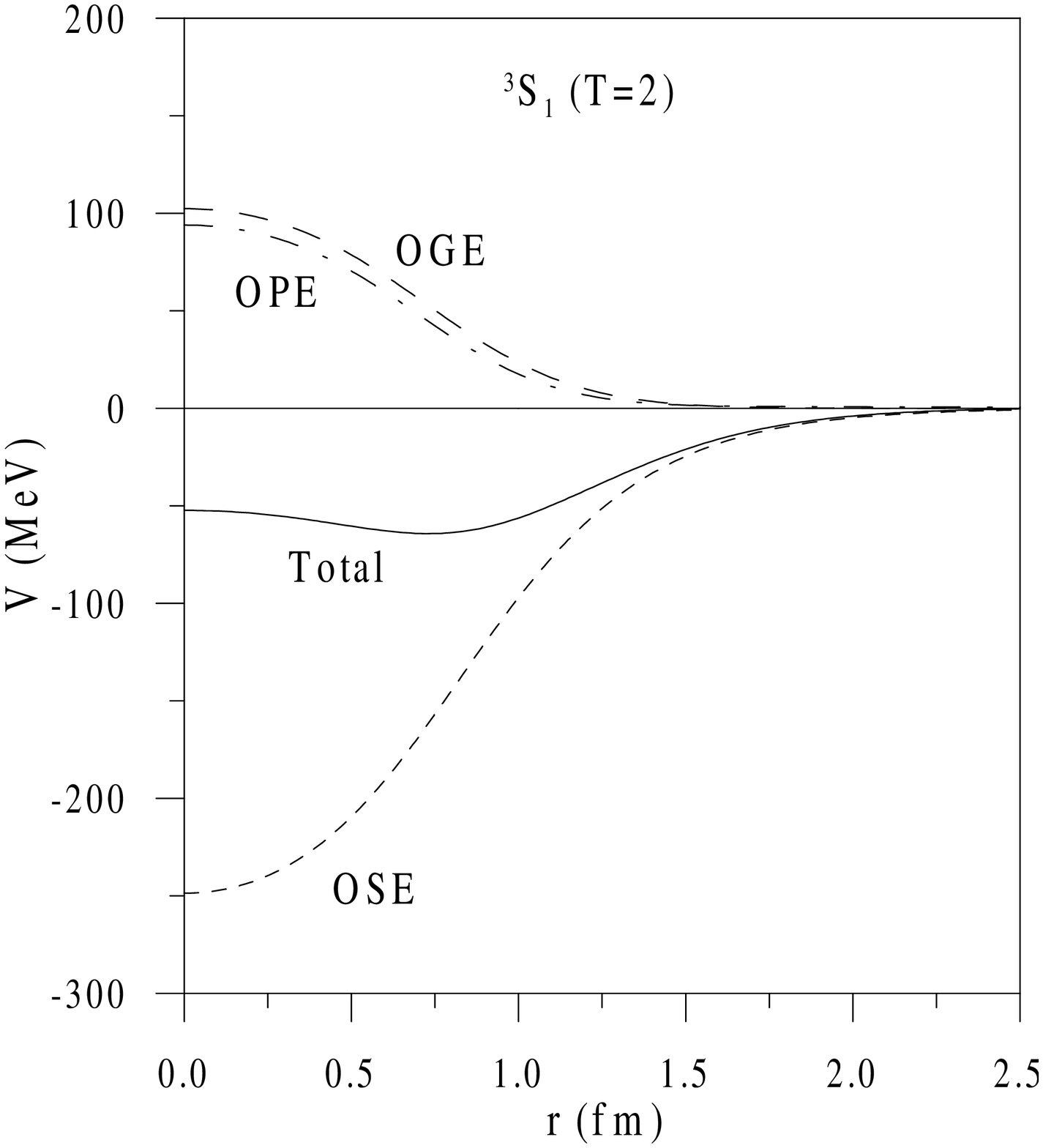,height=2.8in,width=2.4in}}
\hspace*{0.2cm}
\mbox{\psfig{figure=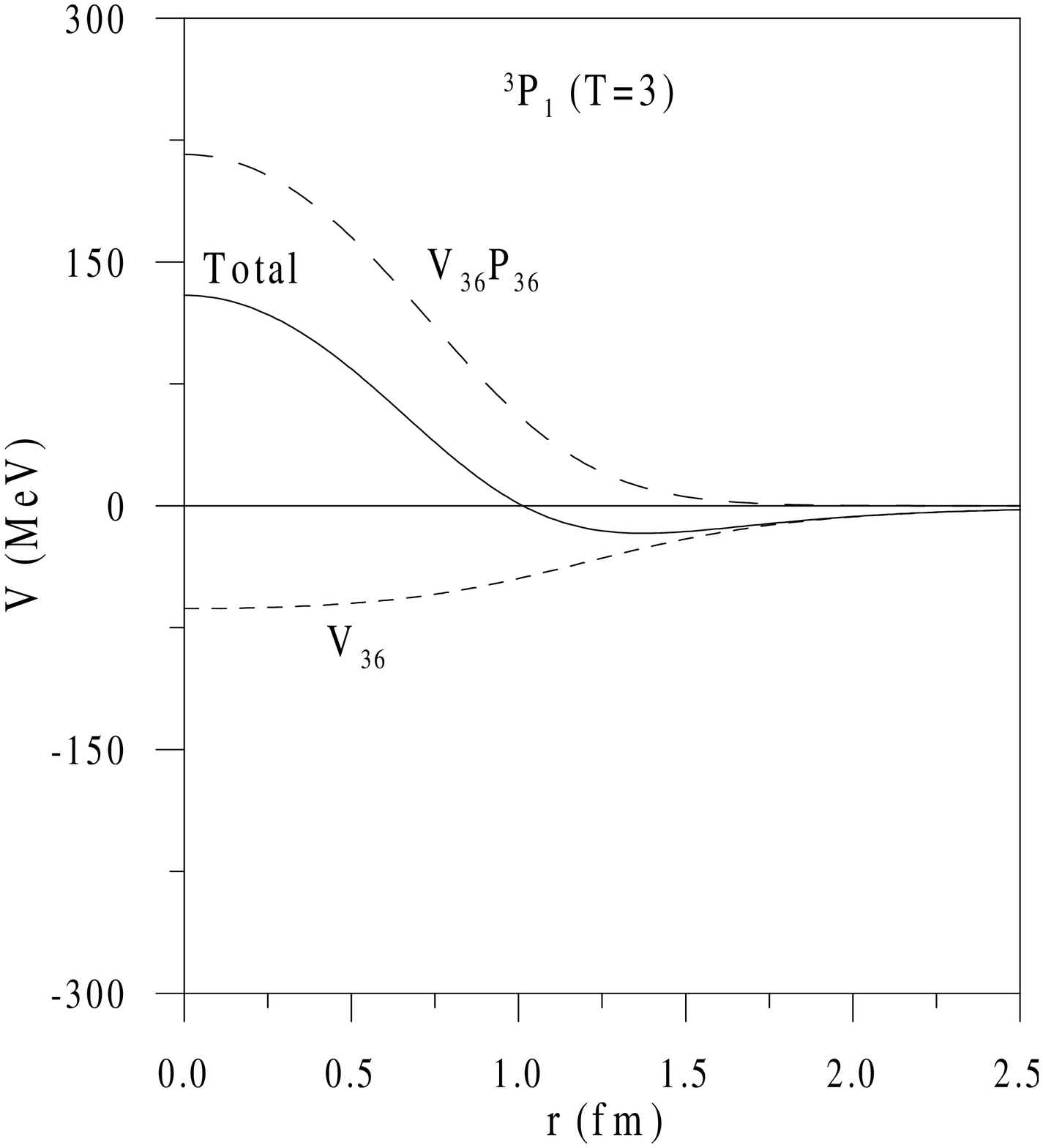,height=2.8in,width=2.4in}}
\vspace*{-1.6cm}
\caption{$^3S_1(T=2)$ and $^3P_1(T=3)$ $\Delta\Delta$ potential.}
\label{fig5}
\end{figure}
\begin{figure}[b]
\vspace*{-0.2cm}
\centerline{\psfig{figure=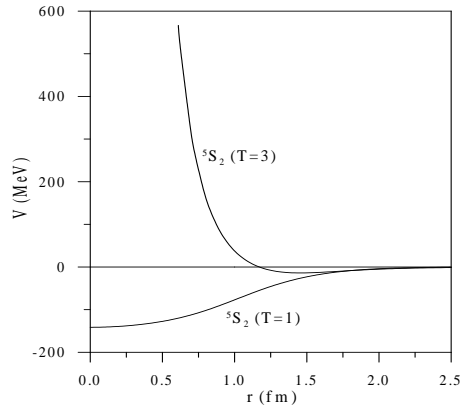,height=2.8in,width=2.4in}}
\vspace*{-1.6cm}
\caption{$^5S_2(T=3)$ and $^5S_2(T=1)$ $\Delta\Delta$ potential.}
\label{fig6}
\end{figure}

The effect of $\Delta\Delta$ components has been considered when
studying NN phase shifts at intermediate energies and deuteron
properties at the baryon level \cite{ALL86,HAI93}. This treatment suffers
from the same shortcomings mentioned in the N$\Delta$ case. 

At the quark level the two identical baryon selection rule,
L+S+T=odd, comes out from quark antisymmetrization.
From the chiral constituent quark model potential
the $\Delta\Delta$ interaction has been derived in reference \cite{VAL97}.
The potential obtained is drawn in figure \ref{fig5} for two partial 
waves of different isospin, $^3S_1(T=2)$ and $^3P_1(T=3)$. 
In one of them the contribution of the different exchanges
has been separated and in the other the contribution of the dominant diagrams
showed in figure \ref{fig0} are depicted. As can be seen, the dominant terms are
diagrams (b) and (d), the others giving almost a negligible contribution.
In both cases the direct contribution, diagram (b), (only generated
by quark-meson exchanges) is attractive, while the effect of
the quark-exchange diagram (d) (due to quark antisymmetry)
is to generate repulsion. Beyond 1.5 fm, the
interaction is driven by the $V_{36}$ term that equals
the total interaction.
\begin{table}[t] 
\caption{\label{models}Character of the
$\Delta \Delta$ interaction obtained in references \cite{OKA84,OKA93},
reference \cite{CVE80}, and reference \cite{VAL97}.
The ($\star$) denotes those channels where a bound
state would be favoured.}
\begin{indented}
\item[]\begin{tabular}{@{}llll}
\br
  $(S,T)$ & Refs. \cite{OKA84,OKA93} & Ref. \cite{CVE80} & Ref. \cite{VAL97} \\
\mr
(0,1) & Attractive       & $-$ & Attractive ($\star$)     \\
(0,3) & Weakly repulsive & Attractive & Weakly repulsive    \\
(1,0) & Attractive       & Repulsive & Attractive ($\star$)     \\
(2,3) & Repulsive        & $-$ & Repulsive           \\
(3,0) & Attractive ($\star$)  & Attractive ($\star$) & Weakly attractive \\
(3,2) & Repulsive        & $-$ & Repulsive           \\
\br
\end{tabular}
\end{indented}
\end{table}

According to the discussion 
in \sref{ch3.1}, the $^5S_2(T=3)$, $^7S_3(T=2)$, 
and $^7P_{2,3,4}(T=3)$ channels correspond to quark Pauli blocked
states. We compare in figure \ref{fig6} the $\Delta\Delta$ potential 
for two $S$ waves, a Pauli blocked channel, $^5S_2(T=3)$, and a
non-Pauli blocked one, $^5S_2(T=1)$. While the last one
presents a soft attractive behaviour at short distances,
the $^5S_2(T=3)$ channel shows a strong repulsive core
coming from the fast decreasing of the
overlapping of the two-baryon wave function (denominator in \eref{BODEF})
together with non-vanishing direct terms in the numerator. 

It is interesting to analyze the possible existence of 
bound states with the chiral constituent quark model since this has been
the object of study of the $\Delta\Delta$ system with quark models
containing confinement plus OGE potentials 
\cite{OKA84,OKA93,WAN92,GOL89,CVE80}.
In table \ref{models} results of reference \cite{VAL97}, based on the chiral
constituent quark model, are compared to those of references 
\cite{OKA84,OKA93,CVE80},
based on a OGE model. With respect to the possibility of having a 
favoured bound state in the
$(S,T)=(3,0)$ channel, as predicted in \cite{OKA84,OKA93,CVE80}, 
reference \cite{VAL97} gets no OGE attraction
but repulsion, although as it is shown in the table the total
interaction is weakly attractive mainly due to the scalar 
exchange potential.
Reference \cite{VAL97} predicts the $(S,T)=(0,1)$ and $(1,0)$ channels to be the
most attractive ones. This attraction, combined with the fact that
they are $S$ waves and therefore the centrifugal barrier is not active,
makes them the best candidates for possible $\Delta \Delta$ bound
states \cite{GAR97b}, as will be discussed in \sref{ch7.2.1}.

\subsubsection{The NN$^*$(1440) system}
\label{ch4.2.4}

The N$^*$(1440) (Roper) couples strongly (60$-$70$\%$) to the $\pi$N
channel and significantly (5$-$10$\%$) to the $\sigma$N channel \cite{PDG}.
Its role in nuclear dynamics as an intermediate state has been
analyzed at the baryon level. The presence of NN$^*$(1440)
configurations in the deuteron was suggested long ago 
\cite{ARE71,HAM62,REI68,ROS75}. 
Graphs involving the excitation of N$^*$(1440)
appear also in other systems, as for example the neutral pion
production in proton-proton reactions \cite{PEN99} or the three-nucleon
interaction mediated by $\pi $ and $\sigma $ exchange contributing to the
triton binding energy \cite{COO95}. The excitation of the Roper resonance has
also been used to explain the missing energy spectra in the 
$p(\alpha,\alpha^{\prime})$ reaction \cite{HIR96} or the $np\to d(\pi\pi)^0$
reaction \cite{ALV99}. 
Pion electro- and photoproduction from nucleons may take
place through the N$^*$(1440) excitation as well \cite{GAR93}. 

At the quark level the involved N$^*$(1440) radial structure
increases the number of diagrams contributing to the interaction. There appear
diagrams generated by the two parts of the wave function, $\phi_1$ and
$\phi_2$ in equation \eref{ropw}. Although involving interactions between
excited and non-excited quarks, they can be classified as in figure \ref{fig0}.
The full set of diagrams for the NN$^*$(1440) potential are explicitly given 
in \cite{JUL01} and those for the ${\rm NN} \to {\rm NN}^*(1440)$ potential 
in \cite{JUL02c}.
\begin{figure}[t]
\mbox{\psfig{figure=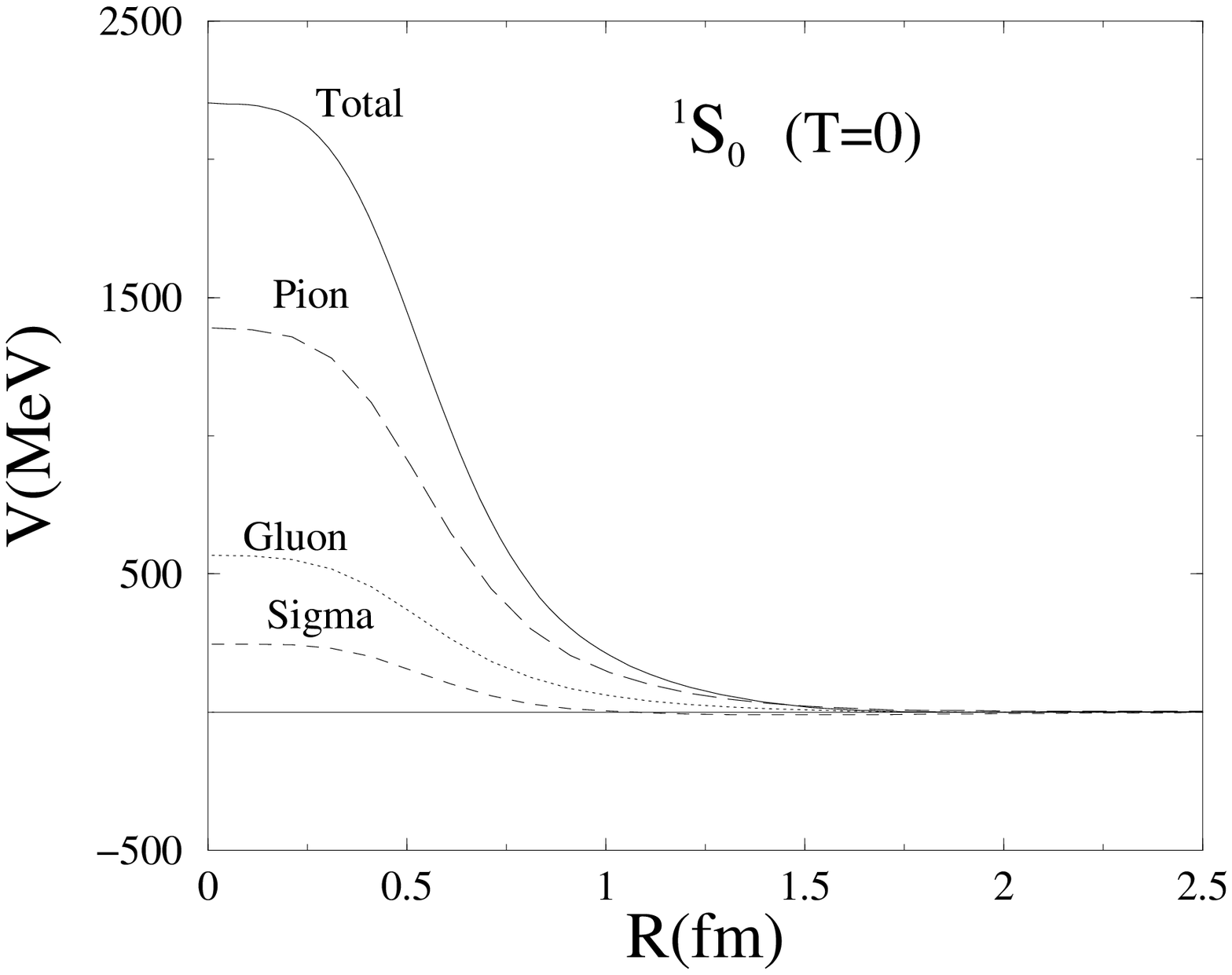,height=2.0in,width=2.4in}}
\hspace*{0.2cm}
\mbox{\psfig{figure=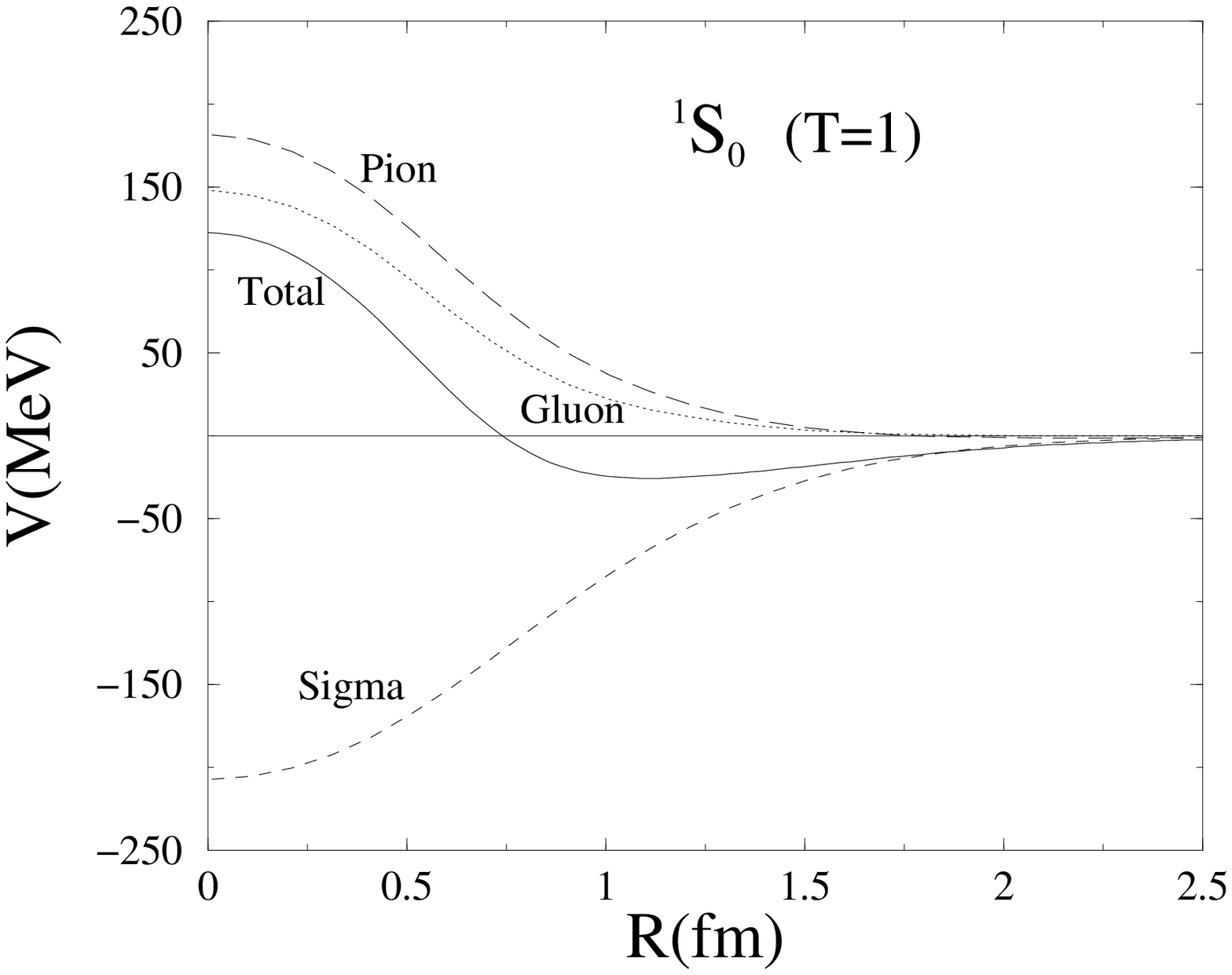,height=2.0in,width=2.4in}}
\vspace*{-0.4cm}
\caption{Contribution of the different terms of the interaction
to the $^1S_0(T=0)$ and $^1S_0(T=1)$ NN$^*$(1440) potentials.}
\label{fig8b}
\end{figure}

\subsubsection{The NN$^*$(1440) interaction}
\label{ch4.2.5}

In this case we shall classify the channels, for the reason that will
be apparent in what follows, as {\it forbidden} channels, i.e.,
allowed in the NN$^*$(1440) system but
forbidden in the NN case, and {\it allowed} channels, i.e.,
allowed in both NN$^*$(1440) and NN systems according to \eref{LST}.
In \fref{fig8b} the potential for a {\it forbidden} channel, 
$^1S_0$ (T=0), and for an {\it allowed} one, $^1S_0$ (T=1), are compared. 
As can be seen, {\it forbidden} channels in $S$ waves 
are much more repulsive than {\it allowed} ones, this repulsion
being mainly driven by the non-vanishing direct terms in the potential.
Moreover, as detailed in reference \cite{JUL01}, 
the potential for the {\it forbidden} $^{1}S_{0}(T=0)$ channel is
very much the same than for the {\it allowed }$^{1}P_{1}(T=0)$
and similarly for $^{3}S_{1}(T=1)$ and $^{3}P_{J}(T=1)$ (in this last case
with small dependence on $J$ due to the tensor interaction). This can be
understood in terms of the Pauli and the centrifugal barrier repulsions. The
Pauli correlations and the centrifugal barrier in the $P$ waves prevent all
the quarks to be in the same spatial state, much the same effect one has due
to Pauli correlations in the $S$ {\it forbidden} waves added to the presence
of the radially excited quark in the N$^*$(1440).

\begin{figure}[t]
\mbox{\psfig{figure=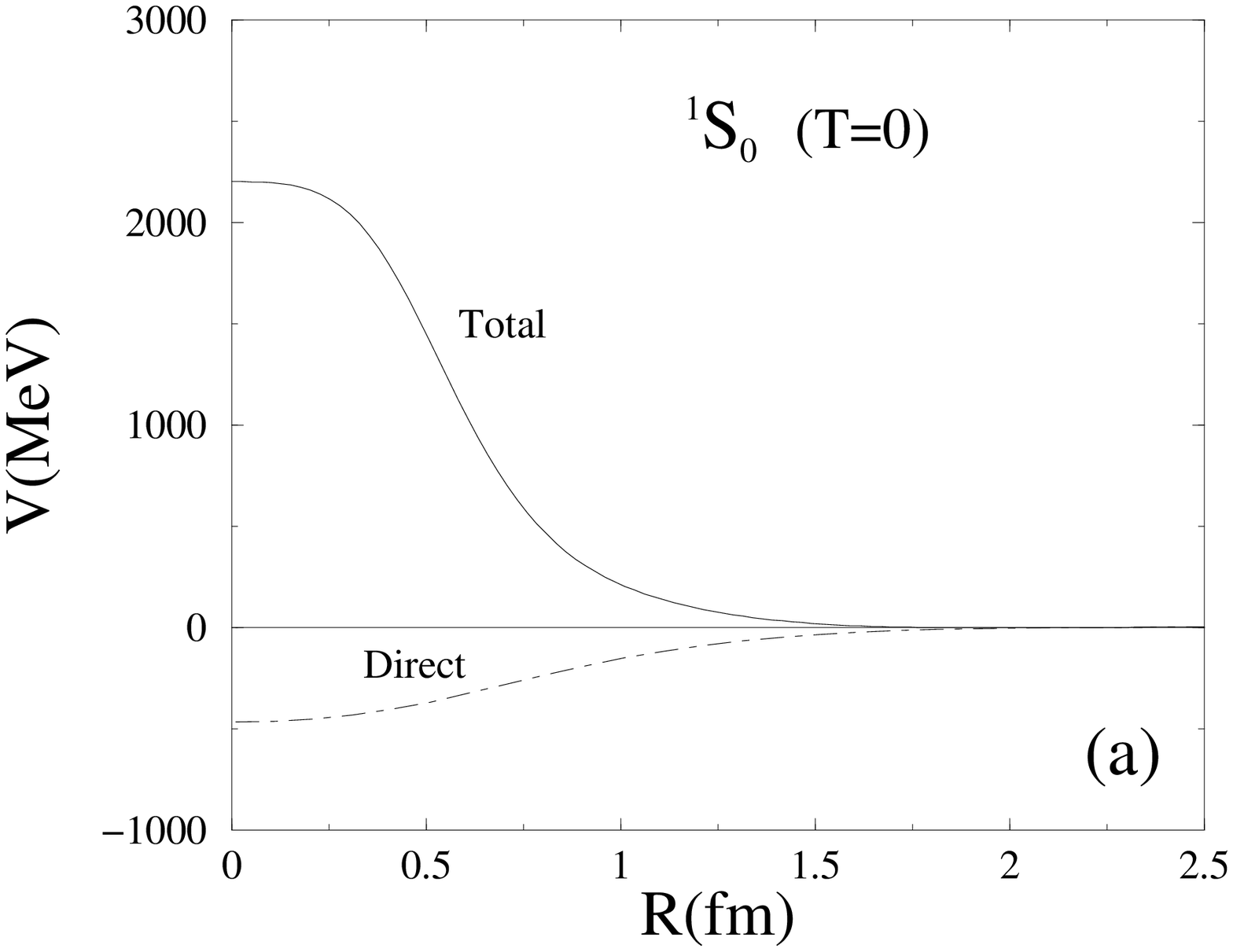,height=2.0in,width=2.4in}}
\hspace*{0.2cm}
\mbox{\psfig{figure=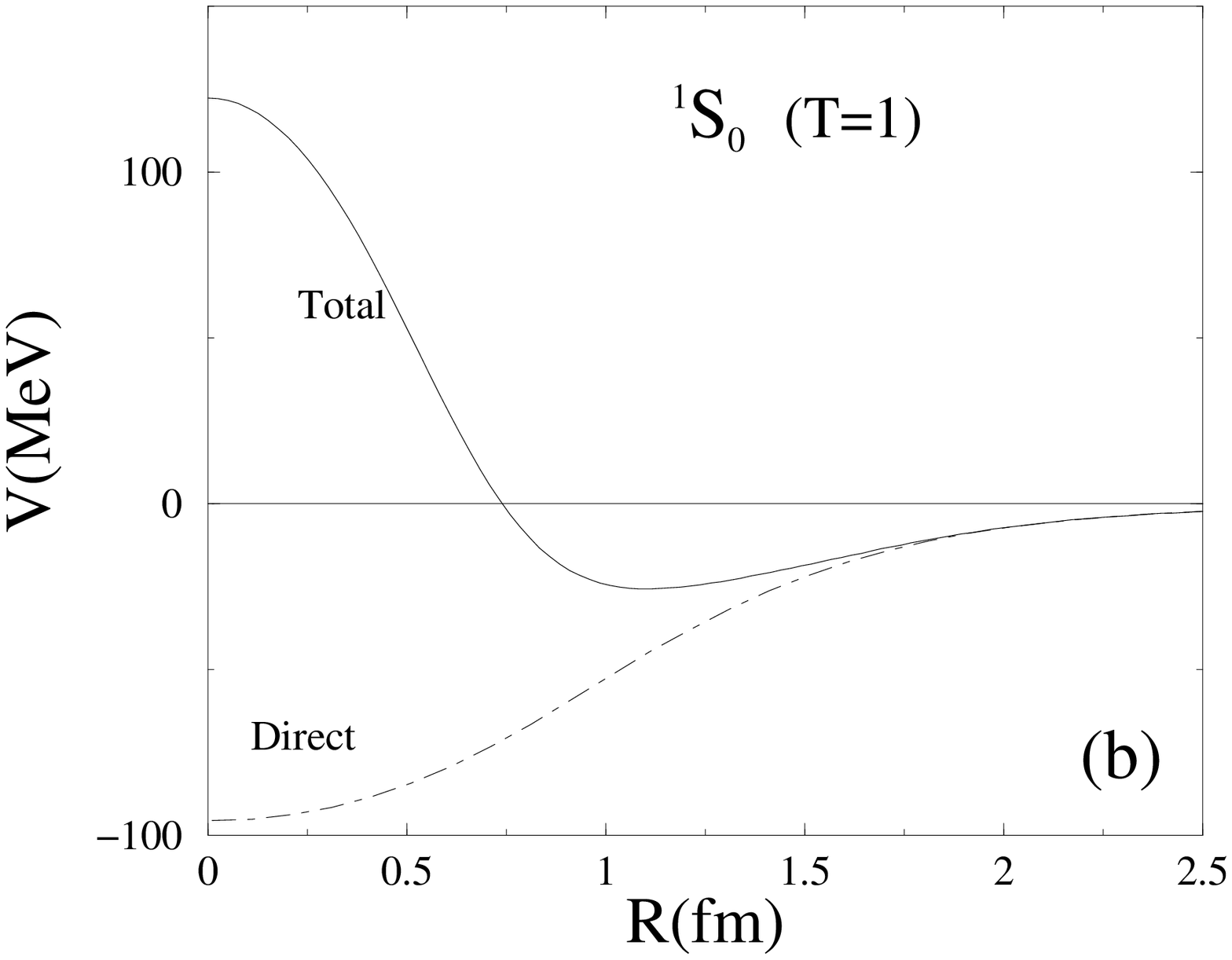,height=2.0in,width=2.4in}}
\vspace*{-0.4cm}
\caption{Comparison between the total and direct potentials for
$^1S_0(T=0)$ and $^1S_0(T=1)$ NN$^*$(1440) potentials.}
\label{fig8bc}
\end{figure}

The lack of available data and the absence of alternative quark
model calculations to compare with makes convenient in this case to
extract a pure baryonic potential whose difference with the total
one emphasizes the effects of the quark substructure.
The dynamical effect of quark antisymmetrization can be estimated by
comparing the total potential with the one arising from diagram (b) 
in figure \ref{fig0}, $V_{36}$,
which is the only significant one that does not include quark exchanges. The 
$V_{36}$ potential turns out to be attractive everywhere. Let us note
however that Pauli correlations are still present in the $V_{36}$ potential,
through the norm, in the denominator of equation (\ref{BODEF}). To eliminate the
whole effect of quark antisymmetrization one should eliminate quark-Pauli
correlations from the norm as well. By proceeding in this way one gets a
genuine baryonic potential, that will be called direct potential. The comparison of
the total and direct potentials reflects the quark antisymmetrization effect
beyond the one-baryon structure, see figure \ref{fig8bc}. 
As $V_{36}$, the direct potential is
attractive everywhere. It becomes then clear that the repulsive
character of the interaction for $S$ and $P$ waves at short distances is due
to dynamical quark-exchange effects. For distances $R\geq$ 2 fm the direct
and total potentials are equal since then the overlap of the N
and the N$^*$(1440) wave functions is negligible and no exchange
diagrams contribute appreciably. These results clearly illustrate
that the use of a NN$^*$(1440) potential as a generalization of the
NN interaction should be taken with great care, 
specially for the {\it forbidden} channels.

\subsubsection{The NN $\rightarrow$ NN$^*$(1440) interaction}
\label{ch4.2.6}

The presence of two identical baryons in the initial state forces
in this case the selection rule L+S+T=odd.
For the ${\rm NN}\rightarrow {\rm NN}^*(1440)$ potential,
most diagrams contributing to the interaction are due to the first term of
the N$^*$(1440) wave function \eref{ropw}, i.e., $|[3](0s)^2(1s)\rangle$.
\begin{figure}[b]
\vspace*{-0.2cm}
\mbox{\psfig{figure=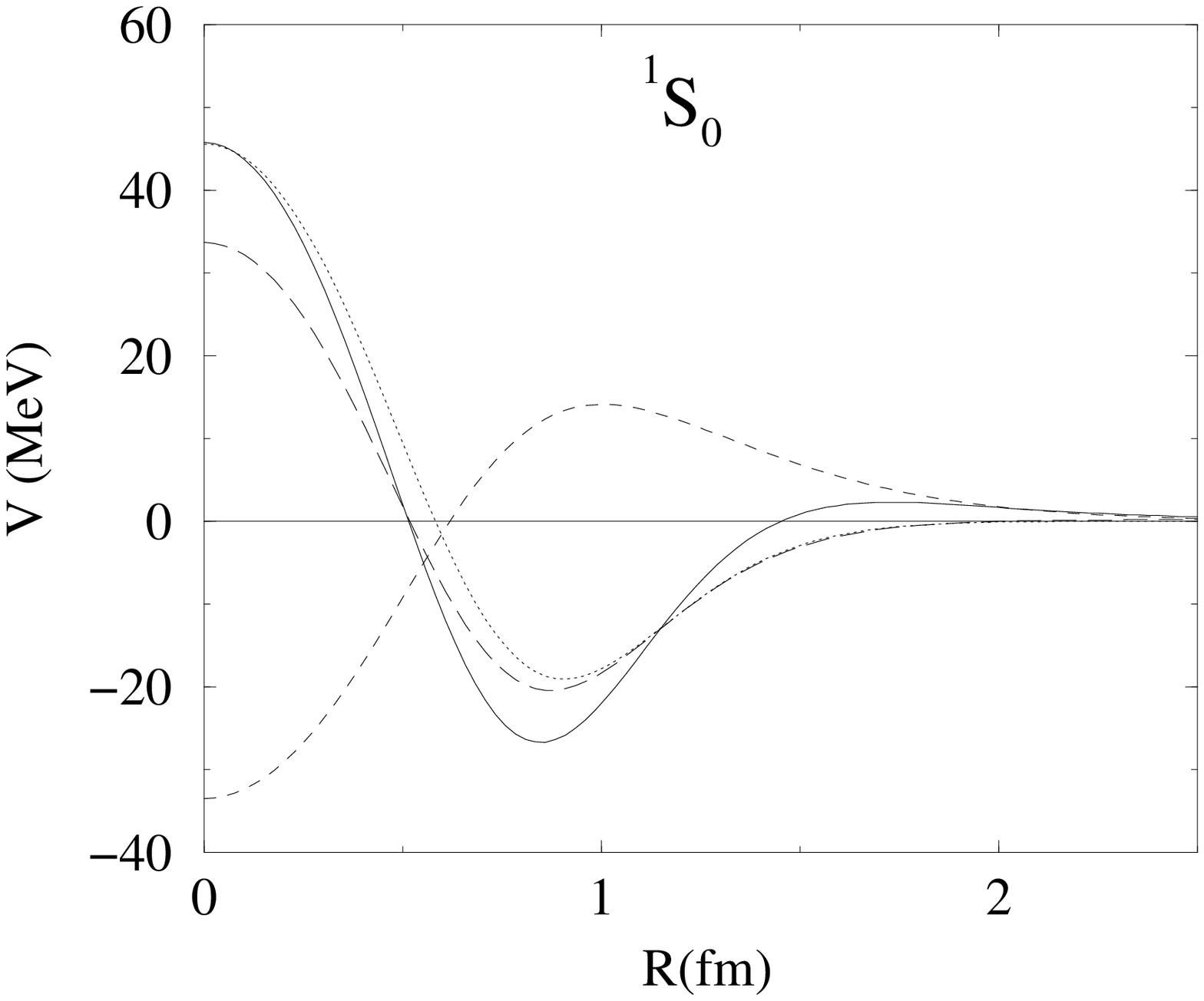,height=2.0in,width=2.4in}}
\hspace*{0.2cm}
\mbox{\psfig{figure=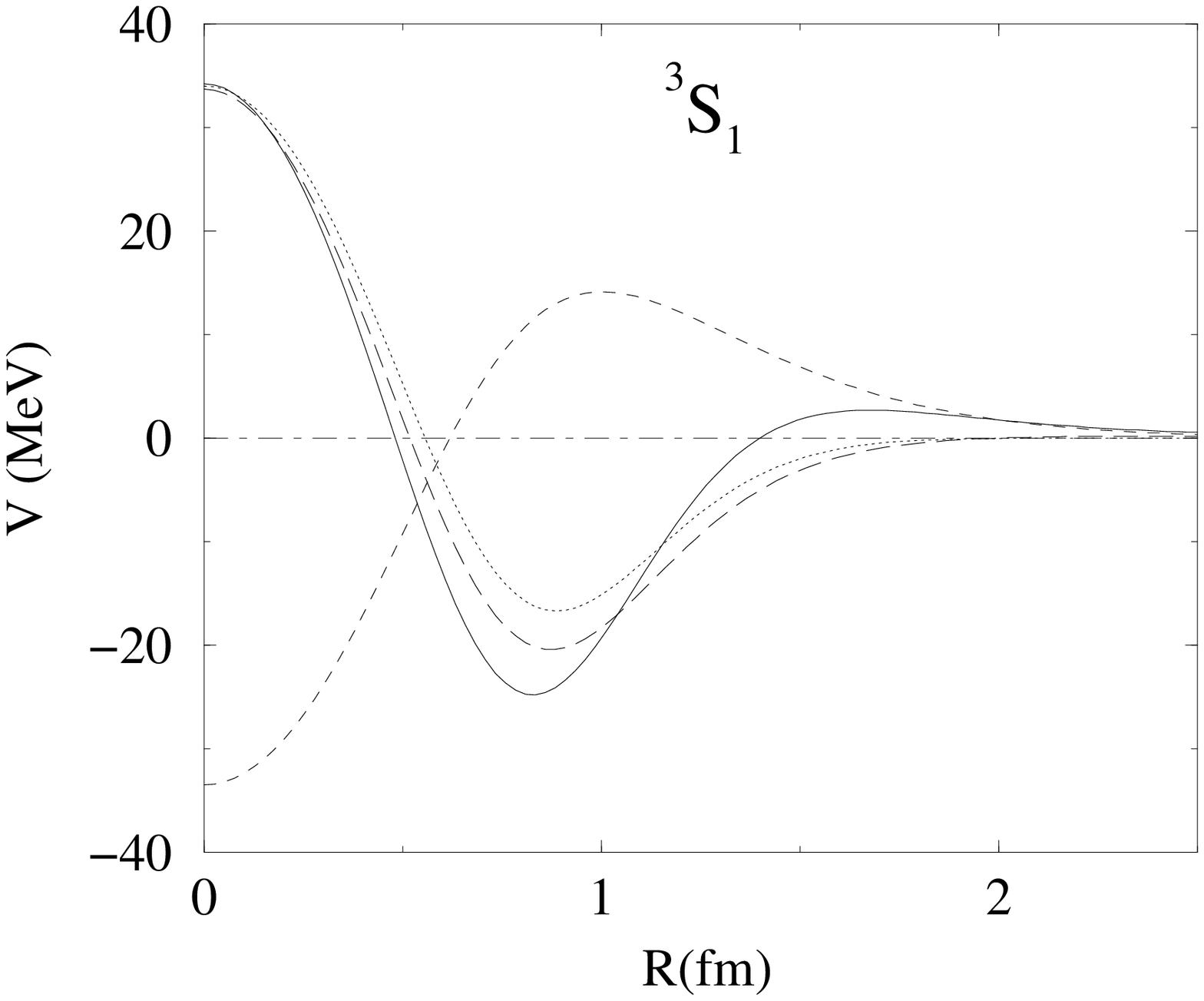,height=2.0in,width=2.4in}}
\vspace*{-0.2cm}
\caption{$^1S_0$ and $^3S_1$ $\rm{NN} \to \rm{NN}^*(1440)$ potentials.
The long-dashed, dashed and dotted lines denote the OPE, OSE and OGE contributions,
respectively.}
\label{fig9}
\end{figure}
In figure \ref{fig9}, the potentials 
for $^{1}S_{0}$ and $^{3}S_{1}$ partial waves are shown (let us note
that an arbitrary global phase between the N and 
N$^*$(1440) wave functions has been chosen). 
Although the long-range part of the interaction ($R>4$ fm ) comes dominated
by the OPE, the asymptotic potential reverses sign with 
respect to both NN and NN$^*$(1440) cases.
Thus for $S$ and $D$ waves the ${\rm NN}\rightarrow {\rm NN}^*(1440)$ interaction
is asymptotically repulsive. This sign reversal is a direct consequence of
the presence of a node in the N$^*$(1440) wave function what implies a
change of sign with respect to the N wave function at long distances 
(for NN$^*$(1440) there are two compensating changes 
of sign coming from the two Ropers). This is also
corroborated by the study of the OSE interaction that is
always asymptotically repulsive at difference to the NN and 
NN$^*$(1440) cases. If 
the opposite sign for the N$^*$(1440) wave function were chosen the
long-range part of the interaction would be attractive but there would also
be a change in the character of the short-range part.

It is worth to remark that no quark antisymmetrization effects survive
either in the numerator or in the 
denominator (norm) of equation (\ref{BODEF}) at
these distances. In other words, the potential corresponds to a direct
baryon-baryon interaction that can be fitted as it is conventional
in terms of a Yukawa function depending on the mass of the meson.

The total potential turns out to be attractive from $R=$ 1.5 fm down to a lower
value of $R$ different for each partial wave. This behaviour, related again
to the node in the Roper wave function, contrasts with the elastic NN
and NN$^*$(1440) cases, where for instance
for $S$ and $D$ waves the scalar part keeps always the same sign and
gives the dominant contribution for $R>$0.8 fm.
Below 0.6 fm the potential becomes repulsive in all partial
waves. Nevertheless there are two distinctive features with respect to 
the elastic NN and NN$^*$(1440) cases:
in $\rm{NN}\rightarrow \rm{NN}^*(1440)$ the intensity of the repulsion at $R=$0
and the value of $R$ at which the interaction becomes repulsive are
significantly lower than in NN and NN$^*$(1440) elastic
potentials. This is a clear effect of the more similarity
(higher overlap) in these cases between initial and final states what makes
the Pauli principle more active.

Let us also mention that at short distances, the
interaction could be fitted in terms of two different Yukawa functions, one
depending on the meson mass, $m$, the other with a shorter range depending on 
$\sqrt{(M_{{\rm N}^*(1440)}-M_{\rm N}+m)m}$. These two Yukawa 
functions could be associated to the two diagrams
with different intermediate states [$m{\rm NN}$ and $m{\rm NN}^*(1440)$] appearing in
time ordered perturbation theory when an effective calculation at the baryonic
level is carried out (let us realize that in a quark calculation the
intermediate state is always $mqq$, the N$^*$(1440)$-$N mass difference
being taken into account through the N and N$^*$(1440) wave functions).

\subsection{$\pi {\rm NN}^*(1440)$ and $\sigma {\rm NN}^*(1440)$ coupling constants}
\label{ch4.3}

A main feature of the quark treatment is its universality in the sense that
all baryon-baryon interactions are treated on an equal footing. This allows a
microscopic understanding and connection of the different baryon-baryon
interactions that is beyond the scope of any analysis based only on
effective hadronic degrees of freedom. 
We will illustrate this discussion by means of the 
${\rm NN} \to {\rm NN}^*(1440)$ transition potential, 
determining the $\pi {\rm NN}^*(1440)$ 
and $\sigma {\rm NN}^*(1440)$ coupling constants.

As has been discussed in \sref{ch4.2.1}, asymptotically
($R\geq$4 fm) the OSE and OPE potentials have at the baryon
level the same spin-isospin structure than at quark level.
Hence one can try to parametrize the asymptotic central interactions as, 
\begin{eqnarray}
\fl V_{{\rm NN}\to {\rm NN}^*(1440)}^{\rm OPE}(R) = \frac{1}{3} \, 
\frac{g_{\pi {\rm NN}}}{%
\sqrt{4\pi }} \, \frac{g_{\pi {\rm NN}^*(1440)}}{\sqrt{4\pi }} \, \frac{%
m_{\pi }}{2M_{N}} \, \frac{m_{\pi }}{2(2M_{r})} \, \frac{\Lambda ^{2}}{%
\Lambda ^{2}-m_{\pi }^{2}}  \nonumber \\
\lo [(\vec{\sigma }_{N}.\vec{\sigma }_{N})(\vec{\tau }%
_{N}.\vec{\tau }_{N})] \, \frac{e^{-m_{\pi }R}}{R} \, ,  \label{lrg}
\end{eqnarray}
\noindent and
\begin{equation}
\fl V_{{\rm NN}\rightarrow {\rm NN}^*(1440)}^{\rm OSE} (R)
=- \, \frac{g_{\sigma NN}}{\sqrt{%
4\pi }} \, \frac{g_{\sigma {\rm NN}^*(1440)}}{\sqrt{4\pi }} \, \frac{\Lambda
^{2}}{\Lambda ^{2}-m_{\sigma }^{2}} \, \frac{e^{-m_{\sigma }R}}{R} \, ,
\label{slrg}
\end{equation}
\noindent where $g_{i}$ stands for the coupling constants at the baryon
level and $M_{r}$ is the reduced mass of the NN$^*$(1440) system. 

By comparing these baryonic potentials with the asymptotic behaviour of the
OPE and OSE previously obtained one can extract
the $\pi {\rm NN}^*(1440)$ and $\sigma {\rm NN}^*(1440)$ coupling constants
in terms of the elementary $\pi qq$ coupling 
constant and the one-baryon model dependent
structure (see \sref{ch2.2}). 
The sign obtained for the meson-NN$^*$(1440) coupling
constants and for their ratios to the meson-NN coupling constants 
is ambiguous since it comes determined by the arbitrarily chosen
relative sign between the N and N$^*$(1440) wave functions. Only
the ratios between $\pi {\rm NN}^*(1440)$ and $\sigma {\rm NN}^*(1440)$
would be free of this uncertainty. This is why only absolute values
will be quoted, except for these cases where the sign comes as a prediction 
of the model. For this study the $^1S_0$ will be used for simplicity. 
This is why only the central
interaction has been written in equation \eref{lrg}.

The $\Lambda ^{2}/(\Lambda ^{2}-m_i^{2})$ vertex factor comes
from the vertex form factor chosen at momentum space as a square root of
monopole $[\Lambda ^{2}/(\Lambda ^{2}+\vec{q}^{\, 2})] ^{1/2}$, 
the same choice taken at the quark level, where chiral
symmetry requires the same form for pion and sigma. A different choice for
the form factor at the baryon level, regarding its functional form as well
as the value of $\Lambda $, would give rise to a different vertex factor and
eventually to a different functional form for the asymptotic behaviour. For
instance, for a modified monopole form, $[(\Lambda^{2}-m^{2})/
(\Lambda^{2}-q^{2})]^{1/2}$, where $m$ is the meson mass ($%
m_{\pi }$ or $m_{\sigma }$), the vertex factor would be $1$, i.e. 
$(\Lambda^{2}-m^{2})/(\Lambda^{2}-m^{2})$, keeping the potential the same
exponentially decreasing asymptotic form. Then it is clear that the
extraction from any model of the meson-baryon-baryon coupling constants
depends on this choice. We shall say they depend on the coupling scheme.
\begin{figure}[b]
\vspace*{-0.2cm}
\mbox{\psfig{figure=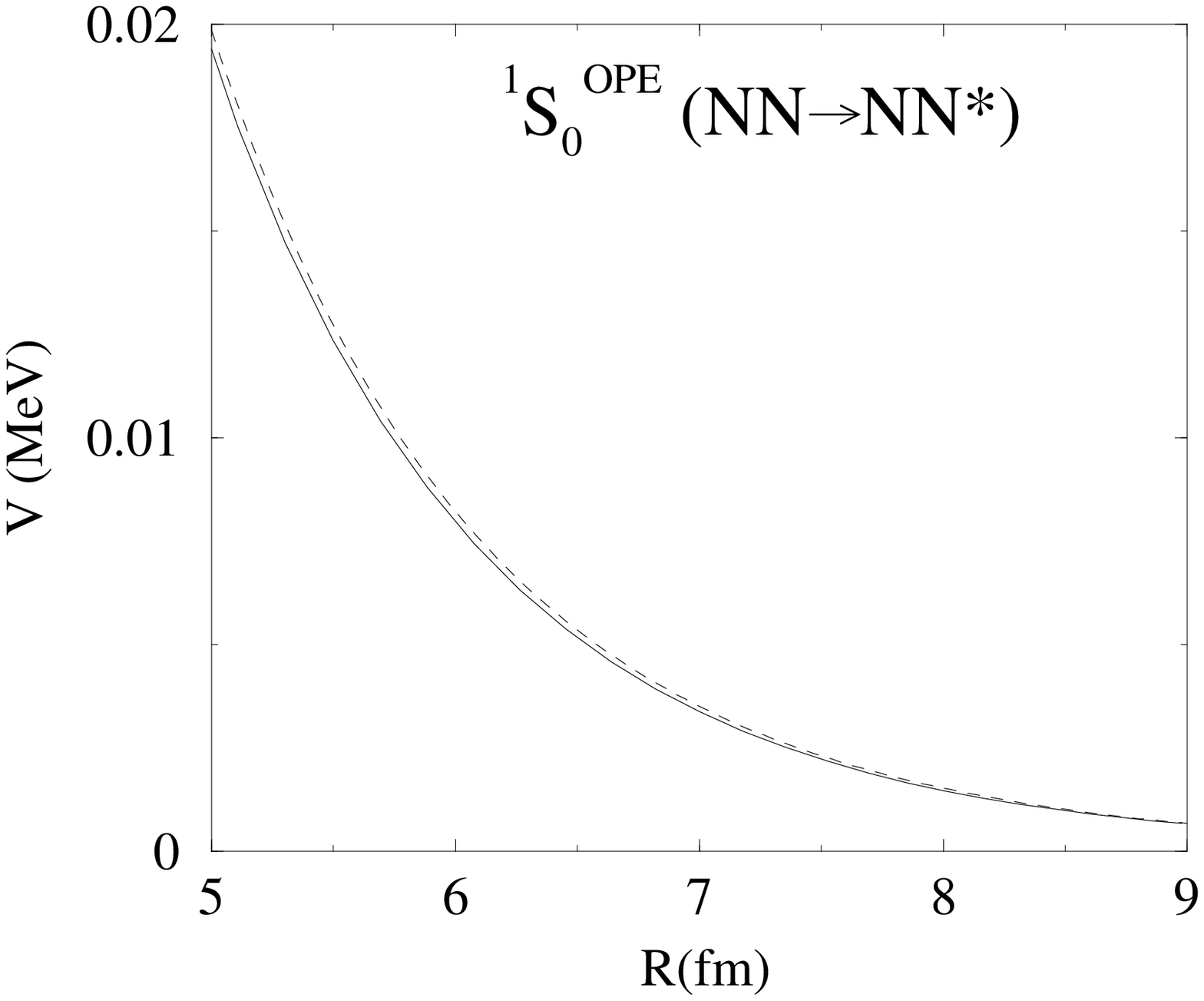,height=2.0in,width=2.4in}}
\hspace*{0.2cm}
\mbox{\psfig{figure=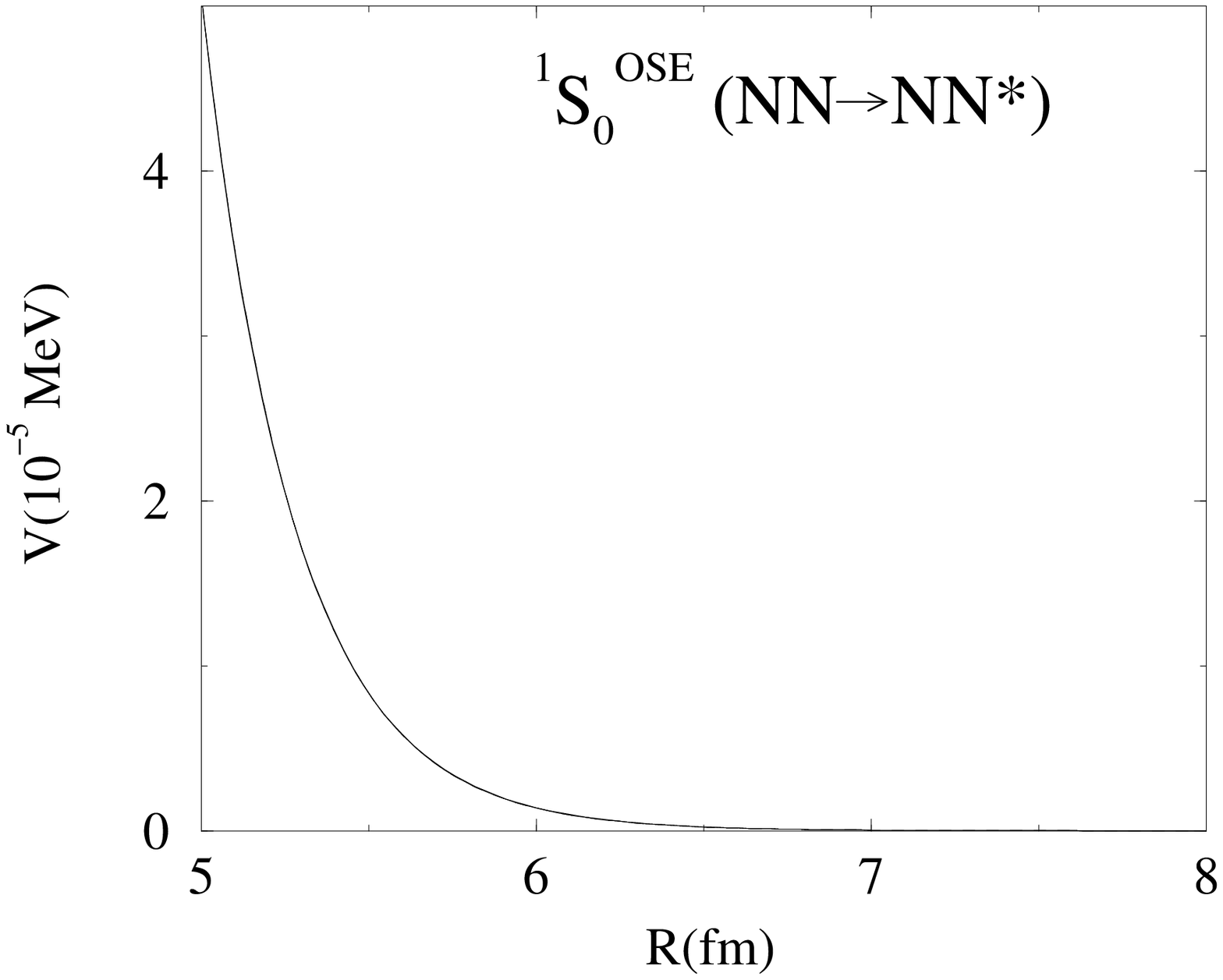,height=2.0in,width=2.4in}}
\vspace*{-0.2cm}
\caption{Asymptotic behaviour of the OPE and OSE $^1S_0$ 
${\rm NN} \to {\rm NN}^*(1440)$ potential
(solid line), the dashed line represents the fitted curves according to \eref{lrg}
and \eref{slrg}. For the OSE potential both lines are indistinguishable.}
\label{fig9c}
\end{figure}

For the OPE with $\Lambda =4.2$ fm$^{-1}$, 
$\Lambda ^{2}/(\Lambda^{2}-m_{\pi}^{2})=1.03$, pretty close to $1$.
As a consequence, in this case the use of this form factor or the modified
monopole form at baryonic level makes little difference in the determination
of the coupling constant. This fact is used when fixing 
$g_{\pi qq}^{2}/{4\pi }$ from the experimental value of 
$g_{\pi {\rm NN}}^{2}/{4\pi }$ extracted from NN data. 
To get $g_{\pi {\rm NN}^*(1440)}/\sqrt{4\pi}$ we turn to the numerical
result for the $^1S_0$ OPE potential and fit
its asymptotic behaviour (in the range $R:5\to$9 fm, see figure
\ref{fig9c}) to equation (\ref{lrg}), obtaining
\begin{equation}
\frac{g_{\pi {\rm NN}}}{\sqrt{4\pi }} \frac{g_{\pi {\rm NN}^*(1440)}}{\sqrt{4\pi }} 
\frac{\Lambda ^{2}}{\Lambda ^{2}-m_{\pi }^{2}}= \, - \, 3.73 \, ,
\end{equation}
i.e. $g_{\pi {\rm NN}^*(1440)}/\sqrt{4\pi}= - 0.94$. As
explained above only the absolute value of this coupling constant is well
defined. In reference \cite{RIS01} a different sign 
is obtained what is a direct consequence of
the different global sign chosen for the N$^*$(1440) wave function.
The coupling scheme dependence can be explicitly eliminated comparing 
$g_{\pi {\rm NN}^*(1440)}$ with $g_{\pi {\rm NN}}$ extracted from the
NN potential within the same quark model,
\begin{equation}
\left | \frac{g_{\pi {\rm NN}^*(1440)}}{g_{\pi {\rm NN}}} \right |=0.25 \,.
\label{eq17}
\end{equation}
By proceeding in the same way for the OSE potential, i.e. by fitting the
potential to equation \eref{slrg}, and following an analogous procedure for 
the NN case one can write
\begin{equation}
\left |\frac{g_{\sigma {\rm NN}^*(1440)}}{g_{\sigma {\rm NN}}} \right |=0.47 \, .
\label{eq18}
\end{equation}

The relative phase chosen for the N$^*$(1440) wave function with respect
to the N wave function is not experimentally relevant in any two step process
comprising N$^*$(1440) production and its subsequent decay. However it
will play a relevant role in those reactions where the same field ($\pi$ or 
$\sigma$) couples simultaneously to both systems, NN and NN$^*$(1440).
In these cases the interference term between both diagrams would
determine the magnitude of the cross section \cite{HIR96}.

The ratio given in (\ref{eq17}) is similar to that obtained in reference 
\cite{RIS01} and a factor 1.5 smaller than the one obtained from the analysis of
the partial decay width. Indeed one can find in the
literature values for $f_{\pi {\rm NN}^*(1440)}$ ranging between 0.27$-$0.47
coming from different experimental analyses with uncertainties
associated to the fitting of parameters \cite{ALV99,HUB94,GAR93}.
Regarding the ratio obtained in (\ref{eq18}), it agrees quite
well with the only experimental available result, obtained in reference 
\cite{HIR96} from the fit of the cross section of the isoscalar Roper excitation
in $p(\alpha,\alpha^{\prime})$ in the 10$-$15 GeV region, where
a value of 0.48 is given.

Furthermore, a very definitive prediction of the magnitude and
sign of the ratio of the two ratios is obtained,
\begin{equation}
\frac{g_{\pi {\rm NN}^*(1440)}}{g_{\pi {\rm NN}}}=0.53 \; \frac{g_{\sigma
{\rm NN}^*(1440)}}{g_{\sigma {\rm NN}}} \, ,
\end{equation}
which is an exportable prediction of the chiral constituent quark model.

For the sake of completeness we give the values of $g_{\sigma {\rm NN}^*(1440)}$
and $g_{\sigma {\rm NN}}$, though one should realize that the corresponding form
factor ${\Lambda ^{2}}/(\Lambda^{2}-m_{\sigma }^{2})=2.97$ differs quite
much from 1. Extracting the quark model factor dependence from the
coupling constant ($e^{m_\sigma^2 b^2/2}$) \cite{JUL01}, that one
may consider included in the baryon form factor, the results
obtained are $g_{\sigma {\rm NN}^*(1440)}^{2} /4\pi=$1.14, that compares quite
well with the value given in reference \cite{HIR96}, $g_{\sigma {\rm NN}^*(1440)}^{2}/
4\pi =1.33$, and $g_{\sigma {\rm NN}}^{2}/{4\pi }=5.06$.
These coupling constants have been also determined in
reference \cite{SOY00}. The results reported there are
sensitive to the decay width of the sigma into two pions and the
mass of the sigma as reflected in the large error bars given. 
Both quantities are highly undetermined in the Particle
Data Book \cite{PDG}, the mass of the sigma 
being constrained between 400$-$1200 MeV
and its width between 600$-$1000 MeV. These values have been fixed
in reference \cite{SOY00} to $m_{\sigma }=$500 MeV and $\Gamma
_{\sigma }=$250 MeV. Varying the mass of the sigma between 400 and 700 MeV
for a fixed width of 250 MeV, the coupling constant according to equation (9) of
reference \cite{SOY00} varies between 0.18$-$2.54. Taking a width of 450 MeV the
resulting coupling is 0.27$-$1.64. In both cases, our values 
may be compatible with the N$^*$(1440) decay and
production phenomenology.

%% file: ch4rev.tex
\section{The NN system}
\label{ch5}

The most accurate description of NN scattering data and
deuteron properties has been done in terms of boson exchanges at 
the baryon level in the framework of the old-fashioned time ordered
perturbation theory, viz., the Bonn potential \cite{MAC89}. 
On a microscopic base different approaches 
have attempted to describe the NN interaction.
Effective theories are very efficient to reproduce the high orbital 
angular momentum $(L)$ partial
waves but for low $L$ they need to introduce a large number of
parameters (more than twenty in reference \cite{ORD96}).
Bag models found severe problems in describing the two-baryon
dynamics due, on the one hand, to the critical difficulty
of separating the center of mass motion and, on the other hand,
to the unphysical sharp boundary of the bag and the problem of how 
to connect six-quark dynamics with external NN dynamics.
In the so-called {\it hybrid model} approaches
short-range dynamics based on
the quark substructure combines with a long-range part 
given by either meson exchanges at the baryonic level 
(for example references \cite{WAK84}
and \cite{VIN91} used the long-range interaction from
the one- and two-pion exchange Paris potential \cite{LAC80})
or phenomenological potentials fitted 
to the experimental data \cite{TAK89}. 
Although they had a relative success in reproducing the phase shifts 
these {\it hybrid models} are not fully consistent.

Chiral constituent quark models 
allow to incorporate the medium and long-range dynamics from the
quark substructure in a natural 
way. They were first applied to the non-strange baryon-baryon
interaction \cite{OBU90,FER93a} and later on to derive 
the potential between all members of the baryon octet \cite{FUJ96}.
In the last case the model parameters for the 
exchange of the two pseudoscalars (pion and kaon) and the 
full scalar and vector octet mesons were independently fitted to the
experimental data. This procedure, although very effective to fit the data,
does explicitly break the chiral construction of the potential and
masks on the model parameters some important physical effects, 
like the coupling to N$\Delta$ channels \cite{VAL95a}. 
In the following we will refer to the model of reference \cite{FER93a},
that has been detailed in section \ref{ch2}, 
although we will use results of the quark model approach of 
Ref. \cite{FUJ96} for comparison.

\subsection{Low-energy scattering parameters and deuteron properties}
\label{ch5.1}

The scattering length and the effective 
range provide a useful way to parametrize 
information on low-energy NN scattering because they are extremely sensitive to 
small variations on the strength of the NN force.
Furthermore, these parameters can be related to other observables 
such as the deuteron binding energy. Therefore, the correct prediction of the 
scattering length, the effective range and the deuteron 
properties should be the first test for any model 
of the NN system to satisfy. 
Table \ref{Lowen} resumes the results obtained by means 
of the chiral constituent quark model \cite{JUL02} 
compared to the quark
model approach of Ref. \cite{FUJ02} and
some standard NN potentials \cite{MAC89,STO94} and
experimental data. For a correct description of the $^1 S_0$ observables 
it is necessary to take into account 
the coupling to the $^5 D_0$ ${\rm N}\Delta$ 
channel, which provides an isospin-dependent mechanism generating additional
attraction. Hence a similitude between the $^1 S_0$ and 
the $^3 S_1$ NN partial waves, both getting the necessary attraction by 
the tensor coupling to a $D$ wave ($^3 D_1$ for $^3 S_1$) \cite{VAL95a}
comes out. Besides it is worth to mention that the
description of the low-energy scattering parameters requires 
a charge symmetry breaking term \cite{MAC89,ENT99} 
that has been taken into account by a 
slight modification of the chiral coupling constant.
In the same table the results for the deuteron properties are
shown, presenting a good agreement with the experimental data.
\begin{table}[t]
\caption{\label{Lowen}NN properties. $a_{s(t)}$ is
the singlet (triplet) scattering length and $r_{s(t)}$ is the
singlet (triplet) effective range. $E_d$
is the deuteron binding energy, $P_D$ the
$D-$state probability, $Q_d$ the quadrupole moment, 
$A_S$ the asymptotic $S$-state normalization, and
$\eta$ the $D/S$-state ratio.}
\begin{tabular}{@{}lllllll}
\br
& & CCQM \protect\cite{JUL02} & Ref. \protect\cite{FUJ02}  &Nijm II \protect\cite{STO94}&
Bonn B\protect\cite{MAC89} &Exp. \\
\mr
\multicolumn{7}{c}{ Low-energy scattering parameters} \\
\mr
 $^1S_0$ & $a_s$ (fm) & $-$23.759  &$-$23.76  &$-$23.739& $-$23.750&
$-$23.74$\pm$0.02  \\
         & $r_s$ (fm) &  2.68     &2.58 & 2.67& 2.71&2.77$\pm$0.05\\
 $^3S_1$ & $a_t$ (fm) & 5.461      &5.399 & 5.418 &5.424&5.419$\pm$0.007 \\
         & $r_t$ (fm) & 1.820      &1.730 &1.753 & 1.761&1.753$\pm$0.008\\
\mr
\multicolumn{6}{c}{ Deuteron properties} \\
\mr
\multicolumn{2}{l}{$E_d$ (MeV)} &$-$2.2242  & $-$2.225 &
$-$2.2246&$-$2.2246&$-$2.224575  \\
\multicolumn{2}{l}{$P_D$ ($\%$)}   & 4.85   & 5.49 & 5.64  & 4.99  &$-$\\
\multicolumn{2}{l}{$Q_d$ (fm$^2$)} &0.276  & 0.270 & 0.271  & 0.278 &
0.2859$\pm$0.0003\\
\multicolumn{2}{l}{$A_S$ (fm$^{-1/2}$)}& 0.891   & $-$ & 0.8845 &
0.8860&0.8846$\pm$0.0009\\
\multicolumn{2}{l}{$\eta $}      &0.0257    & 0.0253 & 0.0252 &
0.0264&0.0256$\pm$0.0004\\
\br
\end{tabular}
\end{table}

\subsection{Deuteron configurations}
\label{ch5.1.n}

Although the deuteron has been usually 
described as a NN isospin singlet in even partial waves,
i.e., $^3 S_1$ and $^3 D_1$, it could also be understood 
as a linear combination of pairs of baryonic resonances provided they
have the adequate quantum numbers. Such a description finds its 
natural framework in quark models for two reasons. First, from 
a quark model point of view baryon resonances are no
more than internally excited nucleons. These excitations may occur even at
low energies due to the virtual process ${\rm NN} \to {\rm NN}^* 
({\rm N}^*{\rm N}^*) \to {\rm NN}$ involving intermediate 
N$^*$'s. Second, all baryon-baryon
interactions can be generated from the same underlying quark-quark potential
and, therefore, described in a consistent way. 
Furthermore, the presence of colour for
the quarks adds the possibility of having new exotic
components such as hidden colour state configurations: two
colour octets coupled to a singlet. Although in some works 
their contribution to the
deuteron wave function has been found to be as large as $5\% $ \cite{MAT77}, 
it is possible to demonstrate that any hidden
colour state can be expressed as a linear 
combination of physical states, i.e., baryon-baryon states \cite{GON87}.

The most prominent low-lying even-parity nucleon resonances are the 
$\Delta$ and the N$^*$(1440). Being the deuteron
isoscalar the N$\Delta$ configuration is forbidden and
therefore $\Delta\Delta$ and NN$^*$(1440) components would be the
relevant non-nucleonic configurations. For the $\Delta\Delta$
configuration, reference \cite{ALL86} established a
significant upper limit of about $0.4\%$. Moreover, although there are no
experimental data on the NN$^*$(1440) configuration, this has been
advocated long ago to understand elastic proton-deuteron backward scattering
at energies above pion threshold \cite{KER69} or the angular distribution of
deuteron photodisintegration at energies above $E_\gamma =$ 100 MeV \cite
{HAD73}. Recent calculations have renewed the interest on these non-nucleonic
components as they could be indirectly observed in several reactions as for
example antiproton-deuteron annihilation \cite{DEN99}, subthreshold
antiproton production \cite{DOR95} or $pd \to dp$ processes \cite{UZI97}.

\begin{table}[t]
\caption{\label{ttt}Probability of the different deuteron components ($\%$)}
\begin{indented}
\item[]\begin{tabular}{@{}llllllll}
\br
\multicolumn{2}{c}{NN}&\multicolumn{2}{c}{NN$^*$(1440)}&
\multicolumn{4}{c}{$\Delta\Delta$}\\
$^3S_1$ & $^3D_1$ & $^3S_1$  & $^3D_1$ &
$^3S_1$ & $^3D_1$ & $^7D_1$ &
$^7G_1$\\
\mr
95.3780&4.6220&$-$&$-$&$-$&$-$&$-$&$-$\\
95.1989&4.5606&$-$&$-$&0.1064&0.0035&0.1243&0.0063\\
95.1885&4.5377& 0.0022&0.0148&0.1224&0.0036&0.1245&0.0063\\
\br
\end{tabular}
\end{indented}
\end{table}

There are several deuteron multichannel calculations
using different approximations. Reference \cite{KUS91} 
proposed a formulation in terms of quark-shell configurations, later on
projected onto physical channels. Reference \cite{GLO94} studied the 
effective numbers for different resonance configurations 
making use of baryon wave functions obtained from the diagonalization of a
quark-quark interaction containing gluon and pion exchanges in a harmonic
oscillator basis including up to 2$\hbar \omega$ excitations, and deuteron
wave functions obtained from the Paris potential. An upper 
limit of $1\%$ for $\Delta\Delta$ components and $0.1\%$ for NN$^*$(1440) 
was obtained. Possible effects of six-quark bags at short-distances
were considered in \cite{ZHA85}. In reference \cite{MAE00} the 
influence of N and $\Delta$ resonances on the NN interaction was studied. 
The most significant contribution was inferred 
from channels involving N and $\Delta$
ground states, although a quantitative calculation of the non-nucleonic
configurations was not performed. 

The most complete multichannel
calculation was done in \cite{ENT00} and \cite{JUL02b} using 
a chiral constituent quark model. The different configurations and
partial waves included in the calculation are shown in table 
\ref{ttt} together with their probabilities (for instance the first
row contains only the NN configurations $^3S_1$ and $^3D_1$).
In all cases the deuteron binding energy is correctly reproduced, being $E_d=-$%
2.2246 MeV. The first remarkable result (the last row of table \ref{ttt})
is that the probabilities
of NN$^*$(1440) channels are smaller than the $\Delta\Delta$ ones being these
later ones compatible with the experimental limits given in \cite{ALL86}.
As compared to other values in the literature, 
the total probability of $\Delta\Delta$ components is three times smaller 
than the one reported in \cite{ARE75b}, although presenting the
same distribution with respect to the different channels. 
Regarding the calculations of reference \cite{DYM90},
a similar distribution is obtained 
but the total probability being twice as much.
Nonetheless, the comparison with these results should be done with care.
For example, reference \cite{ARE75b}
carries out a coupled-channel calculation 
using for the ${\rm NN} \to \Delta\Delta$
transition potential a combination of $\pi$ and $\rho$ exchanges, and the
Reid soft core potential for the rest. 
Such calculations are not fully consistent since 
they use NN potentials (Reid soft core or Paris)
which were designed, in principle, without explicit $\Delta$
degrees of freedom. Therefore, parts of the contribution
from the $\Delta$ degrees of freedom are
implicitly included in the parameterization of the potentials.

Regarding the NN$^*$(1440) probabilities, they show a larger component of the
$^3D_1$ partial wave against the $^3S_1$ partial wave, in agreement with 
the ordering obtained by other calculations \cite{ROS75,MAE00}.
This can be understood taking into account that
the tensor coupling, which is the main responsible for the
presence of non-nucleonic components on the
deuteron, is much stronger
for the $^3S_1 ({\rm NN}) \to \! ^3D_1 [{\rm NN}^*(1440)]$ transition than
for the $^3D_1 ({\rm NN}) \to \! ^3S_1 [{\rm NN}^*(1440)]$ one,
enhancing in this way the $D$-wave influence with respect to the $S$-wave component.
Concerning the absolute value of the probabilities, they are a factor ten
smaller than those reported in reference \cite{ROS75}, where
an estimation of 0.17$\%$ for the NN$^*$(1440)
configuration was obtained (0.06$\%$ for the $^3S_1$
and 0.11 $\%$ for the $^3D_1$ partial wave). 
The difference with the result 
of the chiral constituent quark model may be
understood in the following way. The deuteron is calculated using
the NN Reid hard-core potential. When including NN$^*$(1440)
components, the channel coupling induces an attractive
interaction on the NN system that needs to be subtracted
out. Such a subtraction is done by reducing the intermediate range attraction
of the central part of the Reid hard-core potential without
modifying the tensor part (as done in reference \cite{HAA74}
to calculate the probability of $\Delta\Delta$ components).
So, the strength of the tensor coupling to the
NN$^*$(1440) state is not strongly constrained.
In fact, a variation of this strength may be
compensated by a corresponding modification of the
intermediate range attraction. Thus, the balance of these
two sources of attraction cannot be disentangled in a clearcut way.
This is a similar problem to the one arising in the $^1S_0$ NN partial
wave when the coupling to the N$\Delta$ system is included.
The same attractive effect may be obtained by a central potential or
by a tensor coupling to a state with larger mass, being necessary
other observables to discriminate between the two processes \cite{GRE74}.

\begin{figure}[t]
\vspace*{-0.3cm}
\mbox{\psfig{figure=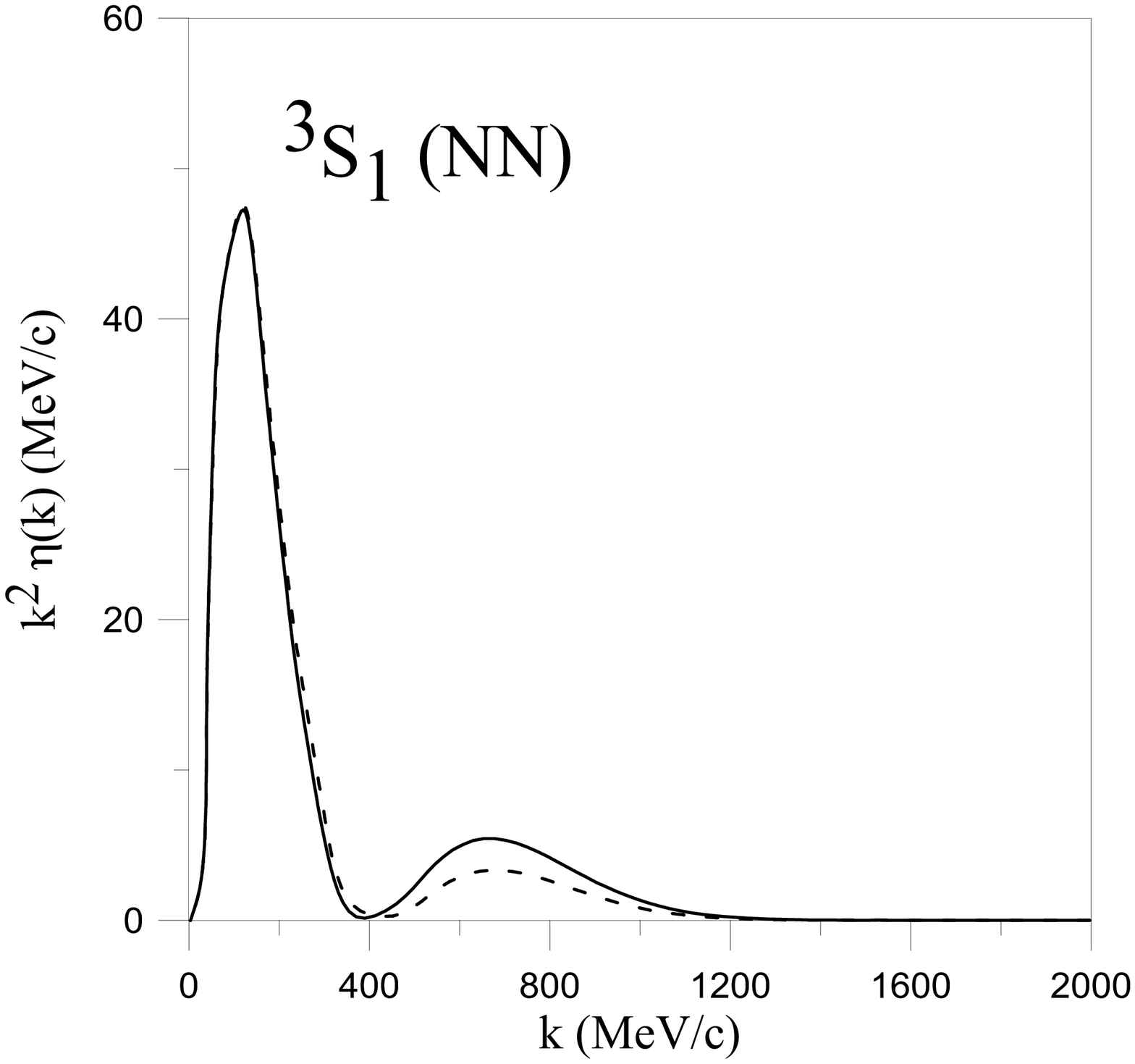,height=2.2in,width=2.4in}}
\hspace*{0.1cm}
\mbox{\psfig{figure=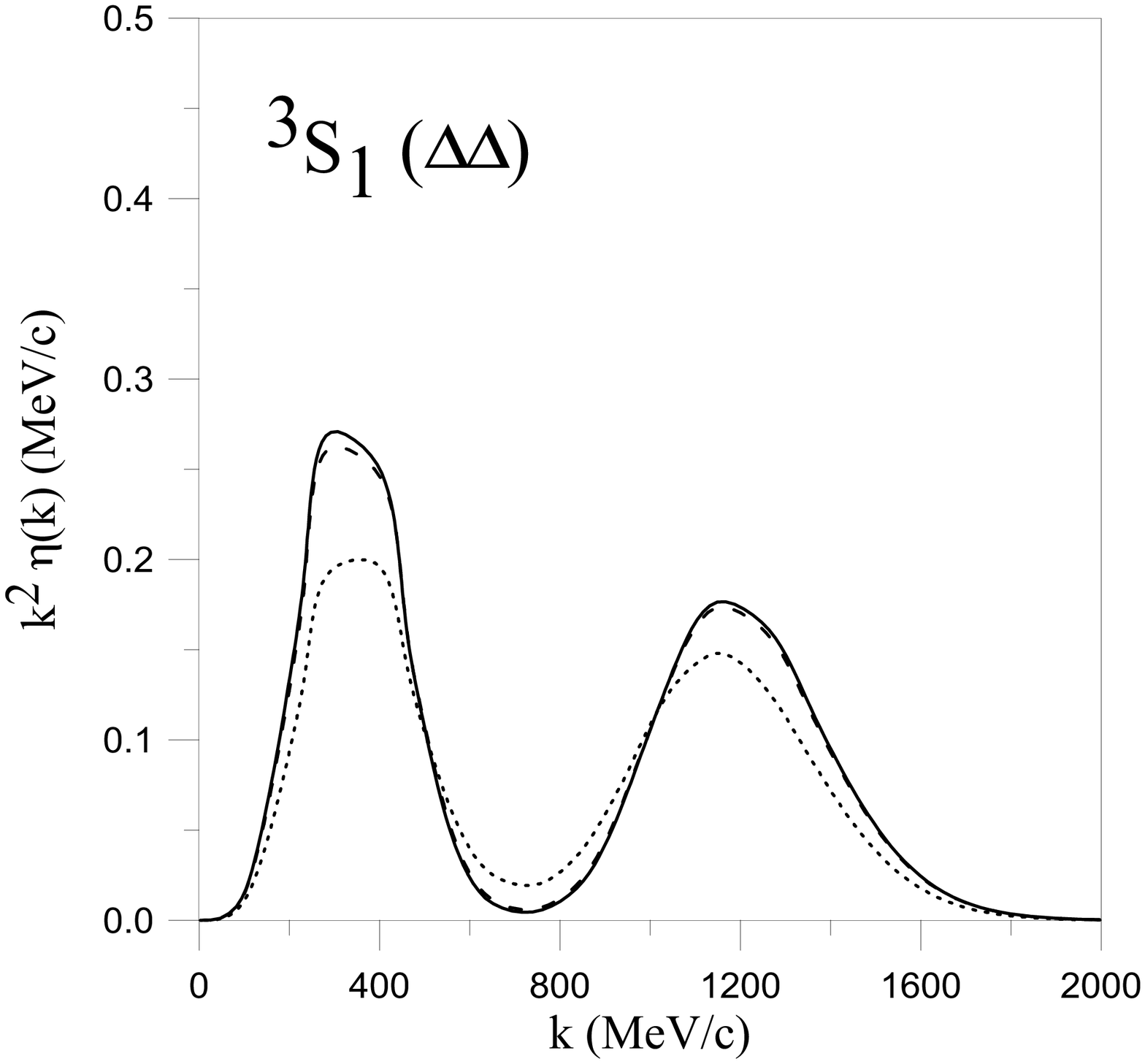,height=2.2in,width=2.4in}}
\vspace*{-0.3cm}
\caption{Momentum distribution for the $^3S_1({\rm NN})$ and the
$^3S_1(\Delta\Delta)$ channels.}  
\label{fig2c4}
\end{figure}
Although the probability of $\Delta\Delta$ components is small, 
its influence for some specific observables can be important.
Figure \ref{fig2c4} shows the momentum distribution $k^2 \eta (k)$ 
($\eta(k)=|\Psi(k)|^2$ is the probability density in momentum
space) for the $S$-wave deuteron components. For the $^3S_1({\rm NN})$
case, the solid line represents
the results including all NN and $\Delta\Delta$ partial waves while the
dashed line refers to the calculation only with NN channels. For
the $^3S_1(\Delta\Delta)$ case the dotted line represents the result
obtained with all NN channels and only the $^3S_1(\Delta \Delta)$
partial wave, while the dashed line includes all NN channels 
and the most important $\Delta\Delta$ partial waves, $^3S_1$ and $^7D_1$, 
being almost identical to the full calculation.
As can be observed, the $\Delta\Delta$
components extend to higher momenta regions. This may influence the
structure function $B(q)$ which presents a zero for momentum around 7.1
fm$^{-1}$. Including only NN components, this result always comes lower,
being the $\Delta\Delta$ component a possible candidate to solve the
problem \cite{DYM90,SIT87}. 
\begin{figure}[t]
\vspace*{-0.3cm}
\mbox{\psfig{figure=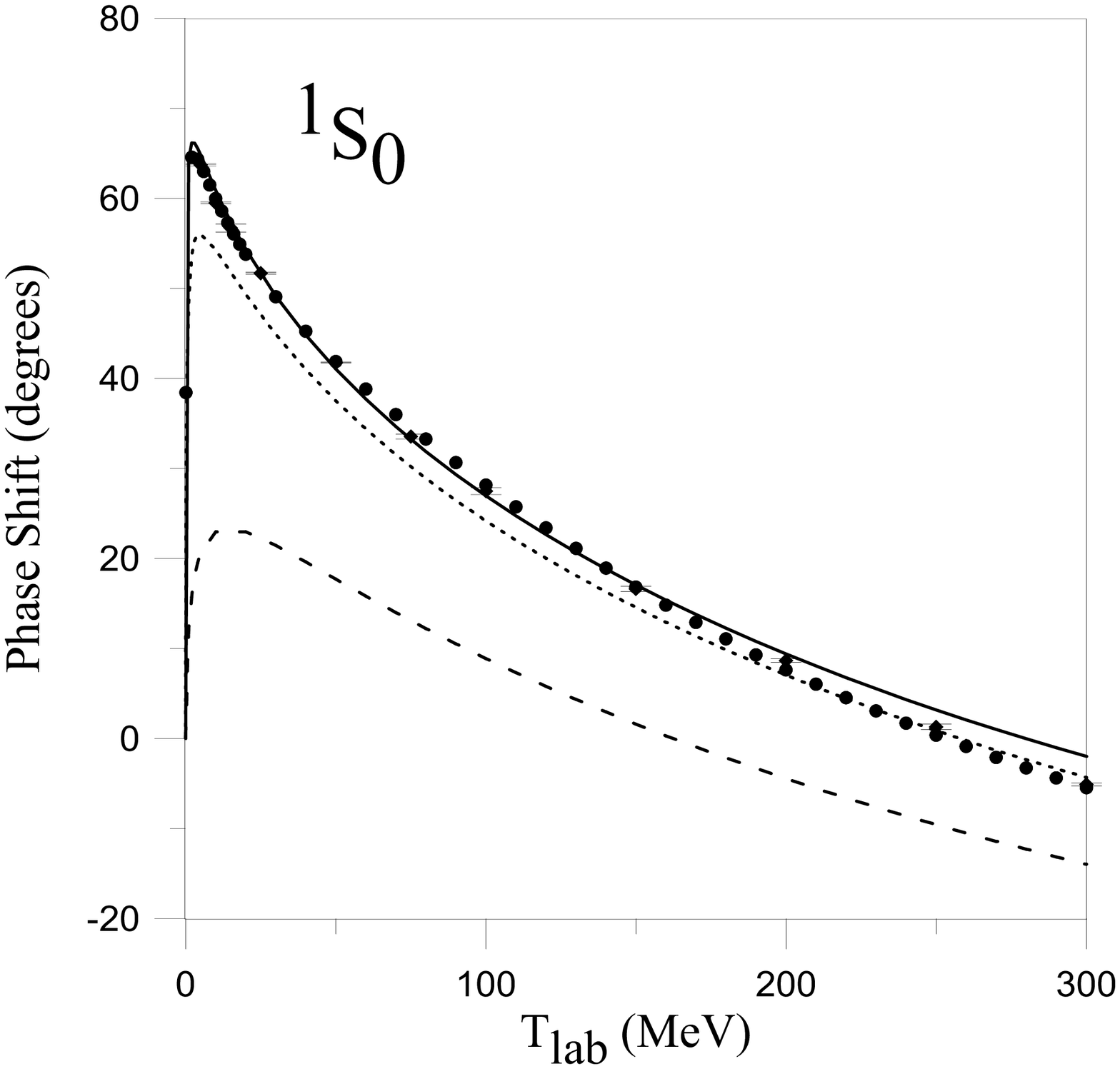,height=2.2in,width=2.4in}}
\hspace*{0.1cm}
\mbox{\psfig{figure=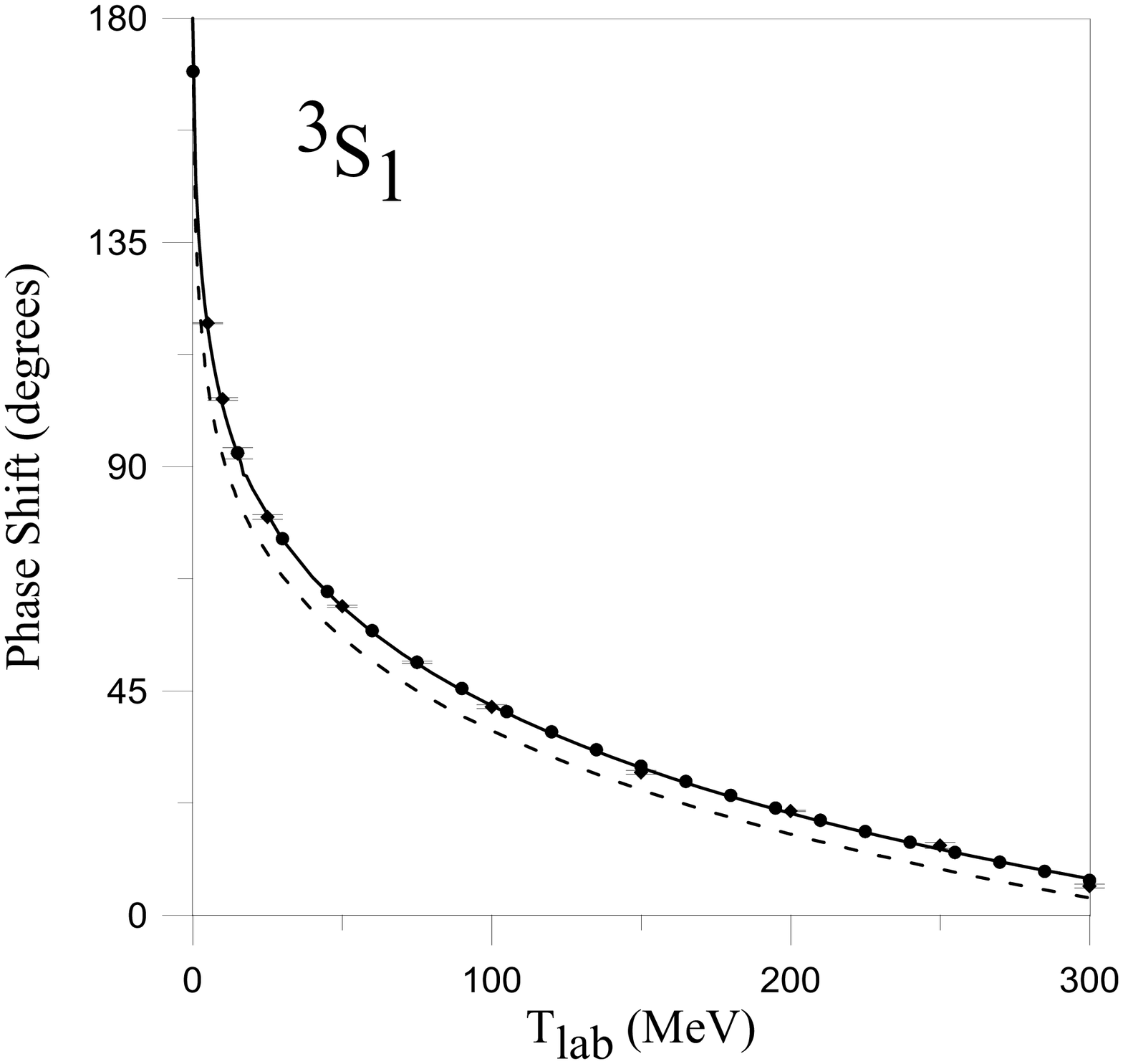,height=2.2in,width=2.4in}}
\vspace*{-0.3cm}
\caption{$^1 S_0$ and $^3 S_1$ NN phase shifts.}
\label{fig3c4}
\end{figure}

\subsection{NN phase shifts}
\label{ch5.2}

A calculation of the NN phase shifts below the pion threshold 
is presented in this section. 
It includes coupling to $\Delta\Delta$ channels for iso-singlet ($T=0$) 
partial waves, and to N$\Delta$ and $\Delta\Delta$ 
channels for iso-triplet ($T=1$) partial waves. 
Experimental data have been obtained through the interactive 
program SAID \cite{SAI00} corresponding to the neutron-proton solution SP98.

For the discussion of the results the phase shifts are divided into
three groups. First, the $L=0$ partial waves. They are the most
sensitive to the short-range part of the interaction and therefore one would
expect that quark-exchange effects played for them
an important role. A second group will be
$P$, $D$ and $F$ waves. They are still sensitive to the short-range part and
therefore to quark exchanges, but the middle-range and spin-orbit terms
will also play a relevant role. Finally, those partial waves with orbital 
angular momentum $L > 3$ depend basically on the long-range 
part of the interaction. Although quark exchanges are 
not expected to be relevant for them, the correct description
of these phase shifts supposes a crucial test to the chiral 
symmetry hypothesis used to derive the basic quark-quark interaction.

\subsubsection{S partial waves}
\label{ch5.2.1}
\begin{figure}[t]
\vspace*{-0.3cm}
\mbox{\psfig{figure=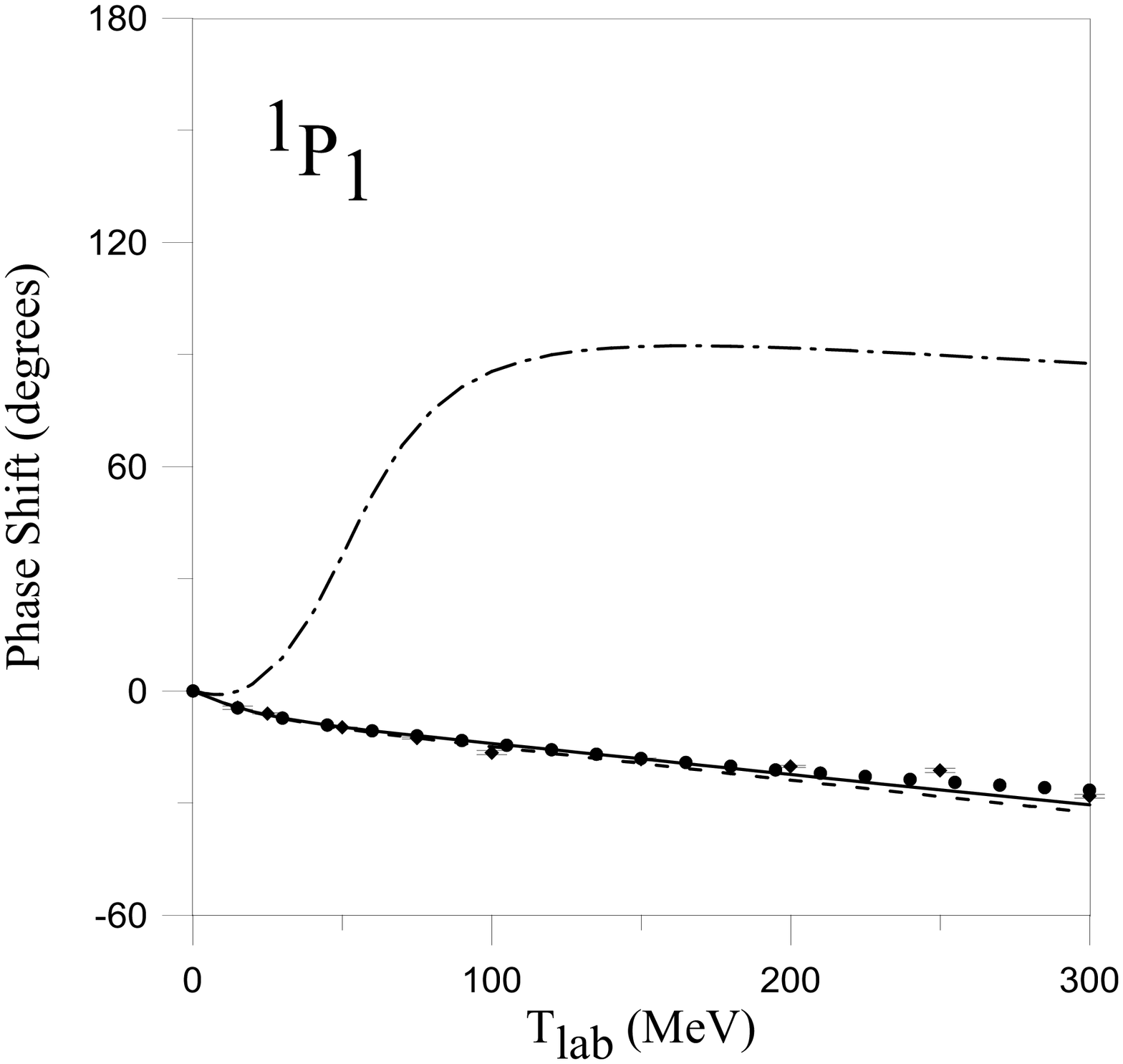,height=2.2in,width=2.4in}}
\hspace*{0.1cm}
\mbox{\psfig{figure=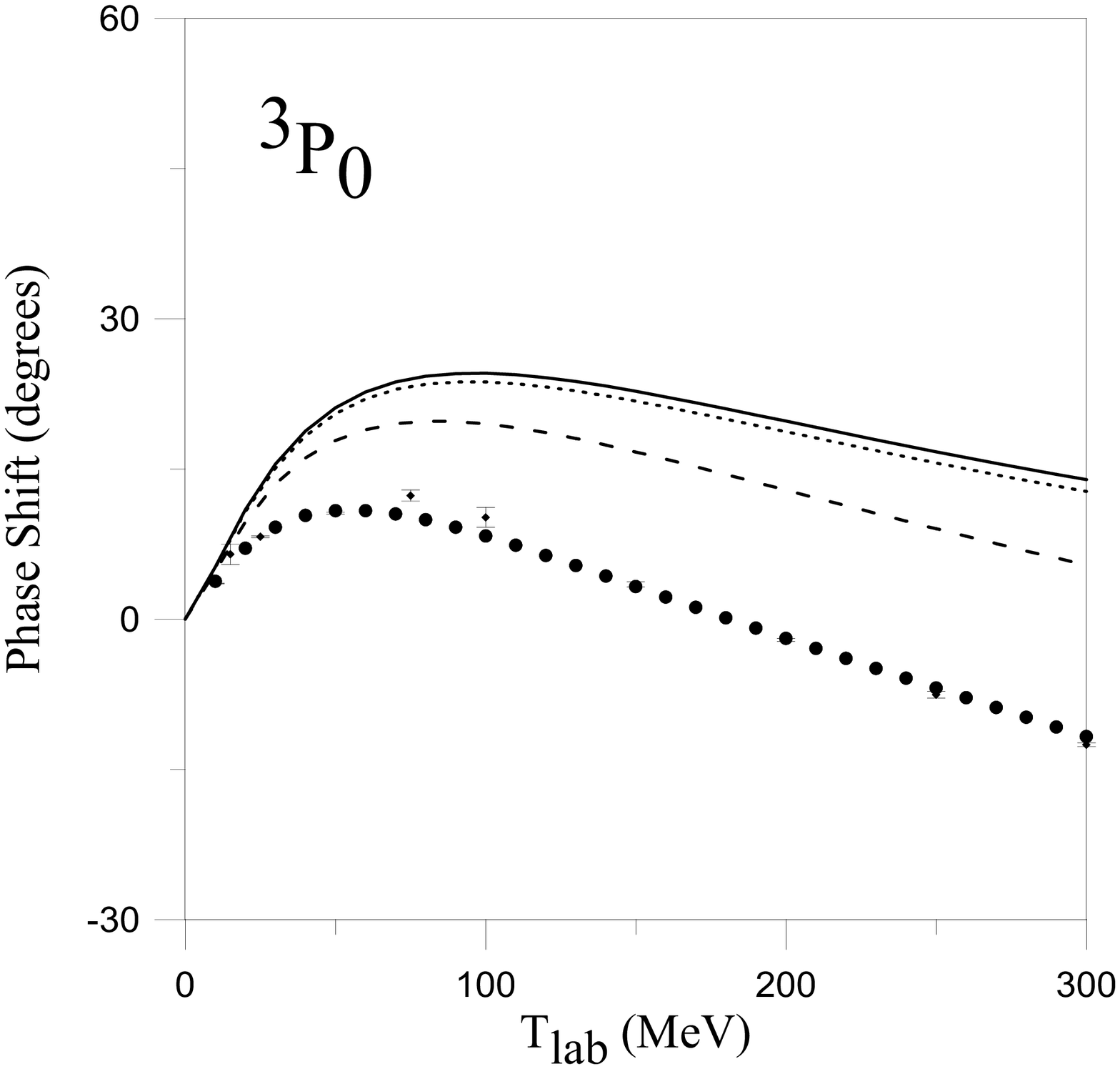,height=2.2in,width=2.4in}}
\mbox{\psfig{figure=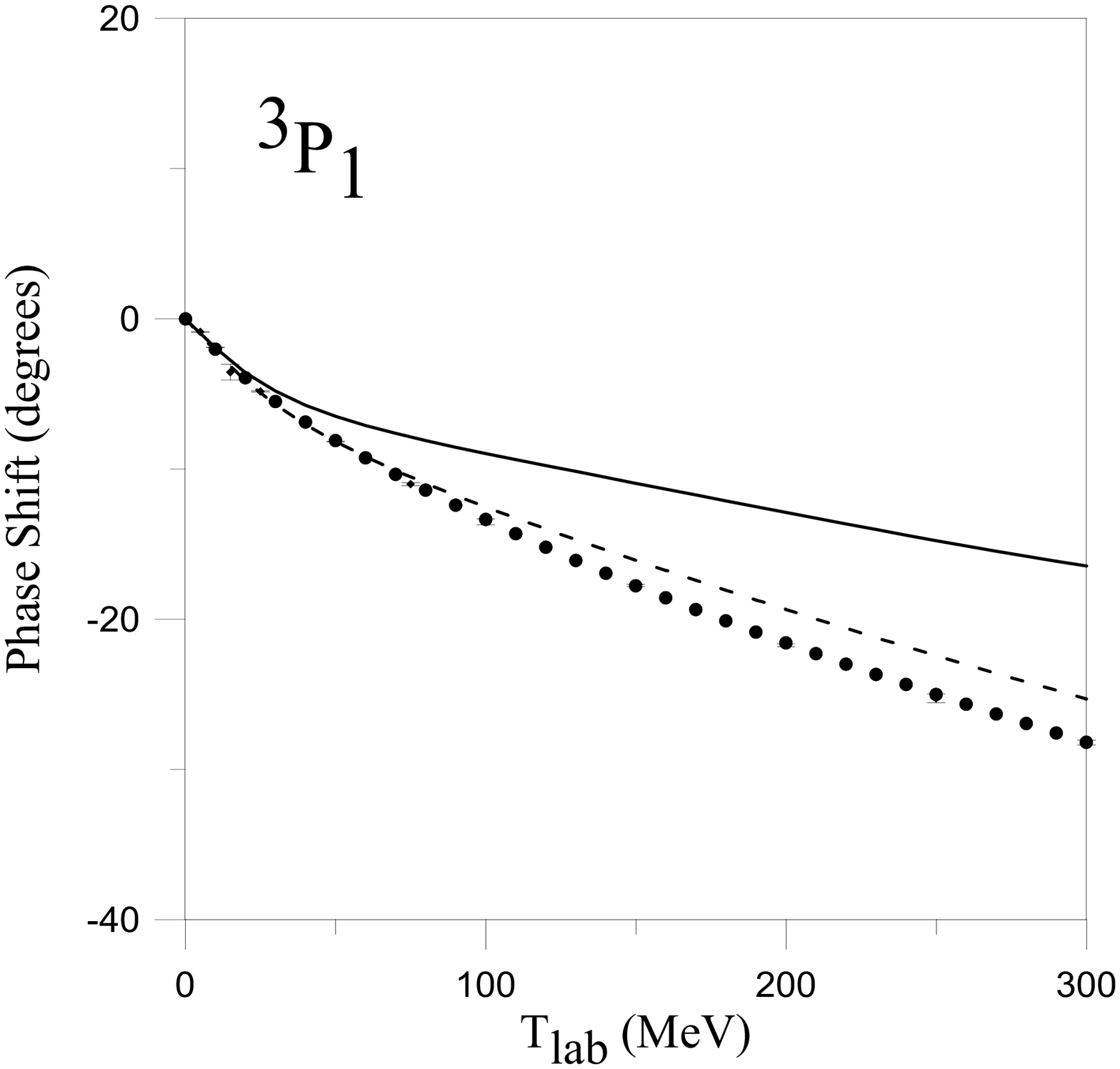,height=2.2in,width=2.4in}}
\hspace*{0.1cm}
\mbox{\psfig{figure=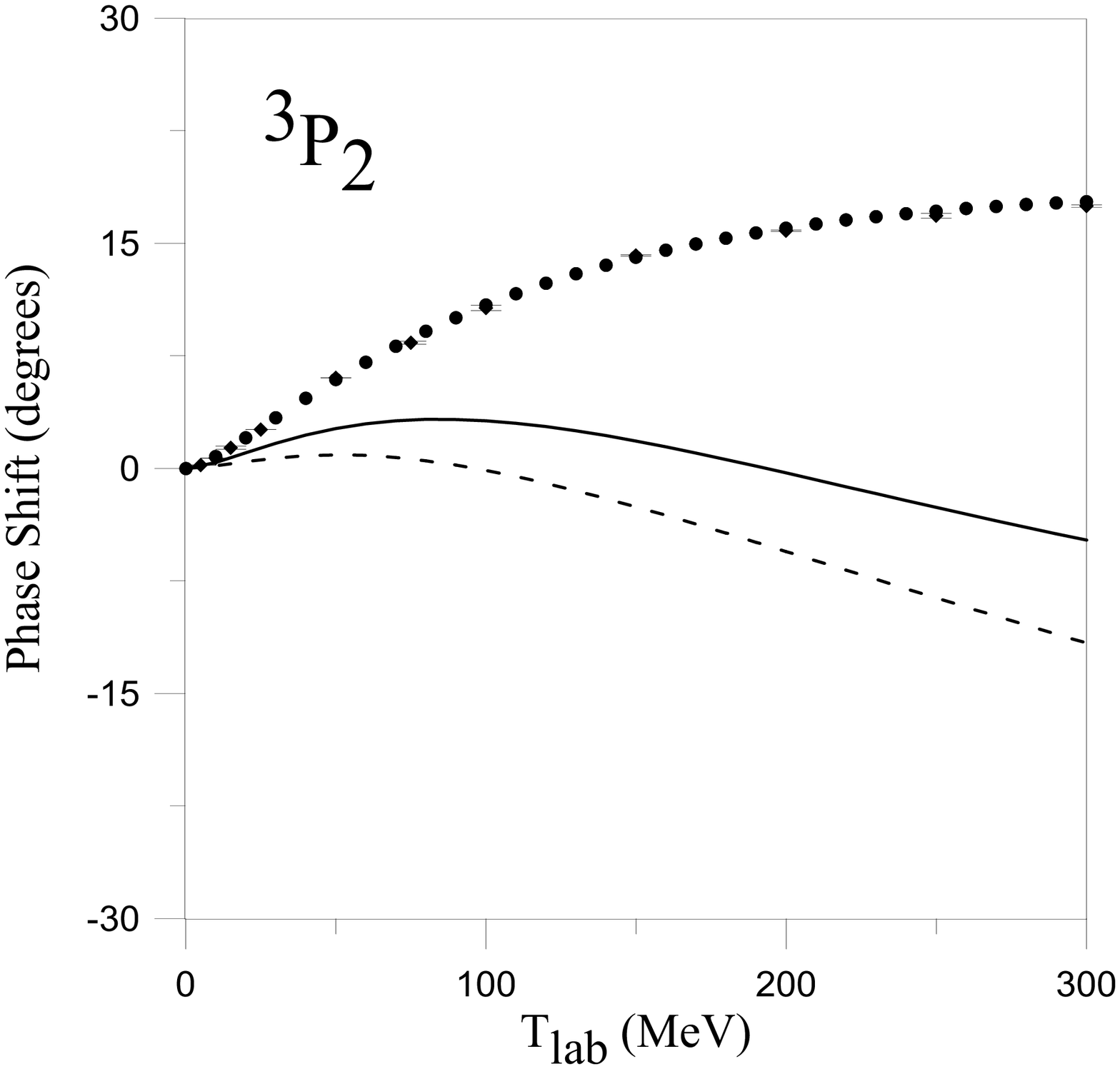,height=2.2in,width=2.4in}}
\vspace*{-0.3cm}
\caption{$^1P_1$ and $^3P_J$ NN phase shifts.}
\label{fig4c4}
\end{figure}

Figure \ref{fig3c4} shows the results for the $^1 S_0$ and $^3S_1$ partial
waves. A perfect fit is obtained for $^1S_0$ up to 250 MeV
laboratory kinetic energy and up to 300 MeV for $^3S_1$.
As can be checked, the chiral components of the NN interaction (dashed line) 
do not provide enough attraction to reproduce the experimental
data. As previously mentioned, in the $^1S_0$ channel 
the required attraction is supplied by addition of a coupling
to the $^5 D_0$ ${\rm N}\Delta$ channel (dotted line). 
A complete agreement with the experimental data is obtained when the coupling 
to $\Delta\Delta$ channels is included (solid line). This result shows 
that the requirements of chiral symmetry (the presence and parameters
of the scalar exchange between quarks) are fulfilled by the data provided
that one includes the necessary additional physics into the problem. 

For the iso-triplet partial wave $^3S_1$,
the coupling to $\Delta\Delta$ channels has a 
smaller influence on the phase shift as can be seen on the figure,
and can be simulated by a fine tuning of the mass of
the scalar boson \cite{ENT00}. 

Let us emphasize that it is the strong spin-isospin 
independent repulsion generated by quark antisymmetry
effects on the OPE what allows to 
reproduce the behaviour of the $S$-wave phase shifts.

\subsubsection{P, D and F partial waves}
\label{ch5.2.2}
\begin{figure}[t]
\vspace*{-0.3cm}
\mbox{\psfig{figure=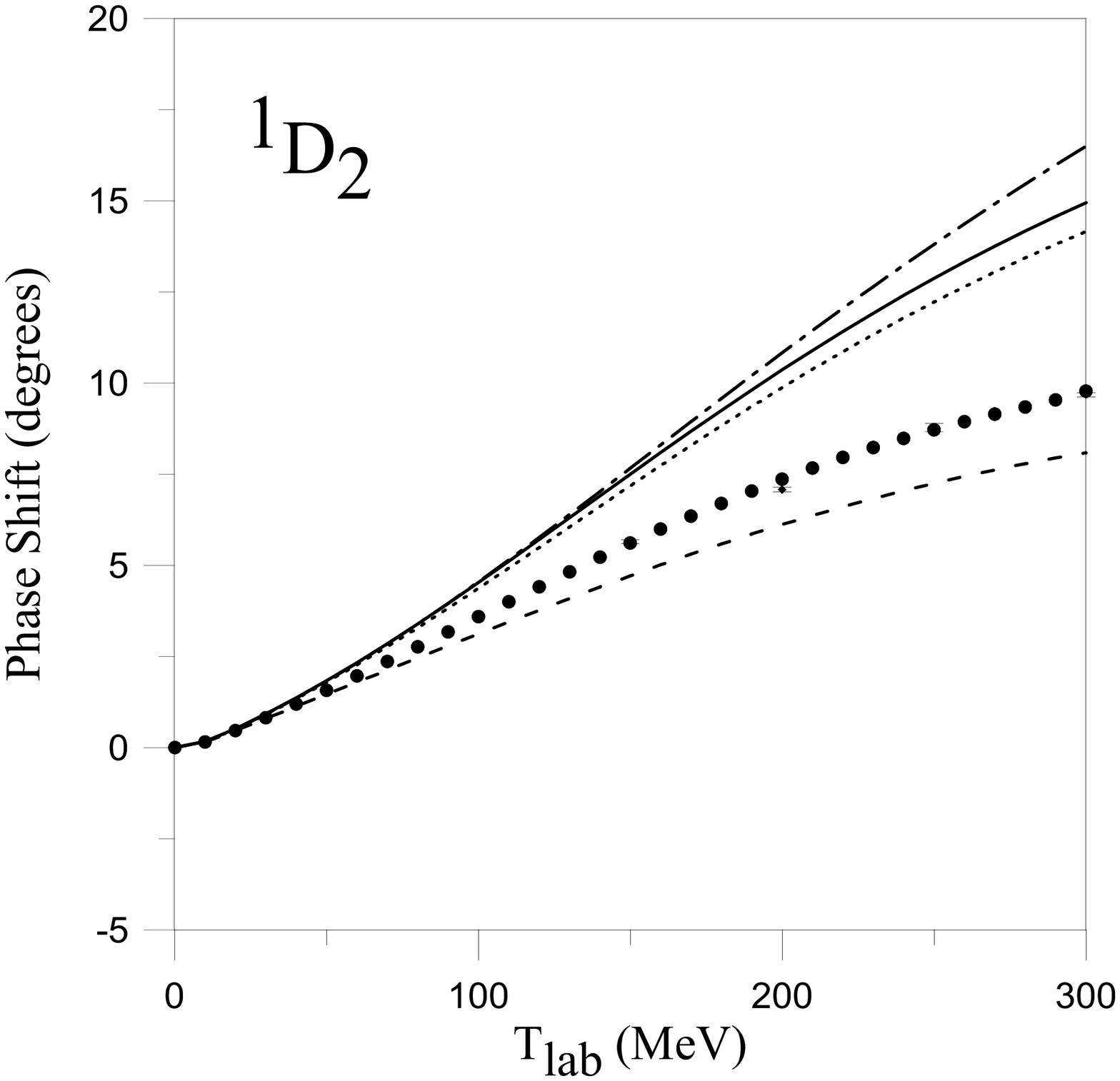,height=2.2in,width=2.4in}}
\hspace*{0.1cm}
\mbox{\psfig{figure=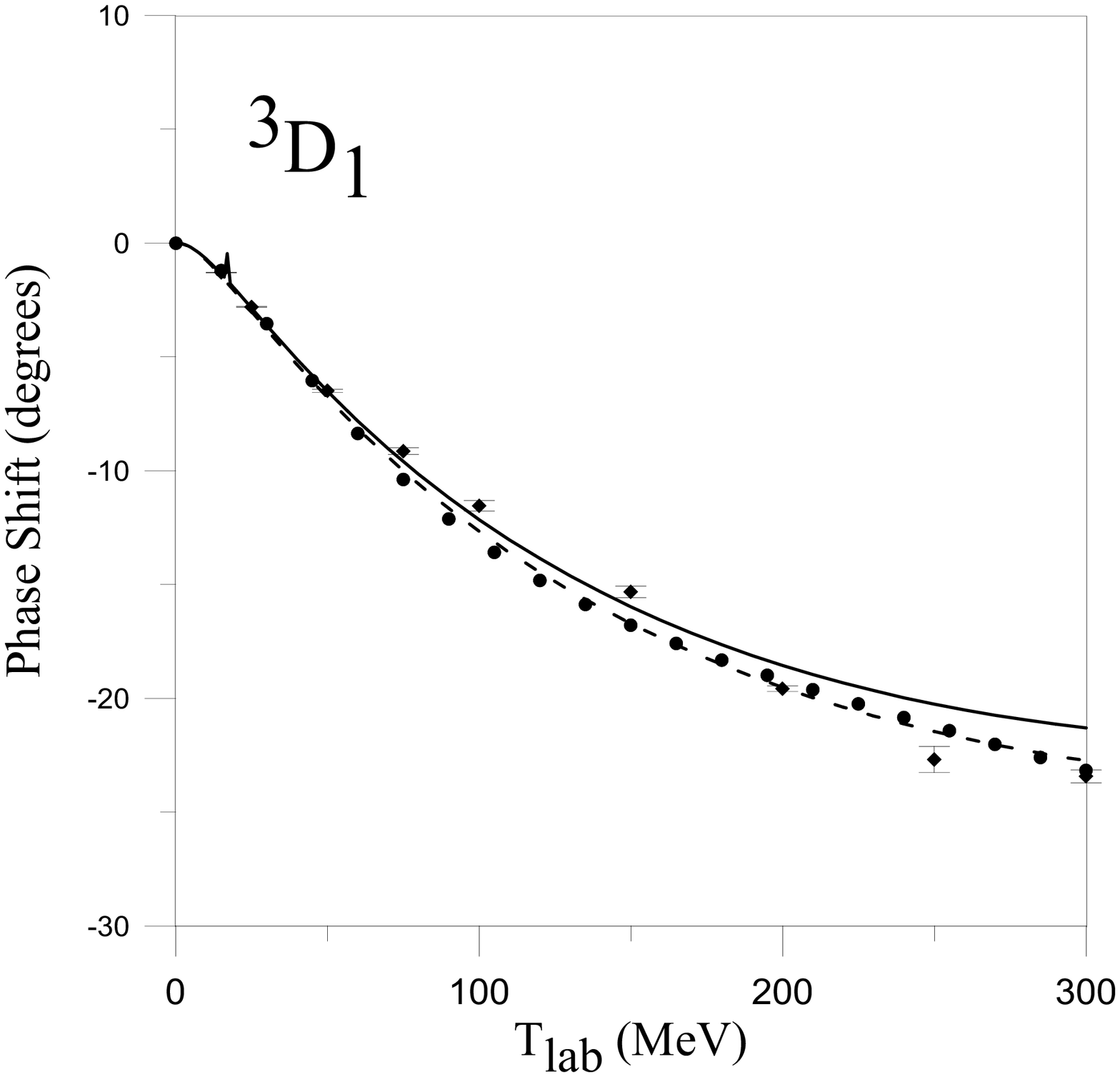,height=2.2in,width=2.4in}}
\mbox{\psfig{figure=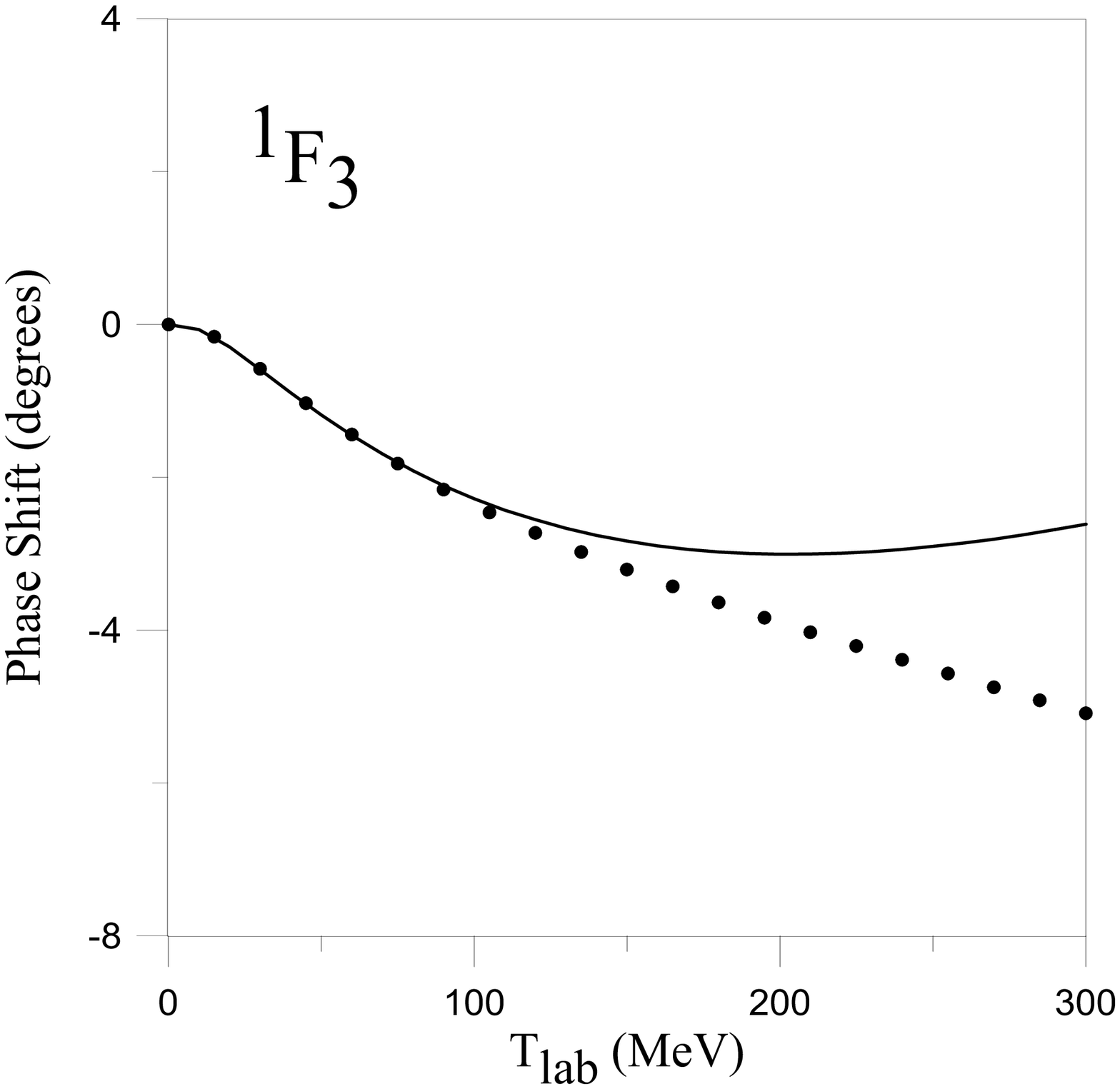,height=2.2in,width=2.4in}}
\hspace*{0.1cm}
\mbox{\psfig{figure=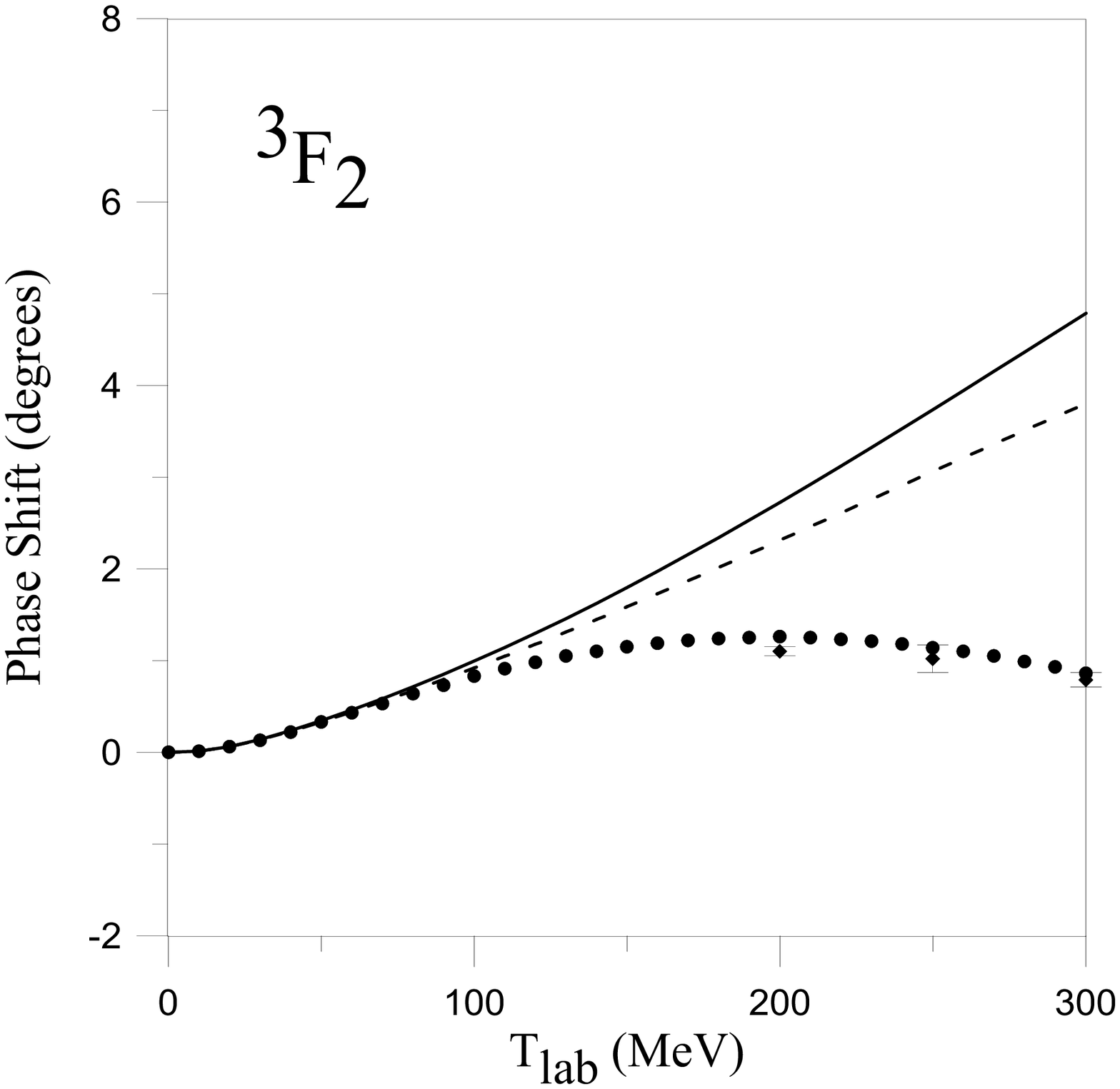,height=2.2in,width=2.4in}}
\vspace*{-0.7cm}
\caption{Some D and F NN phase shifts.}
\label{fig5c4}
\end{figure}
Among the four $L= 1$ partial waves, only the $^1P_1$ is not affected by the
spin-orbit interaction. As seen in figure \ref{fig4c4} the 
chiral constituent quark model result (solid line) is in 
perfect agreement with the experimental data. It is interesting to
notice that for this partial wave quark-exchange effects
still play an important role. Indeed, in the same figure, 
the calculation done by removing the terms coming
from quark antisymmetry is shown (dashed-dotted line), 
and the phase shift becomes attractive.
This result is easily understood in terms of the direct
OPE interaction (the dominant OGE interaction coming from
the spin-spin term does not contribute to $P$-waves 
because of its $\delta$-like behaviour).
At short-range the direct OPE potential is repulsive for
$S$ waves, but it is attractive for the $^1P_1$ wave due to the different sign of
the spin-isospin matrix element. Only when quark antisymmetry terms 
are considered the OPE produces the correct experimental behaviour.
\begin{figure}[t]
\vspace*{-0.3cm}
\mbox{\psfig{figure=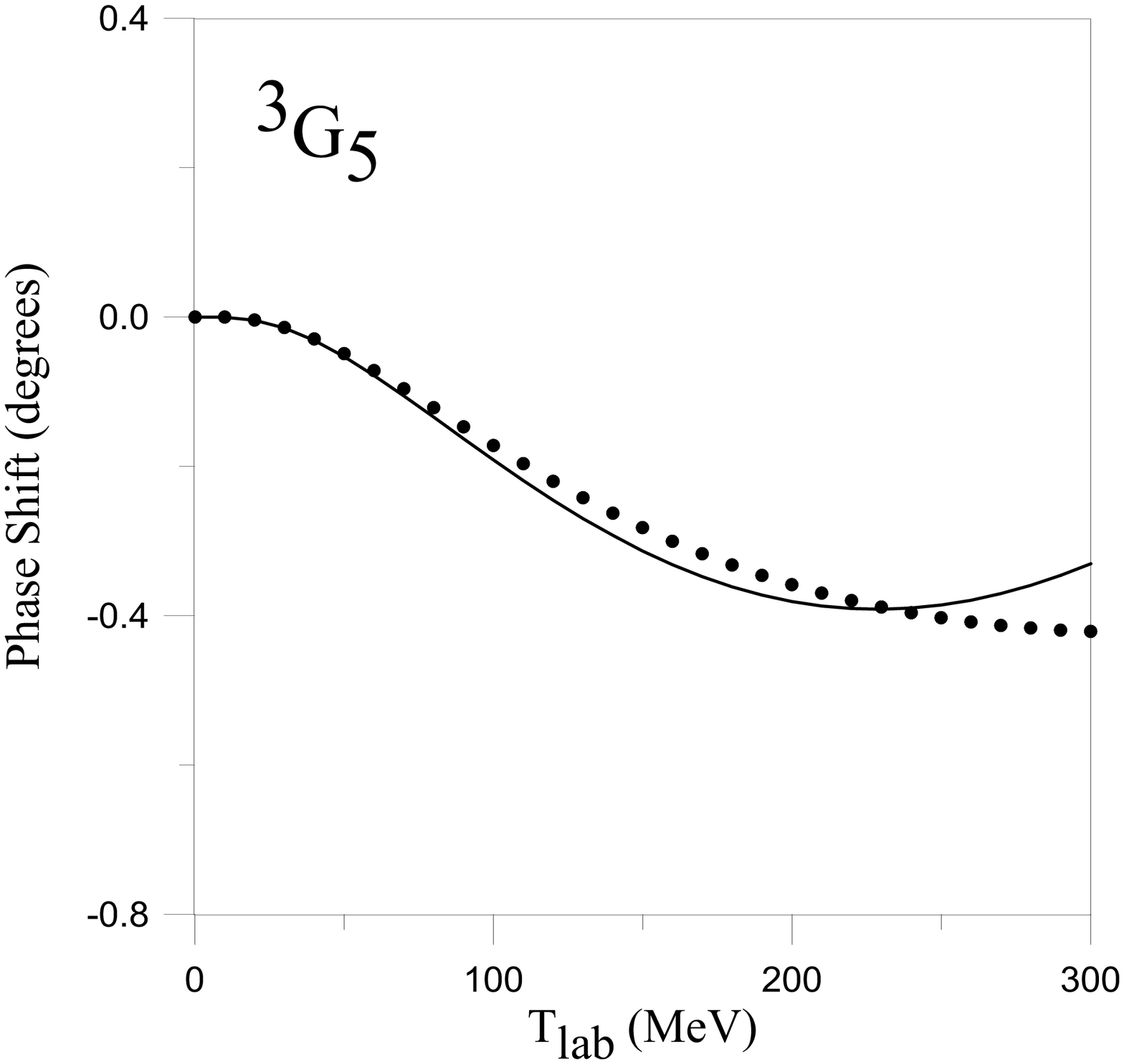,height=2.2in,width=2.4in}}
\hspace*{0.1cm}
\mbox{\psfig{figure=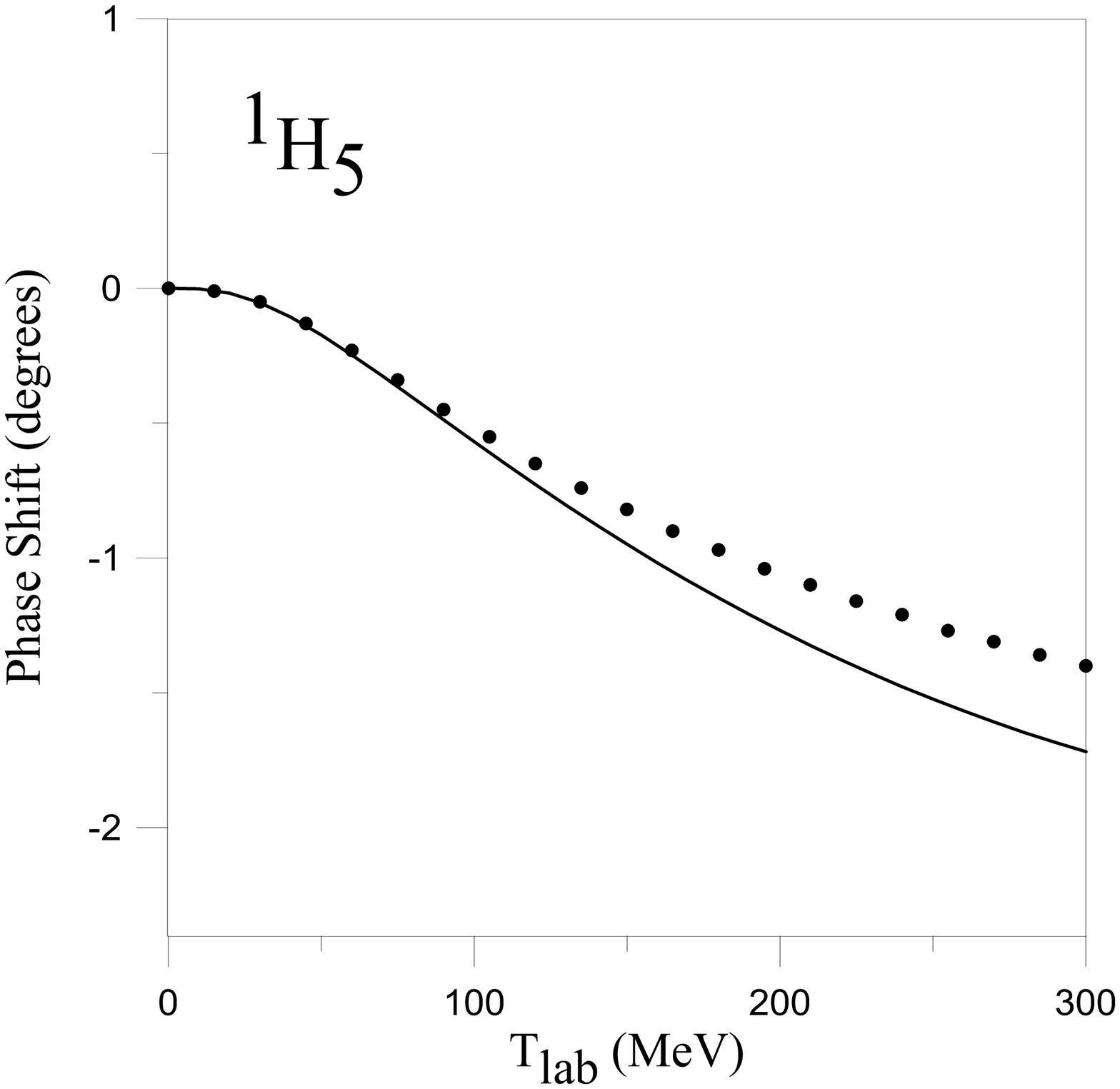,height=2.2in,width=2.4in}}
\mbox{\psfig{figure=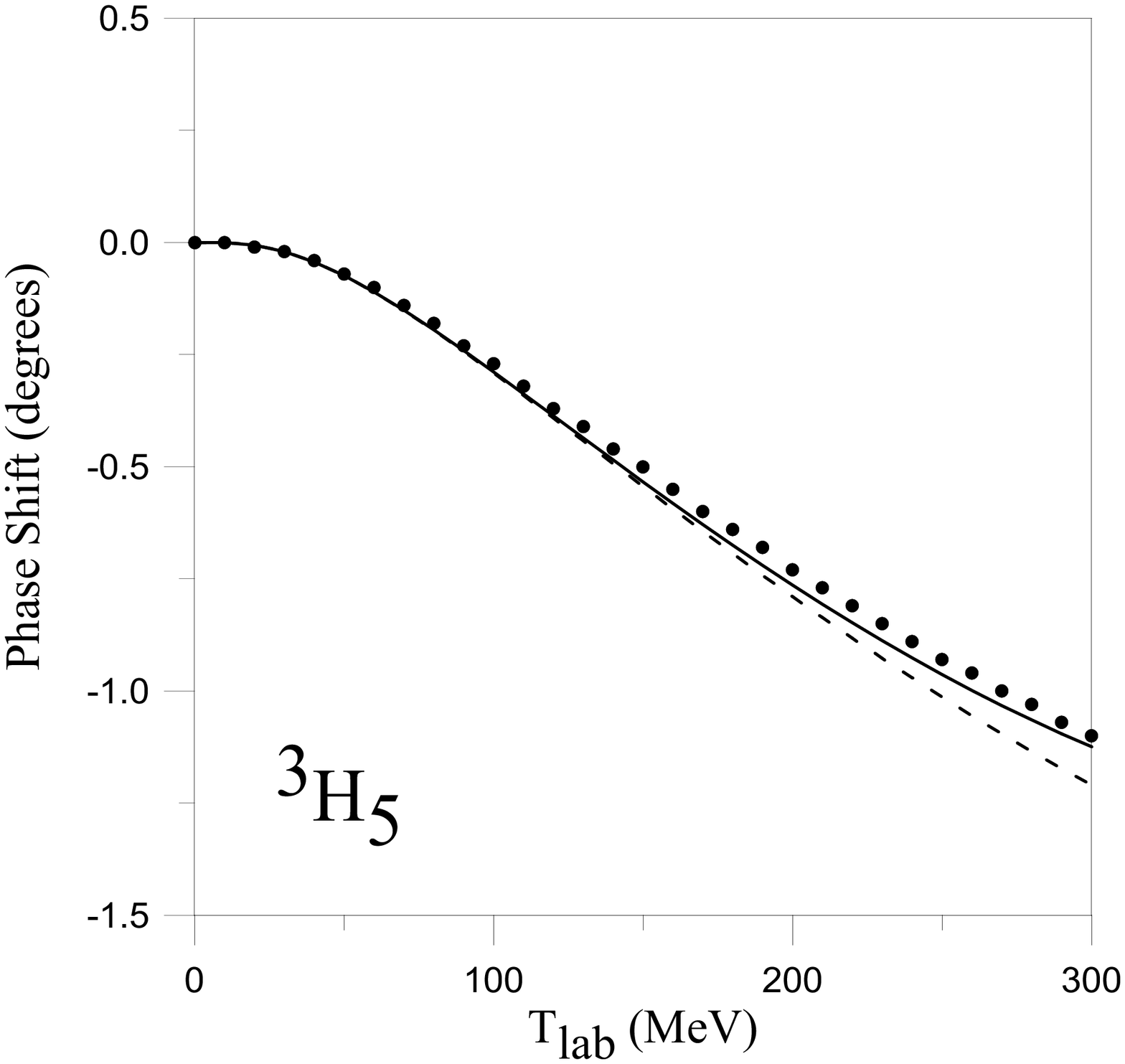,height=2.2in,width=2.4in}}
\hspace*{0.1cm}
\mbox{\psfig{figure=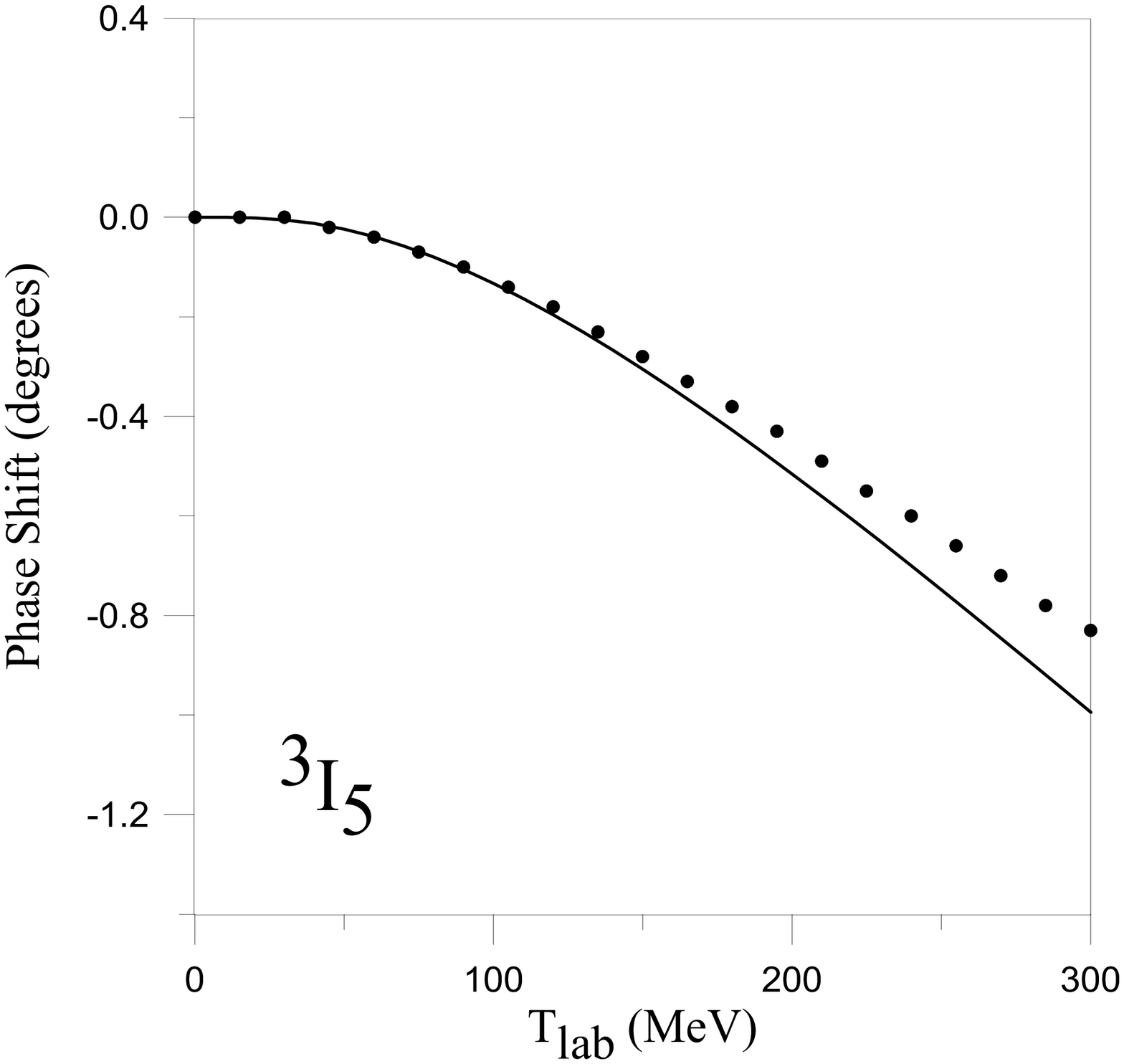,height=2.2in,width=2.4in}}
\vspace*{-0.2cm}
\caption{$J=5$ NN phase shifts.}
\label{fig7c4}
\end{figure}

In the same figure the $^3P_J$ triplet phase shifts are shown 
(notation as in figure \ref{fig3c4}). 
One observes, as has been explained for $S$-waves, how the coupling
to $\Delta\Delta$ channels in higher angular momentum iso-triplet partial
waves is very small. The coupling to N$\Delta$ channels does
not improve the description. The failure to reproduce the data
may be associated to the lack of spin-orbit interaction. 

Some selected $D$ waves are shown in figure \ref{fig5c4}
(notation as in figure \ref{fig3c4}). 
As a general trend, one can say that the model gives too 
much attraction, except for the $^3D_1$ partial
wave where the experimental phase shifts are perfectly reproduced. While for
iso-triplet partial waves the coupling to N$\Delta$ channels is still
important (worsening the quality of the results), the coupling to 
$\Delta\Delta$ channels does not produce any considerable effect. 
For the $^1D_2$ partial wave the
result obtained switching off the antisymmetry effects is again displayed
by a dashed-dotted line, showing clearly how the
influence of antisymmetry diminishes as compared to $S$ and $P$ waves.
It is important to note that the
$D$-waves are correctly reproduced up to about 80-100 MeV, which may be
considered as the characteristic window where the NN interaction is
basically governed by chiral symmetry \cite{KAI98}. Higher energies require an
improvement in the description of short-range effects, as could be the
spin-orbit force which still plays an important role for $D$ waves.
 
The same figure \ref{fig5c4} includes some selected $F$-wave phase shifts. 
In general there is a better agreement than in $D$-waves up to higher energies
of 150-200 MeV. For iso-triplet partial waves the 
coupling to N$\Delta$ channels produces too much attraction, as it was the case
for $P$ and $D$ waves, but the description is
of the same quality as that of effective theories \cite{KAI97,KAI98}.
A comment about the $^3 F_2$ phase shift is in order. This wave is coupled
to the $^3 P_2$ wave, which has a strong influence of the spin-orbit
interaction. The spin-orbit interaction is attractive for the $^3 P_2$ wave
and repulsive for the $^3 F_2$. Therefore, one expects that when the $^3 P_2$
wave approaches the experimental data, the $^3 F_2$ will also do the same.

\subsubsection{G, H and I partial waves}
\label{ch5.2.3}
\begin{figure}[t]
\vspace*{-0.3cm}
\mbox{\psfig{figure=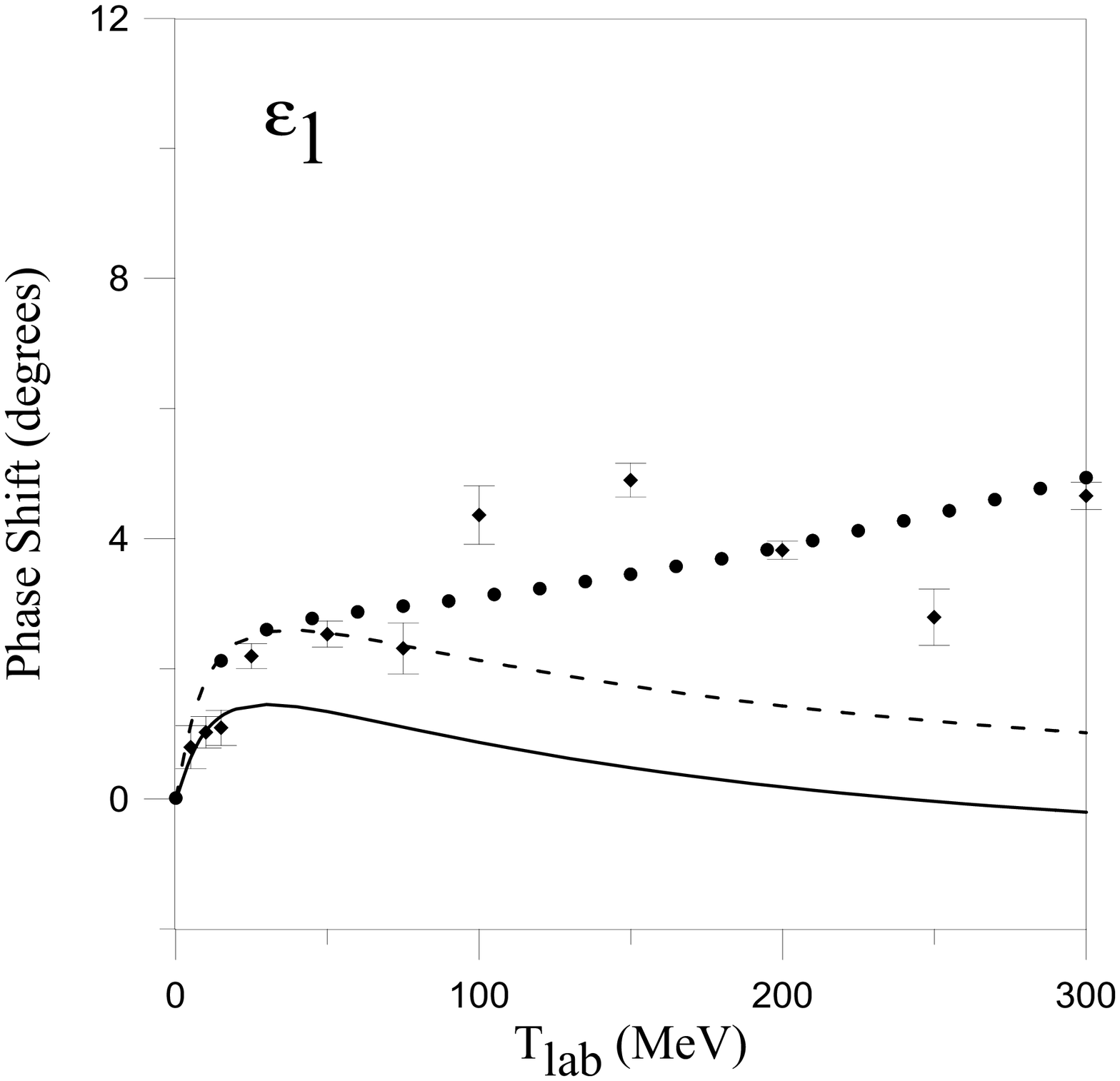,height=2.2in,width=2.4in}}
\hspace*{0.1cm}
\mbox{\psfig{figure=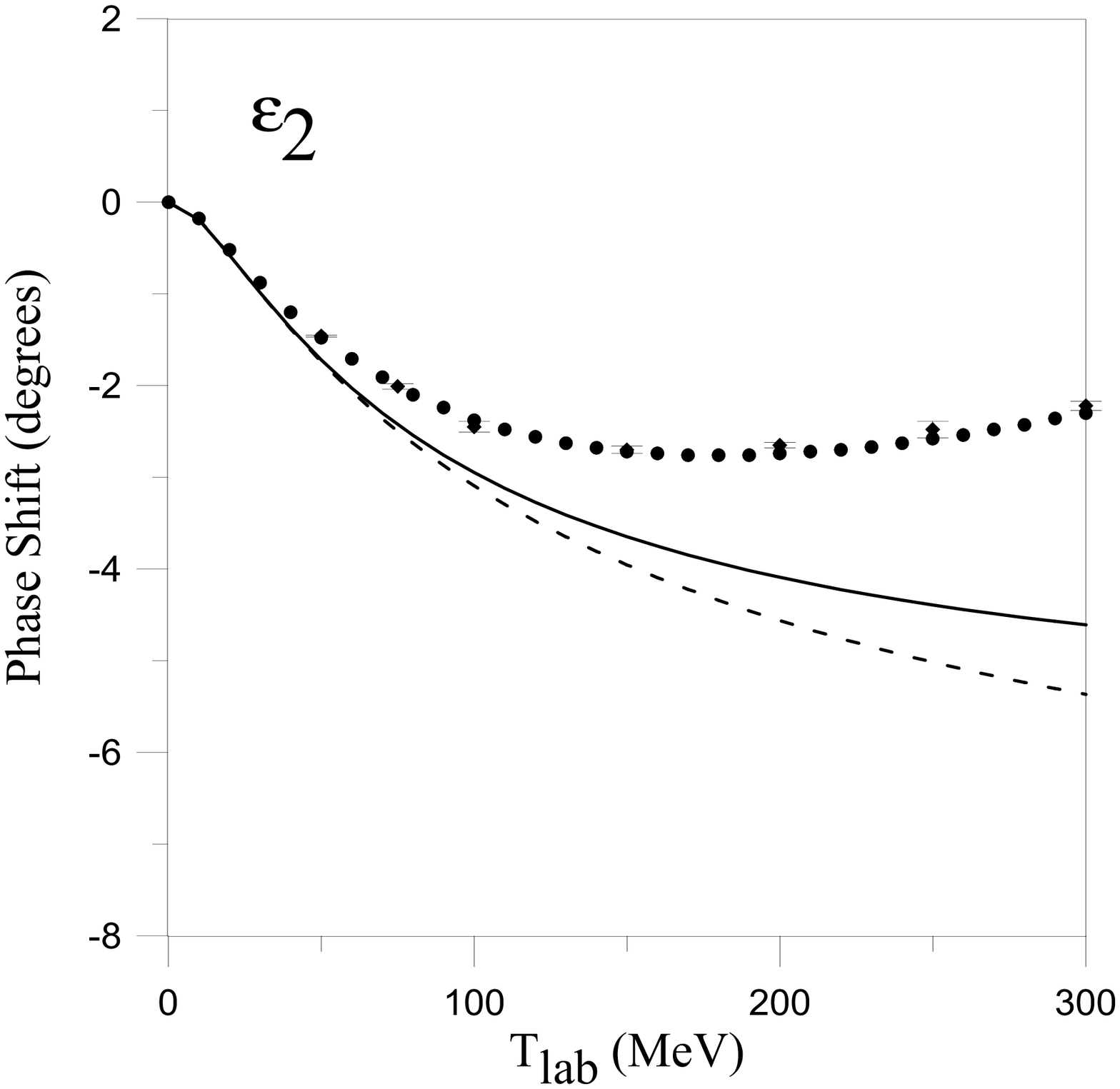,height=2.2in,width=2.4in}}
\mbox{\psfig{figure=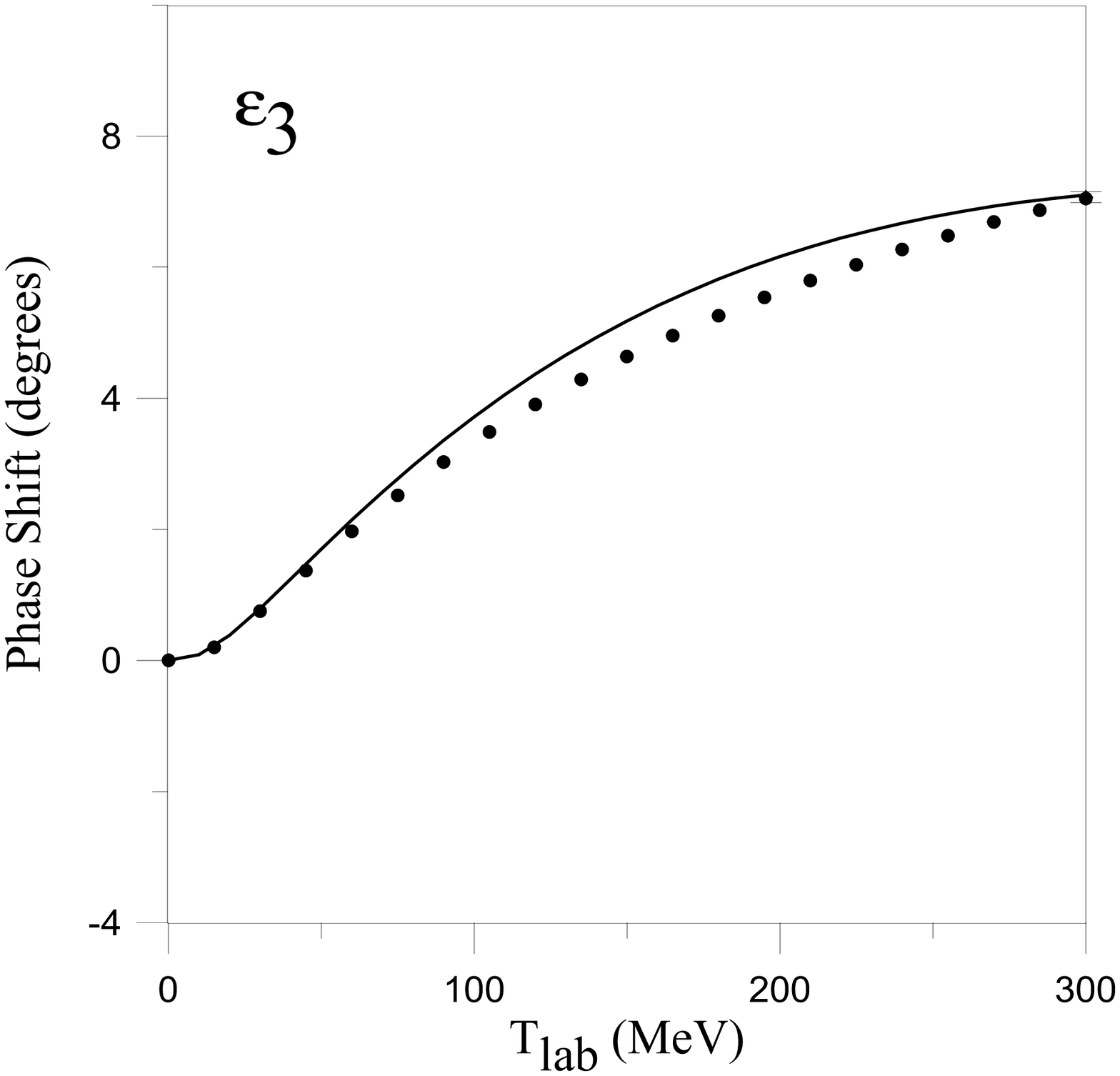,height=2.2in,width=2.4in}}
\hspace*{0.1cm}
\mbox{\psfig{figure=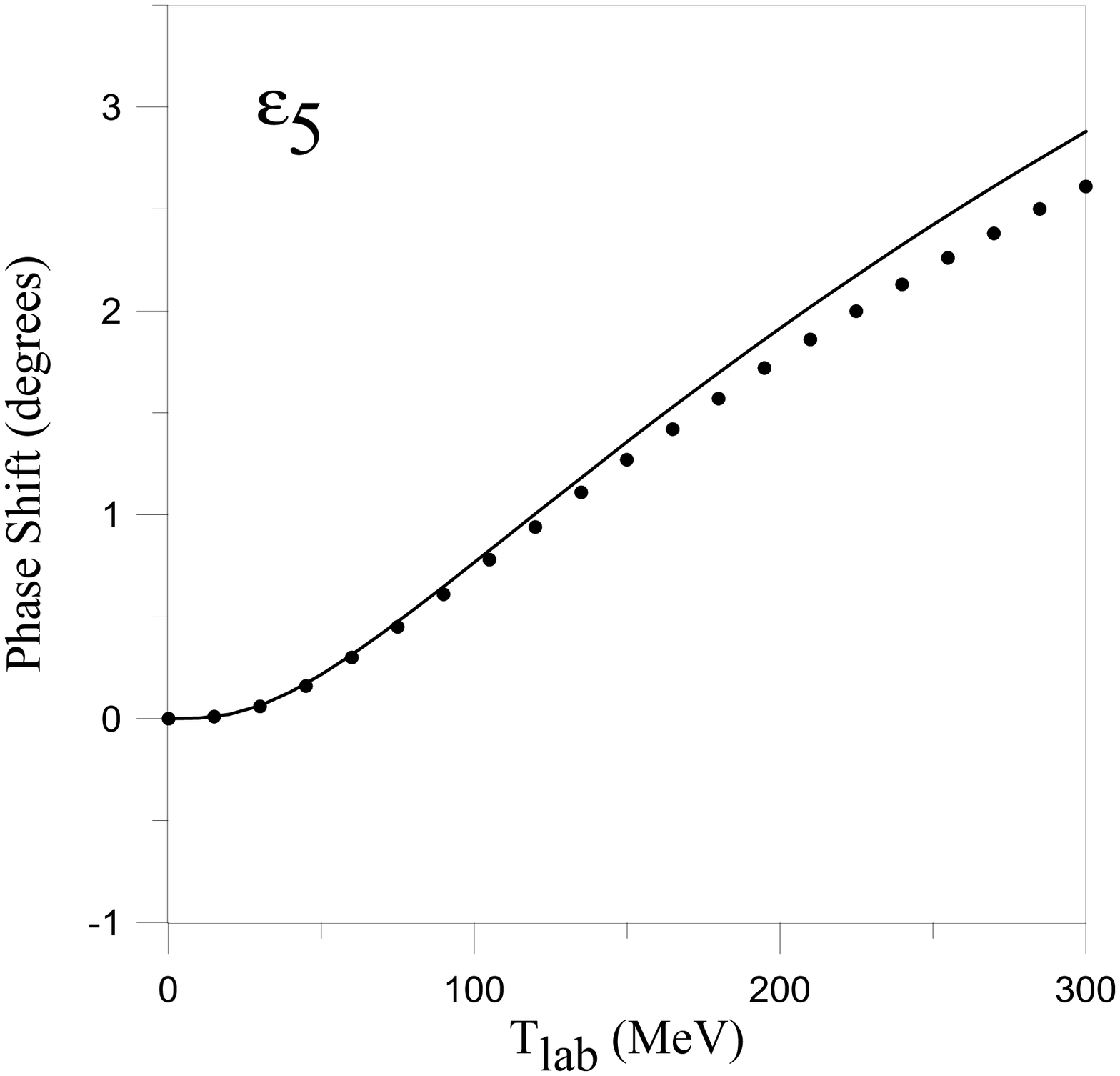,height=2.2in,width=2.4in}}
\vspace*{-0.2cm}
\caption{Mixing parameters.}
\label{fig10c4}
\end{figure}

In figure \ref{fig7c4} $J=5$ $G$, $H$ and $I$ phase shifts are shown
(notation as in figure \ref{fig3c4}). 
The results obtained are in good agreement with the experimental data, being
fundamentally dominated by the interaction of chiral origin. Iso-triplet
partial waves show that the coupling to N$\Delta$ channels, so important for
low orbital angular momenta, becomes smaller but contributes to improve the results.
Once again they are similar to those
obtained by effective theories \cite{KAI97,KAI98}, indicating that these
waves are governed exclusively by chiral symmetry. 

\subsubsection{Mixing parameters}
\label{ch5.2.4}

Figure \ref{fig10c4} shows the mixing parameters for the tensor
coupling between different NN partial waves: $^3S_1-{}^3\! D_1$,
$^3P_2-{}^3\! F_2$, $^3D_3-{}^3\! G_3$ and $^3G_5-{}^3\! I_5$. 
The dashed line shows the results when only NN components are
included. As can be seen, $\epsilon_1$ (which represents the
tensor coupling $^3S_1-{}^3\! D_1$) and $\epsilon_2$ (which 
represents the tensor coupling $^3P_2-{}^ 3\! F_2$)
are not correctly reproduced.
For $\epsilon_2$ one cannot draw any definite conclusion, because the model
does not accurately describe the $^3P_2$ and $^3F_2$ 
phase shifts. However, despite the very good description of the 
$^3 S_1$ and $^3 D_1$ phase shifts,
$\epsilon_1$ has the correct behaviour only at very low
energies. This seems to indicate the need for a stronger
short-range tensor force. Such interaction might be obtained
from a simultaneous exchange of pions and gluons between 
quarks \cite{HEL98}.
Higher angular momentum mixing parameters are governed by the OPE
tensor term and present a good agreement with the experimental data.

\subsubsection{Alternative quark model approaches}
\label{newsec}

Other constituent quark model approaches have 
been used to describe the NN dynamics. Ref. \cite{YUZ95}
used a generalized chiral symmetry restoration scheme
giving rise to four chiral fields instead of the simple image
of the $\pi$ and $\sigma$ mesons. Similar results
to the ones presented here are obtained. 
In Refs \cite{FUJ96,FUJ04,FUJ02} Fujiwara and collaborators
have performed a 
complete analysis of the NN scattering phase shifts and mixing coefficients.
Their quark model approach incorporates a non-relativistic kinetic energy
term, a quadratic confinement potential, the full Fermi-Breit interaction
with explicit quark-mass dependence and the full octet of
scalar and pseudoscalar meson-exchange potentials between quarks. A refined
version of the model considered also vector meson-exchange potentials
and the momentum-dependent Bryan-Scott term in the scalar
and vector meson potentials. 
An impressive good agreement with the experimental
data has been obtained for almost all partial waves and also for the mixing
coefficients. However consistency with the one-body problem is lost (see
also \sref{ch8.1}).
The main difference with respect to the CCQM
used here comes from the presence in the CCQM
of a form factor as a consequence
of the momentum dependent constituent quark mass generated by
chiral symmetry breaking, Eq. \eref{fofa}, 
that allows to regularize the non-central
contributions of the OPE without resorting to multiple
meson exchanges as in Refs. \cite{FUJ96,FUJ04,FUJ02}. 
In fact, in these works an important contribution
from the non-central terms of the octet meson-exchange potential
is obtained. In particular,
the quadratic spin-orbit component cancels the strong
one-pion tensor force. 
The interest reader could find a detailed description in the quoted works.

\subsection{The NN spin-orbit force} 
\label{ch5.3}

The study of $P$ and $D$ phase shifts makes evident
a lack of spin-orbit interaction in chiral constituent quark models.
The origin of the spin-orbit force is a long-standing problem of the
quark model in its description of the NN interaction and
the baryon spectrum
\cite{FAE93,BRA85,SHI89,KOI86,MOR84b,SUZ84,ISG78,HOL84}. 
On the one hand the spin-orbit force is known to be
short ranged and thus a priori it may seem 
easy to understand on a quark model scheme, but on the
other hand, it is a relativistic correction and therefore
difficult a priori to accommodate in a non-relativistic approach. 
In baryonic meson-exchange models \cite{MAC87,MAC89} 
the spin-orbit force has its origin 
on the $\omega$-meson exchange, also responsible 
for the short-range repulsion. In quark models the
inclusion of vector meson exchanges between quarks belonging 
to different clusters provide contributions similar to the quark-exchange 
mechanism and would lead to double counting \cite{YAZ90},
as explained in \sref{ch2.1}. 
Therefore, vector mesons are not included and one has 
to look for a different source of the spin-orbit force.

The non-relativistic reduction of the 
one-gluon exchange diagram provides at $(p_q/m_q)^2$ order
two different types of spin-orbit terms, a Galilei invariant (symmetric)
term and a Galilei non-invariant (antisymmetric) term. The Galilei-invariant
spin-orbit term has received considerable attention in the literature 
\cite{FAE93,BRA85,MOR84b,MOR84a,KOI86}, because it presents the same structure 
as the standard spin-orbit interaction between nucleons. Using this source 
of spin-orbit force it was demonstrated 
in reference \cite{MOR84a} that it is possible
to reproduce the observed spin-orbit splitting in the NN interaction using
a large quark-gluon coupling constant $\alpha_s \sim 1.6$. However, 
an effective meson-exchange potential between nucleons and not between
quarks was used. When quark-meson exchange is incorporated instead,
$\alpha_s$ is considerably reduced and the resulting spin-orbit force is not
large enough to reproduce the experimental $P$-wave splittings. In fact, when
gluon and pion exchange are consistently treated at the quark level, the OGE
spin-orbit force has to be multiplied by some numerical factor (usually
between 5$-$12 depending on the set of parameters \cite{BRA85,FER93a,ZHA94}) in
order to reproduce the experimental splittings in the NN phase
shifts, with the counterpart that it severely
disturbs the description of the excited negative-parity baryon spectrum 
\cite{MOR84a,ISG78}. Regarding the antisymmetric spin-orbit OGE term,
the strength of the full triplet-odd NN spin-orbit 
potential is greatly reduced and its sign may be 
even reversed when considered.
This is the reason why it has not been usually considered
in the literature, although there is no fundamental reason to justify this.

\begin{figure}[t]
\vspace*{-0.2cm}
\mbox{\psfig{figure=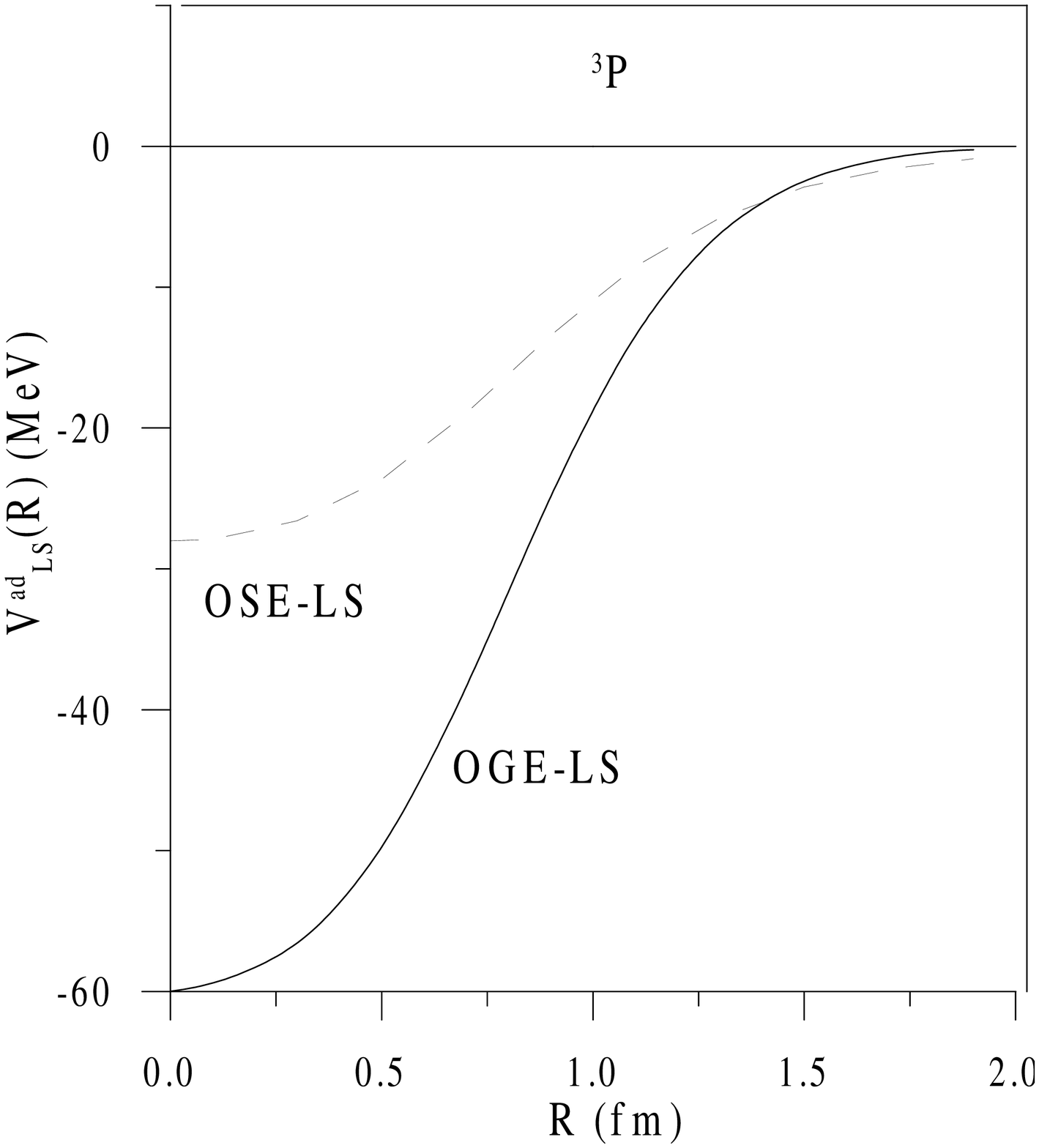,height=2.8in,width=2.2in}}
\hspace*{0.6cm}
\mbox{\psfig{figure=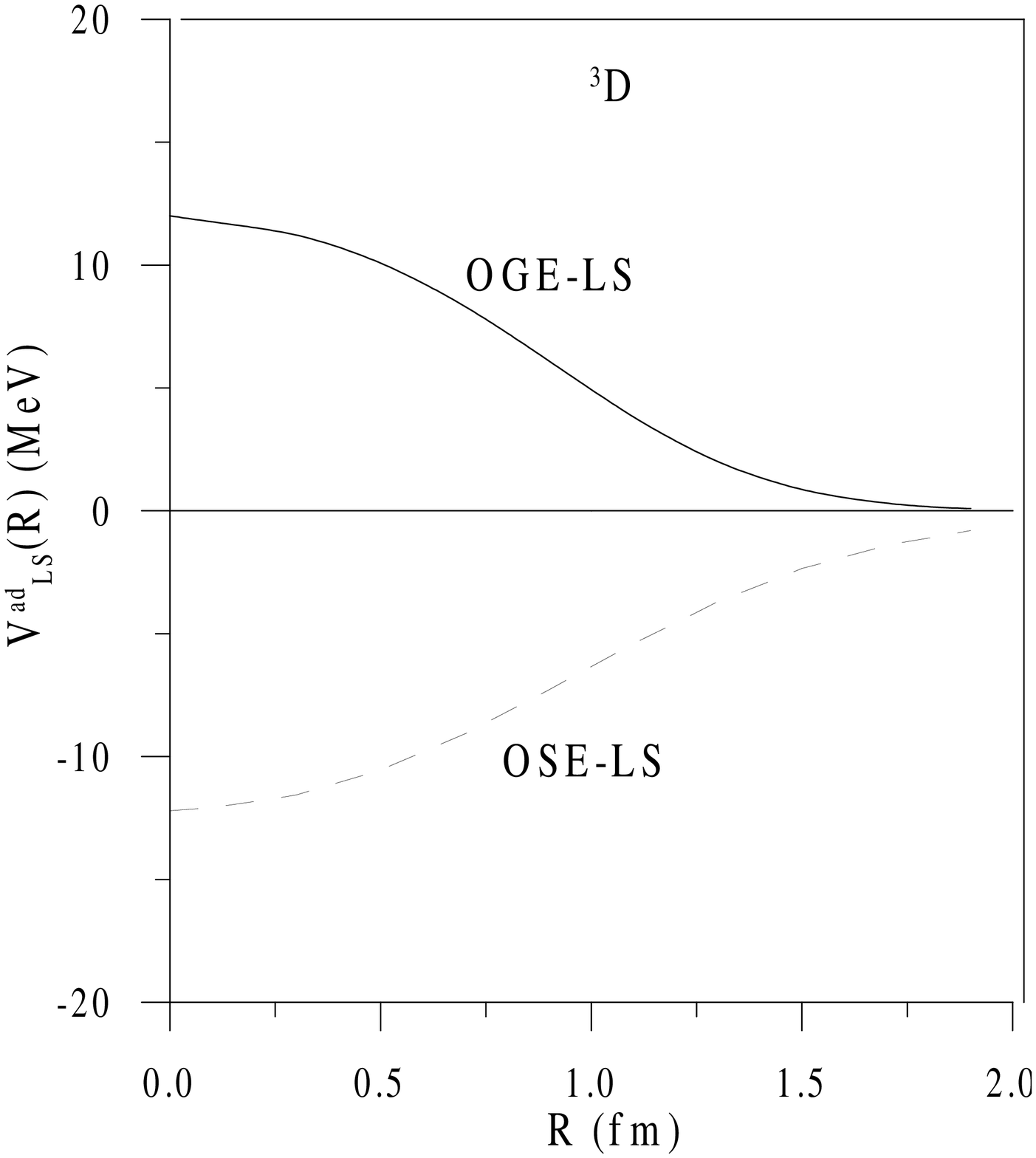,height=2.8in,width=2.2in}}
\vspace*{-1.3cm}
\caption{The adiabatic OGE and OSE spin-orbit potential
of reference \protect\cite{VAL95c} for $^3P_J$ and $^3D_J$ partial waves.}
\label{fig1c}
\end{figure}

Another source of spin-orbit force in chiral constituent quark models is
the scalar exchange. In reference \cite{VAL95c} it was shown that
it is not enough to explain the experimentally observed effects.
The strength of the NN one-sigma exchange spin-orbit force (OSE-LS) can be
compared to the one-gluon exchange symmetric spin-orbit
force (OGE-LS) in the adiabatic approximation. Figure \ref{fig1c} shows
the radial part of the NN adiabatic spin-orbit
potential for $^3P_J$ and $^3D_J$ waves.
The OGE-LS
is generated only by quark-exchange diagrams what makes it highly
non-local, while the OSE-LS has a non-vanishing direct term making it
longer-ranged. The sign of the OSE-LS is the same for $P$ and $D$
waves, because the direct term dominates. However, the OGE-LS force
has a different sign for $P$ and $D$ partial waves,
since only the exchange terms contribute.

Finally, the spin-orbit potential generated by confinement 
is strongly dependent on
the particular confinement model. For example, while the symmetric
spin-orbit term associated with a scalar two-body confinement potential
cancels the OGE symmetric spin-orbit force in the NN interaction \cite{SHI89},
the one coming from a flip-flop confinement interaction interferes
constructively with it in $P$ waves \cite{KOI86}. 
For $D$-waves they almost
cancel each other, which seems to be consistent
with the fact that the NN spin-orbit force
is known to be very small for even orbital angular momentum channels.
In reference \cite{SUZ84}, the uncertainties
associated with spin-orbit terms generated by quark confinement have been
emphasized by studying an alternative model in which confinement is
described through a mass term in the relativistic single particle equation.
The single particle Thomas term of such a model yields a NN
spin-orbit interaction with opposite sign to the one generated by a scalar
two-body confining potential. In summary
contributions which are usually neglected
(antisymmetric spin-orbit terms, Thomas precession terms, etc.)
might be even more important than those considered \cite{SUZ84b}.

\subsection{Charge dependence of nuclear forces}
\label{ch5.4}

The possibility of isospin violation in 
the strong interaction endures as one of the 
intriguing and partially unsolved aspects in the description
of nuclear forces \cite{MIL90}. After electromagnetic effects are removed from
the experimental data, small but significant differences exist among the
neutron-neutron and proton-proton interactions (charge symmetry
breaking, CSB), and among the average of these two interactions and the
neutron-proton one (charge independence breaking, CIB). Explicitly
in terms of the scattering lengths,
\begin{equation}
\eqalign{\Delta a_{\rm CSB} & = a_{\rm pp} - a_{\rm nn} = 1.5 \pm 0.5 \, {\rm fm} \\
\Delta a_{\rm CIB} & = {1 \over 2} \left( a_{\rm pp} + a_{\rm nn} \right) - a_{\rm pn}
= 5.7 \pm 0.5 \, {\rm fm} \, .}
\end{equation}
Most of the theoretical calculations to understand these numbers
have been done using baryonic meson-exchange models 
that incorporate basically
isospin breaking through the charged and neutral pion mass difference, the
proton-neutron mass difference, the $\rho^0 - \omega$ electromagnetic mixing 
amplitude, and the electromagnetic corrections. Whilst the major part of the CIB
has been explained by means of the 
mass difference between the neutral and charged
pions, proton-neutron mass difference 
and $\rho^0 - \omega$ mixing have been used
to account for most part of the CSB effects.

From a more fundamental point of view one would expect that the source of
$\Delta a_{\rm CIB}$ and $\Delta a_{\rm CSB}$ 
could be ultimately traced to CIB and
CSB in QCD. The traditional quark-model approach incorporates only two sources 
of isospin breaking, namely up-down quark mass difference and the 
quark-exchange Coulomb interaction. 
Chiral constituent quark models incorporate two additional mechanisms 
contributing to isospin violation, the mass difference between the charged and 
neutral pion and the leading order electromagnetic loop corrections to the
OGE potential (QED-QCD) \cite{STE91}. 
All these contributions have been evaluated in reference \cite{ENT99}
and the results are shown in table \ref{TCIB}. Taking into account the
pion-mass splitting ($V^C_{\rm OPE}$ and $V^T_{\rm OPE}$), the quark-exchange
Coulomb interaction ($V_\gamma$) and the QED-QCD interference terms
($V_{\rm QED-QCD}$) it is possible to reproduce CIB on the $^1S_0$
scattering length without
introducing any additional parameter to those necessary for describing
the NN interaction. The contribution of the quark-exchange Coulomb
interaction and the QED-QCD interference
terms make also possible to explain CSB on the $^1S_0$ scattering length.
The influence of up-down constituent
quark-mass difference is small due to the cancellation
among different terms.
\begin{table}[t]
\caption{\label{TCIB} Different contributions 
to $\Delta a_{\rm CIB}$ and $\Delta a_{\rm CSB}$. The number 
between parenthesis is calculated including 
the contribution of CSB. $\times$ indicates the contributions
considered.}
\begin{indented}
\item[]\begin{tabular}{@{}llllll}
\br
$V_{\rm OPE}^C$ & $V_{\rm OPE}^T$ & $V_{\rm QED-QCD}$ & $V_\gamma$ &
$\Delta a_{\rm CIB}\,({\rm fm})$ & 
$\Delta a_{\rm CSB}\,({\rm fm})$ \\
\mr
$\times$&   $-$  &   $-$  &   $-$  & 3.07 & $-$  \\
$\times$&$\times$&   $-$  &   $-$  & 3.41  & $-$ \\
$\times$&$\times$&$\times$&   $-$  & 5.59  & $-$ \\
$\times$&$\times$&$\times$&$\times$& 5.97 (5.94) & $-$ \\
\mr
   $-$  &   $-$  &   $-$  &$\times$& $-$ & 0.26 \\
   $-$  &   $-$  &$\times$&   $-$  & $-$ & 1.20 \\
   $-$  &   $-$  &$\times$&$\times$& $-$ & 1.46 \\
\br
\end{tabular}
\end{indented}
\end{table}

\subsection{The NN system above the pion threshold}
\label{ch8.2}
\begin{figure}[t]
\vspace*{-4.5cm}
\begin{center}
\mbox{\epsfxsize=140mm\epsfysize=150mm\epsffile{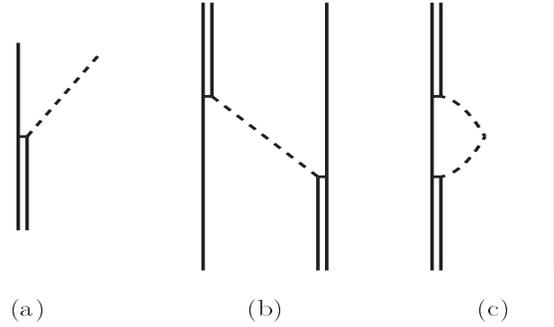}}
\end{center}
\vspace*{-6.4cm}
\caption{Contributions of intermediate $\pi$NN states. The pion is drawn as a
dashed line, the nucleon and $\Delta$ as single and double solid lines,
respectively.}
\label{fig11c4}
\end{figure}

Above the pion production energy the NN experimental data show several
structures which cannot be described by simple extrapolation of 
models explaining the NN data below this energy. To correctly treat the
problem it is necessary to incorporate the description of $\pi$NN states.
One possible way of treating the $\pi$NN system is by means of explicit
nucleon-resonance channels. The three-body problem is then reduced 
to a two-body problem assuming that the NN$\rightarrow \pi$NN transition 
only takes place trough resonances. Once
the decay width of the resonances is incorporated in the propagators, the
equations are able to properly couple physical two (NN) and three ($\pi$NN)
particle states. The effect of the coupling to resonances mainly depends
on the strength and range of the transition potential. Therefore it is
important to use a consistent interaction for the NN and nucleon-resonance 
channels.

Experimentally it is known that the inelasticity in two-nucleon
scattering up to at least 500 MeV in the c.m. system, i.e., far beyond the
two-pion threshold, is predominantly due to single-pion production and it occurs
essentially in the isospin-triplet partial waves, what suggests
that pion production proceeds through single-$\Delta$ excitation. 
This restriction is an enormous technical simplification \cite{POP87} and 
allows to model the $\pi$NN system by means of a NN$-$N$\Delta$ coupled 
channel problem.
\begin{figure}[b]
\hspace*{-1.5cm}
\mbox{\psfig{figure=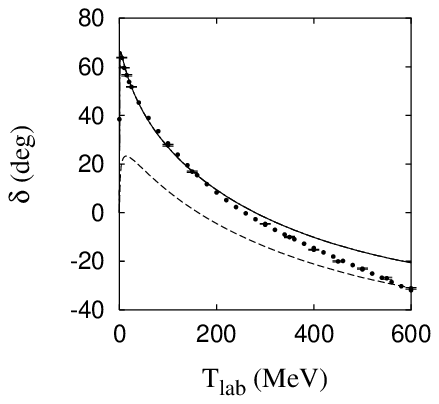,height=2.7in,width=3.7in}}
\hspace*{-3.5cm}
\mbox{\psfig{figure=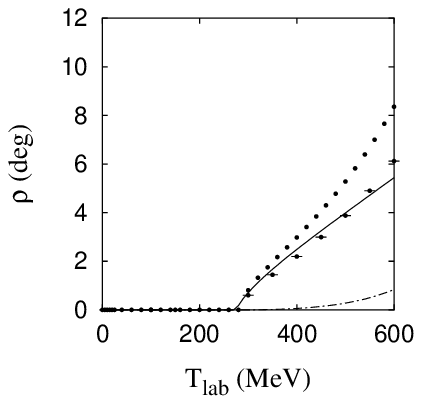,height=2.7in,width=3.7in}}
\vspace*{-1.0cm}
\caption{$^1S_0$ NN phase shift and inelasticity.}
\label{fig12c4}
\end{figure}

The main process for pion production is given by the $\pi$N$\Delta$ 
vertex depicted in figure \ref{fig11c4}(a), in which the $\Delta$ couples 
to a $\pi$N state. The introduction of this diagram has two consequences. 
The first one is that the OPE N$\Delta$ potential has a retarded
interaction, shown in figure \ref{fig11c4}(b), which modifies the $\Delta\Delta$ 
interaction. However, the diagonal $\Delta\Delta$ potential has a minor influence 
on the NN channels and usually this modification is not taken into account. 
The second one is the appearance of the delta self-energy diagram in the 
presence of a spectator nucleon shown in figure \ref{fig11c4}(c), that has a 
contribution to the $\Delta$ width and has to be added to the interaction.
One has then to solve the Lippmann-Schwinger equations for the NN$-$N$\Delta$ system
with the dressed $\Delta$ propagator\cite{GAR90a},
\begin{figure}[t]
\vspace*{-0.2cm}
\hspace*{-1.5cm}
\mbox{\psfig{figure=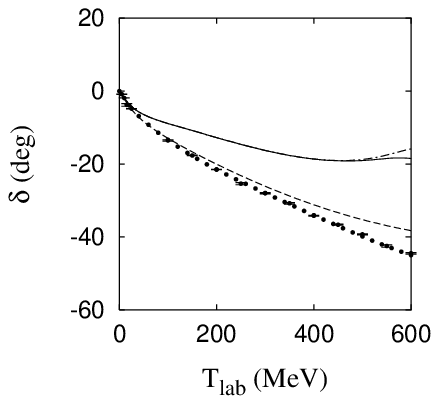,height=2.7in,width=3.7in}}
\hspace*{-3.5cm}
\mbox{\psfig{figure=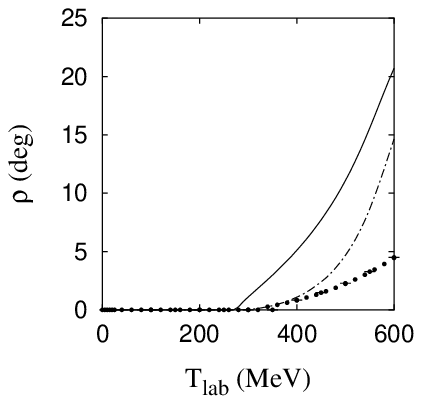,height=2.7in,width=3.7in}}
\vspace*{-1.0cm}
\caption{$^3P_1$ NN phase shift and inelasticity.}
\label{fig12c4b}
\end{figure}
\begin{figure}[b]
\vspace*{-0.2cm}
\hspace*{-1.5cm}
\mbox{\psfig{figure=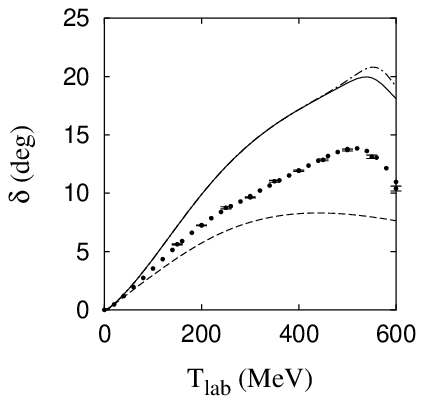,height=2.7in,width=3.7in}}
\hspace*{-3.5cm}
\mbox{\psfig{figure=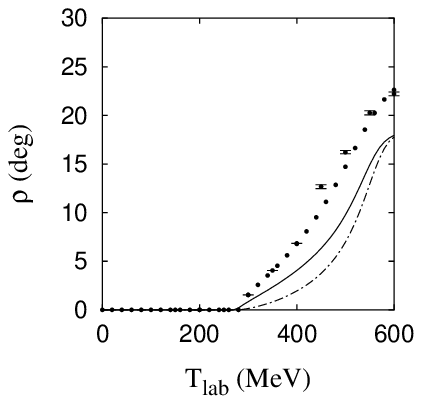,height=2.7in,width=3.7in}}
\vspace*{-1.0cm}
\caption{$^1D_2$ NN phase shift and inelasticity.}
\label{fig12c4c}
\end{figure}
\begin{equation}
G_{\Delta}^{D}(E)=\frac{1}{E-(m_{\Delta}-m_{\rm N})-\frac{q^{2}}{2\mu _{\Delta
}}-v_{\Delta }^{se}(q)+i\varepsilon }  \label{GD}
\end{equation}
where $q$ is the N$-\Delta$ relative momentum,
$\mu_\Delta$ is the N$\Delta$ reduced mass and
$v_{\Delta }^{se}(q)$ is the $\Delta$ self-energy interaction
\begin{equation}
v_{\Delta }^{se}(q)=<\gamma _{\Delta }\mid \frac{1}{E_{\Delta }(q)-\omega
_{\pi }-\omega _{N}+i\varepsilon }\mid \gamma _{\Delta }>
\end{equation}
being $\omega_\pi$ and $\omega_N$ the $\pi$ and N energies and 
$\gamma_{\Delta}$ the structure function of the $\pi$N$\Delta$ vertex.
In equation \eref{GD} only the imaginary part of the $\Delta$ self-energy
diagram, which gives the $\Delta$ width, is used. This is given by
\begin{equation}
\Gamma _{\Delta }=\frac{2}{3}\frac{f_{\pi N\Delta }^{2}(k_{0})}{4\pi
\,m_{\pi }^{2}}\frac{m_{N}}{\omega _{\pi }(k_{0})+m_{N}}\,k_{0}^{3}
\end{equation}
with $\omega _{\pi }(k_0)=\sqrt{m_{\pi }^{2}+k_{0}^{2}}$ the pion energy in the
$\pi$N center of mass system and $k_{0}$ the relative momentum of the $\pi$N system
The effective $f_{\pi {\rm N}\Delta }^{2}(k_{0})$ can be deduced from the $\pi qq$
vertex \cite{YAO88,ENT03},
\begin{equation}
f_{\pi {\rm N}\Delta }(k)=2\sqrt{2}f_{\pi qq}
F(k)e^{-\frac{b^{2}k^{2}}{6}}\left( 1+ \frac{\omega _{\pi }}{6m_{q}}\right) 
\end{equation}
where $F(k)$ is the $\pi qq$ vertex form factor that fixes the chiral
symmetry breaking scale and $b$ the wave function parameter.

\begin{figure}[t]
\vspace*{-0.2cm}
\hspace*{-1.0cm}
\mbox{\psfig{figure=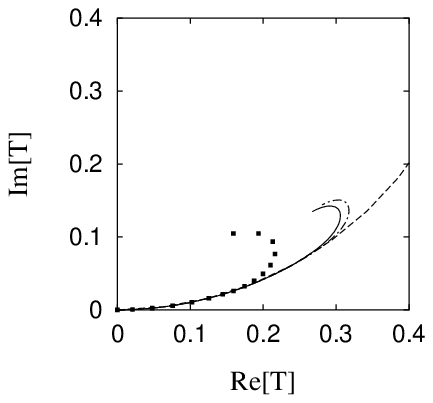,height=2.7in,width=3.5in}}
\hspace*{-2.5cm}
\mbox{\psfig{figure=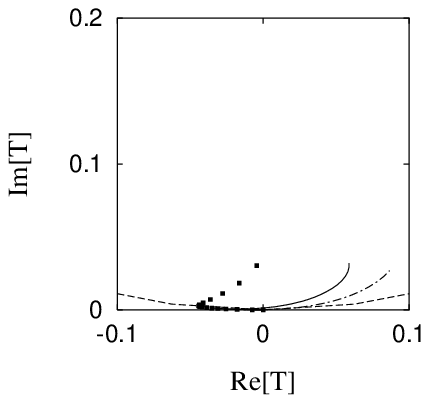,height=2.7in,width=3.5in}}
\vspace*{-1.0cm}
\caption{Argand diagrams for the $^1 D_2$ (left) and 
$^3 F_3$ (right) NN partial waves.}
\label{fig13c4}
\end{figure}

In figures \ref{fig12c4}, \ref{fig12c4b} and \ref{fig12c4c}
some NN phase shifts and inelasticities calculated in 
reference \cite{ENT03} are shown compared to results from the 
energy-dependent (dots without error bars in the figures)
and energy-independent (dots with error bars in the figures) 
partial wave analysis of \cite{ARN97}. 
No parameters are fitted to data above the
pion threshold. As the calculation only includes the inelasticity due to the 
$\Delta$ self-energy only isospin triplet partial waves are affected.
For the phase shifts, the solid line shows results for the coupled-channel
calculation including N$\Delta$ and $\Delta\Delta$ intermediate states
(although, as has been previously explained, the inelasticity is only
included through the N$\Delta$ channel) and dashed lines refer to results
including only the NN channel. For the inelasticities, the solid line
shows the result including the energy dependent $\Delta$ width whereas
dashed-dotted lines include the energy and momentum dependence, as
discussed in reference \cite{ENT03}.
As a general trend the experimental phase shifts are reasonably well
reproduced by the theoretical calculation except for the $P$ waves, 
as it happened below pion threshold. 
The most interesting iso-triplet partial waves are the $^1D_2$ and
$^3F_3$, because in this case the NN$-$N$\Delta$ coupling is very
important. The Argand plot for both cases, figure \ref{fig13c4}, presents a
counterclockwise behaviour with increasing energy, which is considered to be
a signal of a possible resonance. As seen in figure \ref{fig12c4c},
the resonant behaviour observed in the experimental data only appears when
the coupling to the N$\Delta$ channel is considered, although the
calculation overestimates the phase shifts as already known for models
including only N$\Delta$ channels.

%% file: ch5rev.tex
\section{The baryon spectrum}
\label{ch6}

An exciting challenge in contemporary nuclear physics is to achieve a
unified understanding of the baryon-baryon interaction 
and the baryon spectrum in terms of quark degrees of freedom.
Once we have analyzed in detail the description of the
baryon-baryon interaction in chiral constituent quark models
with ansatz harmonic oscillator baryonic wave functions we
should look for a consistency of such ansatz with the solution of the 
baryon spectrum from the same chiral quark-quark potential.
Furthermore we may expect from a detailed description of the
baryon spectrum to deepen the knowledge of some parts of the
quark-quark interaction that are not very relevant
for the baryon-baryon problem as it is the case for the
confining potential.

As we shall see these expectations will be satisfied and a quite
precise description of the non-strange baryon spectrum will finally
be achieved. In the way a long-standing problem in the
prediction of the baryon spectrum from quark models incorporating
a OGE plus confining quark-quark potential, say the level ordering
problem involving the relative position of the first two excited states
of the nucleon (and perhaps also of the $\Delta$) finds a consistent
solution. For the sake of clarity and simplicity and before going
to an extensive revision of the calculational methods and results,
let us establish the level ordering problem in the pure harmonic limit.
The N$^*$(1440) $J^P=1/2^+$ belongs to the $[56,0^+]$
$SU(6)_{Flavour-Spin} \times O(3)$ irreducible representation 
and it appears in the $N=2$ band,
while the N$^*$(1535) $J^P=1/2^-$ belongs to the $[70,1^-]$ appearing in the 
$N=1$ band. As a consequence, the N$^*$(1440) has $2 \hbar \omega$ energy 
excitation while the N$^*$(1535) has only $1 \hbar \omega$ energy excitation,
opposite to the order observed experimentally. 
In fact, the quadratic 
potential places the Roper resonance 1 GeV above the ground state (if its
strength is fixed variationally as it is done in the RGM), while it appears 
in the correct position (but still inverted with respect to N$^*(1535)$)
with a linear potential. This favors a linear 
form for confinement as also suggested by meson spectroscopy and lattice 
calculations \cite{GUT84}. 
We should finally say that the level ordering problem
has been cured by means of appropriate
phenomenological interactions as has 
been the case of anharmonic terms \cite{ISG00},
scalar three-body forces \cite{DES92} or pseudoscalar interactions
\cite{GAR01a,GLO96}, but without any connection to a consistent
simultaneous description of the baryon-baryon interaction 
except for reference \cite{GAR01a}.

\subsection{Formalisms of the three-quark bound state problem}

\subsubsection{Feshbach-Rubinow method}

This method \cite{FES55}, applicable only for the ground state of the system,
assumes that the 
ground state three-particle wave function depends on a single variable
$R={1\over 2}(x_1+x_2+\eta x_3)$, where $x_i=|\vec r_j-\vec r_k|$
with $\vec r_{j,k}$ the particle coordinates and $\epsilon_{ijk}=1$. 
The variational parameter $\eta$ gives account of the
asymmetry in the force bonds and the masses. The three-body Schr\"odinger
equation is then reduced to a single differential equation in the variable $R$
with an effective mass and an effective potential that are dependent on the
variational parameter $\eta$. The ground-state eigenvalue of the equation
is minimized as a function of $\eta$.

\subsubsection{Hyperspherical harmonic method}

This method is specially suited for the case of equal quark masses.
The position of three identical particles with respect to the center of mass 
can be described in terms of the Jacobi coordinates $\vec{\xi_1}$
and $\vec{\xi_2}$, defined in terms of the coordinates $\vec{r_i}$ of the
particles $(i=1,2,3)$ by,
\begin{equation}
\eqalign{ \vec{\xi_1} & = {1 \over \sqrt{2}} \left( \vec{r_1} - \vec{r_2} \right) \\
\vec{\xi_2} & = \sqrt{2 \over 3} \left( - \vec{r_3} + 
{{\vec{r_1} + \vec{r_2}} \over 2} \right) } \,\,\, .
\label{jac}
\end{equation}
From them, the hyperspherical coordinates $(\rho, \Omega)$ are defined
as follows. The hyperradius satisfies $\rho^2 =\xi_1^2 + \xi_2^2$. 
$\Omega$ is a set of five angular coordinates: one 
hyperspherical angle $\phi$ ($\tan \phi = \xi_1 / \xi_2$) and the four
angular coordinates $\hat{\xi_1}$ and $\hat{\xi_2}$ in the three-dimensional
space. 

With this choice the kinetic energy operator is
\begin{equation}
T={\vec{p_{\vec{\xi_1}}}^2 \over {2m}} + {\vec{p_{\vec{\xi_2}}}^2 \over {2m}} =
- {\hbar^2 \over {2m}} \left( {\partial^2 \over {\partial \rho^2}} + {5 \over \rho}
{\partial \over {\partial \rho}} + {1 \over {\rho^2}} K^2(\Omega) \right)
\end{equation}
with
\begin{equation}
K^2(\Omega) = {\partial^2 \over {\partial \phi^2}} + 4 \cot{2\phi}
{\partial \over {\partial \phi}} - 
{{\bf \ell}_1^2(\hat{\xi_1}) \over {\sin^2{\phi}}} -
{{\bf \ell}_2^2(\hat{\xi_2}) \over {\cos^2{\phi}}}  \,\,\, .
\end{equation}
The eigenfunctions of this operator, $K^2(\Omega)$, are the orthonormalized 
hyperspherical harmonics $Y_{[k,\ell_1,m_{\ell_1},\ell_2,m_{\ell_2}]}$. 
The Schr\"odinger equation written in hyperspherical coordinates becomes
\begin{equation}
- \left[ {\hbar^2 \over {2m}} \left\{ {d^2 
\over {d \rho^2}} + {5 \over \rho}
{d \over {d \rho}} + {1 \over {\rho^2}} K^2(\Omega) \right\}
+ V \left( \rho, \Omega \right) - E \right] \Psi (\rho,\Omega)=0
\end{equation}
where $V(\rho,\Omega)$ is the interaction.

\subsubsection{Stochastic variational method}

The variational method is a widely used tool to solve the bound state  
N-particle problem. It is based on 
the choice of a trial wave function and the search of the parameter set 
minimizing the energy of the system. The trial function has to include 
the correlations between the different particles and at the same time
give rise, when calculating observables, to easily calculable
matrix elements. A linear combination of harmonic oscillator wave
functions, as the simplest 
\begin{equation}
f(\vec r_i,\vec r_j)= e^{-\alpha_{ij}(\vec r_i- \vec r_j)^2}
\end{equation}
where $\vec r_{i,j}$ refer to the particle coordinates,
fulfills these two requirements and is one of the most popular
choices in variational calculations.

In the literature there are different strategies to select the trial wave 
function parameters. The stochastic variational method attempts to set up
the most appropriate parameter set by following a stepwise procedure:
one generates a basis function by choosing randomly the nonlinear gaussian 
parameters, $\alpha_{ij}$, while the linear ones are 
determined by diagonalizing
the Hamiltonian matrix. From a set of N randomly chosen functions one
selects those which give the minimal energy. Then one increases the 
number of functions by one with randomly generated nonlinear parameters.
Its utility is judged by the energy lost or gained and so the extra
function is kept or discarded, respectively. 
This trial and error procedure is repeated
until the basis set up leads to convergence. The 
versatility and the efficiency of the stochastic variation 
has been demonstrated for several systems \cite{SUZ98}.

\subsubsection{Coordinate-space Faddeev method}

In this method the Schr\"odinger equation for the 
three-body problem is written as
\begin{equation}
\left( H_0 +V_1 + V_2 + V_3 \right) \Psi (\vec{r}_1, \vec{r}_2, \vec{r}_3)
= E \Psi (\vec{r}_1, \vec{r}_2, \vec{r}_3)
\end{equation}
where $H_0$ is the kinetic energy operator of the three particles
\begin{equation}
H_0= \sum_{i=1}^{3} {k_i^2 \over {2m_i}} \, ,
\label{kin}
\end{equation}
and $V_i=V_i(\vec{r}_j-\vec{r}_k)$ with $\vec r_{j,k}$ the 
particles coordinates and $\epsilon_{ijk}=1$ 
is the potential energy.

In order to guarantee a unique solution of the three-body problem
Faddeev decomposed the wave function of the system as \cite{FAD61}
\begin{equation}
\Psi = \phi_1 +  \phi_2 + \phi_3  \, ,
\label{fad0}
\end{equation}
where the Faddeev components satisfy the coupled equations
\begin{equation}
\left( E- H_0 - V_i \right) \phi_i = V_i \left( \phi_j + \phi_k \right) \, .
\label{fad1}
\end{equation}
This is the form used in coordinate-space calculations where the 
spatial part of $\phi_i$ is decomposed in terms of the 
spherical harmonics relative to the
Jacobi coordinates as defined in equation \eref{jac}.

Due to the presence of confining and constant terms in the potential,
the convergence of these equations with respect to the number
of angular momentum channels is rather slow. Therefore, a modification
of the method has been proposed in references \cite{PAP99,PAP00} in order to 
accelerate the convergence of the solution with respect to the number
of angular momentum channels. The two-body interaction is splitted into 
confining and non-confining parts as,
\begin{equation}
 V_i = V_i^C+V_i^{NC},
\end{equation}
which leads to the modified form of the Faddeev equations 
\begin{equation}
\left( E- H_0 -V_1^C-V_2^C-V_3^C - V_i^{NC} \right) \phi_i = V_i^{NC} \left( \phi_j + \phi_k \right) \, .
\end{equation}
This modified method, however, must be used with some care since the splitting
into confining and non-confining parts is not unique so that it can lead to
spurious solutions \cite{PAP99,PAP00}.

\subsubsection{Momentum-space Faddeev method}
\label{ch6.1}

Formally equation \eref{fad1} can be rewritten as
\begin{equation}
\phi_i = G_0 t_i \left( \phi_j + \phi_k \right)
\end{equation}
where $G_0$ is the propagator of three free particles 
\begin{equation}
G_0 = \left( E - H_0 \right)^{-1}
\end{equation}
and $t_i$ is the two-body T-matrix which obeys the Lippmann-Schwinger 
equation,
\begin{equation}
t_i = V_i + V_i G_0 t_i
\end{equation}

The method to solve the Faddeev equations for three quarks in momentum 
space has been described in detail in reference \cite{GAR03}.
Let us mention here that there is a problem when dealing with
confining potentials of the form
\begin{equation}
V_C(r)=br^n; \;\;\;\;\;\;\;\;\; n=1,2,...
\label{e2c5}
\end{equation}
in momentum space, since the Fourier transform 
does not exist because $V_C\to\infty$ when $r\to\infty$. However, if the 
potential \eref{e2c5} is replaced by the finite potential

\begin{equation}
V(r)=\left\{\matrix{b(r^n-R^n) &; r\le R \cr
              0 &; r>R \cr}\right. 
\label{e3c5}
\end{equation}
for which the Fourier transform is well defined,
in the limit $R\to\infty$ both potentials will give the same 
result provided that the eigenvalues are related by 
\begin{equation}
E_C=E-3bR^n
\label{e4c5}
\end{equation}
where $E_C$ and $E$ are the eigenvalues corresponding to the interactions
\eref{e2c5} and \eref{e3c5}, respectively.

If there are no tensor or spin-orbit forces the Faddeev equations for the
bound-state problem of three quarks can be written as
\begin{eqnarray}
\fl
<p_iq_i;\ell_i\lambda_iS_iT_i|\phi_i^{LST}>  =  
{1\over E-p_i^2/2\eta_i-q_i^2/2\nu_i}
\sum_{j\ne i}
\sum_{\ell_j\lambda_jS_jT_j}
{1\over 2}\int_{-1}^1 dcos\theta\int_0^\infty q_j^2 dq_j \nonumber \\
 \times\, t_i^{\ell_iS_iT_i}(p_i,p_i^\prime;E-q_i^2/2\nu_i) 
A_L^{\ell_i\lambda_i\ell_j\lambda_j}(p_i^\prime q_i p_j q_j) \nonumber \\
 \times  <S_iT_i|S_jT_j>_{ST}\,
<p_jq_j;\ell_j\lambda_jS_jT_j|\phi_j^{LST}>,
\label{e5c5}
\end{eqnarray}
where $S_i$ and $T_i$ are the spin and isospin of the 
pair $jk$ while $S$ and $T$ are the total spin and isospin.
$\ell_i$ ($\vec{p}_i$) is the orbital angular momentum (momentum) 
of the pair $jk$, $\lambda_i$ ($\vec{q}_i$)
is the orbital angular momentum (momentum) 
of particle $i$ with respect to the pair
$jk$, and $L$ is the total orbital angular momentum.
$cos\theta=\vec q_i\cdot\vec q_j/(q_iq_j)$ while
\begin{eqnarray}
\eta_i & = & {m_j m_k \over m_j + m_k},
\label{e6c5} \\
\nu_i & = & {m_i(m_j+m_k) \over m_i+m_j+m_k},
\label{e7c5}
\end{eqnarray}
are the usual reduced masses.
For a given set of values of $LST$ the integral equations \eref{e5c5}
couple the amplitudes of the different configurations
$\{\ell_i\lambda_i S_i T_i\}$ with $(-)^{\ell_i+S_i+T_i}=1$ as required by the
Pauli principle since the wave function is colour antisymmetric.

The spin-isospin recoupling coefficients $<S_iT_i|S_jT_j>_{ST}$ are given by
\begin{eqnarray}
\fl
<S_iT_i|S_jT_j>_{ST}  = 
(-)^{S_j+\sigma_j-S}\sqrt{(2S_i+1)(2S_j+1)} \, W(\sigma_j\sigma_kS\sigma_i;S_iS_j)
 \nonumber \\
 \times 
(-)^{T_j+\tau_j-T}\sqrt{(2T_i+1)(2T_j+1)} \, W(\tau_j\tau_kT\tau_i;T_iT_j),
\label{e8c5}
\end{eqnarray}
with $\sigma_i$ and $\tau_i$ the spin and isospin of particle $i$, and $W$
is the Racah coefficient.
The orbital angular momentum recoupling coefficients 
$A_L^{\ell_i\lambda_i\ell_j\lambda_j}(p_i^\prime q_i p_j q_j)$ are given by 
\begin{eqnarray}
\fl
A_L^{\ell_i\lambda_i\ell_j\lambda_j}(p_i^\prime q_i p_j q_j)  = 
{1\over 2L+1}\sum_{M m_i m_j}C^{\ell_i \lambda_i L}_{m_i,M-m_i,M}
C^{\ell_j \lambda_j L}_{m_j,M-m_j,M}\Gamma_{\ell_i m_i}\Gamma_{\lambda_i
M-m_i}\Gamma_{\ell_j m_j} \nonumber \\  \times
\Gamma_{\lambda_j M-m_j}
cos[-M(\vec q_j,\vec q_i)-m_i(\vec q_i,{\vec p_i}^{\,\prime})
+m_j(\vec q_j,\vec p_j)],
\label{e9c5}
\end{eqnarray}
with $\Gamma_{\ell m}=0$ if $\ell -m$ is odd and
\begin{equation}
\Gamma_{\ell m}={(-)^{(\ell+m)/2}
\sqrt{(2\ell+1)(\ell+m)!(\ell-m)!}
\over 2^\ell((\ell+m)/2)!((\ell-m)/2)!}
\label{e10c5}
\end{equation}
if $\ell-m$ is even. The angles 
$(\vec q_j,\vec q_i)$, $(\vec q_i,{\vec p_i}^{\,\prime})$, and
$(\vec q_j,\vec p_j)$ can be obtained in terms of the magnitudes of the 
momenta by using the relations
\begin{equation}
\vec p_i^{\,\prime} = - \vec q_j - {\eta_i\over m_k}\vec q_i ,
\label{e11c5}
\end{equation}
\begin{equation}
\vec p_j = \vec q_i + {\eta_j\over m_k} \vec q_j,
\label{e12c5}
\end{equation}
where $ij$ is a cyclic pair.
The magnitude of the momenta $p_i^\prime$ and $p_j$, on the other hand,
are obtained in terms of $q_i$, $q_j$, and
$cos\theta$ using equations \eref{e11c5} and \eref{e12c5} as 
\begin{equation}
p_i^\prime=\sqrt{q_j^2+
\left({\eta_i\over m_k}\right)^2q_i^2
+{2\eta_i\over m_k}q_i q_j cos\theta},
\label{e13c5}
\end{equation}
\begin{equation}
p_j=\sqrt{q_i^2+
\left({\eta_j\over m_k}\right)^2q_j^2
+{2\eta_j\over m_k}q_i q_j cos\theta}.
\label{e14c5}
\end{equation}

Finally, the two-body amplitudes 
$t_i^{\ell_iS_iT_i}(p_i,p_i^\prime;E-q_i^2/2\nu_i)$
are given by the solution of the Lippmann-Schwinger equation
\begin{eqnarray}
\fl
t_i^{\ell_iS_iT_i}(p_i,p_i^\prime;E-q_i^2/2\nu_i)  = 
V_i^{\ell_iS_iT_i}(p_i,p_i^\prime) + \int_0^\infty {p_i^{\prime\prime}}^2 
dp_i^{\prime\prime}\, V_i^{\ell_iS_iT_i}(p_i,p_i^{\prime\prime})
 \nonumber \\  \times
{1\over E-{p_i^{\prime\prime}}^2/2\eta_i-q_i^2/2\nu_i}\,
 t_i^{\ell_iS_iT_i}(p_i^{\prime\prime},p_i^\prime;E-q_i^2/2\nu_i),
\label{e15c5}
\end{eqnarray}
with
\begin{equation}
V_i^{\ell_iS_iT_i}(p_i,p_i^\prime) 
= {2\over \pi}\int_0^\infty r_i^2 dr_i\,j_{\ell_i}(p_ir_i)
V_i^{S_iT_i}(r_i)j_{\ell_i}(p_i^\prime r_i).
\label{e16c5}
\end{equation}
and $j_{\ell}$ the spherical Bessel function.

\subsection{Results of different potential models}

\subsubsection{Isgur-Karl model}

In a series of pioneering works
Isgur and Karl \cite{ISG78} performed an extensive 
study of the baryon spectrum. They used an interaction
composed of a confining potential, usually taken to be 
a harmonic oscillator, the hyperfine term 
of the OGE potential proposed by
de R\'ujula, Georgi and Glashow \cite{RUJ75}, and finally
anharmonic terms to treat on a different footing states belonging
to different major shells. 
They worked in first-order perturbation theory, diagonalizing
the Hamiltonian in a harmonic oscillator basis, and they
obtained an overall good agreement with experimental data. 
Nonetheless to get such agreement the zero order masses
of positive- and negative-parity states were
fitted independently, without any sound justification.

\subsubsection{Bhaduri-like potential models}

Bhaduri \etal \cite{BHA81} performed a calculation of the ground-state 
baryon masses using an interaction designed to describe the
meson spectra, composed of a standard OGE interaction besides a linear
confinement. They used the Feshbach-Rubinow method assuming
only $S$-wave interactions. They obtained $M_N=1052$ MeV and $M_\Delta=
1354$ MeV, which are more than 100 MeV above the experimental values.

\begin{figure}[t]
\hspace*{-7.5cm}
\mbox{\psfig{figure=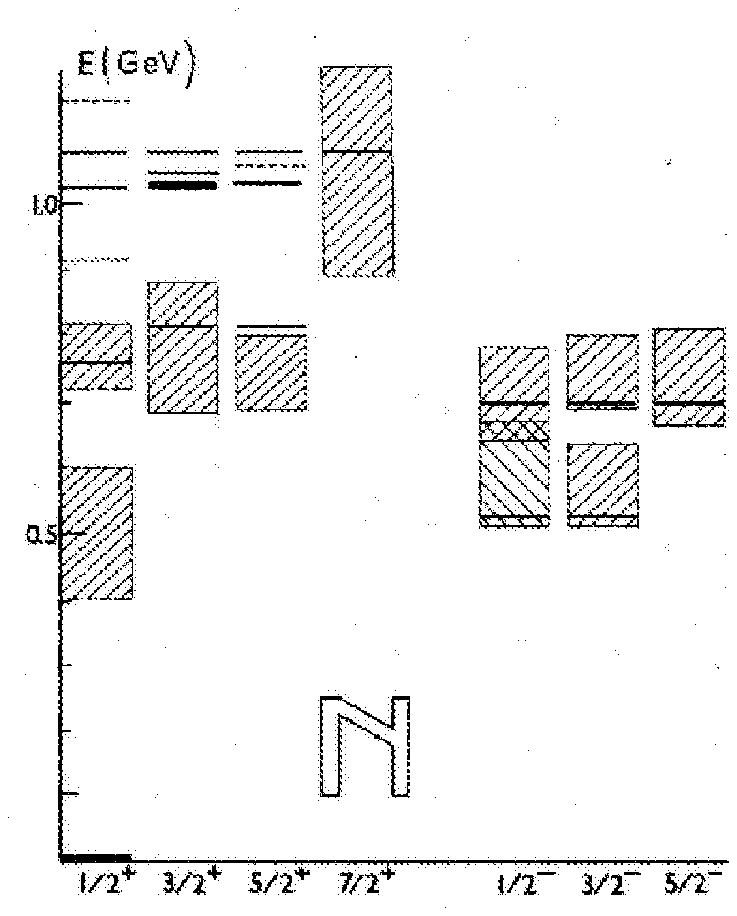,height=7.3in}}
\hspace*{.5cm}
\mbox{\psfig{figure=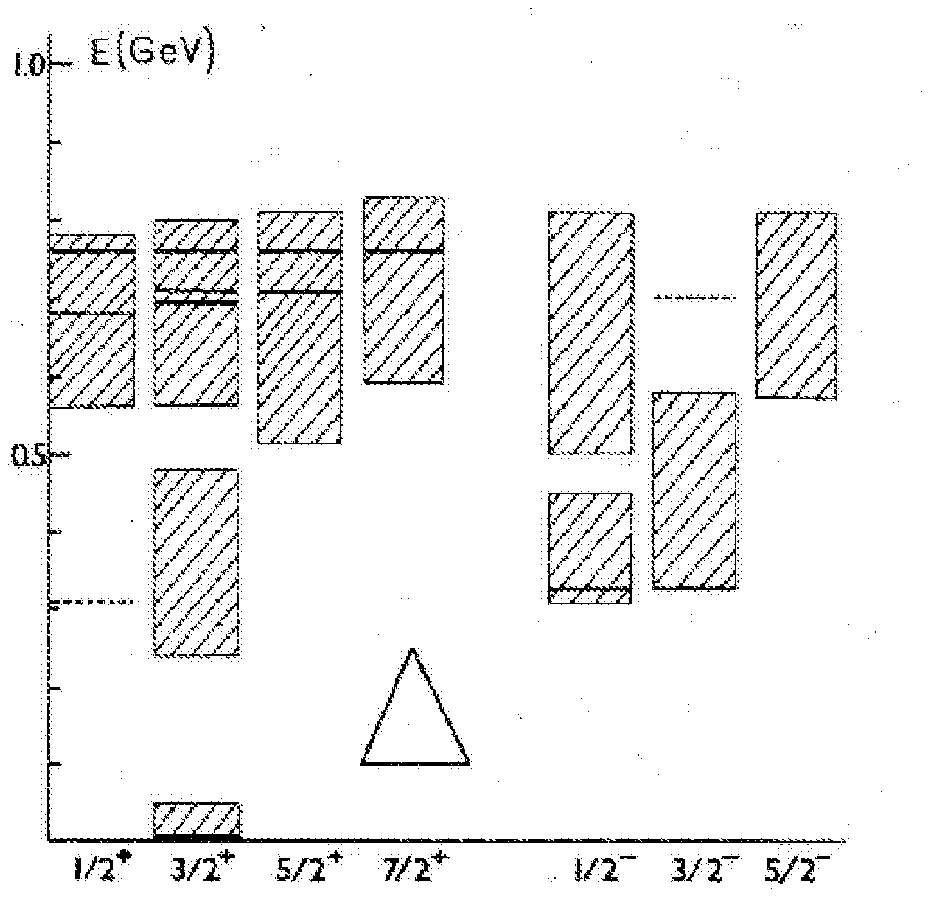,height=7.3in}}
\vspace*{-12.1cm}
\caption{Relative energy N and $\Delta$ spectrum of reference
\protect\cite{SIL85}.}
\label{f2rc5}
\end{figure}

Later on a complete coordinate-space Faddeev method calculation was
carried out with the same interaction \cite{SIL85}. The comparison of 
the results with those of the harmonic
oscillator diagonalization method allowed to establish the 
high degree of accuracy of the last technique.
For the masses of the N and $\Delta$ the values
1024 MeV and 1318 MeV, respectively, were predicted.
These are about 80 MeV above
the experimental values although their splitting is correctly given
around 290 MeV. For the N and $\Delta$ excited spectra (see 
\fref{f2rc5}) they got a reasonable description of the negative
parity excitations while in the positive parity case the 
first radial excitation for both, N and $\Delta$, was above 
the data by about 300 MeV. 

The problem of the relative position of the negative- and positive-parity
excitations of the nucleon was addressed in reference \cite{DES92} 
by implementing the Bhaduri potential with a scalar three-body 
force corresponding to the exchange of two $\sigma$'s. 
Hyperspherical harmonic and coordinate-space Faddeev
calculations were consistently performed. This allowed for a resolution
of the level ordering problem in both cases, N and $\Delta$, and for
a pretty good description of the non-strange baryon spectra.
The question immediately arising was whether this effective
three-body force effect could be obtained or not with a two-body
description once the chiral pseudoscalar exchanges were
incorporated in the quark-quark potential. 

\subsubsection{Goldstone-boson exchange potential model}

In this model the interaction between quarks is assumed to be given
by a linear potential responsible for confinement plus a 
one-boson-exchange potential generated by the exchange
of the octet of low-mass pseudoscalar mesons and a flavour
singlet $\eta'$ \cite{GLO96,GLO99}, i.e., 
\begin{equation}
V = V_\chi + V_{conf} \,
\end{equation}
where
\begin{equation}
\fl \eqalign{V_\chi^{octet} (\vec{r_{ij}}) & = 
\left[ \sum_{a=1}^{3} V_\pi (\vec{r_{ij}}) \lambda_i^a \lambda_j^a +
\sum_{a=4}^{7} V_K (\vec{r_{ij}}) \lambda_i^a \lambda_j^a +
V_\eta (\vec{r_{ij}}) \lambda_i^8 \lambda_j^8 \right] \vec{\sigma_i} \cdot
\vec{\sigma_j} \\
V_\chi^{singlet} (\vec{r_{ij}}) & = {2 \over 3}
V_{\eta'} \vec{\sigma_i} \cdot \vec{\sigma_j}} \, .
\end{equation}
\begin{figure}[t]
\hspace*{-0.5cm}
\centerline{\psfig{figure=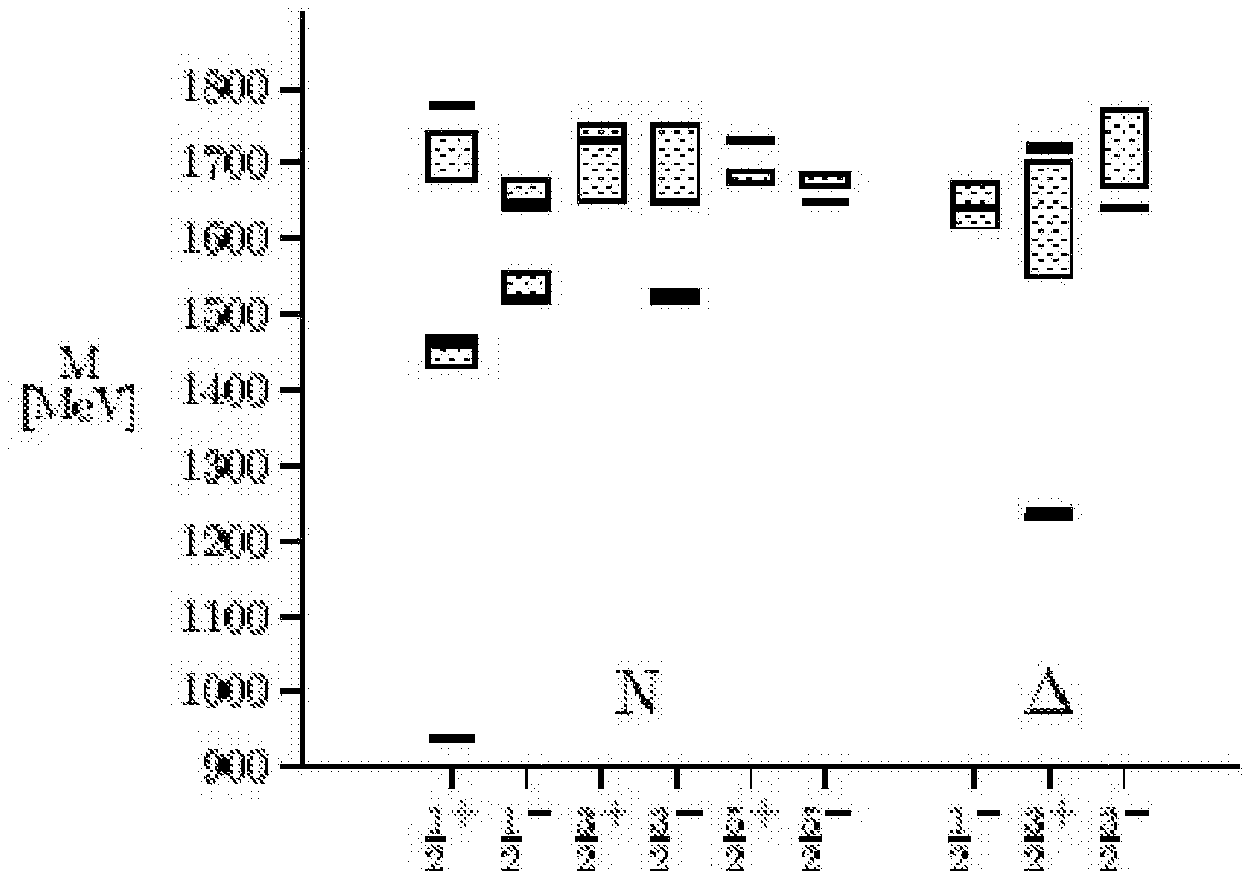,height=7.4in,width=5.4in}}
\vspace*{-11.8cm}
\caption{Relative energy N and $\Delta$ spectrum of reference
\protect\cite{GLO99}.}
\label{f3rc5}
\end{figure}

The model was first applied by Glozman and Riska \cite{GLO96} within
a first-order perturbation theory approach. In the more elaborated 
calculations of the model the one-boson-exchange potentials were
constructed with different meson-quark-quark coupling constants
for the octet and singlet and were regularized at the origin by smearing
the $\delta$-function part with appropriate cut-off parameters
which are different for each meson. 
As the coordinate-space Faddeev method allows to obtain a solution only
in the case of a non-relativistic kinetic energy operator and the effect
of replacing the non-relativistic kinetic energy 
operator \eref{kin} by the relativistic one, i.e.,
\begin{equation}
H_0 \rightarrow \sum_{i=1}^{3} \left( \sqrt{ m_i^2 + q_i^2} - m_i \right) \, ,
\label{kinrel}
\end{equation}
was estimated to be important for the level ordering problem, a
calculation in the stochastic variational formalism
was carried out (see \fref{f3rc5} from \cite{GLO99}). 
The fit to the experimental spectrum was remarkable
up to 900 MeV excitation energy. The first positive- and
negative-parity nucleon states were correctly predicted though 
some caution should be in order to judge the $\Delta$ case. Actually 
the Roper of the $\Delta$ was predicted above the first 
negative-parity state though experimental 
error bars do not allow in this case
to draw any definitive conclusion.

\subsubsection{Chiral constituent quark models}
\label{sec624}

\begin{figure}[t]
\centerline{\psfig{figure=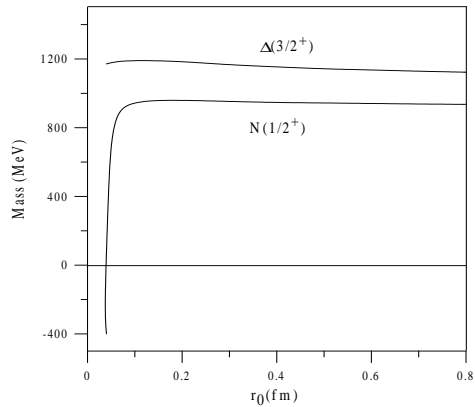,height=3.4in,width=2.6in}}
\vspace*{-3.4cm}
\caption{N$(1/2^+)$ and $\Delta(3/2^+)$ ground state masses as a function
of $r_0$.} 
\label{f1c5}
\end{figure}

The non-strange baryon spectrum has been also studied from the 
chiral quark potential of \sref{ch2}.
Reference \cite{VAL96a} presented a calculation with a small value 
for $r_0$ ($r_0=0.0367$ fm), the regularization parameter for the
contact term of the OGE, and truncating the Hilbert space 
by means of a hyperspherical harmonic approach up to
$k$ = 2 excitations. Similar results were presented in reference \cite{DZI96}.
These results were strongly criticized \cite{GLO98} by arguing that such a
small value of $r_0$ would give rise
to unstable values for the baryon masses. Moreover, it was argued that
the level ordering problem could not be solved with a chiral constituent
quark model due to the presence of the OGE term and the subsequent
reduction of the chiral quark coupling with respect to the Goldstone-boson
exchange potential case. However the results of \cite{VAL96a} and 
\cite{DZI96} can be
alternatively reproduced almost exactly from a momentum-space Faddeev
calculation including only the lowest order configurations
$(\ell_i,\lambda_i,S_i,T_i)$ \cite{GAR01b,GAR01a} for a much bigger
value of $r_0=$ 0.2 fm, lying in the stability region. 
\begin{table}[b]
\caption{\label{t1c5}Different sets of quark model parameters.}
\begin{indented}
\item[]
\begin{tabular}{@{}llllll}
\br
Set & A & B & C & D \\
\mr
  $m_q ( {\rm MeV})$                & 313    & 313   & 313    & 313   \\
  $\alpha_s$                        & 0.72   & 0.65  & 0.485  & 0.50  \\
  $a_c({\rm MeV}\cdot{\rm fm}^{-1})$& 72.518 & 60.12 & 67.0   &110.0  \\
  $\alpha_{ch}$                     & 0.0269 & 0.0269& 0.0269 &0.0269 \\
  $r_0 ({\rm fm})$                  & 0.2    & 0.8   & 0.25   &0.74   \\
  $m_\sigma ({\rm fm}^{-1})$        & 3.42   & 3.42  & 3.42   & 3.42  \\
  $m_\pi ({\rm fm}^{-1})$           &  0.7   &  0.7  & 0.7    &0.7    \\
  $\Lambda_\pi ({\rm fm}^{-1})$     &  4.2   &  5.4  & 4.2    & 2.0   \\
  $\Lambda_\sigma ({\rm fm}^{-1})$  &  4.2   &  4.2  & 4.2    & 2.0   \\
\br
\end{tabular}
\end{indented}
\end{table}
The dependence of the N$(1/2^+)$ and $\Delta(3/2^+)$ ground state masses
on the regularization parameter $r_0$, as given by the Faddeev calculation,
is illustrated in 
figure \ref{f1c5}. While the baryon-baryon interaction does not crucially 
depend on $r_0$, the baryon spectrum does. For values of $r_0$ below 0.1 fm 
the N$(1/2^+)$ mass decreases very quickly while the $\Delta(3/2^+)$ 
remains almost stable. As a consequence, the excitation energy of the whole 
spectrum would become too high. This result can be easily understood 
taken into account that the short-ranged spin-spin force is attractive for 
the nucleon but repulsive for the $\Delta$, lowering the $L=0$ nucleon states. 
It is also worth to notice that the nucleon ground state energy almost does 
not change for $r_0$ values greater than 0.1 fm. 

\begin{figure}[t]
\vspace*{-0.4cm}
\mbox{\psfig{figure=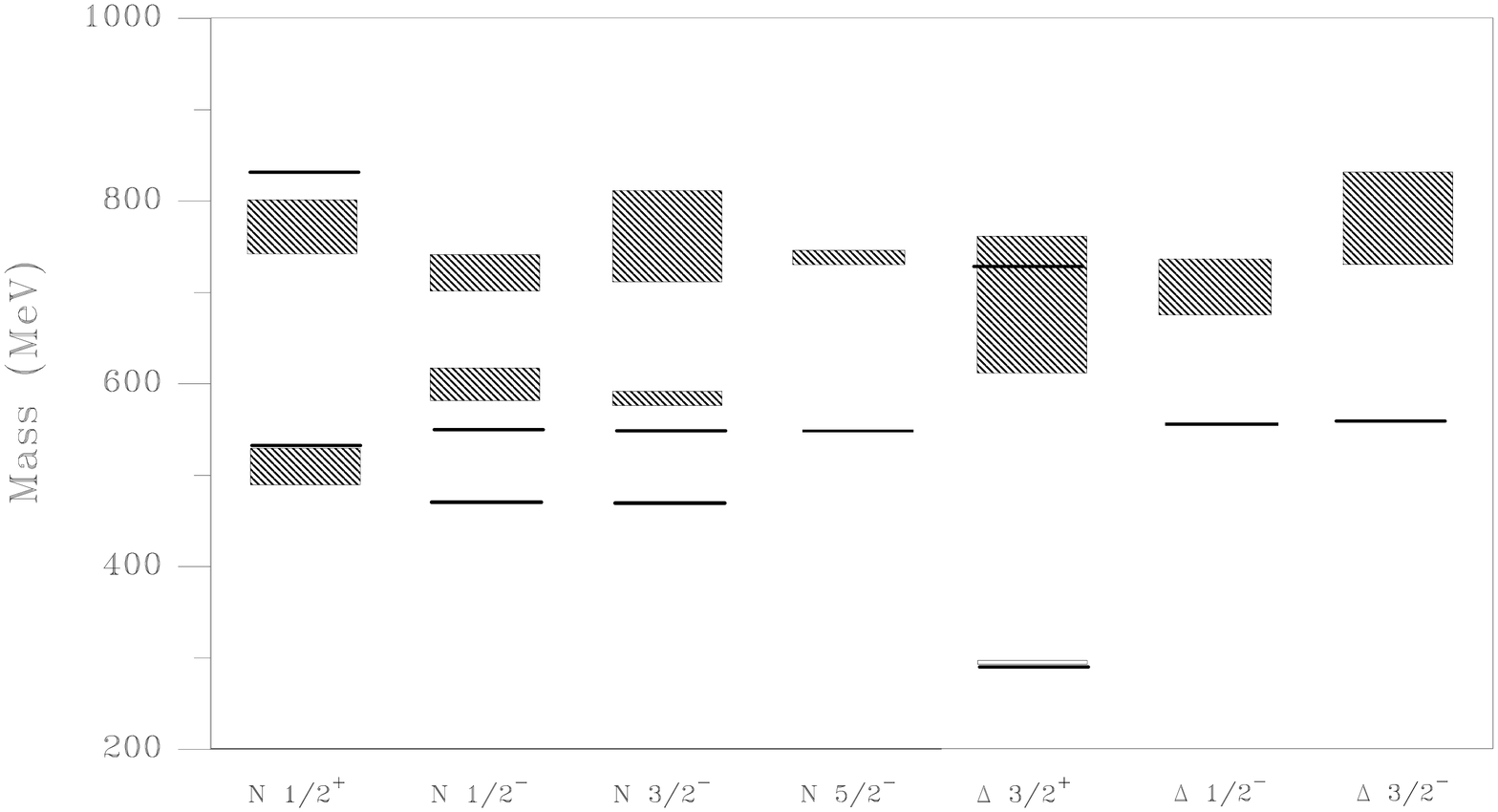,height=2.7in,width=2.55in}}
\hspace*{-0.2cm}
\mbox{\psfig{figure=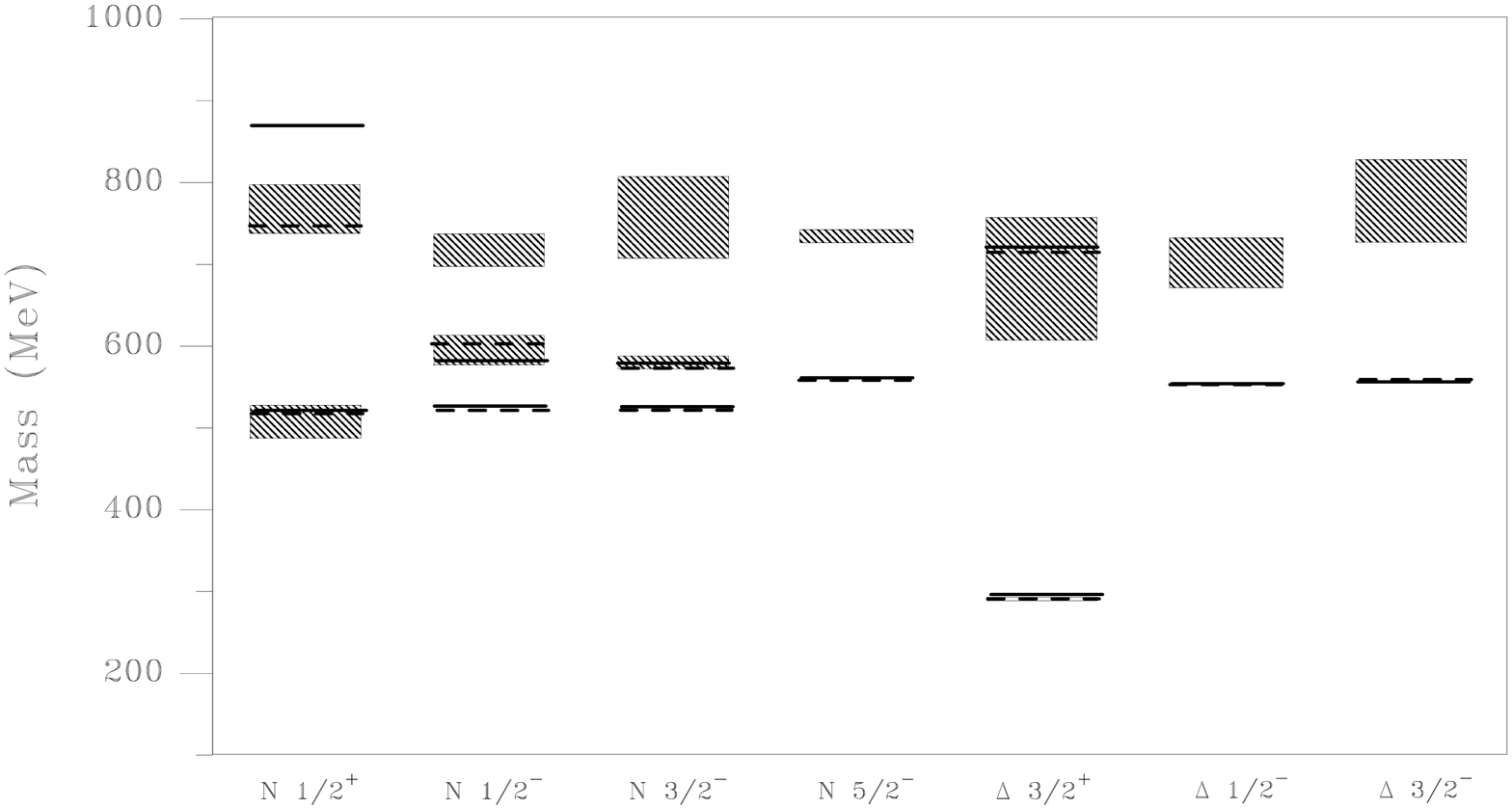,height=2.7in,width=2.55in}}
\vspace*{-1.2cm}
\caption{Relative energy N and $\Delta$ spectrum up to 1 GeV excitation
energy for the set of parameters A (left) and B (right) of table \ref{t1c5}.}
\label{f2c5}
\end{figure}

Increasing the value of $r_0$ the excitation energy
of all the states diminishes, but maintaining a similar ordering.
The excitation energy can be easily increased just by changing the confinement
strength and slightly modifying the strong coupling constant.
The spectrum obtained in a lowest-order Faddeev calculation 
with the set of parameters A of table \ref{t1c5}
is shown in figure \ref{f2c5}, where the solid lines
correspond to the predicted excitation energies 
with respect to the nucleon ground state and the shaded 
areas to the experimental values including their uncertainties.
As can be seen, apart from the energy difference between the positive and negative
parity excitations of the nucleon, the spectrum is
almost the same as those reported in \cite{VAL96a}
and \cite{DZI96}.
These results make clear that the regularization depends on the model space
in which the calculation is done and 
the parameter of this regularization should not
be understood as a true parameter of the model Hamiltonian but
instead as a free parameter to be fitted.

Let us note that the level ordering problem
remains in spite of the fact that the $(\vec \sigma_i
\cdot \vec \sigma_j)(\vec \tau_i \cdot \vec \tau_j)$ structure of the
pseudoscalar potential tends to favor the right location. 
Indeed, the correct ordering
of the Roper and the negative parity excitations of the nucleon can be obtained 
by increasing the strength of the pseudoscalar potential. The strong correlation
among the parameters of the model obliges for a modification of $\alpha_s$
and $a_c$ obtaining the set of parameters B of table \ref{t1c5}.
The spectrum obtained with this set of parameters is shown 
by solid lines in figure \ref{f2c5}.
It can be checked how the correct level ordering 
between the positive and negative parity can be obtained at the expense of having
a stronger OPE (the dashed line in this figure stands for the results of reference
\cite{VAL96a}). However, enhancing artificially the 
pseudoscalar potential the agreement for the two-baryon sector,
specially the binding energy of the deuteron, is destroyed \cite{NAK98}. 
Therefore in the lowest-order Faddeev calculation the 
level ordering problem has no solution if
consistency with the two-baryon system is required.

\begin{table}[t]
\caption{\label{t2c5}
Convergence of baryon masses (in MeV) with respect to the number (Nr.)
of configurations $(\ell_i,\lambda_i,S_i,T_i)$. See text for details.}    
\begin{indented}
\item[]
\begin{tabular}{@{}llllllll}
\br
 Nr. & $M_{\rm N}$ & 
 Nr. & $M_\Delta$ & 
 Nr.  & $ M_{{\rm N}^*(1440)}$ &
 Nr. & $M_{{\rm N}^*(1535)}$  \\
\mr
 2   & 115 & 1 & 423 &  2  & 641 & 4  & 643  \\
 4   & 68  & 2 & 419 &  4  & 622 & 8  & 601  \\
 6   & 21  & 3 & 418 &  6  & 587 & 12 & 570  \\
 8   & 12  &   &     &  8  & 578 & 16 & 551  \\
 10  & 2   &   &     &  10 & 562 & 20 & 540  \\
 12  & 0   &   &     &  12 & 558 &    &      \\
\br
\end{tabular}
\end{indented}
\end{table}
Let us now examine the situation when one extends the calculation to include all
the configurations $(\ell_i,\lambda_i,S_i,T_i)$ with $\ell_i$ and $\lambda_i$ up
to 5. The convergence with respect to the number of 
configurations is shown in table \ref{t2c5} for
N, $\Delta$, N$^*$(1440), and N$^*$(1535) states corresponding to the set
of parameters B of table \ref{t1c5} (figure \ref{f2c5}).
The mass difference with respect to the nucleon ground
state mass as the reference (i.e., $M_{\rm N}=0$) is given.
As can be seen, the convergence with respect to the
number of configurations is different for different states. So,
the $\Delta$ comes down by only 5 MeV while the N, N$^*$(1440), 
and N$^*$(1535) come down by approximately 100 MeV when one
includes the higher orbital angular momentum configurations. It is also
worth to notice that while in the lowest-order calculation the
N$^*$(1440) lies below the N$^*$(1535), when one includes the higher orbital
angular momenta the situation is reversed.
\begin{figure}[b]
\vspace*{-0.5cm}
\centerline{\psfig{figure=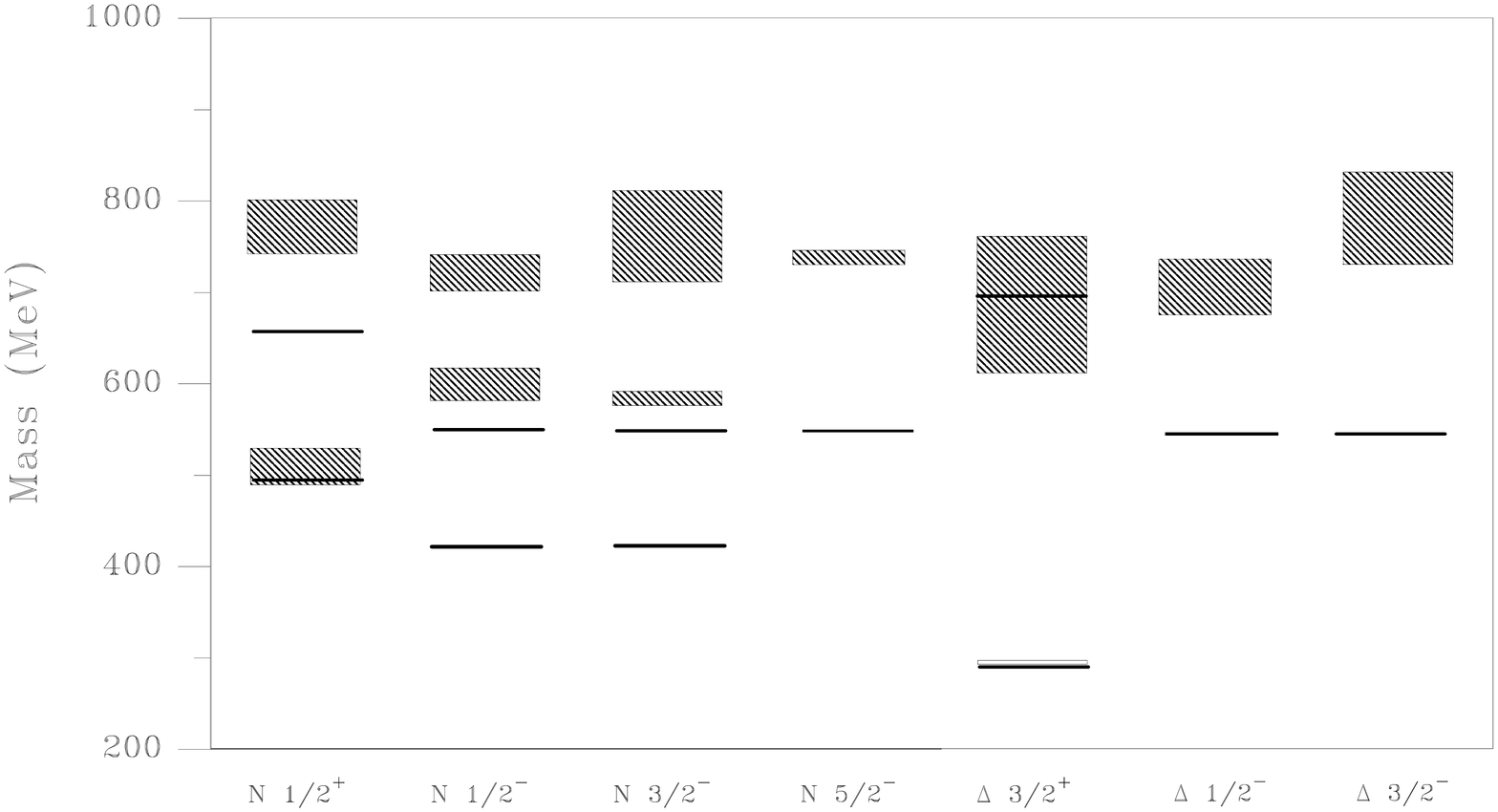,height=2.7in,width=2.6in}}
\vspace*{-0.9cm}
\caption{Relative energy N and $\Delta$ spectra up to 1 GeV excitation
energy including all configurations with $\ell$ and $\lambda$ up to 5
for the set of parameters C.}
\label{f3c5}
\end{figure}

Thus, the higher orbital angular
momentum components play an important role on the description of the
low-energy baryon spectrum. They influence the relative energy
of the nucleonic and $\Delta$ sectors and 
they also affect the relative position of the positive- and negative-parity
excitations of the nucleon. 
In figure \ref{f3c5} we show the baryon spectrum obtained 
with an extended Faddeev calculation for the set of 
parameters C of table \ref{t1c5} that correspond to the ones used
for the description of the baryon-baryon interaction \cite{FER93a,VAL94a}.
The parameter $r_0$ has been taken as 0.25 fm, a typical value
for spectroscopic models, such that the delta function of the OGE
interaction is regularized in the stability region (see figure \ref{f1c5}). 
As one can see, the relative position of the positive-parity states N, 
$\Delta$, and N$^*$(1440) are correctly given, although the negative-parity
states lie below the experimental results. 
It is interesting to check that the spectrum obtained is qualitatively
very similar to the one obtained in \cite{VAL96a}.
The slope of the confining potential, $a_c$, which in set C
of table \ref{t1c5} is 67.0 MeV fm$^{-1}$, is not so different from the 
value 72.5 MeV fm$^{-1}$ used in \cite{VAL96a}
\footnote{One should not forget at this point that consistency with the
NN sector does not impose any restriction to the value
of $a_c$, because the confining interaction does not contribute
appreciably to the NN potential}. 

It is worthwhile to investigate the role played by the OGE and OPE potentials
in the level ordering problem. In reference \cite{GAR01a} 
the energies of the N$^*$(1440) and the N$^*$(1535) 
were calculated starting with the set of parameters 
C and varying the contribution 
of the colour-magnetic interaction by modifying the coupling constant 
$\alpha_s$ of the OGE interaction. It turns out that the
relative position of the N$^*$(1440) and the N$^*$(1535) is not very
much affected by the modification of $\alpha_s$. This behaviour 
indicates that the correct level ordering between the negative
and positive parity excited states of the nucleon 
does not come from a weakening of the OGE interaction. Furthermore,
a suppression of the OGE would imply a stronger pseudoscalar
interaction in order to reproduce the $\Delta-$N mass difference, and
therefore one would obtain a model that it is incompatible with
the understanding of the basic features of the two-nucleon
system \cite{NAK98}.
\begin{figure}[t]
\centerline{\psfig{figure=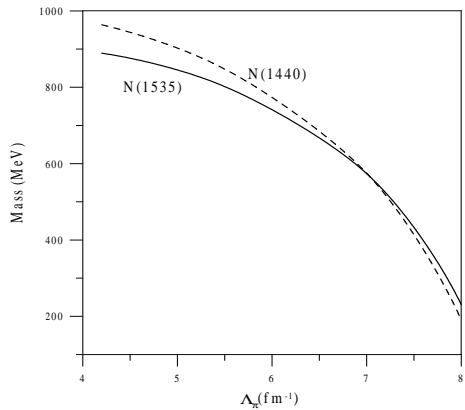,height=3.4in,width=2.6in}}
\vspace*{-3.0cm}
\caption{N$^*$(1440) and N$^*$(1535) masses as a function of the OPE cutoff mass.}
\label{f4c5}
\end{figure}

Regarding the pseudoscalar interaction the $(\vec{\sigma}\cdot \vec{\sigma}) 
(\vec{\tau}\cdot \vec{\tau})$ structure of the OPE gives attraction for 
symmetric spin-isospin pairs and repulsion for antisymmetric ones (a quite distinctive
feature since the colour-magnetic part of the OGE gives similar contributions 
in both cases). This lowers the position of the first nucleon radial excitation,
N$^*$(1440), with regard to the first negative parity state 
solving part of the discrepancy between usual two-body potential models.
To illustrate this point the energy of the N$^*$(1440) and the N$^*$(1535) 
was recalculated starting with the set of parameters C of table \ref{t1c5}, 
but increasing the contribution of the pseudoscalar interaction by
letting the OPE cutoff parameter $\Lambda_\pi$ to increase.
The results are shown in figure \ref{f4c5}. As can be seen, the inversion of
the ordering between the positive and negative parity states
can be achieved if $\Lambda_\pi$ becomes sufficiently large 
(around 7 fm$^{-1}$). A model with such a strong cutoff for 
the OPE is not realistic, the resulting $\Delta-$N mass
difference being around 955 MeV. If one fits again the
$\Delta-$N mass difference by modifying the confinement constant
and suppressing the OGE one loses again the inversion
between the negative and positive parity states. 
\begin{figure}[t]
\vspace*{-0.2cm}
\centerline{\psfig{figure=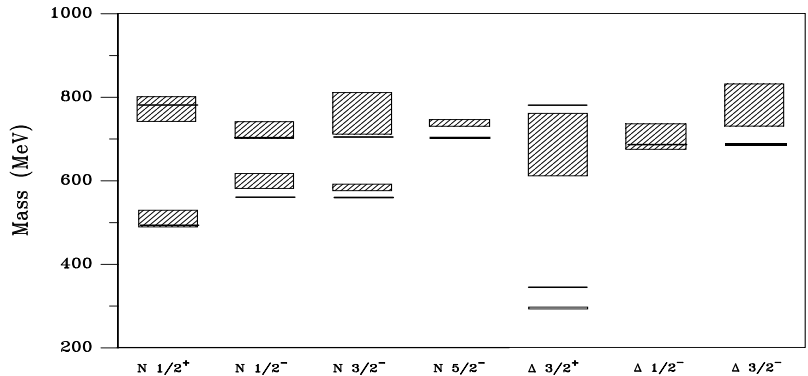,height=5.4in,width=4.4in}}
\vspace*{-8.1cm}
\caption{Relative energy N and $\Delta$ spectra up to 1 GeV excitation
energy with relativistic kinematics 
for the set of parameters D of table \ref{t1c5}.}   
\label{f1rc5}
\end{figure}
Therefore we are forced to conclude that a consistent solution to the
level ordering problem cannot be attained with the pure non-relativistic
chiral constituent quark model.

Recently the baryon spectrum has been studied by means of a chiral constituent
quark model incorporating 
relativistic kinematics \cite{GAR03b}. We show in figure \ref{f1rc5}
the spectrum obtained from a momentum-space Faddeev calculation,
in order to include full three-body relativistic
kinematics using the set of parameters given in column D of table \ref{t1c5}.
As one sees the level-ordering problem that was evident in the spectrum
of figure \ref{f3c5} has now been cured.
As suggested long-time ago the relativistic 
kinematics has an important influence on the N$^*$(1440) mass 
\cite{BAS86,CAR83} and the correct level ordering is obtained
due to the combined effect of relativistic kinematics and one-pion-exchange.
The OGE does not play a major role to this respect although
it should be remarked that
a decreasing of the strength of the OGE goes against the correct level
ordering. 

From a technical point of view, 
the replacement of the non-relativistic by the relativistic 
kinematics for short-range
attractive interactions has to be done together with a careful study of the
short-range part of the potential in the relativistic approach. 
This is due to the fact that the non-relativistic propagator falls down as
$1/k^2$ while the relativistic one does as $1/k$.
This is, for example, the reason why the semirelativistic solutions of
reference \cite{FUR02} collapse when a scalar and an one-gluon-exchange
potential are added to the pseudoscalar interaction. 
This makes evident the risk of replacing
the non-relativistic kinetic energy by the relativistic one
when the details of the short-range part of the potential are not well known.

\subsubsection{Other potential models}

In view of the importance of relativity with regard to the inversion of
positive- and negative-parity nucleon states it 
is also interesting to mention studies
using a fully covariant approach \cite{YUB94,LOR01a,LOR01b,LOR01c}. This 
approach is based on the assumption of instantaneous interactions 
between the quarks which has the consequence that the three-quark
Bethe-Salpeter equation reduces from its four-dimensional form to 
a three-dimensional form, the Salpeter equation, whose 
degree of complexity is similar to the non-relativistic Faddeev method.

A very complete study of the baryon spectrum has been carried out by
the Bonn group \cite{LOR01a,LOR01b,LOR01c} within a covariant formalism
using the Salpeter equation and where the spin has been treated dynamically.
Its model consists of a constant term, a linear confining term, and a 
residual quark-quark interaction. Two models of the part of the 
interaction containing the constant and confining terms (models A and B)
which correspond to different choices for the Dirac structure of each
of these terms are considered. The residual 
quark-quark interaction is based in 't Hooft's
instanton-induced interaction. This determines its Dirac structure
while its radial structure is taken of gaussian type
$e^{-|\vec r_{ij}|^2/\lambda^2}$ with $\lambda=0.4$ fm. This instanton-induced
interaction plays the role of the one-gluon-exchange potential in the more
standard models. Since this interaction acts only in flavour-antisymmetric
states it is responsible for the hyperfine splitting in the baryon spectrum.
Thus, the $\Delta$ spectrum where all the states are flavour-symmetric is
determined only by the constant and linear confining terms. This fact is used
to fit the parameters of the two models (A and B). The strength of
the instanton-induced residual interaction is then fitted to the $\Delta-$N
mass splitting. A reasonable description of the 
$\Delta$ spectrum is obtained for both models. However, in the case
of the nucleon spectrum only model A gives a satisfactory description,
the problem of the inversion of 
positive- and negative-parity nucleon excitations 
being only partially cured since complete inversion is not achieved.
This shows that relativity 
alone is not sufficient to give rise to the inversion of states observed
in the experimental spectrum (as also demonstrated in reference \cite{GAR03}) 
being mandatory to this aim the inclusion of a
pseudoscalar exchange interaction\cite{GAR03b}.

Let us finally mention that in reference \cite{JEN98} the mass spectrum
of the low-lying ground-state baryons has been calculated by solving the Dirac
equation for each quark in a single-particle confining potential and treating
the residual interactions (OGE and OPE potentials) as low-order perturbations. 

\subsubsection{Consistency of the baryon spectrum wave functions with the
ansatz baryon-baryon wave functions}ç
\begin{figure}[t]
\vspace*{-0.2cm}
\mbox{\psfig{figure=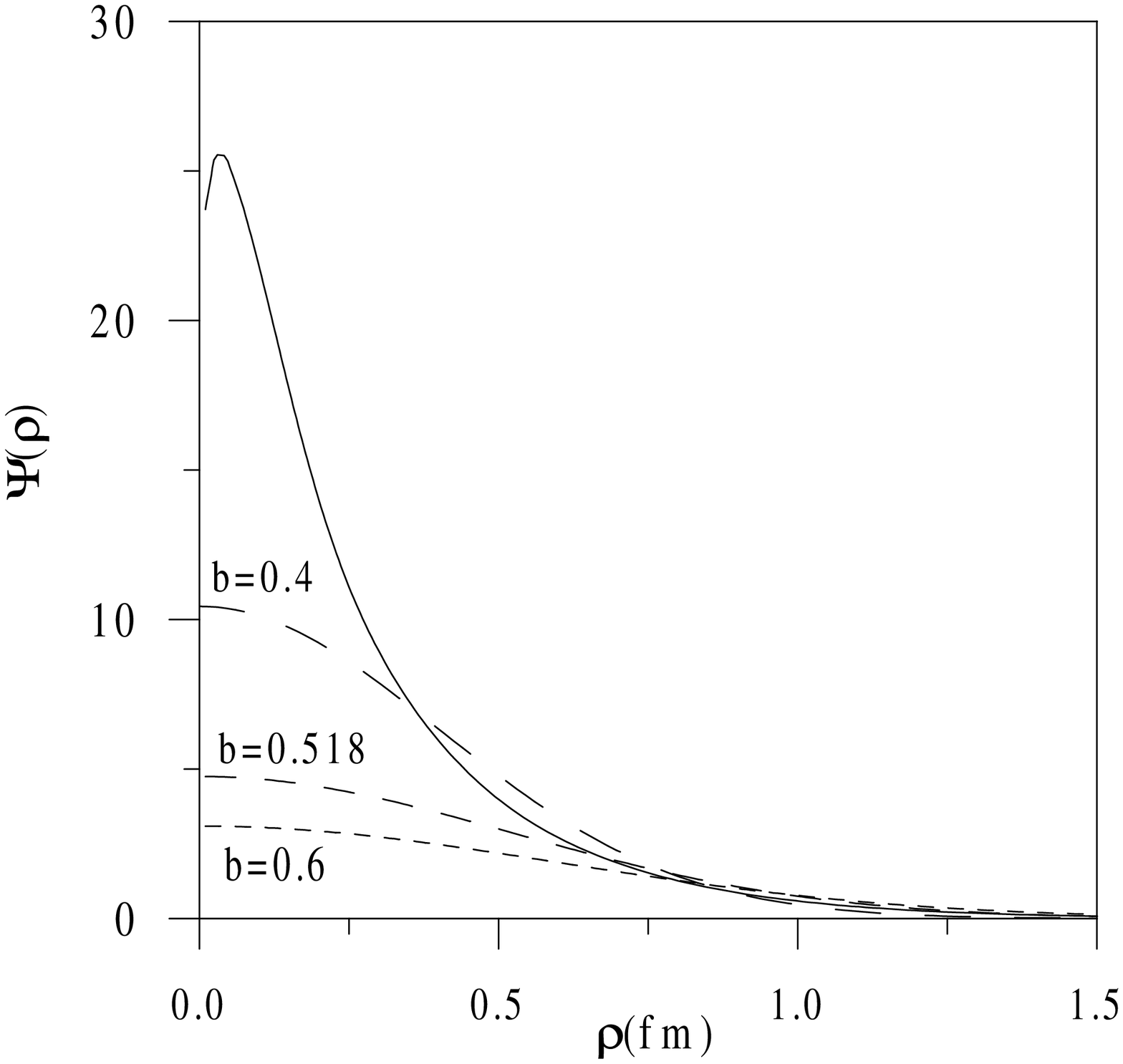,height=3.0in,width=2.6in}}
\hspace*{-0.3cm}
\mbox{\psfig{figure=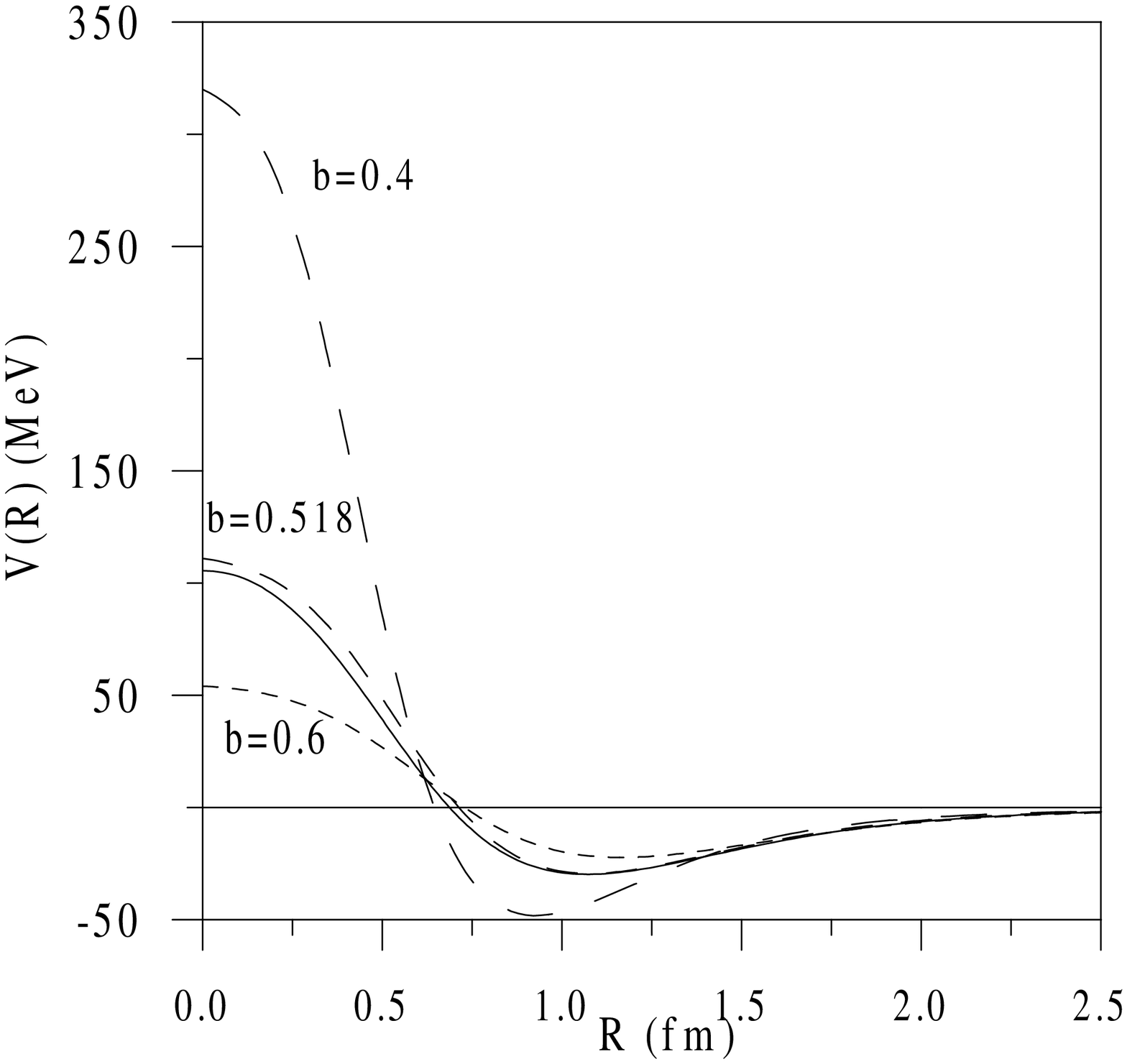,height=3.0in,width=2.6in}}
\vspace*{-2.9cm}
\caption{Main hyperradial component of the nucleon wave function $\Psi (\rho)$
and diagonal kernels of the NN potential $V(R)$ for the $(S,T)=(1,0)$ channel.
The solid line corresponds to the solution of the Sch\"odinger 
equation, the long, medium and short-dashed lines to gaussian
ansatz with values of the parameter $b=$ 0.4, 0.518, and 0.6 fm,
respectively.}
\label{f5c5}
\end{figure}

After discussing the one-baryon problem one question should immediately
arise. The two-baryon system has been studied by means of gaussian-like
wave functions for the spatial part while later on the one-baryon
problem has been solved and therefore one has 
the baryon wave functions predicted by the
quark-quark Hamiltonian. Does it make sense the gaussian hypothesis for the
radial part of the wave function? To answer this question figure
\ref{f5c5} compares the wave function predicted by the quark-quark Hamiltonian 
used for figure \ref{f3c5} to
the ansatz used to study the baryon-baryon interaction for three different 
values of the oscillator parameter. Although at long range all the wave 
functions look very similar, the discrepancy at the origin may reach a factor
four. This effect may be very important for quantities depending on the
value of the wave function at small distances. Hence an evaluation
of the importance of this difference in the calculation of the 
phase shifts should be done. From the technical
point of view a complete RGM calculation would be very difficult to perform
with the full solution of the Schr\"odinger equation, but one can easily
compare the diagonal kernels of the NN potential (which correspond to
the diagonal part used to calculate 
the phase shifts and the deuteron properties)
with both type of wave functions as shown in figure \ref{f5c5}. We see how
the NN potential calculated with the full solution of the Schr\"odinger
equation looks pretty similar to the potential calculated with the harmonic
oscillator wave function with $b=$0.518 fm, this is precisely the value
used in the RGM calculation of the phase shifts and deuteron properties
conferring to it a self-consistent character. Other values of $b$ give
quite different potentials even in the medium-long range part. In this way
the validity of the gaussian ansatz wave function in the RGM calculations
is confirmed, and the old controversy about the possible values of the 
parameter $b$ can be solved.

\subsection{Exotic baryons}

Once the baryon spectrum has been analyzed one may wonder about the
predictions of chiral quark models for exotic baryons, i.e., systems with baryon
number $+1$ whose quantum numbers cannot correspond to a simple three-quark
picture. There has been a renewed interest on this subject from the
announcement of the discovery of the so-called $\Theta^+$ resonance,
with strangeness $+1$, 
in several experiments listed in Table \ref{t3c5}. This resonance,
carrying unit positive charge, shows out as a peak in the invariant mass
distribution of $(K^{+}n)$ or $(K_S^{0}p)$. The location of the peak
corresponds to an invariant mass that ranges from 1520 to 1570 MeV
depending on the experiment and a small width in all cases lesser than 
25 MeV (actually $K$N phase shifts analyses give an upper bound 
of $\sim$ 1 MeV, see reference \cite{NUS04} and references therein).
The angular momentum and parity of
the resonance are still experimentally undetermined whereas there are
experimental indications on its iso-singlet character \cite{BAT03}.
After the initial excitement several recent experiments also listed 
in Table \ref{t3c5} have failed to find evidence for pentaquark signals
(see Ref. \cite{DZI04b} for a detailed description of the experiments).

Different possible explanations for such a peak have been discussed. Leaving
aside its possible origin as a kinematic reflection of the intermediate
production of known meson resonances \cite{DZI04}
what deserves a complete analysis and having been discarded
its $K$N scattering potential nature, the most compelling explanations are
associated to a chiral soliton or a pentaquark $(udud\overline{s})$
structure. The chiral soliton model by
Diakonov, Petrov and Polyakov \cite{DIA97} predicts an
exotic $1/2^{+}$ baryon belonging to the 
$\overline{10}_{F}$ representation of $SU(3)_{Flavour}$ with a mass around
1540 MeV and a width less than 15 MeV (this number has been recalculated as
30 MeV in reference \cite{JAF04}) stimulated its experimental search.
It should be also mentioned that years before Praszalowicz \cite{PRA87}
predicted with a Skyrme model a 1530 MeV mass 
for the iso-singlet member of the antidecuplet.

From a quark model point of view and under the assumption of a 
$J^{P}=1/2^{+}$ assignment, alternative mechanisms, mainly based on colour or
chiral interactions, have been proposed.

\begin{table}[t]
\caption{\label{t3c5}Overview of the positive and null
pentaquark experiments.}
\begin{tabular}{@{}llllll}
\br
\multicolumn{6}{c}{ Positive experiments} \\   
\mr
Exp. & Reaction & Mass (MeV/c$^2$) & Width (MeV) & Significance & Ref. \\
\mr
LEPS(1)  & $\gamma ^{12}C \to K^+ K^- X$ & 1540$\pm$10 &$<25$& 4.6$\sigma$ & \cite{NAK03} \\
LEPS(2)  & $\gamma d \to K^+ K^- X$     & $\sim 1540$ &$-$& $-$ & \cite{NAK03b} \\
DIANA    & $K^+ Xe \to K^0_s p (Xe)'$   & 1539$\pm$2  &$<9$ & 4.4$\sigma$ & \cite{BAR03} \\
CLAS-d   & $\gamma d \to K^+ K^- p(n)$  & 1542$\pm$5  &$<21$& 5.2$\sigma$ & \cite{STE03} \\
SAPHIR   & $\gamma p \to K^+ K^0_S (n)$ & 1540$\pm$6  &$<25$& 4.8$\sigma$ & \cite{BAT03} \\
CLAS-p &$\gamma p \to K^+ K^- \pi^+ (n)$ &1555$\pm$10 &$<26$&7.8$\sigma$ & \cite{KUB03} \\
ITEP     & $\nu A \to K^0_S p X$        & 1533$\pm$5  &$<20$& 6.7$\sigma$ & \cite{ASR03} \\
SVD      & $p A \to p K^0_S X$          & 1526$\pm$3  &$<24$& 5.6$\sigma$ & \cite{ALE04} \\
HERMES & $e^+ d \to K^0_S p X$ & 1526$\pm$3  &13$\pm$9&$\sim 5\sigma$ & \cite{AIR03} \\
ZEUS   & $e p \to K^0_S p X$ & 1522$\pm$3  &8$\pm$4&$\sim 5\sigma$ & \cite{ZEU04} \\
COSY   & $p p \to K^0_S p \Sigma^+$ & 1530$\pm$5  &$<18$&$\sim 5\sigma$ & \cite{ABD04} \\
\br
\multicolumn{6}{c}{Null experiments} \\   
\mr
\multicolumn{2}{l}{Exp.} & \multicolumn{3}{l}{Reaction} & Ref. \\ 
\mr
\multicolumn{2}{l}{HERA-B} & 
\multicolumn{3}{l}{$p A \to \Theta^+ X$} & 
\cite{KNO04} \\ 
\multicolumn{2}{l}{BES} & 
\multicolumn{3}{l}{$e^+ e^- \to J/\psi[\psi(2S)] \to \Theta \overline{\Theta}$} & 
\cite{BAI04} \\ 
\multicolumn{2}{l}{CDF} & 
\multicolumn{3}{l}{$p \overline{p} \to \Theta^+ X$} & 
\cite{LIT04} \\ 
\multicolumn{2}{l}{BaBar} & 
\multicolumn{3}{l}{$e^+ e^- \to \Upsilon(4S)$} & 
\cite{AUB04} \\ 
\multicolumn{2}{l}{ALEPH} & 
\multicolumn{3}{l}{$e^+ e^- \to Z \to q \overline{q}$} & 
\cite{SCH04} \\ 
\multicolumn{2}{l}{DELPHI} & 
\multicolumn{3}{l}{$e^+ e^- \to Z \to q \overline{q}$} & 
\cite{LIN04} \\ 
\multicolumn{2}{l}{FNAL E690} & 
\multicolumn{3}{l}{$p p \to p K^- \pi^+ \Theta^+$} & 
\cite{CHR04} \\ 
\multicolumn{2}{l}{HyperCP} & 
\multicolumn{3}{l}{($\pi^+, K^+, p) Cu \to \Theta^+ X$} & 
\cite{LON04} \\ 
\multicolumn{2}{l}{PHENIX} & 
\multicolumn{3}{l}{$d Au \to \Theta^+ X$} & 
\cite{PIN04} \\ 
\multicolumn{2}{l}{BELLE} & 
\multicolumn{3}{l}{$KN \to \Theta^+ X$} & 
\cite{MIZ04} \\ 
\multicolumn{2}{l}{FOCUS} & 
\multicolumn{3}{l}{$\gamma p \to \Theta^+ X$} & 
\cite{STE04} \\ 
\multicolumn{2}{l}{LASS} & 
\multicolumn{3}{l}{$K^+ p \to K^+ n \pi^+$} & 
\cite{NAP04} \\ 
\multicolumn{2}{l}{L3} & 
\multicolumn{3}{l}{$\gamma \gamma \to \Theta \overline{\Theta}$} & 
\cite{ARM04} \\ 
\multicolumn{2}{l}{SELEX} & 
\multicolumn{3}{l}{$(\pi,p,\Sigma)p \to \Theta^+ X$} & 
\cite{ENG04} \\ 
\multicolumn{2}{l}{SPHINX} & 
\multicolumn{3}{l}{$p C(N) \to \Theta^+ \overline{K^0}C(N)$} & 
\cite{ANT04} \\ 
\br
\end{tabular}
\end{table}

Among the colour based models we shall mention the diquark-diquark and
diquark-triquark models. One should first realize that a simple independent
quark model based on a confinement plus OGE potential would predict the ground
state pentaquark to have negative parity, since the four quarks and the
antiquark would be in $0s$ states and the intrinsic parity of the antiquark
is negative. This difficulty can be overcome by introducing multiquark
correlations. Jaffe and Wilczek \cite{JAF03}, realizing the
presence of strong colour-spin correlations in the baryon spectrum, built a
diquark-diquark-antiquark $((ud)(ud)\overline{s}$) model with the diquark
structures $(3_{Flavour},3_{Colour},S=0)$ favoured by the colour magnetic
(spin-spin) OGE interaction. The assignment of the nucleon Roper resonance
as well as the $\Theta$ to the same $\overline{10}_{F}$ representation
allowed for an understanding not only of the pentaquark mass but also of the
Roper mass on the basis of three-quark $8_{F}-$pentaquark $\overline{10}_{F}$
mixing. On the other hand Karliner and Lipkin \cite{KAR03}
proposed a diquark-triquark model $((ud)(ud\overline{s}))$ with a triquark 
$(\overline{6}_{F},3_{C},S=1/2)$ structure so that the diquark and
triquark have a colour electric OGE interaction (the Coulomb term) between
them which is identical to the quark-antiquark interaction in a meson. Both
models give the right mass for the pentaquark. The small width is attributed
to the weak coupling to the $KN$ continuum due to the difference in colour,
spin and spatial wave functions. Both models predict new stable pentaquarks
in the heavy quark sectors what could serve, if experimentally found, to
disentangle the role of quark correlations.

On the other hand chiral quark models may provide a natural explanation for the positive
parity ground state pentaquark. This is not surprising since they
incorporate Goldstone bosons merging from spontaneous chiral symmetry
breaking and it is precisely this breaking the crucial fact in the
formulation of the chiral soliton model. Explicitly the flavour-spin
interaction may shift the orbitally excited four-quark state below the 
$(0s)^{4}$ since the $L=1$ configuration allows for a completely symmetric
flavour-spin four quark wave function whereas the $(0s)^{4}$ configuration
does not \cite{GLO03}. This parity inversion is similar to the
one giving rise to the correct level ordering in the nucleon spectrum as we
have seen. Stancu and Riska \cite{STA03} have pointed out that pentaquark
stability could be provided by a strong spin-spin interaction between the
light flavour and the strange antiquark. 

These arguments should encourage
more work along these lines in order to get a deeper understanding at the
theoretical level. From the experimental point of view further experiments
are needed to clarify the situation.

%% file: ch6rev.tex
\section{Dibaryons and tribaryons}
\label{ch7}

The possible existence of unstable non-strange two- and three-baryon 
resonances corresponding to bound-state solutions of systems composed 
of nucleons and deltas has been speculated for years
\cite{MUL80,MAL85,WAN95,WON98,BUC98,YUA99,MOT99,GAR99,MOT02,CVE81}.
In the case of two-body
systems (that we shall refer to as dibaryons) 
they will decay mainly into two nucleons and either one
or two pions, while for the three-body case (tribaryons) they will decay
mainly into three nucleons and either one, two, or three pions.
In principle, any nucleus with at least three nucleons
can serve as test system that may be excited 
by forming a dibaryon or a tribaryon \cite{HUB00}.
The lifetime of bound states involving one or more $\Delta$'s should be similar 
to that of the $\Delta$ in the case of very weakly bound systems and larger
if the system is very strongly bound. Therefore, these 
resonances will have widths similar or smaller than the width of 
the $\Delta$ so that, in principle, they may be experimentally observable.

\subsection{Dibaryons in the constituent quark model}

The study of dibaryon resonances within the framework of the constituent
quark model has generally been carried out by considering the baryon-baryon
interaction obtained from a model consisting of confinement and a 
OGE interaction. This baryon-baryon interaction exhibits 
a repulsive core at short distances in most channels due to 
quark antisymmetrization effects and the hyperfine quark-quark 
interaction. Thus, dibaryons may exist in such channels where the strong
short-range repulsion is missing and therefore quark-exchange forces
can produce a tight deep bound state of six quarks.

The detailed investigation of the 
nature (attractive or repulsive) of the different 
channels of the NN, N$\Delta$, and $\Delta\Delta$ systems has been 
an important subject of study ever since the first applications of the 
constituent quark model to derive a baryon-baryon interaction.
Oka and Yazaki \cite{OKA80b} showed
that the Pauli principle and the OGE interaction generate a hard core in
the NN $(j,i)$ 
($j$ stands for the total angular momentum and $i$ for the isospin)
channels (0,1) and (1,0), 
whereas in the $\Delta\Delta$ channels
(0,3) and (3,0) the interaction is strongly attractive such that the 
$(j,i)=(3,0)$ channel is bound. In a later revised analysis by Oka
\cite{OKA93} the $\Delta\Delta$ $(j,i)$ channels (1,0), (0,1),
and (3,0) were found to be attractive while the channel (0,3) was
weakly repulsive.

The $\Delta\Delta$ states $(j,i)=(0,3)$ and (3,0) have also been studied
by Maltman \cite{MAL85} using the standard confinement and OGE interaction
model constructed in \cite{MAL84}. He found that the (3,0) state is bound
by 260 MeV while the (0,3) state is bound by only 30 MeV which is in 
qualitative agreement with the results of Oka and Yazaki \cite{OKA80b,OKA93}
and Cveti\v{c} \etal \cite{CVE80}.

The $\Delta\Delta$ state $(j,i)=(3,0)$ 
has been called the {\it deltaron} by Goldman
and collaborators \cite{GOL89,WAN92} since this state is the analogue of
the deuteron in the case of two $\Delta's$. Their model incorporates the OGE
interaction as well as the quark delocalization (similar to electron
delocalization in nuclear physics) and color 
screening (predicted by unquenched lattice calculations).
Indeed the combined effect of the last two terms produces the 
intermediate-range attraction observed in the NN system. The binding 
energy of the (3,0) state in their model is larger than 200 MeV so that 
the state lies below the $\pi {\rm N}\Delta$ threshold.

\subsection{Dibaryons and tribaryons in the chiral constituent quark model}

This study of the possible existence of such structures has been done
with the help of the baryon-baryon interactions obtained from the
chiral constituent quark model described in \sref{ch2}.
It presents the advantage that once the model parameters are fixed in the study
of the NN system, all the other interacting potentials are parameter-free.
Besides, it allows for a comparison of the results obtained
by means of the local (BO potential) and non-local (RGM potential) interactions
derived from the underlying quark-quark potential.

In order to perform the calculations one can assume that the $\Delta$
is a stable particle, that is, one can neglect its width and the effects of 
retardation in the OPE interaction of the baryon-$\Delta$ subsystems. These 
two effects have been estimated in the case of the simple N$\Delta$
system \cite{GAR97a}. There, it was 
found that the assumption of a stable $\Delta$ 
leads to very reliable predictions for the mass of N$\Delta$ resonances 
since the effects of retardation and width of the $\Delta$ are responsible 
for producing the width of the N$\Delta$ resonance but have almost no effect 
over its mass. Thus, one can be confident that the masses of the bound states 
containing $\Delta$'s will not change very much when its unstable nature
is explicitly taken into account. For the three-baryon systems one can take 
advantage of the experience gained in the three-nucleon bound-state 
problem \cite{HAR72,BET86}, where one knows that the dominant 
configuration is that in which all particles are in $S$-wave states.
However, in order to get reasonable results for the binding energy, the
$S$-wave two-body amplitudes used as input in the Faddeev equations must
also contain the effect of the tensor force. So, for example, in the
case of the Reid soft-core potential if one considers only the $S$-wave
configurations but neglects the tensor force in the two-body subsystems
there is no bound triton. However, 
including the effect of the tensor force in the NN 
$^3S_1-{}^3\! D_1$ channel, but using 
only the $^1S_0$ and $^3S_1$ components of 
the two-body amplitudes in the three-body equations (2-channel calculation),
one gets a triton binding energy of 6.58 MeV against the experimental
value 8.49 MeV. Notice that including the 
remaining configurations (34-channel calculation)
a triton binding energy of 7.35 MeV is obtained \cite{CHE85}. This means that
the $S$-wave truncated T-matrix approximation leads to a binding energy
which differs from the exact value by less than 1 MeV. Hence
this approach may be efficient to look for 
the best candidates for bound states
and the energy ordering of the different tribaryon systems.

\subsubsection{Formalism of two-body bound states}
\label{ch7.1.1}

Let us consider two baryons $B_1$ and $B_2$ in a 
relative $S$-wave interacting through 
a potential that contains a tensor force. 
Then there is a coupling to the $B_1 B_2$
$D$-wave so that the Lippmann-Schwinger equation of the system is of the form
\begin{eqnarray}
\fl
t_{i;j_ii_i}^{l_is_il_i^{\prime\prime}s_i^{\prime\prime}}
(p_i,p_i^{\prime\prime};E)  = 
V_{i;j_ii_i}^{l_is_il_i^{\prime\prime}s_i^{\prime\prime}}
(p_i,p_i^{\prime\prime}) + \sum_{l_i^{\prime}s_i^{\prime}}
\int_0^\infty {p_i^{\prime}}^2dp_i^{\prime}\,
V_{i;j_ii_i}^{l_is_il_i^{\prime}s_i^{\prime}}(p_i,p_i^{\prime}) \nonumber \\
 \times {1 \over E - {p_i^{\prime}}^2/2\eta_i + i\epsilon}
t_{i;j_ii_i}^{l_i's_i'l_i^{\prime\prime}s_i^{\prime\prime}}
(p_i^{\prime},p_i^{\prime\prime};E),
\label{e1c6}
\end{eqnarray}
where $j_i$ and $i_i$ are the total angular momentum and isospin of the
system, while $l_is_i$, $l_i^{\prime}s_i^{\prime}$,
and $l_i^{\prime\prime}s_i^{\prime\prime}$
are the initial, intermediate, and final orbital angular momentum and
spin of the system, respectively. $p_i$ and $\eta_i$ are, respectively, the
relative momentum and reduced mass of the two-body system.
Tables \ref{t1c6} and \ref{t3c6} show
the corresponding NN, N$\Delta$ and $\Delta\Delta$
two-body channels in a relative $S$-wave that are coupled together for the
possible values of $j$ and $i$
(since the NN state is the one with the lowest mass, in this case
we have considered also the possibility of transitions to
larger mass states like N$\Delta$ and $\Delta\Delta$).
In the NN and $\Delta\Delta$ cases the Pauli principle
requires $(-)^{l_i+s_i+i_i} = -1$.
\begin{table}[t]
\caption{\label{t1c6} NN channels $(l_{\rm NN},s_{\rm NN})$,
N$\Delta$ channels $(l_{{\rm N}\Delta},s_{{\rm N}\Delta})$, and
$\Delta\Delta$ channels $(l_{\Delta\Delta},s_{\Delta\Delta})$
that are coupled together in the $^3S_1-{}^3\! D_1$ and $^1S_0$ NN states.}
\begin{indented}
\item[]
\begin{tabular}{@{}llllll}
\br
NN state & $j$ & $i$ & $(l_{\rm NN},s_{\rm NN})$ & 
$(l_{{\rm N}\Delta},s_{{\rm N}\Delta})$
& $(l_{\Delta\Delta},s_{\Delta\Delta})$ \\
\mr
$^3S_1-{}^3\! D_1$   & 1   &  0   &(0,1),(2,1)&   $-$  & (0,1),(2,1),(2,3) \\
$^1S_0$           & 0   &  1   &(0,0)      &(2,2) & $-$ \\
\br
\end{tabular}
\end{indented}
\end{table}
\begin{table}[b]
\caption{\label{t3c6} Coupled channels $(l,s)$ that contribute to a given
N$\Delta$ or $\Delta\Delta$ state with total angular momentum $j$ and isospin $i$.}
\begin{indented}
\item[]
\begin{tabular}{@{}llll}
\br
    &       & N$\Delta$ & $\Delta\Delta$ \\
$j$ & $i$   & $(l,s)$   & $(l,s)$  \\
\mr
0   &  1    & $-$ & (0,0),(2,2)        \\
1   &  0    & $-$ & (0,1),(2,1),(2,3)  \\
1   &  1    & (0,1),(2,1),(2,2) &       $-$          \\
1   &  2    & (0,1),(2,1),(2,2) & (0,1),(2,1),(2,3)  \\
2   &  1    & (0,2),(2,1),(2,2) & (0,2),(2,0),(2,2)  \\
2   &  2    & (0,2),(2,1),(2,2) &       $-$          \\
0   &  3    & $-$ & (0,0),(2,2)        \\
3   &  0    & $-$ & (0,3),(2,1),(2,3) \\
2   &  3    & $-$ & (0,2),(2,0),(2,2) \\
3   &  2    & $-$ & (0,3),(2,1),(2,3) \\
\br
\end{tabular}
\end{indented}
\end{table}

As mentioned before, for the solution of the three-body system one can use
only the component of the T-matrix obtained from the 
equation \eref{e1c6} with $l_i=l_i^{\prime\prime}=0$,
so that for that purpose one can define the $S$-wave truncated amplitude
\begin{equation}
t_{i;s_ii_i}(p_i,p_i^{\prime\prime};E) \equiv
t_{i;s_ii_i}^{0s_i0s_i}(p_i,p_i^{\prime\prime};E).
\label{e2c6}
\end{equation}

\subsubsection{Formalism of three-body bound states}
\label{ch7.1.2}
\begin{table}[b]
\caption{\label{t5c6}
Two-body $\Delta\Delta$ (NN) channels $(j,i)$ that contribute to a given
$\Delta\Delta\Delta$ (NNN) state with total spin $S$ and isospin $I$.}
\begin{indented}
\item[]
\begin{tabular}{@{}llll}
\br
    &       & NN & $\Delta\Delta$ \\
$S$ & $I$   & $(j,i)$ & $(j,i)$ \\
\mr
 1/2    &  1/2    & (1,0),(0,1) & (1,2),(2,1) \\
 1/2    &  3/2    & (0,1) & (1,0),(1,2),(2,1),(2,3) \\
 3/2    &  1/2    & (1,0) & (0,1),(1,2),(2,1),(3,2)  \\
 1/2    &  5/2    & $-$ & (1,2),(2,1),(2,3) \\
 5/2    &  1/2    & $-$ & (1,2),(2,1),(3,2)  \\
 1/2    &  7/2    & $-$ & (1,2),(2,3) \\
 7/2    &  1/2    & $-$ & (2,1),(3,2) \\
 1/2    &  9/2    & $-$ & (2,3) \\
 9/2    &  1/2    & $-$ & (3,2) \\
 3/2    &  3/2    & $-$ & (0,1),(0,3),(1,0),(1,2),
(2,1),(2,3),(3,0),(3,2) \\
 3/2    &  5/2    & $-$ & (0,1),(0,3),(1,2),(2,1),
(2,3),(3,2) \\
 5/2    &  3/2    & $-$ & (1,0),(1,2),(2,1),(2,3),
(3,0),(3,2) \\
 3/2    &  7/2    & $-$ & (0,3),(1,2),(2,3),(3,2) \\
 7/2    &  3/2    & $-$ & (2,1),(2,3),(3,0),(3,2) \\
 3/2    &  9/2    & $-$ & (0,3),(2,3) \\
 9/2    &  3/2    & $-$ & (3,0),(3,2) \\
 5/2    &  5/2    & $-$ & (1,2),(2,1),(2,3),(3,2) \\
 5/2    &  7/2    & $-$ & (1,2),(2,3),(3,2) \\
 7/2    &  5/2    & $-$ & (2,1),(2,3),(3,2) \\
 5/2    &  9/2    & $-$ & (2,3) \\
 9/2    &  5/2    & $-$ & (3,2) \\
 7/2    &  7/2    & $-$ & (2,3),(3,2) \\
 7/2    &  9/2    & $-$ & (2,3) \\
 9/2    &  7/2    & $-$ & (3,2) \\
 9/2    &  9/2    & $-$   & $-$ \\
\br
\end{tabular}
\end{indented}
\end{table}

If one restricts to configurations where all three particles
are in $S$-wave states, the Faddeev equations for the bound-state problem
in the case of three particles with total spin $S$ and
isospin $I$ are (see equation \eref{e5c5})
\begin{eqnarray}
\fl
T_{i;SI}^{s_ii_i}(p_iq_i)  =  \sum_{j\ne i} \sum_{s_ji_j}
<s_ii_i|s_ji_j>_{SI}{1 \over 2}
\int_0^\infty q_j^2 dq_j \int_{-1}^1 dcos\theta\,
t_{i;s_ii_i}(p_i,p^\prime_i;
E - q_i^2/2\nu_i) \nonumber \\
 \times {1 \over E - p_j^2/2\eta_j -q_j^2/2\nu_j}\,
T_{j;SI}^{s_ji_j}(p_jq_j),
\label{e3c6}
\end{eqnarray}
where $p_i$ and $q_i$ are the usual momentum-space 
Jacobi coordinates (see \sref{ch6.1}), $\eta_i$ and
$\nu_i$ are the reduced masses defined by equations \eref{e6c5} and \eref{e7c5},
and $<s_ii_i|s_ji_j>_{SI}$ are the spin-isospin coefficients given by
equation \eref{e8c5}.
\begin{table}[t]
\caption{\label{t6c6}
Two-body N$\Delta$ $(j_{{\rm N}\Delta},i_{{\rm N}\Delta})$ and 
NN channels $(j_{\rm NN},i_{\rm NN})$ that contribute to a given
NN$\Delta$ state with total spin $S$ and isospin $I$.}
\begin{indented}
\item[]
\begin{tabular}{@{}llll}
\br
 $S$ & $I$   & $(j_{{\rm N}\Delta},i_{{\rm N}\Delta})$ &  $(j_{\rm NN},i_{\rm NN})$ \\
\mr
 1/2    &  1/2    & (1,1) &  \\
 1/2    &  3/2    & (1,1),(1,2) & (1,0) \\
 1/2    &  5/2    & (1,2) &  \\
 3/2    &  1/2    & (1,1),(2,1) & (0,1) \\
 3/2    &  3/2    & (1,1),(1,2),(2,1),(2,2) & (1,0),(0,1) \\
 3/2    &  5/2    & (1,2),(2,2) & (0,1) \\
 5/2    &  1/2    & (2,1) &  \\
 5/2    &  3/2    & (2,1),(2,2) & (1,0) \\
 5/2    &  5/2    & (2,2) &  \\
\br
\end{tabular}
\end{indented}
\end{table}
\begin{table}[b]
\caption{\label{t7c6}
Two-body N$\Delta$ $(j_{{\rm N}\Delta},i_{{\rm N}\Delta})$ and
$\Delta\Delta$ channels $(j_{\Delta\Delta},i_{\Delta\Delta})$
that contribute to a given N$\Delta\Delta$ state with
total spin $S$ and isospin $I$.}
\begin{indented}
\item[]
\begin{tabular}{@{}llll}
\br
 $S$ & $I$ & $(j_{{\rm N}\Delta},i_{{\rm N}\Delta})$ & 
$(j_{\Delta\Delta},i_{\Delta\Delta})$ \\
\mr
 1/2    &  1/2    & (1,1),(1,2),(2,1),(2,2) & (1,0),(0,1) \\
 1/2    &  3/2    & (1,1),(1,2),(2,1),(2,2) & (0,1),(1,2) \\
 1/2    &  5/2    & (1,1),(1,2),(2,1),(2,2) & (0,3),(1,2) \\
 1/2    &  7/2    & (1,2),(2,2) & (0,3) \\
 3/2    &  1/2    & (1,1),(1,2),(2,1),(2,2) & (1,0),(2,1) \\
 3/2    &  3/2    & (1,1),(1,2),(2,1),(2,2) & (1,2),(2,1) \\
 3/2    &  5/2    & (1,1),(1,2),(2,1),(2,2) & (1,2),(2,3) \\
 3/2    &  7/2    & (1,2),(2,2) & (2,3) \\
 5/2    &  1/2    & (1,1),(1,2),(2,1),(2,2) & (2,1),(3,0) \\
 5/2    &  3/2    & (1,1),(1,2),(2,1),(2,2) & (2,1),(3,2) \\
 5/2    &  5/2    & (1,1),(1,2),(2,1),(2,2) & (2,3),(3,2) \\
 5/2    &  7/2    & (1,2),(2,2) & (2,3) \\
 7/2    &  1/2    & (2,1),(2,2) & (3,0) \\
 7/2    &  3/2    & (2,1),(2,2) & (3,2) \\
 7/2    &  5/2    & (2,1),(2,2) & (3,2) \\
 7/2    &  7/2    & (2,2) & \\
\br
\end{tabular}
\end{indented}
\end{table}

The three amplitudes $T_{1;SI}^{s_1i_1}(p_1q_1)$, $T_{2;SI}^{s_2i_2}(p_2q_2)$,
and $T_{3;SI}^{s_3i_3}(p_3q_3)$ in equation \eref{e3c6} are coupled together.
The number of coupled equations can be reduced since some of the 
particles are identical. In the case of three identical particles (NNN and
$\Delta\Delta\Delta$ systems) all three amplitudes are
equal and therefore equation \eref{e3c6} becomes,
\begin{eqnarray}
\fl
T_{i;SI}^{s_ii_i}(p_iq_i)  =   \sum_{s_ji_j}
<s_ii_i|s_ji_j>_{SI}
\int_0^\infty q_j^2 dq_j \int_{-1}^1 dcos\theta\,
t_{i;s_ii_i}(p_i,p^\prime_i;
E - q_i^2/2\nu_i) \nonumber \\
 \times {1 \over E - p_j^2/2\eta_j -q_j^2/2\nu_j}\,
T_{j;SI}^{s_ji_j}(p_jq_j).
\label{e4c6}
\end{eqnarray}
Table \ref{t5c6} details the three NNN states characterized by total spin
and isospin $(S,I)$ that are possible as well as the two-body
NN channels that contribute to each state. 
The 25 possible $\Delta\Delta\Delta$ states as well as the two-body $\Delta\Delta$
channels that contribute to each state are also given.

When two particles are identical and one different
(NN$\Delta$ and N$\Delta\Delta$ systems) two of the amplitudes are equal.
With the assumption that particle 1 is the different one and
particles 2 and 3 are the two identical, only the amplitudes
$T_{1;SI}^{s_1i_1}(p_1q_1)$ and $T_{2;SI}^{s_2i_2}(p_2q_2)$ are
independent from each other. The reduction procedure for the case where one has two
identical fermions has been described in \cite{GAR90a} and \cite{AFN74}
and will not be repeated here. Tables \ref{t6c6}
and \ref{t7c6} resume the possible NN$\Delta$ and N$\Delta\Delta$
states characterized by its total spin and isospin $(S,I)$ 
as well as the two-body N$\Delta$ and NN, or N$\Delta$ and 
$\Delta\Delta$, channels that contribute to each state. 

\subsubsection{Results}
\label{ch7.2}

Before going to the numbers obtained two technical
precisions are in order. 
It has been already discussed in \sref{ch3.1} that the
$(j,i)=$ (1,1) and (2,2) N$\Delta$ states, and the
$(j,i)=$ (2,3) and (3,2) $\Delta\Delta$ states present 
quark Pauli blocking for $L=0$, what translates into
a strong repulsive core in the baryon-baryon potential.
The reason for such a repulsive core 
is the presence of a forbidden state, that in the RGM potential
should be eliminated from the relative motion
wave function for each partial wave. 
This procedure is tedious both
from the conceptual and numerical point of view \cite{SAIT68,OTS65}.
It has been demonstrated \cite{TOK80} that for the Pauli blocked channels
the local N$\Delta$ and $\Delta\Delta$ potentials reproduce
the qualitative behaviour of the RGM kernels after the subtraction of 
the forbidden states. This is why reference \cite{MOT02}
used the local version of the quark Pauli blocked channels.

For the three-body systems the binding-energy spectrum, say, the energy of 
the states measured with respect to the three-body threshold, as well as the 
separation-energy spectrum, the energy of the states measured with 
respect to the threshold of one-free particle and a bound state of the 
other two, will be calculated. The deepest bound three-body state is not 
the one with the largest binding energy but the one with the largest separation 
energy, since such a state is the one that requires more energy in order to become 
unbound, that is, to move it to the nearest threshold. The numerical results
for two-body and three-body systems appear in tables \ref{t8c6} and
\ref{t11c6} respectively.

\subsubsection{The NN system}
\label{ch7.2.3}

Out of the two states of table \ref{t1c6} only 
the $(j,i)=(1,0)$, that is the deuteron, 
is bound. The non-local model gives a deuteron binding energy of
2.14 MeV, while the local version gives 3.13 MeV.
The exact chiral constituent quark model NN potential gives a deuteron binding 
energy of 2.225 MeV \cite{ENT00}. This value was 
obtained by taking into account the 
$\Delta\Delta$ partial wave $(l_{\Delta\Delta},s_{\Delta\Delta})=(4,3)$
coupled together in addition to those given in table \ref{t1c6}. 
The present calculation only considers $S$- and $D$-waves
obtaining for the non-local potential a deuteron 
binding energy of 2.14 MeV, which differs 
less than 0.1 MeV from the exact result.

\subsubsection{The N$\Delta$ system}
\label{ch7.2.2}

Out of the four possible N$\Delta$ states of table \ref{t3c6}
only one, the $(j,i)=(2,1)$, has a bound state which lies exactly
at the N$\Delta$ threshold for the local model, and with a small binding
of 0.141 MeV for the non-local one.
The states $(j,i)=(1,1)$ and $(2,2)$ are unbound because they present
quark Pauli blocking that, as in the case of the $\Delta\Delta$ system,
will play an important role in the three-body spectrum.
The state $(j,i)=(2,1)$ can also exist in the NN system and there 
it corresponds to the $^1D_2$ partial 
wave which has a resonance at an invariant 
mass of 2.17 GeV \cite{HOS78,YOK80,ARN00}. This means that the N$\Delta$ bound 
state may decay into two nucleons and appear in the NN system as a resonance.
As the N$\Delta$ bound state has for both local and non-local models
energies very close to the N$\Delta$ threshold, 
the invariant mass of the system should be very close to 2.17 GeV. Thus, 
the chiral constituent quark model predicts the NN $^1D_2$ resonance 
as being a N$\Delta$ bound state.

\subsubsection{The $\Delta\Delta$ system}
\label{ch7.2.1}

Out of the eight possible $\Delta\Delta$ states given in table \ref{t3c6}
with the non-local interaction five have a bound state, whereas the
local interaction binds six of them (in both cases there 
are no excited states in any channel). The channels $(j,i)=(2,3)$ and $(3,2)$ 
are unbound as expected, due to the strong repulsive barrier 
at short distances in the $S$-wave central interaction.
These repulsive cores will largely determine the three-body spectrum.
\begin{table}[b]
\caption{\label{t8c6} Binding energies $B_2$ (in MeV) of the NN, 
N$\Delta$ and $\Delta\Delta$ states. 'ub' denotes unbound channels.
$B_2^L$ ($B_2^{NL}$) refers to the local (nonlocal) potential.}
\begin{indented}
\item[]
\begin{tabular}{@{}|ll|ll|ll|ll|}
\br
 & & \multicolumn{2}{|c|}{\rm NN} & \multicolumn{2}{|c|}{N$\Delta$} & 
\multicolumn{2}{|c|}{$\Delta\Delta$} \\
$j$ & $i$  & $B_2^L$ & $B_2^{NL}$ & $B_2^L$ & $B_2^{NL}$ & $B_2^L$ & $B_2^{NL}$ \\
\mr
0   &  1    & ub    & ub   & $-$  & $-$  & 108.4 & 159.5  \\
1   &  0    & 3.13  & 2.14 & $-$  & $-$  & 138.5 & 190.3   \\
1   &  1    & $-$   & $-$  & ub   & ub   &  $-$  &  $-$       \\
1   &  2    & $-$   & $-$  & ub   & ub   & 5.7   & ub  \\
2   &  1    & $-$   & $-$  & 0.0  & 0.141& 30.5  & 7.4  \\
2   &  2    & $-$   & $-$  & ub   & ub   &  $-$  & $-$     \\
0   &  3    & $-$   & $-$  & $-$  & $-$  & 0.4   & 0.2  \\
3   &  0    & $-$   & $-$  & $-$  & $-$  & 29.9  & 7.8  \\
2   &  3    & $-$   & $-$  & $-$  & $-$  & ub    & ub  \\
3   &  2    & $-$   & $-$  & $-$  & $-$  & ub    & ub  \\
\br
\end{tabular}
\end{indented}
\end{table}

The predicted bound states $(j,i)$ = (1,0), (0,1), (2,1) and (3,0) appear 
also in the NN system. One can see that the 
deepest bound state is $(j,i)$ = (1,0), the second is $(j,i)$ = (0,1), the
third is $(j,i)$ = (2,1), and the fourth is $(j,i)$ = (3,0). 
The first three states appear also, and precisely in the same order, in the NN system.
The $(j,i)$ = (1,0) state has the quantum numbers of the deuteron, the $(j,i)$ = (0,1) is the
$^1S_0$ virtual bound state, and the $(j,i)$ = (2,1) is the $^1D_2$ resonance that 
lies at $\sim$ 2.17 GeV \cite{HOS78} (notice that the $^3F_3$ NN resonance has 
no counterpart in table \ref{t8c6} because only even parity states have been
calculated and $^3F_3$ has odd parity). Thus, the $(j,i)$ = (3,0) state which 
is also allowed in the NN system could correspond 
to a new NN resonance found in reference \cite{VAL00}. 

\begin{figure}[t]
\vspace*{-0.3cm}
\hspace*{-0.4cm}
\mbox{\psfig{figure=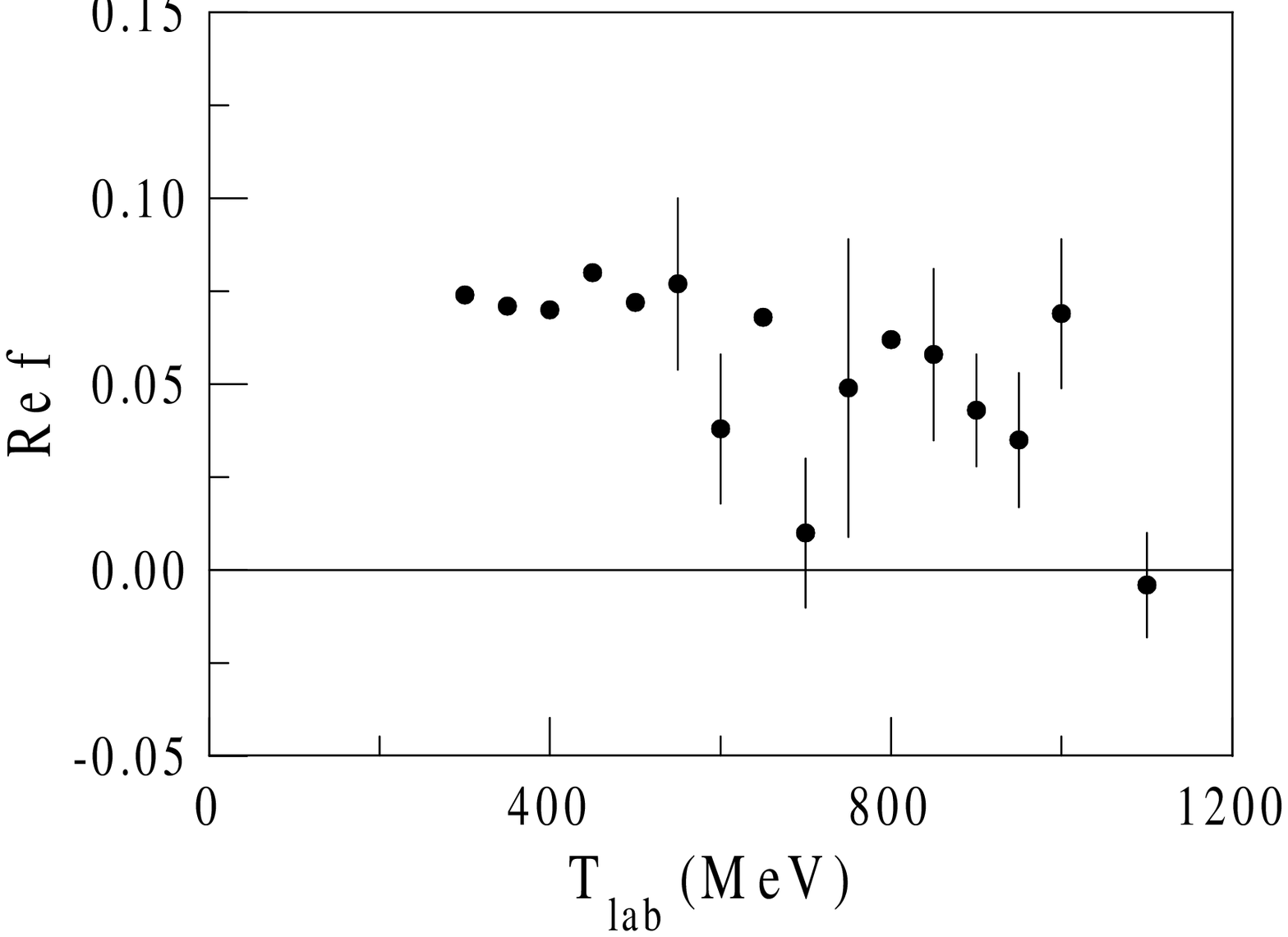,height=4.0in,width=2.8in}}
\hspace*{-0.7cm}
\mbox{\psfig{figure=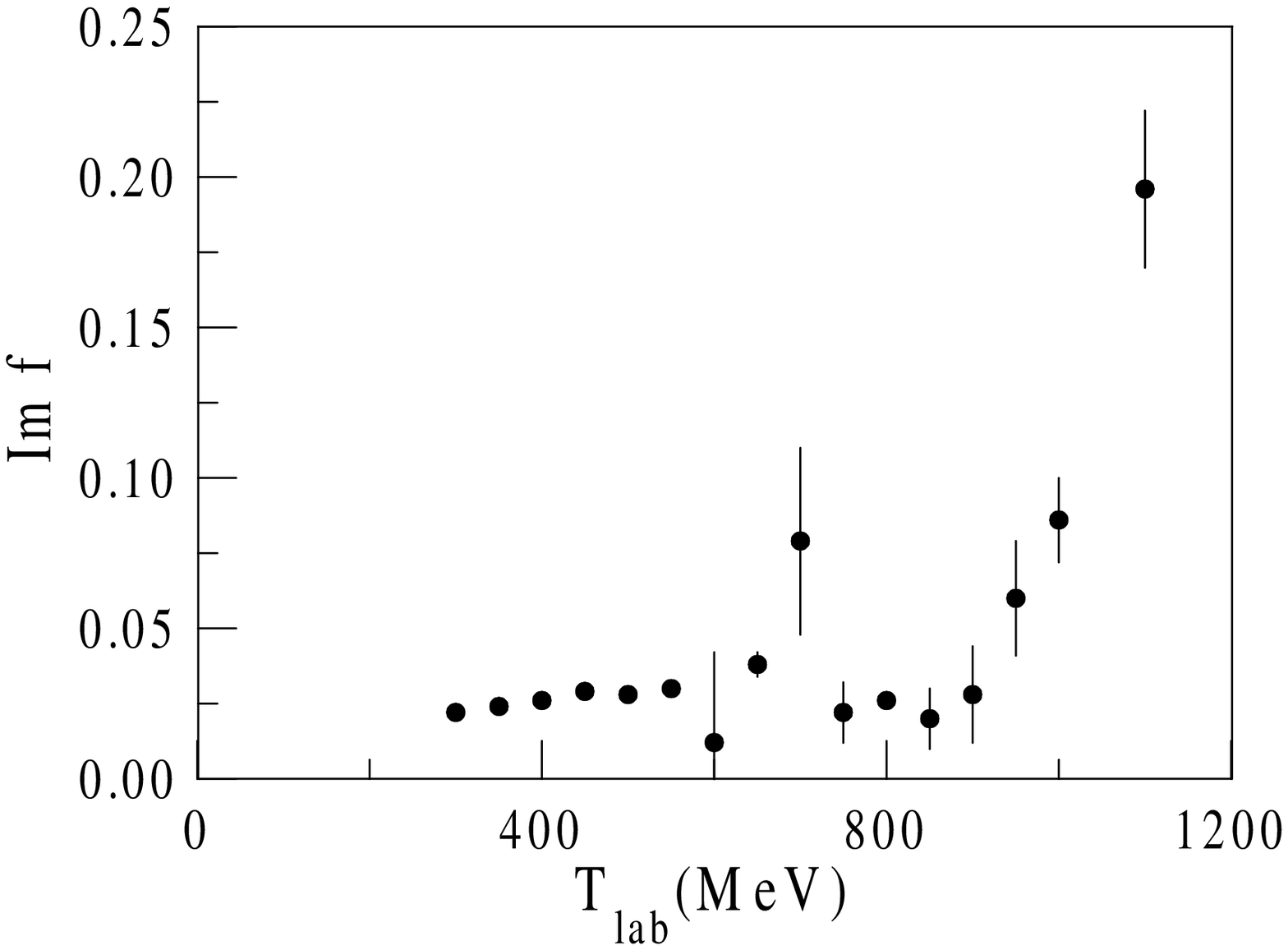,height=4.0in,width=2.8in}}
\vspace*{-5.7cm}
\caption{Real and imaginary parts of the the $^3D_3$ amplitude \cite{ARN00}.} 
\label{f1c6}
\end{figure}

It is interesting  to note that some indication of a $(j,i)$ = (3,0) 
resonance can already be seen 
in the recent analyses of the NN data by Arndt \etal \cite{ARN00}.
This channel corresponds to the NN $^3D_3$ partial wave. 
The most distinctive feature of a resonance is that as the energy increases
the real part of the scattering amplitude changes sign from positive to negative while 
the imaginary part becomes large, so that the amplitude describes a 
counterclockwise loop in the Argand diagram. The energy at which
this change of sign occurs corresponds to the mass of the resonance.
Figure \ref{f1c6} shows the real and imaginary parts of the $^3D_3$ amplitude
obtained from the single-energy analysis of reference \cite{ARN00}. As one 
can see a resonance-like behaviour seems to be present at about 700 and 1100 MeV.
These kinetic energies correspond to invariant masses of 2.20 and 2.37 GeV so that in 
either case the ordering of the state agrees with that predicted in table \ref{t8c6}.
The $\Delta\Delta$ bound state in the $(j,i)$ = (3,0) channel has been
predicted also by other models 
\cite{KAM97,WAN92,GOL89,OKA84,OKA93,CVE80}, and
a method to search for it experimentally has been proposed \cite{WON98}.

\subsubsection{The NNN system}
\label{ch7.2.7}

As a reliability test of the calculating scheme for the
three-baryon system, the NNN bound-state problem has been studied.
Of the three NNN states of table \ref{t5c6} only the channel
$(S,I)=(1/2,1/2)$, that is the triton, has a bound
state. By using the local potential a binding energy of 5.76
MeV is obtained, and 6.52 MeV with the non-local one. 
For comparison, we notice that the triton binding energy for 
the Reid-soft-core potential in the truncated T-matrix approximation is 6.58 MeV.
Since the experimental value is $B_{\rm exp}=8.49$ MeV, the difference with
the theoretical result, of about 3 MeV, is a measure of the uncertainty
of the two-channel calculation in the case of the three-baryon system.
A detailed study of the triton binding energy will be presented 
in the next section.
\begin{table}[t]
\caption{\label{t11c6}
Binding energies $B_3$ (in MeV) of the NNN, N$\Delta\Delta$, NN$\Delta$, 
and $\Delta\Delta\Delta$ states. 'ub' denotes unbound channels.
For the bound channels the separation energy is given below
in parenthesis. $B_3^L$ ($B_3^{NL}$) refers to the local (nonlocal) potential.}
\begin{tabular}{@{}|ll|p{1.1cm}p{1.1cm}|p{0.5cm}p{1.1cm}|p{0.5cm}p{1.2cm}|p{1.1cm}p{1.1cm}|}
\br
& &\multicolumn{2}{|c|}{NNN} & \multicolumn{2}{|c|}{N$\Delta\Delta$}&
\multicolumn{2}{|c|}{NN$\Delta$} &\multicolumn{2}{|c|}{$\Delta\Delta\Delta$} \\
$S$ & $I$ & $B_3^L$ & $B_3^{NL}$ & $B_3^L$ & $B_3^{NL}$ 
& $B_3^L$  & $B_3^{NL}$ & $B_3^L$  & $B_3^{NL}$ \\
\mr
$1/2$&$1/2$&$\begin{array}{l}5.76 \\(2.63)\end{array}$
&$\begin{array}{l} 6.52 \\(4.38)\end{array}$
&$-$&$-$        &  $-$  &   $-$      
&$\begin{array}{l} 84.0 \\(53.5)\end{array}$
&$\begin{array}{l} 16.6 \\(9.2)\end{array}$ \\
1/2&3/2&ub        &ub        &$-$&$-$        
&  $-$  &   $-$      
&$\begin{array}{l} 139.2 \\(0.7)\end{array}$
& ub         \\
3/2&1/2&ub        &ub        &$-$&$-$        & ub    
&$\begin{array}{l} 0.14 \\(0.002)\end{array}$
&$\begin{array}{l} 109.5 \\(1.1)\end{array}$
& ub         \\
3/2&3/2&$-$       &$-$       &$-$&$-$        & ub    
&$\begin{array}{l} 2.28 \\(0.14)\end{array}$
& $-$       &  $-$       \\
1/2&5/2&$-$       &$-$       &ub 
&$\begin{array}{l} 0.63 \\(0.43)\end{array}$
&  $-$  &   $-$      & $-$       &  $-$       \\
5/2&1/2&$-$       &$-$       &ub 
&$\begin{array}{l} 8.16 \\(0.36)\end{array}$
&  $-$  &   $-$      
&$\begin{array}{l} 39.1 \\(8.6)\end{array}$
&$\begin{array}{l} 9.3 \\(1.9)\end{array}$ \\
5/2&5/2&$-$       &$-$       &ub 
&$\begin{array}{l} 0.18 \\(0.04)\end{array} $
&  $-$  &   $-$      & $-$       &  $-$       \\
1/2&7/2&$-$       &$-$       &$-$&$-$        &  $-$  &   $-$      
&$\begin{array}{l} 6.3 \\(0.6)\end{array}$
& ub           \\
7/2&1/2&$-$       &$-$       &$-$&$-$        &  $-$  &   $-$      
&$\begin{array}{l} 31.7 \\(1.2)\end{array}$
&$\begin{array}{l} 7.8 \\(0.4)\end{array}$ \\
7/2&3/2&$-$       &$-$       &$-$&$-$        &  $-$  &   $-$      
&$\begin{array}{l} 35.1 \\(4.6)\end{array}$ 
&$\begin{array}{l} 9.8 \\(2.0)\end{array}$ \\
\br
\end{tabular}
\end{table}

\subsubsection{The NN$\Delta$ system}
\label{ch7.2.6}

Though the presence of bound states in the N$\Delta$
$(j,i)=(2,1)$ and NN $(j,i)=(1,0)$ channels could induce to 
expect several bound states in the N$\Delta\Delta$ system,
this is not the case indeed. As a matter of fact, 
with the non-local potential only two out of nine possible three-body states
given in table \ref{t6c6} are bound. Because of the attractive
contribution of the N$\Delta$ $(j,i)=(2,1)$ bound state with the
non-local model, the three-body state $(3/2,1/2)$
turns out to be very weakly bound, with an energy of 0.14 MeV, and a
separation energy different hardly from zero.
This means that the $(S,I)=(3/2,1/2)$ state is very near
the NN$\Delta$ threshold and therefore it represents the tribaryon
resonance with the lowest possible mass since it can decay into
three nucleons and one pion. Also, for this case
the three-body state $(3/2,3/2)$ is bound. As it can be
seen from table \ref{t6c6} this state has the contribution of all the
two-body N$\Delta$ and NN channels. In spite of the fact that
the N$\Delta$ two-body channels $(j,i)=(1,1)$ and
$(2,2)$ present Pauli blocking, the attractive contribution of
the N$\Delta$ $(j,i)=(2,1)$ and NN $(j,i)=(1,0)$ channels
results to be enough to weakly bound this state.
We note that neither one of the three-body states
$(S,I)=(3/2,1/2)$ and $(3/2,3/2)$ is bound with the local interaction.

The predicted NN$\Delta$ states with
$(S,I)=(3/2,1/2)$ and $(S,I)=(3/2,3/2)$
which correspond to $M \approx 3.4$ GeV are the
tribaryon resonances with the lowest mass and therefore the ones that
would be more easy to detect experimentally.

\subsubsection{The N$\Delta\Delta$ system}
\label{ch7.2.5}

In the non-local case one finds
that three of the sixteen possible N$\Delta\Delta$ states given in table 
\ref{t7c6} are bound. They are the $(S,I)=(1/2,5/2)$, $(5/2,1/2)$,
and $(5/2,5/2)$ states and their bound
state energies are 0.63 MeV, 8.16 MeV, and 0.18 MeV, respectively.
In the case of the states $(S,I)=(1/2,5/2)$ and $(S,I)=(5/2,1/2)$ the 
repulsive barrier due to the quark Pauli blocked N$\Delta$ states
$(j,i)=(1,1)$ and $(2,2)$ is less strong than the
attraction due to the state $(j,i)=(2,1)$, so that they result
to be bound in the non-local model. The state $(S,I)=
(5/2,5/2)$ is the weakest bound state of this system,
since in addition to the contribution of the N$\Delta$ quark Pauli
blocked channels, there exists that of the
$\Delta\Delta$ quark Pauli blocked channels $(j,i)=(2,3)$
and $(3,2)$. This confirms what has been mentioned
that it is the structure of the interaction of the
two-body system the one which largely determines the three-body
spectrum.  Thus, the non-local interaction predicts the bound states
$(S,I)=(1/2,5/2)$, $(5/2,1/2)$ and $(5/2,5/2)$, which in principle may
be observed as tribaryon resonances which decay into three
nucleons and two pions with masses close to the N$\Delta\Delta$ threshold.

\subsubsection{The $\Delta\Delta\Delta$ system}
\label{ch7.2.4}

While the non-local model presents four bound states the local interaction
finds seven. From table \ref{t11c6} one observes that 
the three states which are missing in the non-local version
are barely bound in the local version, i.e., they have very small
separation energies. Since the non-local interaction tends to lower
the attraction in all the $\Delta\Delta\Delta$ channels it is not
surprising that those which were barely bound have disappeared.
The more strongly bound three-body state
(that is, the one with the largest separation energy) is the
$(S,I)=(1/2,1/2)$ state which has precisely
the triton quantum numbers. This shows again, like in the $\Delta\Delta$
and NN systems, the similarity between 
the $\Delta\Delta\Delta$ and NNN systems.

The reason why the $(S,I)=(1/2,1/2)$ state is the more
strongly bound is very simple. As shown in table \ref{t5c6} this is the only
state where none of the Pauli blocked two-body channels
$(j,i)=(2,3)$ and $(3,2)$ contribute. In all the other three-body states the
strong repulsion of these channels either
completely destroys the bound state or allows just a barely bound one.
The state $(S,I)=(7/2,3/2)$ comes next with respect to the
separation energy. This state has
a somewhat anomalous behaviour since it has a relatively large separation 
energy. This behaviour is sort of accidental and it
can be understood as follows. 
As seen in table \ref{t5c6}, there are four two-body
channels contributing to the $(S,I)=(7/2,3/2)$ state, the
two attractive ones $(j,i)=(2,1)$ and $(3,0)$ and the two repulsive ones
$(j,i)=(2,3)$ and $(3,2)$. However, as seen in table \ref{t8c6} 
the attractive channels $(2,1)$ and $(3,0)$
have bound states at $E=-7.4$ MeV and $E=-7.8$ MeV, respectively,
for the non-local model, and $E=-30.5$ MeV and $E=-29.9$ MeV,
respectively, for the local one, so that
the poles in the scattering amplitude of these two channels are very
close together and therefore there is a reinforcement between
them, giving rise to the anomalously large separation energy

\subsection{The triton binding energy}
\label{ch8.1}

The relevance and/or necessity of considering the non-local part 
of NN potentials in realistic interactions is still under debate. 
Over the past few years several studies have appeared
in the literature which stress the potential importance
of non-local effects for the quantitative understanding of few-body 
observables and, specifically, for the triton binding energy 
\cite{TAK92,FUJ02,GIB95,HAI96,MAC96,ELS96,DOL00}. However, 
the majority of these investigations \cite{GIB95,HAI96,MAC96,ELS96,DOL00}
explore only non-localities arising from the baryonic meson-exchange
picture of the NN interaction.

The non-localities generated in a quark-model derivation of baryonic potentials
may be relevant for the three-nucleon bound state. It has been
argued that the assumptions associated with meson-exchange
models sharply limit the nature of the off-shell
properties of those potentials, once the on-shell matrix
elements are constrained to fit the NN data \cite{RED83}.
Therefore, it is very interesting to investigate the off-shell
features of potentials derived from a quark-model. Some
studies in this direction were done in reference \cite{TAK92},
but only the short-range part of the interaction was obtained
by quark-model techniques, the intermediate and long-range parts being 
described by standard meson-exchanges between baryons.
Accordingly, that work allowed only very limited conclusions with regard
to effects of the quark substructure.

The triton binding energy has been studied in detail by means of
the chiral constituent quark model \cite{JUL02}. The calculation
has been done in a momentum-space Faddeev formalism 
restricted to the $^1S_0$ and $^3S_1-{}^3\! D_1$ NN partial waves, 
those which provide the bulk contribution to the three-nucleon binding energy. 
To solve the three-body Faddeev equations in momentum space
a separable finite-rank expansion of the
NN(coupled to N$\Delta)$ sector utilizing 
the EST method \cite{ERN73} is first performed. 
Such a technique has been extensively studied in other works 
for various realistic NN potentials \cite{SCH00} and
specifically for a model including the coupling to the
N$\Delta$ system \cite{NEM98}. In these works it was shown that,
with a separable expansion of sufficiently high rank, reliable and
accurate results on the three-body level can be achieved. 
In the present case it turns out that separable 
representations of rank 6 (for $^1S_0-(^5D_0)$ and for
$^3S_1-{}^3\! D_1$) are sufficient to get converged 
results. We refer the reader to references \cite{SCH00,NEM98,HAI84} for 
technical details on the expansion method.
The quality of the separable expansion on the NN sector 
is very good. The phase shifts for the original non-local potential and 
for the corresponding separable expansion are almost 
indistinguishable \cite{JUL02}.

\begin{table}[t]
\caption{\label{t15c6} Triton binding energy, $E_B$ ($E_{\rm Exp}=-8.49$ MeV), 
and wave function components: $P_S$ is the $L=0$ symmetric spatial
probability, $P_{S'}$ the $L=0$ mixed symmetric spatial probability,
$P_P$ the $L=1$ probability and $P_D$ the $L=2$ probability.}
\begin{indented}
\item[]
\begin{tabular}{@{}lllll}
\br
  & CCQM \protect\cite{JUL02}   &
Nijm II \protect\cite{FRI93} & Bonn B \protect\cite{SCH00}\\
\mr
$E_B$ (MeV)    & $-$7.72     &$-$7.65 & $-$8.17   \\ 
$P_S$ ($\%$)   & 91.49       &90.33   &  91.35  \\ 
$P_{S'}$ ($\%$)& 1.430       &1.339   &   1.368  \\ 
$P_P$ ($\%$)   & 0.044       &0.064   &  0.049  \\ 
$P_D$ ($\%$)   & 7.033       &8.267   &  7.235  \\ 
\br
\end{tabular}
\end{indented}
\end{table}

Results for the triton are summarized in table \ref{t15c6}. 
It is reassuring to see that the predicted value of the triton binding energy 
is comparable to the values obtained from conventional NN
potentials, such as the Bonn or Nijmegen models. 
Thus, these calculations show that quark model based NN interactions are 
definitely able to provide a realistic description of the triton. The
results also give support to the use of such interaction model for
further few-body calculations.
One should not forget at this point that the number of free parameters
is greatly reduced as compared, for example, to effective theories, 
and, in addition, they are strongly correlated by 
the requirement of describing the baryon spectrum.

One can find in the literature other quark-inspired analysis of the
triton binding energy. Some preliminary studies were done in Ref. \cite{TAK92}
with a hybrid model as mentioned above.
Recently the triton has been studied with the same quark model approach
of Ref. \cite{FUJ96}, used for the study of the NN phase shifts (see
\sref{newsec}). A 34-channel calculation is done.
The result obtained,
$-$8.390 MeV \cite{FUJ02}, is pretty close to experiment.
Although this model is only intended for the study of the
baryon-baryon interaction, i.e., the connection with the one-body
problem is lost, the quality of its predictions is remarkable.

%% file: ch7rev.tex
\section{Summary}
\label{ch9}

We have reviewed recent calculations of few-body systems 
within a chiral constituent quark model framework. 
As for the moment an exact solution 
of QCD is not attainable, models incorporating its basic symmetries 
and some dynamics reveal as an important
alternative to study low-energy hadron phenomenology. 
In this energy regime, confinement and chiral symmetry breaking appear as the most
important properties of QCD. When properly implemented by means of chiral
constituent quark models, the interplay between gluons, Goldstone bosons and
constituent quarks is able to provide a consistent description of
the baryon structure and properties and the baryon-baryon phenomenology.
Several aspects of chiral constituent quark models in their 
application to study few-body systems deserve to be emphasized. 

First, their universality in a double sense: on the one hand they allow to 
describe a great variety of phenomena (regarding N, NN and NNN systems)
in terms of the same interacting potential, on the other hand
once the very few parameters of the models are determined in the 
well-known NN system they allow to predict in a parameter-free 
way the spectrum and the interaction involving any other baryonic resonances. 

Second, quark antisymmetry is properly implemented showing up as
cluster antisymmetry and quark exchange between clusters.
Its interplay with quark
dynamics provides an elegant explanation of the
observed hard core of the NN force. Similarly  a precise description
of the quark Pauli short-range repulsion appearing in other systems
is obtained.

Third, these models at their current status 
have raised a degree of accuracy that allows to reproduce
fine properties of two and
three-baryon systems, like deuteron properties, 
the triton binding energy or charge symmetry
breaking, with similar quality to baryonic meson-exchange models. 
Furthermore, the non-strange baryon spectrum
is nicely reproduced once some relativistic effects are taken into account.

Despite this success a lot of work is still to be done.
The confining interaction lacks a fundamental description and the
phenomenological constraints are not enough to determine its properties.
This has an important influence on the spin-orbit
force which for the moment is not correctly described.
Moreover, a complete relativistic treatment should be properly implemented and its
consequences analyzed. Finally, other systems like mesons or multiquark
systems should be studied within the same scheme. 
Meanwhile one can safely conclude that chiral 
constituent quark models have contributed in the last twenty years 
to a better understanding of the baryonic phenomenology.

\ack This work has been partially funded by 
Junta de Castilla y Le\'{o}n under Contract No. SA-104/04, 
by Ministerio de Educaci\'on y Ciencia under Contract No. FPA2004-05616 
by COFAA-IPN (M\'exico).

%% file: paper.bbl
\section*{References}
\begin{thebibliography}{999}

\bibitem{FRI73} Fritzsch H, Gell-Mann M and Leutwyler H 1973
\PL B {\bf 47} 365
\nonum Marciano W and Pagels H 1978
\PRP {\bf 36C} 137

\bibitem{MON94} Montvay I and Munster G 1994 
{\it Quantum Fields on a Lattice} (UK: Cambridge)

\bibitem{WEI79} Weinberg S 1979 
\PH A {\bf 96} 327

\bibitem{GAS84} Gasser J and Leutwyler H 1984
\APNY {\bf 158} 142

\bibitem{ISG89} Isgur N and Wise M B 1989
\PL B {\bf 232} 113

\bibitem{THA91} Thacker B A and Lepage G P 1991
\PR D {\bf 43} 196

\bibitem{THO74} 't Hooft G 1974
\NP B {\bf 72} 461

\bibitem{WIT79} Witten E 1979
\NP B {\bf 160} 57

\bibitem{ZAH86} Zahed I and Brown G 1986
\PRP {\bf 142} 2

\bibitem{DIA97} Diakonov D, Petrov V and Polyakov M 1997
\ZP A {\bf 359} 305

\bibitem{GEL64} Gell-Mann M 1964
\PL {\bf 8} 214

\bibitem{ZWE64} Zweig G 1964
CERN Preprints 411 and 412 (unpublished)

\bibitem{DAL65} Dalitz R H 1965
{\it High Energy Physics} (New York: Gordon and Breach) p~253

\bibitem{GRE64} Greenberg O W 1964
\PRL {\bf 13} 598

\bibitem{HAN65} Han M Y and Nambu Y 1965
\PR {\bf 139} 1006

\bibitem{RUJ75} de R\'ujula A, Georgi H and Glashow S L 1975
\PR D {\bf 12} 147

\bibitem{APP75} Appelquist T, de R\'ujula A, Politzer H D and Glashow S L 1975
\PRL {\bf 34} 365

\bibitem{EIC75} Eichten E, Gottfried K, Kinoshita T, 
Kogut J B, Lane K D and Yan T M 1975
\PRL {\bf 34} 369

\bibitem{ISG78} Isgur N and Karl G 1978 
\PR D {\bf 18} 4187 
\nonum Isgur N and Karl G 1979
\PR D {\bf 19} 2653 
\nonum Isgur N and Karl G 1979
\PR D {\bf 20} 1191

\bibitem{BHA81} Bhaduri R K, Cohler L E and Nogami Y 1981
\NC A {\bf 65} 376

\bibitem{SIL85} Silvestre-Brac B and Gignoux C 1985
\PR D {\bf 32} 743

\bibitem{CHO74} Chodos A, Jaffe R L, Johson K, Thorn C B and Weisskopf V F 1974
\PR D {\bf 9} 3471
\nonum DeGrand T, Jaffe R L, Johnson K and Kiskis J E 1975
\PR D {\bf 12} 2060

\bibitem{CHO75} Chodos A and Thorn C B 1975
\PR D {\bf 12} 2733
\nonum Brown G E and Rho M 1979
\PL B {\bf 82} 177
\nonum Brown G E, Rho M and Vento V 1979
\PL B {\bf 84} 383
\nonum Th\'eberge S, Thomas A W and Miller G A 1980
\PR D {\bf 22} 2838

\bibitem{GOD85} Godfrey S and Isgur N 1985
\PR D {\bf 32} 189
\nonum Capstick S and Isgur N 1986
\PR D {\bf 34} 2809

\bibitem{LIB77} Liberman D A 1977,
\PR D {\bf 16} 1542

\bibitem{NEU77} Neudatchin V G, Smirnov Yu F and Tamagaki R 1977
\PTP {\bf 58} 1072

\bibitem{TOK80} Toki H 1980
\ZP A {\bf 294} 173

\bibitem{RIB80} Ribeiro J E T 1980
\ZP C {\bf 5} 27

\bibitem{OKA80a} Oka M and Yazaki K 1980
\PTP {\bf 66} 556

\bibitem{OKA80b} Oka M and Yazaki K 1980 
\PL B {\bf 90} 41

\bibitem{HARV81} Harvey M 1981
\NP A {\bf 352} 326

\bibitem{FAE82} Faessler A, Fern\'andez F, L\"ubeck G and Shimizu K 1982 
\PL B {\bf 112} 201
\nonum Faessler A, Fern\'andez F, L\"ubeck G and Shimizu K 1983 
\NP A {\bf 402} 555

\bibitem{CVE83} Cvetic M, Golli B, Mankoc-Borstnik N and Rosina M 1983
\NP A {\bf 395} 349

\bibitem{FUJ86} Fujiwara Y and Hecht K T 1986
\PL B {\bf 171} 17

\bibitem{OKA83} Oka M and Yazaki K 1983
\NP A {\bf 402} 477 

\bibitem{FUJ86b} Fujiwara Y and Hecht K T 1986
\NP A {\bf 462} 621

\bibitem{MAN84} Manohar A and Georgi H 1984
\NP B {\bf 234} 189

\bibitem{DIA84} Diakonov D and Petrov V Yu 1984 
\NP B {\bf 245} 259

\bibitem{DIA96} Diakonov D 1996 
\PP hep-ph/9602375

\bibitem{SHI84} Shimizu K 1984
\PL B {\bf 148} 418

\bibitem{BRA85} Br\"auer K, Faessler A, Fern\'andez F and Shimizu K 1985
\ZP A {\bf 320} 609 
\nonum Br\"auer K, Faessler A, Fern\'andez F and Shimizu K 1990
\NP A {\bf 507} 599 

\bibitem{OBU90} Obukhovsky I T and Kusainov A M 1990
\PL B {\bf 238} 142

\bibitem{FER93a} Fern\'andez F, Valcarce A, Straub U and Faessler A 1993
\jpg {\bf 19} 2013

\bibitem{VAL96a} Valcarce A, Gonz\'alez P, Fern\'andez F and Vento V 1996 
\PL B {\bf 367} 35

\bibitem{GLO96} Glozman L Ya and Riska D O 1996 
\PRP {\bf 268} 263

\bibitem{NAK98} Nakamoto C and Toki H 1998
\PTP {\bf 99} 1001

\bibitem{FUJ96} Fujiwara Y, Nakamoto C and Suzuki Y 1996
\PR C {\bf 54}, 2180
\nonum Fujiwara Y, Fujita T, Kohno M, Nakamoto C and Suzuki Y 2002
\PR C {\bf 65}, 014002

\bibitem{FUJ04} Fujiwara Y, Miyagawa K, Khono M, Suzuki Y and Nakamoto C 2004
\NP A {\bf 737} 243
\nonum Fujiwara Y, Fujita T, Nakamoto C, Suzuki Y and Khono M 1998
\NP A {\bf 639} 41c
\nonum Fujiwara Y, Khono M, Nakamoto C and Suzuki Y 2001
\PR C {\bf 64} 054001 

\bibitem{GAR03b} Garcilazo H and Valcarce A 2003
\PR C {\bf 68} 035207 

\bibitem{DES04} Theussl L, Amghar A, Desplanques B and Noguera S 2003
\FBS {\bf 14} 393
\nonum Amghar A, Desplanques B and Theussl L 2003
\PL B {\bf 574} 201
\nonum Glozman L Ya, Radici M, Wagenbrunn R F, Boffi S, Klink W and Plessas W 2001
\PL B {\bf 516} 183
\nonum Desplanques B 2004
\PP nucl-th/0405059
\nonum Desplanques B 2004
\PP nucl-th/0405060
\nonum Desplanques B 2004
\PP nucl-th/0407074
\nonum  Melde T, Canton L, Plessas W and Wagenbrunn R F 2004
\PP hep-ph/0411322

\bibitem{BAL01} Bali G S 2001
\PRP {\bf 343} 1

\bibitem{SHI89} Shimizu K 1989
\RPP {\bf 52} 1

\bibitem{YND99} Yndurain F J 1999 {\it The Theory of Quark and Gluon Interactions}
(Berlin: Springer)

\bibitem{PIC95} Pich A 1995 
\PP hep-ph/9505231

\bibitem{GOD61} Goldstone J 1961 
\NC {\bf 19} 154

\bibitem{BHA80} Bhaduri R K, Cohler L E and Nogami Y 1980
\PRL {\bf 44} 1369 

\bibitem{YAZ90} Yazaki K 1990 
\PPNP {\bf 24} 353

\bibitem{FER94} Fern\'andez F, Valcarce A, Gonz\'alez P and Vento V 1992
\PL B {\bf 287} 35
\nonum Fern\'andez F, Valcarce A, Gonz\'alez P and Vento V 1994
\NP A {\bf 567} 741

\bibitem{SCA93} Scadron M D 1993
\PAN {\bf 56} 1595

\bibitem{LIU93} Liu G Q, Swift M, Thomas A W and Holinde K 1993
\NP A {\bf 556} 331

\bibitem{ERI92} Ericson T E O 1992
\NP A {\bf 543} 409c

\bibitem{VAL94a} Valcarce A, Buchmann A, Fern\'andez F and Faessler A 1994 
\PR C {\bf 50} 2246 

\bibitem{STA96} Stancu F 1996 {\it Group Theory in Subnuclear Physics}
(Oxford: Clarendon Press)

\bibitem{HOL84} Holinde K 1984
\NP A {\bf 415} 477

\bibitem{SUZ84b} Suzuki Y and Hecht K T 1984 
\NP A {\bf 420} 525

\bibitem{VAL95b} Valcarce A, Fern\'andez F, Gonz\'alez P and Vento V 1995 
\PR C {\bf 52} 38

\bibitem{JUL01} Juli\'a-D\'{\i}az B, Fern\'andez F, Gonz\'alez P and Valcarce A 2001
\PR C {\bf 63} 024006

\bibitem{FERR89} Ferreira E and Dosch H G 1989 
\PR C {\bf 40} 1750

\bibitem{MYH86} Myhrer F and Wroldsen J 1986
\PL B {\bf 174} 366
\nonum Myhrer F and Wroldsen J 1988
\RMP {\bf 60} 629

\bibitem{SAIT68} Saito S 1969 
\PTP {\bf 41} 705

\bibitem{VAL97} Valcarce A, Fern\'andez F and Gonz\'alez P 1997
\PR C {\bf 56} 3026

\bibitem{HET69} Hecht K T and Pang S C 1969 
\JMP {\bf 10} 1571

\bibitem{VAL95c} Valcarce A, Buchmann A, Fern\'andez F and Faessler A 1995 
\PR C {\bf 51} 1480

\bibitem{YUZ95} Yu Y W, Zhang Z Y, Shen P N and Dai L R  1995
\PR C {\bf 52} 3393 

\bibitem{WHE37} Wheeler J A 1937
\PR {\bf 52} 1107

\bibitem{TAN77} Tang Y C, LeMere M and Thompsom D R 1978
\PRP {\bf 47} 167

\bibitem{KAM78} Kamimura M 1977
\PTPS {\bf 62} 236

\bibitem{TAR78} DeTar C 1978
\PR D {\bf 17} 323

\bibitem{OKA84} Oka M and Yazaki K 1984
\IRNP 1 489

\bibitem{JUL02} Juli\'a-D\'{\i}az B, Haidenbauer J, Valcarce A and Fern\'andez F 2002 
\PR C {\bf 65} 034001

\bibitem{BUC89} Buchmann A, Yamauchi Y and Faessler A 1989
\NP A {\bf 496} 621

\bibitem{JUL03} Juli\'a-D\'{\i}az B, Valcarce A, Gonz\'alez P and Fern\'andez F 2003
\EPJ A {\bf 19} s01,99 

\bibitem{JAI93} Jain B K and Santra A B 1993
\PRP {\bf 230} 1

\bibitem{SAU86} Sauer P U 1986 
\PPNP {\bf 16} 35

\bibitem{LOM82} Lomon E L 1982
\PR D {\bf 26} 576

\bibitem{PEN92} Pe\~na M T, Garcilazo H, Oelfke U and Sauer P U 1992
\PR C {\bf 45} 1487
 
\bibitem{PEN90} Pe\~na M T, Henning H and Sauer P U 1990
\PR C {\bf 42} 855

\bibitem{ALE90} Alexandrou C and Blankleider B 1990
\PR C {\bf 42} 517 

\bibitem{TAKA88} Takaki T and Thies M 1988
\PR C {\bf 38} 2230

\bibitem{GAR90b} Garcilazo H, Mizutani T, Pe\~na M T and Sauer P U 1990 
\PR C {\bf 42} 2315

\bibitem{VAL94b} Valcarce A, Fern\'andez F, Garcilazo H, Pe\~na M T and Sauer P U 1994
\PR C {\bf 49} 1799 

\bibitem{SHY89} Shypit R L, Bugg D V, Sanjari A H, Lee D M, McNaughton M W, 
Silbar R R, Hollas C L, McNaughton K H, Riley P and Davis C A 1989
\PR C {\bf 40} 2203
\nonum Shypit R L, Bugg D V, Lee D M, McNaughton M W, Silbar R R, Stewart N M, 
Clough A S, Hollas C L, McNaughton K H, Riley P and Davis C A 1988
\PRL {\bf 60} 901 

\bibitem{ALL86} Allasia D \etal 1986
\PL B {\bf 174} 450 

\bibitem{HAI93} Haidenbauer J, Holinde K and Johnson M B 1993
\PR C {\bf 48} 2190

\bibitem{OKA93} Oka M 1993 {\it Frontiers Science Series} No. 6, p~950

\bibitem{WAN92} Wang F, Wu G, Teng L and Goldman T 1992
\PRL {\bf 69} 2901

\bibitem{GOL89} Goldman T, Maltman K, Stephenson Jr. G J, Schmidt K E and Wang F 1989
\PR C {\bf 39} 1889

\bibitem{CVE80} Cveti\v{c} M, Golli B, Manko\v{c}-Bor\v{s}tnik N and Rosina M 1980
\PL B {\bf 93} 489

\bibitem{GAR97b} Garcilazo H, Fern\'andez F, Valcarce A and Mota R D 1997
\PR C {\bf 56} 84

\bibitem{PDG} Hagiwara K \etal 2002
\PR D {\bf 66} 010001

\bibitem{ARE71} Arenh\"ovel H, Danos M and Williams H T 1971
\NP A {\bf 162} 12

\bibitem{HAM62} Hamada T and Johnston I D 1962
\NP {\bf 34} 382

\bibitem{REI68} Reid R V 1968
\APNY {\bf 50} 411

\bibitem{ROS75} Rost E 1975
\NP A {\bf 249} 510

\bibitem{PEN99} Pe\~na M T, Riska D O and Stadler A 1999
\PR C {\bf 60} 045201

\bibitem{COO95} Coon S A, Pe\~{n}a M T and Riska D O 1995
\PR C {\bf 52} 2925

\bibitem{HIR96} Hirenzaki S, Oset E and Fern\'andez de Cordoba P 1996 
\PL B {\bf 378} 29
\nonum Hirenzaki S, Fern\'andez de Cordoba P and Oset E 1996 
\PR C {\bf 53} 277

\bibitem{ALV99} \'Alvarez-Ruso L 1999 
\PL B {\bf 452} 207 

\bibitem{GAR93} Garcilazo H and Moya de Guerra E 1993
\NP A {\bf 562} 521

\bibitem{JUL02c} Juli\'a-D\'{\i}az B, Valcarce A, Gonz\'alez P and Fern\'andez F 2002
\PR C {\bf 66} 024005

\bibitem{RIS01} Riska D O and Brown G E 2001
\NP A {\bf 679} 577

\bibitem{HUB94} Huber S and Aichelin J 1994 
\NP A {\bf 573} 587

\bibitem{SOY00} Soyeur M 2000
\NP A {\bf 671} 532

\bibitem{MAC89} Machleidt R 1989 
\ANP {\bf 19} 189

\bibitem{ORD96} Ord\'o\~nez C, Ray L and van Kolck U 1996
\PR C {\bf 53} 2086

\bibitem{WAK84} Wakamatsu M, Yamamoto R and Yamauchi Y 1984
\PL B {\bf 146} 148

\bibitem{VIN91} Vinh Mau R, Semay C, Loiseau B and Lacombe M 1991
\PRL {\bf 67} 1392

\bibitem{LAC80} Lacombe M, Loiseau B, Richard J M, Vinh Mau R, C\^ot\'e J. 
Pir\`es P and de Tourreil R 1980
\PR C {\bf 21} 861

\bibitem{TAK89} Takeuchi S, Shimizu K and Yazaki K 1989
\NP A {\bf 504} 777

\bibitem{VAL95a} Valcarce A, Faessler A and Fern\'andez F 1995
\PL B {\bf 345} 367

\bibitem{FUJ02} Fujiwara Y, Miyagawa K, Kohno M, Suzuki Y and Nemura H 2002
\PR C {\bf 66} 021001

\bibitem{STO94} Stoks V G J, Klomp R A M, Terheggen C P F and de Swart J J 1994
\PR C {\bf 49} 2950

\bibitem{ENT99} Entem D R, Fern\'andez F and Valcarce A 1999
\PL B {\bf 463} 153

\bibitem{MAT77} Matveev V A and Sorba P 1977
\LNC {\bf 20} 435

\bibitem{GON87} Gonz\'alez P and Vento V 1987
\FB {\bf 2} 145

\bibitem{KER69} Kerman A K and Kisslinger L S 1969
\PR {\bf 180} 1483

\bibitem{HAD73} Hadjimichael E 1973
\PL B {\bf 46} 147

\bibitem{DEN99} Denisov O Yu, Kuksa S D and Lykasov G I 1999
\PL B {\bf 458} 415

\bibitem{DOR95} Dorodnykh Yu L, Lykasov G I, Rzyanin M V and Cassing W 1995
\PL B {\bf 346} 227

\bibitem{UZI97} Uzikov Yu N 1997
\PAN {\bf 60} 1458 
\nonum Uzikov Yu N 1997 
\PAN {\bf 60} 1616

\bibitem{KUS91} Kusainov A M, Neudatchin V G and Obukhovsky I T 1991
\PR C {\bf 44} 2343

\bibitem{GLO94} Glozman L Ya and Kuchina E I 1994
\PR C {\bf 49} 1149

\bibitem{ZHA85} Zhang Z Y, Br\"auer K, Faessler A and Shimizu K 1985
\NP A {\bf 443} 557

\bibitem{MAE00} Maeda I, Arima M and Masutani K 2000
\PL B {\bf 474} 255

\bibitem{ENT00} Entem D R, Fern\'{a}ndez F and Valcarce A 2000
\PR C {\bf 62} 034002

\bibitem{JUL02b} Juli\'a-D\'{\i}az B, Entem D R, Valcarce A and Fern\'andez F 2002 
\PR C {\bf 66} 047002

\bibitem{ARE75b} Arenh\"ovel H 1975
\ZP A {\bf 275} 189

\bibitem{DYM90} Dymarz R and Khanna F C 1990
\NP A {\bf 516} 549

\bibitem{HAA74} Haapakoski P and Saarela M 1974
\PL B {\bf 53} 333

\bibitem{GRE74} Green A M and Haapakoski P 1974
\NP A {\bf 221} 429

\bibitem{SIT87} Sitarski W P, Blunden P G and Lomon E L 1987
\PR C {\bf 36} 2479

\bibitem{SAI00} SAID program. Available on 
the web site {\it http://gwdac.phys.gwu.edu}

\bibitem{KAI98} Kaiser N, Gerstend\"orfer S and Weise W 1998
\NP A {\bf 637} 395

\bibitem{KAI97} Kaiser N, Brockmann R and Weise W 1997
\NP A {\bf 625} 758 

\bibitem{HEL98} Helminen C and Riska D O 1998
\PR C {\bf 58} 2928

\bibitem{SUZ84} Suzuki Y and Hecht K T 1984 
\PR C {\bf 29} 1586

\bibitem{FAE93} Faessler A, Buchmann A and Yamauchi Y 1993 
\IJMP E {\bf 2} 39

\bibitem{MOR84b} Morimatsu O, Yazaki K and Oka M 1984
\NP A {\bf 424} 412

\bibitem{KOI86} Koike Y 1986 
\NP A {\bf 454} 509
\nonum Koike Y, Morimatsu O and Yazaki K 1986
\NP A {\bf 449} 635

\bibitem{MAC87} Machleidt R, Holinde K and Elster C 1987 
\PRP {\bf 149} 1

\bibitem{MOR84a} Morimatsu O, Ohta S, Shimizu K and Yazaki K 1984
\NP A {\bf 420} 573

\bibitem{ZHA94} Zhang Z Y, Faessler A, Straub U and Glozman L Ya 1994 
\NP A {\bf 578} 573

\bibitem{MIL90} Miller G A, Nefkens B M K and Slaus I 1990
\PRP {\bf 194} 1

\bibitem{STE91} Stephenson Jr. G J, Maltman K and Goldman T 1991
\PR D {\bf 43} 860

\bibitem{POP87} P\"opping H, Sauer P U and Xi-Zhen Z 1987
\NP A {\bf 474} 557

\bibitem{GAR90a} Garcilazo H and Mizutani T 1990
{\it $\pi NN$ Systems} (Singapore: World Scientific)

\bibitem{YAO88} Le Yaouanc A, Oliver L, P\`ene O and Raynal J C 1988
{\it Hadron Transitions in the Quark Model} (New York: Gordon and Breach)

\bibitem{ENT03} Entem D R, Fern\'andez F and Valcarce A 2003
\PR C {\bf 67} 014001

\bibitem{ARN97} Arndt R A, Oh C H, Strakovsky I I and Workman R L 1997
\PR C {\bf 56} 3005 

\bibitem{GUT84} Gutbrod F and Montvay I 1984
\PL B {\bf 136} 411

\bibitem{ISG00} Isgur N 2000 
\PR D {\bf 62} 054026
\nonum Isgur N 2000 
\PP nucl-th/0007008

\bibitem{DES92} Desplanques B, Gignoux C, Silvestre-Brac B, Gonz\'alez P, 
Navarro J and Noguera S 1992
\ZP A {\bf 343} 331

\bibitem{GAR01a} Garcilazo H, Valcarce A and Fern\'andez F 2001
\PR C {\bf 64} 058201

\bibitem{FES55} Feshbach H and Rubinow S I 1955
\PR {\bf 98} 188

\bibitem{SUZ98} Suzuki Y and Varga K 1998
{\it Stochastic Variational Approach to Quantum-Mechanical Few-Body Problems} 
(Berlin: Springer)

\bibitem{FAD61} Faddeev L D 1961 
\SPJ  {\bf 12} 1014

\bibitem{PAP99} Papp Z 1999 
\FB {\bf 26} 99

\bibitem{PAP00} Papp Z, Krassnigg A and Plessas W 2000 
\PR C {\bf 62} 044004

\bibitem{GAR03} Garcilazo H 2003
\PR C {\bf 67} 055203

\bibitem{GLO99} Glozman L Ya, Plessas W, Varga K and Wagenbrunn R F 1998 
\PR D {\bf 58} 094030

\bibitem{DZI96} Dziembowski Z, Fabre de la Ripelle M and Miller G A 1996
\PR C {\bf 53} 2038

\bibitem{GLO98} Glozman L Ya, Papp Z, Plessas W, Varga K and Wagenbrunn R F 1998 
\PR C {\bf 57} 3406

\bibitem{GAR01b} Garcilazo H, Valcarce A and Fern\'andez F 2001
\PR C {\bf 63} 035207

\bibitem{BAS86} Basdevant J L and Boukraa S 1986
\ZP C {\bf 30} 103

\bibitem{CAR83} Carlson J, Kogut J and Pandharipande V R 1983
\PR D {\bf 27} 233

\bibitem{FUR02} Furuichi M and Shimizu K 2002
\PR C {\bf 65} 025201

\bibitem{YUB94} Dong Y, Su J and Wu S 1994
\jpg {\bf 20} 73

\bibitem{LOR01a} L\"oring U, Kretzschmar K, Metsch B Ch and Petry H R 2001 
\EPJ A {\bf 10} 309

\bibitem{LOR01b} L\"oring U, Metsch B Ch and Petry H R 2001 
\EPJ A {\bf 10} 395

\bibitem{LOR01c} L\"oring K, Metsch B Ch and Petry H R 2001 
\EPJ A {\bf 10} 447

\bibitem{JEN98} Jena S N, Behera M R and Panda S 1998
\jpg {\bf 24} 1089

\bibitem{NAK03} Nakano T \etal 2003
\PRL {\bf 91} 012002

\bibitem{NAK03b} Nakano T 2004
Presented at QNP2004

\bibitem{BAR03} Barmin V V \etal 2003
\PAN {\bf 66} 1715

\bibitem{STE03} Stepanyan S \etal 2003
\PRL {\bf 91} 252001

\bibitem{BAT03} Barth J \etal 2003
\PL B {\bf 572} 127

\bibitem{KUB03} Kubarovsky V 2003
\PRL {\bf 92} 032001

\bibitem{ASR03} Asratyan A E 2004
\PAN {\bf 67} 682 

\bibitem{ALE04} Aleev A 2004
\PP hep-ex/0401024

\bibitem{AIR03} Airapetian A 2004
\PL B {\bf 585} 213 

\bibitem{ZEU04} ZEUS Collaboration 2004
\PL B {\bf 591} 7

\bibitem{ABD04} Abdel-Bary M 2004
\PL B {\bf 595} 127

\bibitem{KNO04} Kn\"opfle K T \etal 2004
\jpg {\bf 30} S1363

\bibitem{BAI04} Bai J Z \etal 2004
\PR D {\bf 70} 012004

\bibitem{LIT04} Litvintsev D 2004
\PP hep-ex/0410024

\bibitem{AUB04} Aubert B \etal 2004
\PP hep-ex/0408064

\bibitem{SCH04} Schael S \etal 2004
\PL B {\bf 599} 1

\bibitem{LIN04} Lin C 2004
Presented at ICHEP2004

\bibitem{CHR04} Christian D \etal 2004
Presented at QNP2004

\bibitem{LON04} Longo M \etal 2004
\PR D {\bf 70} 111101

\bibitem{PIN04} Pinkenburg C \etal 2004
\jpg {\bf 30} S1201

\bibitem{MIZ04} Mizuk R \etal 2004
\PP hep-ex/0411005

\bibitem{STE04} Stenson K 2004
\PP hep-ex/0412021

\bibitem{NAP04} Napolitano J, Cummings J and Witkowski M 2004
\PP hep-ex/0412031

\bibitem{ARM04} Armstrong S R 2004
\PP hep-ex/0410080

\bibitem{ENG04} Engelfried J 2004
Presented at Quark Confinement 2004

\bibitem{ANT04} Antipov Yu M \etal 2004
\EPJ A {\bf 21} 455

\bibitem{NUS04} Nussinov S 2004
\PP hep-ph/0403028

\bibitem{DZI04b} Dzierba A R, Meyer C A and Szczepaniak A P 2004
\PP hep-ex/0412077

\bibitem{DZI04} Dzierba A R, Krop D, Swat M, Szczepaniak A P and Teige S 2004
\PR D {\bf 69} 051901

\bibitem{JAF04} Jaffe R L 2004
\EPJ C {\bf 35} 221

\bibitem{PRA87} Praszalowicz M 1987
{\it Skyrmions and Anomalies} (Singapore: World Scientific) p~112

\bibitem{JAF03} Jaffe R L and Wilczek F 2003
\PR L {\bf 91} 232003

\bibitem{KAR03} Karliner M and Lipkin H J 2003
\PP hep-ph/0307243

\bibitem{GLO03} Glozman L Ya 2003
\PL B {\bf 575} 18

\bibitem{STA03} Stancu Fl and Riska D O 2003
\PL B {\bf 575} 242 

\bibitem{MUL80} Mulders P J, Aerts A T and de Swart J J 1980
\PR D {\bf 21} 2653

\bibitem{MAL85} Maltman K 1985
\NP A {\bf 438} 669

\bibitem{WAN95} Wang F, Ping J and Goldman T 1995
\PR C {\bf 51} 1648

\bibitem{WON98} Wong C W 1998
\PR C {\bf 57} 1962

\bibitem{BUC98} Buchmann A J, Wagner G and Faessler A 1998
\PR C {\bf 57} 3340

\bibitem{YUA99} Yuan X Q, Zhang Z Y, Yu Y W and Shen P N 1999
\PR C {\bf 60} 045203

\bibitem{MOT99} Mota R D, Valcarce A, Fern\'andez F and Garcilazo H 1999 
\PR C {\bf 59} 46

\bibitem{GAR99} Garcilazo H, Valcarce A and Fern\'andez F 1999
\PR C {\bf 60} 044002

\bibitem{MOT02} Mota R D, Valcarce A, Fern\'andez F, Entem D R and Garcilazo H 2002
\PR C {\bf 65} 034006

\bibitem{CVE81} Cveti\v{c}-Krivec M, Golli B, Manko\v{c}-Bor\v{s}tnik N and Rosina M 1981
\PL B {\bf 99} 486

\bibitem{HUB00} Huber G M \etal 2000
\PR C {\bf 62} 044001

\bibitem{MAL84} Maltman K and Isgur N 1984
\PR D {\bf 29} 952

\bibitem{GAR97a} Garcilazo H 1997
\jpg {\bf 23} 1101

\bibitem{HAR72} Harper E P, Kim Y E and Tubis A 1972
\PRL {\bf 28} 1533

\bibitem{BET86} Berthold G H, Zankel H, Mathelitsch L and Garcilazo H 1986 
\NC A {\bf 93} 89 

\bibitem{CHE85} Chen C R, Payne G L, Friar J L and Gibson B F 1985
\PRL {\bf 55} 374

\bibitem{AFN74} Afnan I R and Thomas A W 1974
\PR C {\bf 10} 109

\bibitem{OTS65} Otsuki S, Tamagaki R and Yasuno M 1965
\PTPS {\it Number Extra} 578

\bibitem{HOS78} Hoshizaki N 1978
\PTP {\bf 60} 1796 
\nonum Hoshizaki N 1979
\PTP {\bf 61} 129

\bibitem{ARN00} Arndt R A, Strakovsky I I and Workman R L 2000
\PR C {\bf 62} 034005

\bibitem{YOK80} Yokosawa A 1980
\PRP {\bf 64} 47

\bibitem{VAL00} Valcarce A, Garcilazo H, Mota R D and Fern\'andez F 2001
\jpg {\bf 27} L1

\bibitem{KAM97} Kamae T and Fujita T 1977
\PRL {\bf 38} 471

\bibitem{DOL00} Doleschall P and Borb\'ely I 2000 
\PR C {\bf 62} 054004

\bibitem{ELS96} Elster Ch, Evans E E, Kamada H and Gl\"ockle W 1996  
\FB {\bf 21} 25

\bibitem{GIB95} Gibson B F, Kohlhoff H and von Geramb H V 1995
\PR C {\bf 51} R465

\bibitem{HAI96} Haidenbauer J and Holinde K 1996
\PR C {\bf 53} R25

\bibitem{MAC96} Machleidt R, Sammarruca F and Song Y 1996
\PR C {\bf 53} R1483

\bibitem{TAK92} Takeuchi S, Cheon T and Redish E F 1992
\PL B {\bf 280} 175

\bibitem{RED83} Redish E F and Stricker-Bauer K 1983
\PL B {\bf 133} 1

\bibitem{ERN73} Ernst D J, Shakin C M, Thaler R M and Weiss D L 1973
\PR C {\bf 8} 2056

\bibitem{FRI93} Friar J L, Payne G L, Stoks V G J and de Swart J J 1993
\PL B {\bf 311} 4

\bibitem{SCH00} Schadow W, Sandhas W, Haidenbauer J and Nogga A 2000
\FB {\bf 28} 241

\bibitem{NEM98} Nemoto S, Chmielewski K, Schellingerhout N W, Sauer P U, 
Haidenbauer J and Oryu S 1998
\FB {\bf 24} 213

\bibitem{HAI84} Haidenbauer J and Plessas W 1983
\PR C {\bf 27} 63 
\nonum Haidenbauer J and Plessas W 1984
\PR C {\bf 30} 1822

\end{thebibliography}
